\documentclass[a4paper,12pt]{book}

\usepackage[spanish,british]{babel}

\usepackage{float}

\usepackage{epsfig}

\usepackage{cite}

\usepackage{appendix}

\usepackage{amssymb,amsmath,amsthm,fixmath, bm}
\usepackage{acronym}

\usepackage{cases}

\usepackage{fancyhdr,fancybox}

\usepackage{tocbibind}
\usepackage[titles]{tocloft}

\renewcommand*{\headrulewidth}{0.6pt}

\addto\captionsspanish{}
\spanishdecimal{.}

\font\bigrm= cmr12   scaled \magstep 2
\font\bigbf= cmbx12  scaled \magstep 2
\font\Bigbf= cmbx12  scaled \magstep 4

\DeclareMathAlphabet{\mathitbf}{T1}{cmr}{bx}{it}

\begin{document}

\fancyhf{}
\fancyhead[RE]{\slshape \nouppercase \leftmark}
\fancyhead[LO]{\slshape \nouppercase \rightmark}
\fancyfoot[C]{\thepage}

\renewcommand*{\headrulewidth}{0.6pt}

\normalsize

\pagestyle{empty}

\enlargethispage*{2cm}

\thisfancyput(-1.3cm,-7.5cm){\vbox{\includegraphics[width=1.22\columnwidth]{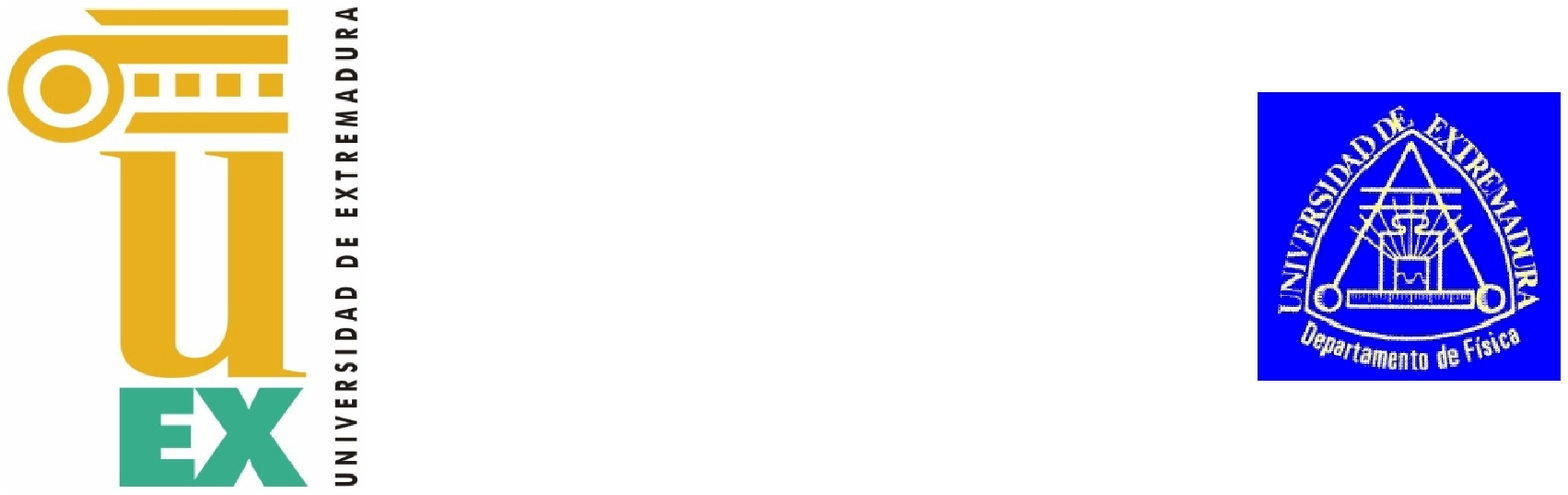}}}

\vspace*{1.2cm}

\begin{center}
\bigrm
Universidad de Extremadura\\[7mm]	
Facultad de Ciencias\\[6mm]
Departamento de F\'{\i}sica
\vskip 3.2 cm

\Bigbf
Transiciones de Fase en\\[10mm] Sistemas Desordenados\\[10mm]
\vskip 0.5 cm
---------------
\vskip 1 cm
Phase Transitions in \\[10mm] Disordered Systems

\end{center}

\vskip 3.9 cm

\bigrm
\rightline{Memoria de tesis doctoral}
\quad\\[5mm]
\rightline{presentada por}
\quad\\[5mm]
\rightline{\bigrm Antonio Gordillo Guerrero}\\[8mm]

\newpage	
\quad

\newpage
\pagenumbering{arabic}
\font\bigrm= cmr12   scaled \magstep 2
\font\bigbf= cmbx12  scaled \magstep 2
\font\Bigbf= cmbx12  scaled \magstep 3

\begin{center}

\vskip 1 cm

\bigrm
Universidad de Extremadura\\[7mm]	
Facultad de Ciencias\\[6mm]
Departamento de F\'{\i}sica
\vskip 4 cm

\Bigbf
  Transiciones de Fase en \\[9mm]  Sistemas Desordenados 
\vskip 0.5 cm
---------------
\vskip 1 cm
Phase Transitions in \\[10mm] Disordered Systems

\end{center}

\vskip  3 cm

\bigrm
\centerline{\bigbf Antonio Gordillo Guerrero}
\vskip 2.5 cm
\centerline{Dirigida por}
\quad\\[4mm]
\centerline{{ Juan Jes\'us Ruiz Lorenzo} }

\newpage
\quad

\newpage
\normalsize

\pagestyle{empty}

\vspace*{2.5cm}

\selectlanguage{spanish}

\font\bigbf= cmbx12  scaled \magstep 3
\font\bigrm= cmr12   scaled \magstep 1
\font\bigit= cmti12   scaled \magstep 2
{\bigbf Agradecimientos}
\bigrm
\quad\\[0.5cm]
{
\baselineskip 15pt
\parskip 6 pt
\hskip 20pt

La elaboraci\'on de esta tesis doctoral ha sido \'unicamente posible gracias al apoyo prestado
por muy diversas personas e instituciones, es por ello muy importante para m\'{\i}
hacer una menci\'on a su labor.

Soy muy afortunado al formar parte del grupo investigaci\'on 
Statistical Physics in Extremadura (SPHINX), formado por
cient\'{\i}ficos de primer nivel adem\'as de excelentes personas. Su acogida
ha sido impecable desde el primer d\'{\i}a y he recibido su apoyo en todo momento.
Gracias a Andr\'es Santos, Vicente Garz\'o, Santos Bravo, Antonio Astillero,
Francisco Vega, Mariano L\'opez, Enrique Abad, Rafael Borrego, Rene Rohrmann y por supuesto
Juan Jes\'us Ruiz.

Juan Jes\'us Ruiz tiene m\'as que merecido un p\'arrafo aparte, ha sido el director de este trabajo
y es mi principal mentor. Es una persona muy trabajadora, con una formaci\'on excelente 
y grandes habilidades pedag\'ogicas. Adem\'as es un director comprensivo y muy accesible.
No imagino otro director de tesis mejor. Ha sido un placer trabajar con \'el 
durante estos a\~nos y espero poder seguir haci\'endolo.

Agradezco la ayuda prestada durante estos a\~nos por Luis Antonio Fern\'andez 
y V\'{\i}ctor Mart\'{\i}n, de la Universidad Complutense de Madrid.
Luis Antonio supone todo un ejemplo de maestr\'{\i}a
en F\'{\i}sica Computacional y de capacidad de trabajo, mientras que V\'{\i}ctor
nunca deja de sorprenderme por su capacidad de s\'{\i}ntesis. Estoy tambi\'en en deuda con Ralph Kenna, de
la Universidad de Coventry; trabajar con \'el ha sido agradable y fruct\'{\i}fero.

Estoy muy agradecido por el apoyo computacional ofrecido por el Instituto de 
Biocomputaci\'on y F\'{\i}sica de Sistemas Complejos (BIFI) de Zaragoza.
Resalto la labor de Alfonso Taranc\'on, su secretario, 
siempre atento por el buen desarrollo de nuestras investigaciones.
El equipo de soporte t\'ecnico ha sido excelente; gracias en particular a Guillermo Losilla, Arturo Giner
y Ferm\'{\i}n Serrano. Los desarrolladores de Ibercivis han hecho un
esfuerzo importante en la adaptaci\'on y soporte de nuestra aplicaci\'on; gracias en
concreto a Alejandro Rivero y David Benito.

Mis compa\~neros del \'area de Electr\'onica en la Escuela Polit\'ecnica de C\'aceres
me han facilitado desde el primer d\'{\i}a la integraci\'on en su equipo docente,
siendo siempre comprensivos con mis continuos viajes.
Ellos son Horacio Gonz\'alez, Ram\'on Gallardo y Antonio Garc\'ia.

Env\'{\i}o tambi\'en mi agradecimiento al Departamento de F\'{\i}sica y al Departamento
de Ingenier\'{\i}a El\'ectrica, Electr\'onica y Autom\'atica de la Universidad de Extremadura.
Agradezco en particular la labor de sus directores Miguel Jurado y
Miguel \'Angel Jaramillo, as\'{\i} como de su personal de administraci\'on.

%

Estas p\'aginas de agradecimientos estar\'{\i}an del todo incompletas si no mencionara el apoyo continuo
que tengo de mi familia. En especial quiero agradecer a mis padres la educaci\'on
que me dan a trav\'es de su ejemplo. Dedico tambi\'en estas p\'aginas a mis abuelas y abuelos ya que
suponen para m\'{\i} importantes modelos de buena conducta.
Quiero adem\'as mencionar a mi hermano Fernando, me gustar\'{\i}a
que este trabajo sirviera de motivaci\'on para su carrera cient\'{\i}fica, tiene
sobradas capacidades que creo que a\'un no ha descubierto.

Para la realizaci\'on de esta tesis ha sido imprescindible la ayuda de
mi pareja: Ana Chac\'on Chamorro. Ha sido todo lo comprensiva que se puede ser con
un marido y padre que realiza una tesis doctoral. Sin su apoyo
no hubiera podido finalizar este trabajo. Ella cuida y protege con toda su alma a nuestro principal objetivo vital:
nuestra hija Julia. A Julia le ruego que me perdone por el tiempo irrecuperable de su primera infancia 
que no hemos podido disfrutar juntos. Estar con Julia y Ana
produce en m\'{\i} la felicidad m\'as profunda que nunca he sentido.

No puedo dejar de mencionar la ayuda que me ha prestado mi prima Raquel Acosta durante los
a\~nos vividos entre C\'aceres y Badajoz. Desde el primer d\'{\i}a me ha 
abierto las puertas de su casa sin esperar nada a cambio. Me ha dado la posibilidad de
aprovechar m\'as mi tiempo y me ha acompa\~nado en mis escasos momentos
de ocio. \textexclamdown Muchas gracias Prima!

Finalmente mencionar a mis amigos de Badajoz, Zafra y Madrid, as\'{\i} como al Club de Pirag\"uismo de Badajoz.
Todos hab\'eis estado ah\'{\i} cuando os necesitaba.
}

\vskip 1.5cm
\rightline{C\'aceres, Septiembre de 2009}
\newpage
\pagestyle{empty}
\bigit
\vglue 6 cm
\rightline{A mis padres, responsables de mi educaci\'on.}

\selectlanguage{british}


\newpage

\tableofcontents

\pagestyle{fancy}

\clearpage{\thispagestyle{empty}\cleardoublepage}

\normalfont

\selectlanguage{spanish}

\vspace{4cm}
\chapter{Introducci\'on}
\label{introduccion}

Una transici\'on de fase se define como un cambio brusco en la estructura interna
y las propiedades de un sistema debido a variaciones en su entorno. Este entorno se caracteriza generalmente
por cantidades tales como temperatura, presi\'on, campos electromagn\'eticos, etc.
Los ejemplos m\'as comunes de transiciones de fase son la transici\'on 
de l\'{\i}quido a gas, de conductor normal a superconductor, o de
material paramagn\'etico a ferromagn\'etico. El estudio de las transiciones de fase es de indiscutible
inter\'es tanto te\'orico como tecnol\'ogico.

Los estudios te\'oricos microsc\'opicos de las transiciones de fase
implican el estudio de un fen\'omeno producido por la interacci\'on simult\'anea de un n\'umero enorme
($\sim10^{23}$) de componentes individuales. Esto forz\'o el desarrollo de teor\'{\i}as aproximadas que
proporcionaban soluciones exactas solo en algunos casos simplificados. Un ejemplo es la
teor\'{\i}a de campo medio para transiciones de segundo orden, v\'ease~\cite{GOLDEN} \'o~\cite{ZJ-BOOK},
introducida por L.~D.~Landau al final de la d\'ecada de 1950.
La explicaci\'on m\'as satisfactoria de los fen\'omenos cr\'{\i}ticos fue proporcionada por el
\emph{Grupo de Renormalizaci\'on} (GR), en primer lugar intuido por L.~P.~Kadanoff~\cite{RG-KADANOFF}
y finalmente desarrollado alrededor de 1970 en los importantes art\'{\i}culos de K.~G.~Wilson~\cite{RG-WILSON, RG-WILSON-KONDO},
ver~\cite{REV-FISHER} para una interesante revisi\'on hist\'orica de los logros del GR.

La transici\'on de fase en un sistema puede ser descrita como una discontinuidad en
las derivadas de su energ\'{\i}a libre respecto a alguna de las variables termodin\'amicas y pueden ser
clasificadas de acuerdo a esto, utilizando la llamada clasificaci\'on de Ehrenfest.
Si la discontinuidad se presenta en la primera derivada, se denomina transici\'on de fase de primer orden,
mientras que si es en la segunda derivada, se denomina transici\'on de fase de segundo orden.

De forma general, las transiciones de fase de primer orden son casi siempre las que involucran un calor latente.
Durante dichas transiciones, el sistema absorbe o libera una cantidad fija (y por lo general grande)
de energ\'{\i}a. Durante el proceso, la temperatura del sistema permanece constante conforme se
absorbe o se libera calor. Adem\'as, las transiciones de primer orden est\'an asociadas a \emph{reg\'{\i}menes mixtos}
en los que algunas partes del sistema han completado la transici\'on, mientras que otras no. Un ejemplo t\'{\i}pico de este fen\'omeno
es la coexistencia del r\'egimen de baja temperatura del agua (hielo) y el de alta temperatura
(agua l\'{\i}quida); el agua y el hielo pueden coexistir (existen los icebergs).

Las transiciones de fase de segundo orden son continuas en la primera derivada de la
energ\'{\i}a libre, pero presentan
discontinuidades en su segunda derivada. Estas incluyen la 
transici\'on a la fase ferromagn\'etica en materiales como el hierro, donde la magnetizaci\'on,
que es la primera derivada de la energ\'{\i}a libre con respecto a la fuerza del campo magn\'etico aplicado,
aumenta de forma continua desde cero conforme la temperatura desciende por debajo de
la temperatura de Curie. La susceptibilidad magn\'etica, la segunda derivada de la energ\'{\i}a libre 
respecto al campo, diverge.
Este tipo de transiciones no tiene calor latente asociado pero presenta longitudes 
de correlaci\'on infinita. Ejemplos t\'{\i}picos de transiciones de fase de segundo orden
son las transiciones paramagn\'etica-ferromagn\'etica y conductor-superconductor. Este tipo de transiciones
tambi\'en se caracteriza por comportamientos en forma de leyes de potencia en el punto de transici\'on 
(tambi\'en llamado \emph{punto cr\'{\i}tico}) con exponentes no enteros, llamados \emph{exponentes
cr\'{\i}ticos\footnote{ Seguiremos la nomenclatura habitual (ver por ejemplo~\cite{VICTORAMIT})
  para los exponentes cr\'{\i}ticos: $\nu$ es el exponente de la longitud de correlaci\'on,
  $\alpha$ es el del calor espec\'{\i}fico, $\beta$ el del par\'ametro de orden, 
  mientras que $\omega$ es el exponente (universal) de correcciones de escala de orden dominante.
  Un exponente ligeramente distinto, la dimensi\'on an\'omala $\eta$, se define en Ec.~(\ref{anom_dim}).}}.
Los exponentes cr\'{\i}ticos est\'an relacionados entre s\'{\i} por
\emph{relaciones de hiperescalado} -- conociendo dos de los exponentes, los otros pueden ser deducidos.
Sistemas muy diferentes pueden compartir exactamente el mismo conjunto de exponentes cr\'{\i}ticos y
se dice entonces que pertenecen a la misma \emph{Clase de Universalidad} (CU). La CU de un sistema
se define por propiedades muy generales tales como la simetr\'{\i}a de la interacci\'on microsc\'opica, la dimensionalidad
espacial, o la dimensionalidad del par\'ametro de orden, ver~\cite{PELI-REP} para
una revisi\'on exhaustiva de las CU m\'as habituales.

La existencia de un par\'ametro de orden es tambi\'en caracter\'{\i}stica de las transiciones de fase.
\'Este puede ser definido como una cantidad que es nula en una de las fases 
y no nula en la otra. Refleja el proceso de ruptura de simetr\'{\i}a que normalmente tiene lugar 
a trav\'es del punto de transici\'on. Por ejemplo, para la
transici\'on paramagneto-ferromagneto un par\'ametro de orden v\'alido es la magnetizaci\'on neta (cero
en la fase de alta temperatura y no cero en la de baja temperatura), mientras que
para la transici\'on l\'{\i}quido-gas es la diferencia de densidad de los dos reg\'{\i}menes que coexisten.
Otros tipos de transiciones de fase deben ser descritos por par\'ametros de orden m\'as complejos.

Simulaciones de Monte Carlo (MC) han resultado ser muy \'utiles en esta rama de la Mec\'anica Estad\'{\i}stica,
ver~\cite{MONTEBOOK} para una revisi\'on de los m\'etodos m\'as populares.
Con ellos, se puede simular la evoluci\'on temporal de cada constituyente del sistema
para un determinado Hamiltoniano. En nuestro caso, los sistemas se definen en redes
de dimensionalidad espacial $D$, con tama\~no lineal $L$ y
condiciones de contorno peri\'odicas. En cada nodo de la red se define una variable, llamada \emph{esp\'{\i}n},
que toma ciertos valores (dependientes del modelo) que evolucionan con el tiempo.

Un m\'etodo de MC puede actualizar o bien un solo esp\'{\i}n por iteraci\'on, como es el caso
del algoritmo de Metr\'opolis o del de ba\~no t\'ermico~\cite{MONTEBOOK}, o bien un grupo de espines,
como en el caso de los algoritmos de Wolff~\cite{Wolff89} o de Swendsen-Wang~\cite{SW}.
Estos \'ultimos son llamados \emph{m\'etodos de clusters}.  
En las proximidades del punto cr\'{\i}tico, se produce el llamado
\emph{Decaimiento Cr\'{\i}tico}, ver por ejemplo~\cite{MONTEBOOK}.
Los tiempos de relajaci\'on del par\'ametro de orden divergen como 
una potencia de la longitud de correlaci\'on, $\tau \sim \xi^z$,
siendo $z$ el denominado \emph{exponente cr\'{\i}tico din\'amico}.
Esto implica que el tiempo necesario para producir configuraciones estad\'{\i}sticamente 
independientes diverge cerca del punto cr\'{\i}tico para un sistema finito como $\tau \sim L^z$.
Los m\'etodos de MC de actualizaci\'on de un \'unico esp\'{\i}n tienen un exponente $z\gtrsim 2$.
Por lo tanto es muy complicado obtener datos de alta precisi\'on muy cerca del punto cr\'{\i}tico en sistemas
grandes. Por el contrario, se obtiene un comportamiento din\'amico mucho mejor
utilizando m\'etodos de MC de actualizaci\'on de clusters. Con estos \'ultimos,
dependiendo del modelo y de la dimensionalidad, se obtienen valores de $z$ entre 0 y 1~\cite{Sokal-Dyn}.
En este trabajo hemos utilizado casi siempre algoritmos de clusters.

Los m\'etodos de actualizaci\'on de espines incluyen simulaciones dentro del colectivo can\'onico (a una temperatura fija)
y dentro del colectivo microcan\'onico (a energ\'{\i}a fija).
Respecto a las simulaciones dentro del colectivo microcan\'onico, hemos utilizado un 
m\'etodo de simulaci\'on microcan\'onico propuesto recientemente que permite la simulaci\'on de sistemas de
un tama\~no nunca antes alcanzado que realizan transiciones de fase de primer orden~\cite{VICTORMICRO}.

Incluso con los recursos de computaci\'on de hoy en d\'{\i}a,
estamos restringidos a simular sistemas con m\'as de $10^{15}$ \'ordenes de magnitud menos componentes que los
presentes en un sistema macrosc\'opico real (con $\sim10^{23}$ part\'{\i}culas).
Lo \'unico que podemos hacer es simular sistemas con diferentes tama\~nos y
tratar de extrapolar nuestros resultados al L\'{\i}mite Termodin\'amico ($L\to\infty$).
El estudio del comportamiento de escala de los diferentes observables con el tama\~no del sistema
es llamado Finite-Size Scaling (FSS) y es fundamental para el estudio de las transiciones de fase,
v\'ease por ejemplo~\cite{VICTORAMIT}. En este trabajo hemos usado continuamente
t\'ecnicas de FFS, adem\'as hemos realizado un estudio novedoso del FSS
dentro del colectivo microcan\'onico, ve\'ase el Cap\'{\i}tulo~\ref{chap:FSSmicro}.

Nuestro objetivo principal es el estudio de los efectos del desorden sobre las transiciones de fase.
En concreto, deseamos estudiar el efecto de las impurezas congeladas en la transici\'on
de material paramagn\'etico a ferromagn\'etico. La presencia de desorden aleatorio en un sistema 
produce muchos fen\'omenos interesantes y f\'{\i}sicamente relevantes, lo que ha motivado extensos estudios
te\'oricos y experimentales. Los tipos m\'as caracter\'{\i}sticos de sistema con
desorden aleatorio son: vidrios de esp\'{\i}n~\cite{YOUNG-BOOK, PARISI-SG}, sistemas
aleatoriamente anisotr\'opicos~\cite{DUDKA, HARRIS-ANISOT, STINCH},
sistemas diluidos~\cite{STINCH2, JANKE-PoS, FOLK} y sistemas con los campos aleatorios~\cite{YOUNG-BOOK}.
En todos estos casos, existen variables aleatorias que caracterizan el comportamiento del
sistema.

Al modelar un sistema con desorden aleatorio se pueden utilizar dos enfoques diferentes.
Por un lado, se puede considerar que las variables aleatorias est\'an en equilibrio termodin\'amico
las otras variables din\'amicas del sistema. Por lo que las variables aleatorias tambi\'en ser\'an
``din\'amicas''. Este es el llamado \emph{desorden annealed} y debe ser la elecci\'on
si deseamos modelar un sistema en el que los tiempos caracter\'{\i}sticos de la din\'amica del desorden
sean comparables con los tiempos caracter\'{\i}sticos de la din\'amica de las variables originales,
como ser\'{\i}a el caso por ejemplo de una disoluci\'on de dos l\'{\i}quidos. Por otro lado,
se puede considerar que las variables aleatorias no evolucionan en el tiempo, sino que est\'an congeladas.
Es el llamado \emph{desorden congelado}.
\'Esta \'ultima es una alternativa perfectamente v\'alida por ejemplo si queremos modelar sistemas magn\'eticos con
impurezas. En este caso, el comportamiento magn\'etico se debe a los espines de los electrones no apareados
en las capas at\'omicas exteriores, mientras que las impurezas son \'atomos sin electrones no apareados. Se
sabe que la din\'amica de los electrones es \'ordenes de magnitud m\'as veloz que la din\'amica 
de los n\'ucleos de modo que se puede considerar a los \'atomos de impurezas congelados en el tiempo.
En Ref.~\cite{GIORGIO} se presenta un an\'alisis m\'as detallado de esta cuesti\'on.
Cuando se considera el desorden congelado se generan diferentes configuraciones espaciales aleatorias
del desorden (llamadas \emph{muestras}). Los espines de cada muestra evolucionar\'an independientemente
mientras que las impurezas permanecen fijas.
Para extraer la informaci\'on de un determinado observable, en primer lugar realizamos el promedio de su evoluci\'on
temporal en el interior de cada muestra (en lo sucesivo denominado \emph{promedio termal} y denotado por brackets)
y luego realizamos el promedio de lo anterior entre todas las muestras (denominado \emph{promedio muestral} y denotado por un suprarrayado),
el promedio doble es entonces denotado por $\overline{\langle\cdots\rangle}$.

Uno de los resultados de mayor importancia en el estudio de sistemas desordenados
es el criterio de Harris~\cite{critharris}, v\'ease el Ap\'endice~\ref{Appendix_Harris}.
El criterio se\~nala que si el calor espec\'{\i}fico en
el sistema puro diverge (el exponente cr\'{\i}tico $\alpha_\mathrm{puro}$ es mayor que cero),
el desorden cambiar\'a el comportamiento cr\'{\i}tico del modelo, es decir, aparecer\'a una nueva CU.
En este caso, se dice que el desorden es \emph{relevante}. Por el contrario, si el
calor espec\'{\i}fico no diverge en el sistema puro ($\alpha_\mathrm{puro}<0$) los 
exponentes cr\'{\i}ticos del sistema desordenado no cambiar\'an. Es este caso,
se dice que el desorden es \emph{irrelevante}. En el presente trabajo recomprobaremos
la validez del criterio para el modelo de Heisenberg tridimensional con diluci\'on por sitios.

Otra cuesti\'on muy interesante que surge cuando se estudian sistemas diluidos es la cuesti\'on del
\emph{autopromedio}. El valor medio de un observable $\cal O$ en una red de tama\~no lineal $L$ 
es diferente para cada realizaci\'on del desorden (en nuestro caso, para cada distribuci\'on espacial 
de los sitios no magn\'eticos), por lo que es una variable estoc\'astica caracterizada por un promedio
sobre el desorden $\overline{\cal O}$ y una varianza $(\Delta {\cal O})^2 \equiv \overline{{\cal O}^2} - \overline{{\cal O}}^2$.
Se dice que un sistema exhibe autopromedio para el observable $\cal O$ si
$\Delta{\cal O} / {\cal O}$ tiende a cero cuando $L\to\infty $.
Cuando un sistema diluido no autopromedia los estudios num\'ericos se hacen
muy dif\'{\i}ciles: incluso fijando la temperatura cr\'{\i}tica al valor correcto para 
$L\to\infty$, hacer el sistema m\'as grande no proporciona una gran
mejora estad\'{\i}stica. El autopromedio de las propiedades de los sistemas desordenados genera
gran inter\'es, reflejado en numerosos trabajos tanto num\'ericos~\cite{wise, derou,fytas} como
anal\'{\i}ticos~\cite{harris, korut, harris2}. En esta tesis se estudiar\'a el
autopromedio tanto de la susceptibilidad del modelo de Heisenberg tridimensional,
como del calor latente y la tensi\'on superficial del modelo de Potts tridimensional, ambos
modelos con diluci\'on por sitios.

Hemos estudiado num\'ericamente varios modelos con desorden aleatorio que presentaban
importantes cuestiones abiertas. Nuestra colaboraci\'on ha producido las
publicaciones recogidas en~\cite{OUR_O3, POTTS3D, BIFI2008, OUR_FSSMICRO, OUR_RSIM4D, RSIM4D-PROC, JANUS1, JANUS2,
JANUS3, JANUS4, JANUS-POTTS-SG, JANUS-EA-PRL, JANUS-EA-JSP} aunque en el presente trabajo
s\'olo se presentan los resultados de las referencias~\cite{OUR_O3, POTTS3D, BIFI2008, OUR_FSSMICRO, OUR_RSIM4D, RSIM4D-PROC}.

La disposici\'on del resto de esta tesis doctoral es la siguiente.
En el Cap\'{\i}tulo~\ref{chap:FSSmicro} estudiamos las propiedades de escala, tanto del 
modelo de Potts puro con cuatro estados ($Q=4$) en $D=2$ como del modelo de Ising puro en $D=3$.
Estos modelos realizan transiciones de fase de segundo orden bien conocidas
con calores espec\'{\i}ficos divergentes. Hemos simulado ambos modelos
utilizando el m\'etodo de simulaci\'on microcan\'onico presentado en~\cite{VICTORMICRO}. Obtuvimos 
fuertes evidencias de la bondad de dicho m\'etodo a trav\'es de la comparaci\'on con los
resultados m\'as recientes~\cite{PELI-REP, Salas}. El Cap\'{\i}tulo~\ref{chap:potts3D}
lo dedicamos al estudio
de los efectos de la diluci\'on en un sistema que realiza una transici\'on de fase de primer orden fuerte:
el modelo de Potts tridimensional con $Q=4$ y $Q=8$ estados.
Utilizando el m\'etodo de simulaci\'on comprobado en el Cap\'{\i}tulo~\ref{chap:FSSmicro},
fuimos capaces de simular sistemas con m\'as de $10^6$ componentes, multiplicando por un factor de 100
el n\'umero de componentes de los trabajos m\'as recientes~\cite{Ball00, Chat01, Chat05}.
En el Cap\'{\i}tulo~\ref{chap:O3}
estudiamos las propiedades de autopromedio del modelo de Heisenberg tridimensional diluido por sitios,
donde existen dos escenarios en conflicto que afirman que la susceptibilidad
es~\cite{harris} o no es~\cite{korut} autopromediante.
Adem\'as obtendremos informaci\'on acerca de la validez del criterio de Harris. 
En el Cap\'{\i}tulo~\ref{chap:RSIM4D} exponemos nueva informaci\'on acerca de los exponentes cr\'{\i}ticos
de los t\'erminos logar\'{\i}tmicos del modelo de Ising en cuatro dimensiones con diluci\'on por sitios.
Hemos logrado discriminar entre cinco diferentes teor\'{\i}as~\cite{Ah76,Boris,Jug,GeDe93,ISDIL4D}
mediante simulaciones num\'ericas de gran envergadura. Presentamos nuestras conclusiones generales en el Cap\'{\i}tulo~\ref{chap:conclusions}.
Tambi\'en se exponen varios ap\'endices tratando de ampliar la informaci\'on sobre
algunas de las herramientas utilizadas m\'as importantes o innovadoras.
En el Ap\'endice~\ref{Appendix_Harris} explicamos con detalle 
el criterio de Harris. En el Ap\'endice~\ref{Appendix_Quotient} presentamos brevemente
las t\'ecnicas populares del FSS, as\'{\i} como el M\'etodo de los Cocientes, una t\'ecnica 
que permite el c\'alculo de los exponente cr\'{\i}ticos partiendo de los datos obtenidos en sistemas finitos.
En el Ap\'endice~\ref{Appendix_dataanalysis} se describen dos cuestiones muy importantes para la simulaci\'on 
de sistemas din\'amicos con m\'etodos de MC: los tiempos de autocorrelaci\'on y la estimaci\'on de errores.
El Ap\'endice~\ref{Appendix_Extrapolations} est\'a dedicado a describir los diferentes m\'etodos
para la extrapolaci\'on en temperatura de los resultados obtenidos en una simulaci\'on can\'onica mientras que
en el Ap\'endice~\ref{Appendix_Max} se describe la obtenci\'on de la construcci\'on de Maxwell,
muy \'util para el estudio de transiciones de fase de primer orden. En el Ap\'endice~\ref{Appendix_LY}
se describe el enfoque introducido por Lee y Yang para describir las transiciones de fase,
formulando su importante teorema. Tambi\'en discutimos en este ap\'endice de la distribuci\'on
de los ceros de Lee-Yang sobre el c\'{\i}rculo unidad. Por \'ultimo, dedicamos el Ap\'endice~\ref{Appendix_Ibercivis}
a la descripci\'on de la infraestructura de supercomputaci\'on IBERCIVIS, que ha sido crucial para completar algunas
partes de este trabajo.

\selectlanguage{british}

\clearpage{\thispagestyle{empty}\cleardoublepage}

\normalfont

\addtocounter{chapter}{-1}

\selectlanguage{british}

\vspace{4cm}
\chapter{Introduction}
\label{introduction}

A phase transition is defined as a sharp change in the
internal structure and properties of a system due to
variations in its environment. This environment is usually
characterised by quantities such as temperature, pressure,
electromagnetic fields, etc.  Common examples of phase
transitions are the transition from liquid to gas, from
normal conductor to superconductor, and from paramagnet to
ferromagnet. The study of phase transitions is of major
interest both theoretically and technologically.

Theoretical microscopic studies of phase transitions
involve the study of a phenomenon produced by the simultaneous interaction of an enormous number
($\sim10^{23}$) of individual components. This forced the development of approximate
theories that produce exact solutions only for some simplified cases. An example is the
mean-field theory for second-order phase transitions, see for instance~\cite{GOLDEN} or~\cite{ZJ-BOOK},
introduced by L.~D.~Landau in the late 1950s.
The most satisfying explanation of critical phenomena was provided by the
\emph{Renormalization Group} (RG) picture, first intuited by L.~P.~Kadanoff~\cite{RG-KADANOFF}
and finally developed around 1970 in the landmark papers of K.~G.~Wilson~\cite{RG-WILSON, RG-WILSON-KONDO},
see~\cite{REV-FISHER} for an interesting historical review of RG achievements.

A phase transition in a system can be described as a discontinuity in
the derivatives of its free energy with respect to some thermodynamic variable, and can be
classified according to this by using the so-called Ehrenfest classification.
If the discontinuity is in the first derivative it is called a first-order
phase transition, while if it is in the second derivative it is called a second-order
phase transition.

More generally, first-order phase transitions are usually those involving a latent heat.
During such a transition, the system either absorbs or releases a fixed (and typically large)
amount of energy. During the process, the temperature of the system remains constant as heat is added or released.
In addition, first-order transitions are associated with \emph{mixed-phase regimes} in which some parts of the system
have completed the transition while others have not. A typical example of this phenomenon is 
the coexistence of the low temperature regime of water (ice) and the high temperature one 
(liquid water); water and ice can and do coexist (there exist icebergs).

Second-order phase transitions are
continuous in the first derivative but exhibit
discontinuities in a second derivative of the free
energy. These include the ferromagnetic phase transition in
materials such as iron, where the magnetisation, which is
the first derivative of the free energy with respect to the
applied magnetic field strength, increases continuously from
zero as the temperature is lowered below the Curie
temperature. The magnetic susceptibility, the second
derivative of the free energy with respect to the field,
diverges.  These have no associated latent heats, but present
infinite correlation lengths. Examples of second-order phase
transitions are the paramagnetic-ferromagnetic and the
conductor-superconductor transitions. This kind of
transition is also characterised by power-law behaviour
at the transition point (also-called \emph{critical point})
with non-integer exponents, thus called \emph{critical
exponents\footnote{We follow the
  standard terminology (see e.g~\cite{VICTORAMIT}) for the critical
  exponents: $\nu$ is the exponent for the correlation length,
  $\alpha$ that of the specific heat, $\beta$ that of the order
  parameter, while $\omega$ is the (universal) leading-order
  scaling-corrections exponent. A slightly different exponent, the
  anomalous dimension $\eta$, is defined in Eq.~(\ref{anom_dim}).}}. The
critical exponents are related to each other by
\emph{hyperscaling relations} -- knowing two of the exponents the others can be deduced.
Quite different systems can share exactly the same set of
critical exponents and are said to belong to the same
\emph{Universality Class} (UC). The UC of a system is defined
by very general properties such as symmetry of the
microscopic interaction, dimensionality of the space, or
dimensionality of the order parameter, see~\cite{PELI-REP} for
an exhaustive review of the most usual UC's.

The existence of an order parameter is also characteristic
of phase transitions. It can be defined as a quantity that
is null in one of the phases and non-null in the other, and
reflects the symmetry-breaking process that usually takes
place across the transition point. For example, for the
ferromagnetic-paramagnetic transition a valid order
parameter is the net magnetisation (zero in the high
temperature phase and non-zero in the low-temperature one),
while for the liquid-gas transition it is the density
difference of the two co-existing regimes. Other kinds of
phase transition must be described by more complex order
parameters.

Monte Carlo (MC) simulations have proved very useful in this branch of Statistical Mechanics,
see~\cite{MONTEBOOK} for a review of the most popular methods.
With them, one can simulate the evolution in time of each constituent of the system
for a given Hamiltonian. In our case the systems are defined on lattices
of spatial dimensionality $D$, with linear size $L$ and periodic boundary
conditions. On each node of the lattice we define a variable, called \emph{spin},
that takes on values (depending on the model) that evolve in time.

An MC spin update method can change either just one spin per iteration, as is the case
of the Metropolis or heat-bath algorithms~\cite{MONTEBOOK}, or a cluster of spins, as
is the case of the Wolff~\cite{Wolff89} and Swendsen-Wang~\cite{SW} algorithms.
The latter are thus called \emph{cluster methods}. 
At the critical point is presented the so-called \emph{Critical
Slowing Down}, see for example~\cite{MONTEBOOK}.
The relaxation time of the order parameter diverges as a power
of the correlation length, $\tau \sim \xi^z$, with $z$ being the \emph{dynamic critical exponent}.
This roughly implies that the time needed to produce statistically independent configurations
diverges at the critical point for a finite system as $\tau \sim L^z$.
Single-spin MC update methods have an exponent $z\gtrsim 2$. Therefore
is very hard to obtain high-precision data very close
to the critical point on large systems. However, cluster MC update
methods produce a much better dynamic behaviour. Depending
on the model and dimensionality, cluster methods have
$z$ values between 0 and 1~\cite{Sokal-Dyn}. In this work we
have used basically cluster methods.

Spin update methods include simulations within the canonical ensemble (at fixed temperatures)
and within the microcanonical ensemble (at fixed energies).
With respect to simulations in the microcanonical ensemble, we have exploited a recently proposed microcanonical
simulation method that allows the simulation of systems with a size never reached before that undergo
first-order phase transitions~\cite{VICTORMICRO}.

Even with today's computing resources, we are restricted to
simulating systems with more than $10^{15}$ orders of
magnitude fewer components than the real macroscopic system
(with $\sim10^{23}$ particles). The only thing we can do is
to simulate systems with different sizes and try to
extrapolate the results to the Thermodynamic Limit
($L\to\infty$).  The study of the scaling behaviour of the
different observables with system size is called Finite-Size
Scaling (FSS) and is fundamental for the study of phase
transitions, see for example~\cite{VICTORAMIT}. In this work
we have continually used FSS techniques, as well as
performing a novel study of FSS within the
microcanonical ensemble, see Chap.~\ref{chap:FSSmicro}.

Our main objective is the study of the effects of disorder
in phase transitions. In particular, we study the effect of
quenched impurities on the paramagnet-ferromagnet
transition.  The presence of random disorder in a system
produces many interesting and physically relevant phenomena
which have motivated extensive theoretical and experimental
studies. The most typical types of system with random
disorder are: spin glasses~\cite{YOUNG-BOOK, PARISI-SG},
random anisotropic systems~\cite{DUDKA, HARRIS-ANISOT,
STINCH}, dilute systems~\cite{STINCH2, JANKE-PoS, FOLK}, and
systems with random fields~\cite{YOUNG-BOOK}. In all these
cases there exist random variables characterising the
behaviour of the system.

When modelling a randomly disordered system one can use either of two approaches. On
the one hand, one can consider the random variables in thermodynamic equilibrium with
the other dynamic variables of the system. Thus the random variables will also be ``dynamic''. This is the
so-called \emph{annealed disorder} and should be the choice if we model a system in which
the characteristic times for the dynamics of the disorder are comparable with the characteristic
time of the original dynamic variables, as would be the case for example of
a solution of two liquids. On the other hand, one can consider that the
random variables do not evolve in time, but are frozen. This is the so-called \emph{quenched disorder}.
The latter is a perfectly valid alternative for example if we want to model magnetic systems with
impurities. In this case the magnetic behaviour is due to the spins of the unpaired electrons
in the outer atomic shells while the impurities are whole atoms with no unpaired electrons. It
is known that the dynamics of the electrons is orders of magnitude faster than the dynamics of
the nuclei, so that we can perfectly consider the impurity atoms as frozen in time. In Ref.~\cite{GIORGIO}
there is presented a more detailed discussion of this issue.
When considering quenched disorder we will generate different random spatial configurations
of the disorder (called \emph{samples}). Within each sample the spins evolve independently
but the disorder is fixed. To extract information
of a given observable, first we perform the average of its temporal evolution within each sample
(in the following called \emph{thermal average} and denoted by angle brackets)
and afterwards we perform the \emph{sample average} (denoted by an overline),
the double average is then denoted by $\overline{\langle\cdots\rangle}$.

One of the main results in disordered systems is the Harris criterion~\cite{critharris}, see Appendix~\ref{Appendix_Harris}. 
It states that if the specific heat diverges in
the pure system (the critical exponent, $\alpha_\mathrm{pure}$, is greater than zero),
the disorder will change the critical behaviour of the
model, i.e a new UC will appear. In this case
it is said that the disorder is \emph{relevant}. Conversely, if the
specific heat does not diverge in the pure system ($\alpha_\mathrm{pure}<0$) the critical
exponents of the disordered system will not change. In this case
it is said that the disorder is \emph{irrelevant}. We will recheck in the present work
the validity of the criterion for the three-dimensional site-diluted Heisenberg model.

Another very interesting question arising when studying dilute systems is the issue of
\emph{self-averaging}. The mean value of a quantity $\cal O$ on a lattice of linear size $L$ is different for each realization of
the disorder (in our case, for each spatial distribution of
the non-magnetic sites). Therefore it is a stochastic
variable characterised by an average over the disorder
$\overline{\cal O}$ and a variance $(\Delta{\cal
O})^2\equiv\overline{{\cal O}^2}-\overline{{\cal O}}^2$.  It
is said that a system is self-averaging for the quantity
$\cal O$ if $\Delta{\cal O}/{\cal O}$ goes to zero when
$L\to\infty$.  When a dilute system is not self-averaging,
numerical studies become very difficult: even fixing the
critical temperature to the correct value for
$L\longrightarrow\infty$ making the system larger does not
much improve the statistics. The self-averaging properties
of disordered systems have generated much interest,
reflected in numerous works both
numerical~\cite{wise,derou,fytas} and
analytical~\cite{harris,korut,harris2}. In this work we will
study the self-averaging properties both of the
susceptibility of the three-dimensional site-diluted Heisenberg
model and of the latent heat and surface tension of the
three-dimensional site-diluted Potts model.

We have numerically studied randomly disordered models presenting important open issues. 
Our collaboration has produced the 
papers of Refs.~\cite{OUR_O3, POTTS3D, BIFI2008, OUR_FSSMICRO, OUR_RSIM4D, RSIM4D-PROC, JANUS1, JANUS2, 
JANUS3, JANUS4, JANUS-POTTS-SG, JANUS-EA-PRL, JANUS-EA-JSP}, although in the present work
we only present the results of Refs.~\cite{OUR_O3, POTTS3D, BIFI2008, OUR_FSSMICRO, OUR_RSIM4D, RSIM4D-PROC}.

The organisation of the rest of this PhD thesis is as
follows. In Chapter~\ref{chap:FSSmicro} we study the scaling
properties both of the four-state ($Q=4$) pure Potts model
in $D=2$ and of the pure Ising model in $D=3$. These models
undergo well-known second-order phase transitions with
diverging specific heats. We have simulated them using the
microcanonical simulation method presented
in~\cite{VICTORMICRO} obtaining strong evidence for the
goodness of our approach by comparing it with the most
recent results~\cite{PELI-REP,
Salas}. Chapter~\ref{chap:potts3D} is devoted to the study
of the effects of dilution on a system performing a strong
first-order phase transition: the three-dimensional Potts
model with $Q=4$ and $Q=8$ states.  Using the simulation
method studied in Chapter~\ref{chap:FSSmicro}, we will be
able to simulate systems with more than $10^6$ components,
multiplying by a factor of 100 the number of components
reached in the most recent work~\cite{Ball00, Chat01,
Chat05}. In Chapter~\ref{chap:O3} we study the
self-averaging properties of the three-dimensional
site-diluted Heisenberg model, where there exist two
conflicting results stating that the susceptibility
is~\cite{harris} or is not~\cite{korut} a self-averaging
quantity. We will also obtain information about the validity
of the Harris criterion. In Chapter~\ref{chap:RSIM4D} we
report novel information about the critical exponents of the
logarithmic terms of the four-dimensional site-diluted Ising
model, we try to discriminate between five different
theories~\cite{Ah76,Boris,Jug,GeDe93,ISDIL4D} by using
high-statistics MC simulations. We present our conclusions
in Chapter~\ref{chap:conclusions}.  We also present some
appendices with the aim of extending some of the most
important or innovative tools used.  In
Appendix~\ref{Appendix_Harris} we explain in detail the
Harris criterion. In Appendix~\ref{Appendix_Quotient} we
briefly present the popular FSS techniques and the Quotient
Method, a technique that allows the computation of critical
exponents from data obtained in finite systems. In
Appendix~\ref{Appendix_dataanalysis} we describe two
important issues when simulating dynamical systems with MC
methods -- autocorrelation times and error estimates.
Appendix~\ref{Appendix_Extrapolations} is devoted to
describing the different methods to temperature-extrapolate
the results obtained in a canonical MC simulation, and
Appendix~\ref{Appendix_Max} describes the derivation of the
Maxwell construction, which is very useful in the study of
first-order phase transitions. In Appendix~\ref{Appendix_LY}
we describe the approach introduced by Lee and Yang to
describe phase transitions, formulating their landmark
theorem.  We also discuss in this appendix the distribution
of the LY-zeros on the unit circle. Finally
Appendix~\ref{Appendix_Ibercivis} describes the IBERCIVIS
computing infrastructure, which has been crucial for the
completion of some parts of this work.

\selectlanguage{british}

\clearpage{\thispagestyle{empty}\cleardoublepage}

\normalfont

\vspace{4cm}

\chapter{Microcanonical Finite-Size Scaling}
\label{chap:FSSmicro}


\section{Introduction}
\label{FSSmicro:intro}

The canonical ensemble enjoys a predominating position in
Theoretical Physics due to its many technical advantages
(convex effective potential in finite systems, easily
derived fluctuation-dissipation theorems,
etc.). This somewhat arbitrary choice of ensemble is
justified by the ensemble equivalence property, which holds
in the Thermodynamic Limit (TL) for systems with short-range
interactions.

However, in spite of this long-standing bias in favour of
the canonical ensemble, the canonical analysis of phase
transitions is {\em not} simpler. The advantages of
microcanonical analyses of first-order phase transitions
have long been known~\cite{VICTORMICRO,FIRST-ORDER}, and
indeed become overwhelming in the study of disordered
systems~\cite{POTTS3D}.  Furthermore, the current interest
in mesoscopic or even nanoscopic systems, where ensemble
equivalence does not hold, provides ample motivation to
study other statistical ensembles, in particular the
microcanonical ensemble~\cite{GROSS}. Besides,
microcanonical Monte Carlo~\cite{LUSTIG} is now as simple
and efficient as its canonical counterpart (even
microcanonical cluster algorithms are
available~\cite{VICTORMICRO}). Under such circumstances, it
is of interest to extend Finite-Size Scaling
(FSS)~\cite{VICTORAMIT,BINDER,BARBER,PRIVMAN} to the
microcanonical framework for systems undergoing a continuous
phase transition.

The relation between the microcanonical and the canonical critical
behaviour is well understood only in the TL. A global
constraint modifies the critical exponents, but only if the specific heat
of the unconstrained system diverges with a
positive critical exponent $\alpha>0$~\cite{FISHER-RENORM} (however, see~\cite{DOHM}).
This fact is explained in detail in Sec.~\ref{FSSmicro:Fisher}.
The modification of the
critical exponents, termed Fisher renormalization, is very simple.
Let $L$ be the system size,
and consider an observable $O$ (for instance, the susceptibility)
whose diverging behaviour in the infinite-volume canonical system is
governed by the critical exponent $x_O$
\begin{equation}
\langle O \rangle_{L=\infty,T}^\text{canonical} \propto |t|^{-x_O}\,,\quad
t=\frac{T-T_\text{c}}{T_\text{c}}\,.\label{EXPO-CANONICO}
\end{equation}
Now, let $e$ be the internal energy density and $e_\text{c}=\langle
e\rangle^\text{canonical}_{L=\infty,T_\text{c}}$. Consider the
microcanonical expectation value of the {\em same} observable $O$ in
Eq.~(\ref{EXPO-CANONICO}), but now at fixed energy $e$. The scaling behaviour~(\ref{EXPO-CANONICO})
translates to
\footnote{In the particular case of the fixed-energy constraint,
Eq.~\eqref{EXPO-MICRO-CANONICO} follows from \eqref{EXPO-CANONICO} and from
the ensemble equivalence property
$$\langle O\rangle_{L=\infty,e}=\langle
O\rangle_{L=\infty,T}^\text{canonical}\,,\quad\text{if }\quad e=\langle
e\rangle_{L=\infty,T}^\text{canonical}\,.$$ Indeed, it suffices to
notice that ($C(T)$ is the canonical specific heat, $C\propto
|t|^{-\alpha}$),
$$e-e_\text{c}=\int_{T_\text{c}}^T\,\text{d}T\, C(T)\propto
|t|^{1-\alpha}\ \Longrightarrow\ |t|\propto
|e-e_\text{c}|^{\frac{1}{1-\alpha}}\,.$$ The only exponent whose
renormalization is not clear at this point, is $\alpha$ itself, for
the energy is not a dynamical variable but a parameter in this
ensemble.  If one chooses to define $\alpha_\text{m}$ as the critical
exponent corresponding to $\text{d} t/\text{d} e$, the correspondence
with Fisher renormalization, $\alpha_\text{m}=-\alpha/(1-\alpha)$
becomes complete. In fact, see concluding paragraph in
Sect.~\ref{SUBSECT-ENSEMBLE-EQUIVALENCE}, the microcanonical
$\text{d} t/\text{d} e$ behaves as the canonical $1/[\text{d}
  e/\text{d} t]\,$.
Note that in the above expressions we disregarded subdominant 
terms such as the contribution of the analytical background in the specific heat. Such terms are
subdominant only if $\alpha>0$. In case $\alpha$ were negative, the asymptotic dominance is different.
The specific heat at $T_\text{c}$ is dominated by the analytical background.
As a consequence $|t|\sim |e-e_\text{c}|$ and none of the exponents (not even $\alpha$) gets renormalised.}
\begin{equation}
\langle O\rangle_{L=\infty, e} \propto |e-e_\text{c}|^{-x_{O,\text{m}}}\,,\quad
x_{O,\text{m}}=\frac{x_O}{1-\alpha}\,.
\label{EXPO-MICRO-CANONICO}
\end{equation}
We will denote the microcanonical exponents with the subindex ``m''.
Hence, the Fisher renormalization of the correlation length exponent
$\nu$, is $\nu\rightarrow \nu_\text{m}=\nu/(1-\alpha)$, that of the
order parameter exponent is $\beta\rightarrow
\beta_\text{m}=\beta/(1-\alpha)$, etc. On the other hand, the
anomalous dimension, defined in Eq.~(\ref{anom_dim}), is invariant under
Fisher renormalization~\cite{FISHER-RENORM}, i.e. $\eta=\eta_\text{m}\,$. See also~\cite{KENNA}
for a recent extension of Fisher renormalization to the case of {\em logarithmic} scaling corrections.

As for systems of finite size, the microcanonical
FSS~\cite{MFSS1,MFSS2,MFSS3} is at the level of an ansatz. This ansatz
is obtained from the canonical one merely by replacing the free-energy
density by the entropy density, and using Fisher renormalised critical
exponents.  The microcanonical ansatz reproduces the canonical
one~\cite{KASTNER2001}, and has been the subject of some numerical
testing~\cite{MFSS3,BEHRINGER2006}. Furthermore, systems undergoing
Fisher renormalization (due to some global constraint other than the
energy) do seem to obey FSS as well~\cite{TROSTER08}.

A difficulty lies in the fact that the current forms of the microcanonical FSS
ansatz (FSSA)~\cite{MFSS1,MFSS2,MFSS3} are in a somewhat old-fashioned form.
Indeed, they are formulated in terms of quantities such as
$e_\text{c}$ or the critical exponents, which are not accessible in
the absence of an analytical solution. In this respect, a great step
forward was achieved in a canonical context~\cite{FSS-PISA} when it
was realized that the finite-lattice correlation length~\cite{COOPER}
allows one to formulate the FSSA in terms of quantities computable
in a finite-lattice. This formulation made it practical to extend
Nightingale's phenomenological renormalization~\cite{NIGHTINGALE} to
space dimensions $D>2$ (the so-called quotient
method~\cite{QUOTIENTS}).

Here, we will extend the microcanonical FSSA to a modern form,
allowing us to use the quotient method. We will test numerically this
extended FSSA in two models with $\alpha>0$, hence undergoing
non-trivial Fisher renormalization, namely the $D=3$ ferromagnetic
Ising model, and the $D=2$ four-state ferromagnetic Potts model. The Potts
model has the added interest of undergoing, in its canonical form,
quite strong {\em logarithmic} corrections to scaling that are
nevertheless under relatively strong analytical
control~\cite{Salas}. It will therefore be quite a challenge
to control the logarithmic corrections in the microcanonical setting.

\section{Analytical Framework}
\label{FSSmicro:analytics}

\subsection{Fisher Renormalization of Critical Exponents}
\label{FSSmicro:Fisher}

In 1976, an important paper of M. E. Fisher~\cite{FISHER-RENORM} established a set of elegant
relationships describing the effects of constrained hidden variables on the critical exponents.
The original theory was developed to explain the significant deviations of the
theoretical predictions (basically from the Ising model) from the experimental
measurements of critical exponents. These deviations were attributed to some extra ``hidden'' degrees of freedom, present
in the \emph{real} system but not in the oversimplified \emph{ideal} theoretical
model. Models (like Ising or Potts) are somewhat gross
idealisations of real fluids or magnets, and can be said to lack sufficient
internal degrees of freedom. In addition, some experiments are unavoidably different
from the ideal system, for example due to the presence of defects or impurities (such as quenched
magnetic impurities or non-uniform isotopic composition).

Apart from the exactly soluble models introduced in the original work~\cite{FISHER-RENORM},
the formalism introduced by Fisher has provided explanations of numerous phenomena
and behaviours, both theoretical and experimental. To cite some examples, there are studies of the phase
transition of constrained uniaxial dipolar ferromagnets~\cite{AHA-DIP} and
of the random-field antiferromagnet with competing interactions~\cite{WO-COMP}, efforts made to
distinguish between the Random Field Ising Model (RFIM)
and the Dilute Antiferromagnetic Model (DAFM) under an applied field~\cite{CAL-RF}, and the study of
the tricritical point of the Blume-Capel model in three dimensions~\cite{DEN-BC} related with
the superfluid $\lambda$ transition in $^3$He-$^4$He mixtures in confined films~\cite{KIM-3HE}.
We would  also point to the agreement of
Fisher theory with the results for compressible systems, theoretically for both the Ising~\cite{BAKER-COMPR} and
the $\phi^4$\cite{TRO-COMPR} models~ and experimentally for ammonium chloride at high
pressures~\cite{GARLAND-NHCL}, see also~\cite{IMRY-COMP} for a study of the tricritical point in compressible systems.

The situation was described as follows: firstly, there is an ``ideal'' system
with known variables characterised by the ideal critical exponents $\alpha, \beta, \gamma, \ldots$;
secondly, the ``real'' system has some ``hidden'' degrees of freedom which fluctuate
but remain in equilibrium with the known variables; finally, the hidden variables are
subject to some form of constraint (for example, the total number of impurity atoms
must remain fixed). The critical exponents of the real system are denoted by
$\alpha_X, \beta_X, \gamma_X, \ldots$

In the following we will describe the relationships between the two
sets of critical exponents (now called Fisher renormalization), following in some points the 
recent work of Kenna et al.~\cite{KENNA} in which there also can be found the corresponding set of relationships
for the logarithmic correction exponents. We will focus the discussion on describing the temperature transition 
in a ferromagnet at the Curie point, although the results are perfectly valid for other kinds of phase transitions
(antiferromagnets, gas-liquid transitions, binary fluids, etc.).

The starting point is the description of the ideal system in terms of the free energy:
\begin{equation}
f= f_0(t,h)\,,
\label{f_ideal}
\end{equation}
where $t=(T-T_\text{c})/T_\text{c}$ is the reduced temperature and $h$ is the ``field''. The field is related
in general with the order parameter, $\sigma$, describing the transition under study through
\begin{equation}
\sigma= \sigma(t,h)=-\left(\frac{\partial f}{\partial h}\right)_t \,,
\label{order_gener}
\end{equation}
that in the ideal case becomes
\begin{equation}
\sigma= \sigma_0(t,h)=-\left(\frac{\partial f_0}{\partial h}\right)_t \,.
\label{order_ideal}
\end{equation}
The order parameter can be the magnetisation in a ferromagnet, the sublattice magnetisation
in an antiferromagnet, the density in a gas-liquid transition, etc. The ideal system
will undergo a phase transition at $T_\text{c}^0$. Thus below this temperature there will be
a non-vanishing order parameter (spontaneous magnetisation)
\begin{equation}
\varDelta \sigma_0= \lim_{h\to 0^+} \frac{1}{2} [\sigma_0(t,h)-\sigma_0(t,-h)] \,,
\label{order_below_ideal}
\end{equation}
which vanishes at the critical point as:
\begin{equation}
\varDelta \sigma_0 \sim |t|^\beta \quad \quad,\quad \quad (t\to 0^-) \,.
\label{order_ideal_critical}
\end{equation}
The second field derivative is also described by its corresponding critical exponent
\begin{equation}
\chi_0(t)= \lim_{h\to 0} \left(\frac{\partial \sigma_0}{\partial h}\right)_t \sim |t|^{-\gamma} \quad\quad , \quad\quad (t\to 0) \,.
\label{sus_ideal}
\end{equation}
Finally the critical behaviour of the specific heat
\begin{equation}
c_0(t) \sim |t|^{-\alpha} \quad \quad ,\quad \quad (t\to 0) \,,
\label{cesp_critical}
\end{equation}
means that, in the absence of a field, the ideal free energy behaves for small $t$ as
\begin{equation}
f_0(t,0)={\cal A}_{0\pm}+{\cal A}_{1\pm}|t|+{\cal A}_{2\pm}|t|^2+{\cal B}_{\pm}|t|^{2-\alpha}+{\cal O}(|t|^3) \,,
\label{f_ideal_small_t}
\end{equation}
where the $\pm$ depends on the sign of $t$.

We assume that the real system is derived from the ideal system by the introduction of
a new ``hidden'' thermodynamic variable $x$ which is the conjugate of a force $u$, such
that the thermodynamic potential becomes $f=f(t,h,u)$, with
\begin{equation}
x(t,h,u)=\left(\frac{\partial f}{\partial u}\right)_{t,h} \,.
\label{x_from_f}
\end{equation}
The constraint in the hidden variable is written as
\begin{equation}
x(t,h,u)=X(t,h,u) \,,
\label{constraint}
\end{equation}
where $X(t,h,u)$ is assumed to be an analytical function. Let us introduce the basic hypothesis
(suggested in part by the analytical results of some soluble models~\cite{FISHER-RENORM}) that
the free energy of the constrained system can be written in terms of the ideal free energy $f_0$ as
\begin{equation}
f(t,h,u)= f_0(t^*(t,h,u),h^*(t,h,u))+g(t,h,u)\,,
\label{f_constrained}
\end{equation}
where $t^*$, $h^*$, and $g$ are analytic functions of their arguments. I.e., we assume
that the total free energy consists of a ``regular background'' contribution $g(t,h,u)$ plus a ``singular
contribution'' derived from the ideal free energy $f_0(t,h)$ by a smooth transformation
of the temperature, $t$, and the field, $h$, to the modified versions $t^*$ and $h^*$.
Furthermore, the transition must remain ideal if observed at fixed force $u$, and the ideal
free energy $f_0(t,h)$ is recovered when $u=0$.

To simplify the discussion we assume that the hidden degrees of freedom are neutral in the
sense that they do not bias the value of the external field at the transition. I.e., the
transition still occurs at $h=0$ and
\begin{equation}
h^*(t,h,u)=h{\cal J}(t,h,u)\,.
\label{h_notbiased}
\end{equation}

With the previous assumptions, one can obtain from Eqs.~(\ref{order_below_ideal}),~(\ref{f_constrained}),
and~(\ref{h_notbiased}) that the order parameter of the constrained system behaves as 
\begin{equation}
\varDelta \sigma = \varDelta \sigma_0 (t^*(t,0,u)){\cal J}(t,0,u)\,,
\label{order_constrained_critical}
\end{equation}
where we have made use of the continuity of all the analytical functions for $h\to0$ including the internal
energy of the ideal system, $e_0=\partial f_0 / \partial t$.

We can calculate the internal energy for zero field using Eq.~(\ref{f_ideal_small_t}) as
\begin{equation}
e_0(t,0)=\frac{\partial f_0(t,0)}{ \partial t} = A_0+A|t|+B|t|^{1-\alpha}+\cdots\,,
\label{e_internal}
\end{equation}
with $A_0=\pm{\cal A}_{1\pm}$, $A_1=2{\cal A}_{2\pm}$ and $B=\pm(2-\alpha){\cal B}_{\pm}$ and where
$\cdots$ represents higher-order terms.

We can also obtain from Eqs.~(\ref{x_from_f}) and~(\ref{f_constrained}) for $h\to0$
\begin{eqnarray}
x(t,0,u) & = & \frac{\partial f(t,0,u)}{ \partial u} =  \frac{\partial f}{ \partial t^*}\frac{\partial t^*}{ \partial u}
   +\frac{\partial f}{ \partial h^*}\frac{\partial h^*}{ \partial u}+\frac{\partial g}{ \partial u} \\
& = & e_0(t^*,0)\frac{\partial t^*(t,0,u)}{ \partial u}+\frac{\partial g(t,0,u)}{ \partial u} \,,
 \label{x_arrayl}
\end{eqnarray}
where we used the fact that
\begin{equation}
\frac{\partial h^*}{ \partial u}=h\frac{\partial {\cal J}}{ \partial u}={\cal O}(h)\,.
\label{h_orden}
\end{equation}

Provided that $t^*$ is a smooth function, it can be expanded for $h=0$ around the 
real critical point $T=T_\text{c}$ and $u=u_\text{c}$,
\begin{equation}
t^*(t,0,u)=a_1\mu+a_2\tau+\cdots \,,
\label{t_expanded}
\end{equation}
with $\mu=u-u_\text{c}$ and $\tau=T-T_\text{c}$. The absence of constant terms in the
above expansion is due to Eq.~(\ref{order_constrained_critical}) because 
$\varDelta \sigma (t^*(t_\text{c},0,u_\text{c}))=0$ fixes the temperature in the constrained system
so that $t^*(t_\text{c},0,u_\text{c})=0$.

Then to first order
\begin{equation}
\frac{\partial t^*(t,0,u)}{\partial u}=a_1+\cdots \,,
\label{deriv_t}
\end{equation}
which inserted into Eq.~(\ref{x_arrayl}) with Eq.~(\ref{e_internal}) gives
\begin{equation}
x(t,0,u)=a_1 A_0+a_1A|t^*|+a_1 B|t^*|^{1-\alpha}+\frac{\partial g(t,0,u)}{ \partial u}+\cdots\,.
\label{x_firstorder}
\end{equation}

We can also expand the constraint, Eq.~(\ref{order_constrained_critical}), about the real
critical point $u=u_\text{c}$ and $T=T_\text{c}$
\begin{equation}
X(t,0,u)=X(t_\text{c},0,u_\text{c})+d_1\mu+d_2\tau + \cdots\,.
\label{X_firstorder}
\end{equation}
In addition, from Eq.~(\ref{t_expanded})
\begin{equation}
\mu=\frac{1}{a_1}t^*(t,0,u)-\frac{a_2}{a_1}\tau+\cdots\,.
\label{mu_firstorder}
\end{equation}
Using the above equation, we can insert Eqs.~(\ref{x_firstorder}) and~(\ref{X_firstorder}) into~(\ref{constraint})
to obtain the main result
\begin{equation}
a_1^2(A|t^*|+B|t^*|^{1-\alpha})=d_1t^*+(a_1 d_2-d_1 a_2)\tau\,.
\label{main_equation}
\end{equation}
If $\alpha<0$, the regular term dominates and $|t^*|\propto|\tau|$, resulting in the absence of Fisher renormalization, in
which case the critical exponents of the transition remain unchanged. On the contrary, if $\alpha>0$, one
obtains the central result
\begin{equation}
|t^*|\propto|\tau|^{1/(1-\alpha)} \,.
\label{main_result}
\end{equation}
That means that a deviation from the critical point of
$\tau$ in the real system is equivalent to a deviation
$t^*$ in the ideal system, these deviations being related by
Eq.~(\ref{main_result}).  Then the real system approaches the transition more
slowly than the ideal one. The internal energy of the real
system for $h=0$ is
\begin{eqnarray}
e(t,0,u) & = & \frac{\partial f(t,0,u)}{\partial t} = e_0(t^*,0)\frac{\partial t^*(t,0,u)}{ \partial t}+ \frac{\partial g(t,0,u)}{ \partial t} \\
 & = & (A_0+A|t^*|+B|t^*|^{1-\alpha}) \frac{\partial t^*(t,0,u)}{ \partial t} + \frac{\partial g(t,0,u)}{ \partial t} \\
 & = & (A_0+A|\tau|^{1/(1-\alpha)}+B|\tau|) \frac{\partial t^*(t,0,u)}{ \partial t} + \frac{\partial g(t,0,u)}{ \partial t}\,,
\label{e_real}
\end{eqnarray}
where the only possible singular contributions to the specific heat stem from the terms within parentheses.
The specific heat of the constrained system is then
\begin{equation}
c(t,0,u)=\frac{\partial e(t,0,u)}{\partial t} \sim |\tau|^{\alpha/(1-\alpha)} \,,
\label{c_renorm}
\end{equation}
and therefore the singular behaviour of the specific heat has been replaced for a cusp-like one,
i.e., it remains finite due to the presence of the constrained hidden variable. Then
the Fisher renormalization for the exponent of the specific heat is
\begin{equation}
\alpha_X=-\frac{\alpha}{1-\alpha} \,.
\label{alpha_renorm}
\end{equation}
We can obtain analogously the renormalization for other critical exponent as:
\begin{eqnarray}
\sigma & \sim & |t^*|^\beta \sim |\tau|^{\beta/(1-\alpha)} \quad \quad, \quad \beta_X=\frac{\beta}{1-\alpha} \,, \\
\chi & \sim & |t^*|^{-\gamma} \sim |\tau|^{-\gamma/(1-\alpha)} \quad, \quad \gamma_X=\frac{\gamma}{1-\alpha} \,, \\
\xi & \sim & |t^*|^{-\nu} \sim |\tau|^{-\nu/(1-\alpha)} \quad, \quad \nu_X=\frac{\nu}{1-\alpha} \,.
\label{exp_renorm}
\end{eqnarray}
The critical exponent $\eta$ is defined at just the critical point and therefore is invariant~\footnote{
This can also be obtained from the scaling relationship $(2-\eta)\nu=\gamma$. If $\nu$ and $\gamma$
are renormalised, $\eta$ remains unchanged.
}
\begin{equation}
\eta_X=\eta \,.
\label{eta_renorm}
\end{equation}

If the standard power-law scaling behaviour of the main quantities is modified by multiplicative logarithmic correction,
i.e,
\begin{eqnarray}
c_0(t) & \sim & |t|^{-\alpha}|\log|t||^{\hat{\alpha}} \,, \\
m_0(t) & \sim & |t|^{\beta}|\log|t||^{\hat{\beta}} \quad \text{for } t<0 \,, \\
\chi_0(t)& \sim & |t|^{-\gamma}|\log|t||^{\hat{\gamma}} \,, \\
\xi_0(t)& \sim & |t|^{-\nu}|\log|t||^{\hat{\nu}} \,, \\
m_0(h)& \sim & |h|^{\frac{1}{\delta}}|\log|h||^{\hat{\delta}}\quad  \text{for } t=0\,, \\
{\cal G}_0(x,t)& \sim & x^{-(D-2+\eta)}(\log x)^{\hat{\eta}} G \left( \frac{x}{\xi(t)} \right)\quad \text{for } t\ll1\,,
\label{log_behaviour}
\end{eqnarray}
then Eq.~(\ref{f_ideal_small_t}) is naively changed to give:
\begin{equation}
f_0(t,0)={\cal A}_{0\pm}+{\cal A}_{1\pm}|t|+{\cal A}_{2\pm}|t|^2+{\cal O}(|t|^3)+{\cal B}_{\pm}|t|^{2-\alpha}|\log|t||^{\hat{\alpha}}
\left\{1+{\cal O} \left( \frac{\log|\log|t||}{\log|t|} \right) \right\} \,.
\label{f_ideal_small_t_log}
\end{equation}
And therefore it can be obtained the equivalent of Eq.~(\ref{main_equation}) for the logarithmic case
\begin{equation}
a_1^2(A|t^*|+B|t^*|^{1-\alpha}|\log|t^*||^{\hat{\alpha}})=d_1t^*+(a_1 d_2-d_1 a_2)\tau\,.
\label{main_equation_log}
\end{equation}
The above equation produces the following results:
\begin{itemize}
\item If $\alpha<0$, or $\alpha=0$ and $\hat{\alpha}<0$, the regular term dominates and $|t^*|\propto|\tau|$,
leading to the absence of Fisher renormalization.

\item If $\alpha>0$, or $\alpha=0$ and $\hat{\alpha}>0$, one obtains the modified central result
\begin{equation}
|t^*|\propto|\tau|^{1/(1-\alpha)} |\log|\tau||^{-\hat{\alpha}/(1-\alpha)}\,,
\label{main_result_log}
\end{equation}
which produces the renormalization of the individual exponents of the logarithms:
\begin{eqnarray}
\hat{\alpha}_X & = & -\frac{\hat{\alpha}}{1-\alpha} \,, \\
\hat{\beta}_X & = & \hat{\beta}-\frac{\beta\hat{\alpha}}{1-\alpha} \,, \\
\hat{\gamma}_X & = & \hat{\gamma}+\frac{\gamma\hat{\alpha}}{1-\alpha} \,, \\
\hat{\nu}_X & = & \hat{\nu}+\frac{\nu\hat{\alpha}}{1-\alpha} \,.
\label{exp_renorm_log}
\end{eqnarray}
Again, no renormalization takes place for the exponents $\hat{\eta}_X = \hat{\eta}\,$ and $\hat{\delta}_X = \hat{\delta}\,$.

\end{itemize}

\subsection{The Microcanonical Ensemble}
\label{FSSmicro:MFSSA}

The first step in the construction of the ensemble is an extension of
the configuration space.  We add $N(=L^D)$ real momenta, $p_i$, to our
$N$ original variables, $\sigma_i$ (named spins
here)~\cite{LUSTIG,VICTORMICRO}.  Note that this extended
configuration, $\{\sigma_i,p_i\}$, appears in many numerical schemes
(consider, for instance, Hybrid Monte Carlo~\cite{HYBRID} simulations
in Lattice Gauge Theory). We shall  work in
the {\em microcanonical} ensemble for the $\{\sigma_i,p_i\}$ system.

Let ${\cal U}$ be the original spin Hamiltonian
(i.e., Eq.~(\ref{SPIN-HAMILTONIAN}) in our case).  Our total energy
is~\footnote{Note that this microcanonical ensemble exactly matches
  the conditions in the original Fisher work~\cite{FISHER-RENORM}: the
  momenta are some {\em hidden} degrees of freedom in thermal
  equilibrium with the spins, and a global constraint is imposed. It
  is also curious to rederive the results in Sec.~\ref{FSSmicro:MFSSA}
  considering $\varGamma$ momenta per spin (in this work $\varGamma=1$, while Lustig~\cite{LUSTIG} always considered $\varGamma=3$). If
  one takes the limit $\varGamma\to\infty$, at fixed $N,$ the canonical
  probability is recovered for the spins.}
\begin{equation}
{\cal E}= \sum_{i=1}^{N} \frac{p_i^2}{2}\ +\ {\cal U}\, \quad (e\equiv {\cal
E}/N\,,\ u\equiv {\cal U}/N)\,.\label{ENERGIATOTAL}
\end{equation}
The momenta contribution,
\begin{equation}
N\kappa \equiv \sum_{i=1}^{N} \frac{p_i^2}{2}\,,
\end{equation}
is necessarily positive, and it is best thought of as a ``kinetic''
energy.  In this mechanical analogue, the original spin Hamiltonian
${\cal U}$ can be regarded as a ``potential'' energy.

The canonical partition function is ($\beta\!\equiv\!1/T$)
\begin{equation}
Z_N(\beta)=\int_{-\infty}^\infty\prod_{i=1}^N\text{d}p_i\sum_{\{\sigma_i\}}\
\text{e}^{-\beta{\cal E}}=\left(\frac{2\pi}{\beta}\right)^{\frac{N}{2}} 
\sum_{\{\sigma_i\}}\ \text{e}^{-\beta{\cal U}}\,,
\end{equation}
where $\sum_{\{\sigma_i\}}$ denotes summation over
spin configurations. Hence, the $\{p_i\}$ play the role of a Gaussian thermostat. The
$\{p_i\}$ are statistically uncorrelated with the spins. Since
$\langle \kappa \rangle^\text{canonical}_{L,\beta}=1/(2\beta)$, one has
$\langle e\rangle^\text{canonical}_\beta= \langle
u\rangle^\text{canonical}_\beta + 1/(2\beta)\,$.

Furthermore, given the statistical independence of $\kappa$ and $u$,
the canonical probability distribution function for $e$,
$P_\beta^{(L)}(e)$, is merely the convolution of the distributions for
$\kappa$ and $u$:
\begin{equation}
P_\beta^{(L)}(e)=\int_0^\infty\,\text{d}\,\kappa\  P_\beta^{(L),\kappa}(\kappa)\,
P_\beta^{(L),u}(e-\kappa)\,.
\end{equation}
In particular, note that for spin systems on a finite lattice,
$P_\beta^{(L),u}(u)$ is a sum of (order $N$) Dirac $\delta$
functions.  Now, since the canonical variance of $\kappa$ is $1/(\beta
\sqrt{2 N})$, roughly $\sqrt{N}$ discrete $u$-levels, with $u\sim
e-1/(2\beta)$, give the most significant contribution to
$P_\beta^{(L)}(e)$. We see that the momenta's kinetic energy  
provides a natural smoothing of the comb-like
$P_\beta^{(L),u}(u)$. Once we have a conveniently smoothed
$P_\beta^{(L)}(e)$, we may proceed to the definition of the entropy.

In a microcanonical setting, the crucial role is played by the entropy
density, $s(e,N)$, given by 
\begin{equation}
\exp[N s(e,N)]=
\int_{-\infty}^\infty\prod_{i=1}^N\text{d}p_i\sum_{\{\sigma_i\}}\ 
\delta(Ne-\cal{E})\,.\label{MICRO1}
\end{equation}

Integrating out the $\{p_i\}$ using the Dirac delta function in (\ref{MICRO1})
we get
\begin{eqnarray}
\exp[N s(e,N)] &=&\displaystyle \frac{(2\pi N)^{\frac{N}{2}}}{N \Gamma(N/2)} \sum_{\{\sigma_i\}}
\omega(e,u,N)\,,\label{MICRO2}\\
\omega(e,u,N)&\equiv&(e-u)^{\frac{N-2}{2}}\theta(e-u)\;,\label{MICROPESO}
\end{eqnarray}
where $\Gamma$ is the gamma function and the step
function, $\theta(e-u)$, enforces $e>u$.  Equation~(\ref{MICRO2}) suggests
defining the microcanonical average at fixed $e$ of any function of $e$
and the spins, $O(e,\{\sigma_i\})$, as~\cite{LUSTIG}
\begin{equation}
\langle O\rangle_e\equiv
\frac{\sum_{\{\sigma_i\}}\,O(e,\{\sigma_i\})\,\omega(e,u,N)}{{\sum_{\{\sigma_i\}}\omega(e,u,N)}}\,.\label{MICROPROB}
\end{equation}

We use Eq.~(\ref{MICRO2}) to compute $\text{d}s/\text{d}e$~\cite{VICTORMICRO}:
\begin{eqnarray}
\frac{\text{d} s(e,N)}{\text{d} e}&=
&\langle\hat\beta(e;\{\sigma_i\})\rangle_e\,,\label{FD1}\\
\hat\beta(e;\{\sigma_i\})&\equiv&\frac{N-2}{2N (e -u)}\label{beta_micro}\,.
\end{eqnarray}

Bearing in mind the crucial role of the generating functional in
Field Theory (see e.g.~\cite{VICTORAMIT}), we extend the definition
(\ref{MICRO1}) by considering a linear coupling between the spins and
a site dependent source field $h_i$:
\begin{equation}
\exp[N s(e,\{h_i\},N)]=
\int_{-\infty}^\infty\prod_{i=1}^N\text{d}p_i\sum_{\{\sigma_i\}}\ \text{e}^{\sum_i\, h_i \sigma_i}
\delta(Ne-\cal{E})\,,\label{MICRO-FUENTE}
\end{equation}
where ${\cal E}=Ne$ is still given by Eq.~(\ref{ENERGIATOTAL}),
without including the source term.
In this way, the microcanonical spin correlation functions follow from derivatives of $s(e,\{h_i\},N)$:
\begin{eqnarray}
\left.\frac{\partial [N\, s]}{\partial h_k}\right|_{e,\{h_i\},N}&=&\langle \sigma_k \rangle_{e, \{h_i\}}\,,\\\nonumber
\left.\frac{\partial^2 [N\, s]}{\partial h_k\partial h_l}\right|_{e,\{h_i\},N}&=&\langle \sigma_k \sigma_l \rangle_{e, \{h_i\}}- 
\langle \sigma_k\rangle_{e, \{h_i\}}\, \langle\sigma_l \rangle_{e, \{h_i\}}\,.
\end{eqnarray}
In particular, if the source term is uniform $h_i=h$ we observe that
the microcanonical susceptibility is given by standard
fluctuation-dissipation relations, see Ref.~\cite{VICTORAMIT} and
Eq.~(\ref{chi-micro-def}) below.

\subsubsection{Ensemble equivalence}
\label{SUBSECT-ENSEMBLE-EQUIVALENCE}

Equation~(\ref{MICRO1}) ensures that the {\em
  canonical} probability density function for $e$ is
\begin{equation}
P_{\beta}^{(L)}(e)=\frac{N}{Z_N(\beta)}\exp[N(s(e,N)-\beta e)]\,,\label{LINK0}
\end{equation}
hence, Eq.~(\ref{FD1}),
\begin{equation}
\log P_{\beta}^{(L)}(e_2)-\log P_{\beta}^{(L)}(e_1)=
N\int_{e_1}^{e_2}\text{d}e\, \left(
\langle\hat\beta\rangle_e -\beta\right)\,,\label{LINK1}
\end{equation}
where $\log$ means natural logarithm everywhere in this work.

The relation between the canonical and the microcanonical spin-values is given
by 
\begin{equation}
\langle O \rangle^\text{canonical}_\beta=\int_{-\infty}^\infty\,\text{d}e\ 
\langle O\rangle_e\, P_{\beta}^{(L)}(e)\,.\label{LINK2}
\end{equation}
Now, Eqs.~(\ref{LINK0}) and~(\ref{LINK2}) imply that the canonical mean value will
be dominated by a saddle-point at $e^\text{SP}$,
\begin{equation}
\langle \hat\beta \rangle_{e^\text{SP}_{L,\beta}}=\beta\,,\label{LINK3}
\end{equation}
which can be read as yet another expression of the Second Law of Thermodynamics,
$T\text{d}s=\text{d}e$.

The condition of thermodynamic stability (namely that $\langle
\hat\beta \rangle_e $ be a monotonically decreasing function of $e$)
ensures that the saddle point is unique and that $e^\text{SP}$ is a maximum of
$P_\beta(e)$. Under the thermodynamic stability condition and if, in
the large $L$ limit, 
\begin{equation}
\left.\frac{{\text d} \langle \hat\beta\rangle_e}{{\text d} e}\right|_{e^\text{SP}_{L,\beta}} <0\,,\label{LINK4}
\end{equation}
the saddle point approximation becomes exact:
\begin{equation}
e^\text{SP}_{L=\infty,\beta} = \langle e\rangle_{L=\infty,\beta}^\text{canonical}\,,\label{LINK5}
\end{equation}
and we have ensemble equivalence:
\begin{equation}
\langle O\rangle_{L=\infty,e^\text{SP}_{L=\infty,\beta}}=\langle O\rangle^\text{canonical}_{L=\infty,\beta}\,.\label{LINK6}
\end{equation}

It follows that the microcanonical estimator
\begin{equation}
C_\text{m}(L,e)= \frac{1}{{\text d} \langle \hat\beta \rangle_{e,L}/{\text
    d} e}\,,\label{LINK7}
\end{equation}
evaluated at $e^\text{SP}_{L=\infty,\beta}$ will tend in the
large-$L$ limit to minus the canonical specific heat. Thus, if
the critical exponent $\alpha$ is positive, Eq.~\eqref{LINK4} will fail
precisely at $e_\text{c}$. Hence, Eq.~\eqref{LINK6} can be expected to hold
for all $e$ but $e_\text{c}$ (or for all $\beta$ but $\beta_\text{c}$).

\subsubsection{Double-peaked histogram}{\protect\label{SECT-DOS-PICOS}}

The situation can be slightly more complicated if
$P_{\beta_\text{c}}(e)$ presents two local maxima, remindful of phase
coexistence. This is actually the case for one of our models -- the
$D\!=\!2$, four-state Potts model~\cite{Fukugita}. From
Eq.~(\ref{LINK1}) it is clear that the solution to the saddle-point
equation~(\ref{LINK3}) will no longer be unique. We borrow the
following definitions from the analysis of first-order phase
transitions (where true phase coexistence takes
place)~\cite{VICTORMICRO}:
\begin{itemize}
\item The rightmost root of Eq.~(\ref{LINK3}), $e_{L,\beta}^\text{d}\,$, is a local
maximum of $P_{\beta}^{(L)}$ corresponding to the ``disordered phase''.
\item The leftmost root of Eq.~(\ref{LINK3}), $e_{L,\beta}^\text{o}\,$, is a local
maximum of $P_{\beta}^{(L)}$ corresponding to the ``ordered phase''.
\item The second rightmost root of Eq.~(\ref{LINK3}), $e_{L,\beta}^*\,$, is a local
minimum of $P_{\beta}^{(L)}$.
\end{itemize}
Maxwell construction yields the finite-system  critical
point, $\beta_{\text{c},L}$, see Fig.~\ref{beta_e_L} and Appendix~\ref{Appendix_Max}:
\begin{equation}
0=\int_{e^\text{o}_{L,{\beta_{\text{c},L}}}}^{e^\text{d}_{L,{\beta_{\text{c},L}}}}
\text{d}e\, \left(\langle\hat\beta\rangle_e -\beta_{\text{c},L}\right)\,,\label{MAXWELL}
\end{equation}
and the finite-system estimator of the ``surface tension''
\begin{equation}
\Sigma^L=\frac{N}{2L^{D-1}} \int_{e^*_{L,{\beta_{\text{c},L}}}}^{e^\text{d}_{L,{\beta_{\text{c},L}}}}
\text{d}e\, \left(
    \langle\hat\beta\rangle_e -\beta_{\text{c},L}\right)\,.\label{SIGMAEQ}
\end{equation}
Of course, in the large-$L$ limit and for a continuous transition,
$\Sigma^L\to 0$, $\beta_\text{c}^L\to \beta_\text{c}$ and
$e^\text{d}_{L,{\beta_{\text{c},L}}},e^\text{o}_{L,{\beta_{\text{c},L}}}\to
e_\text{c}\,$, as we will see.

\subsection{Our Microcanonical Finite-Size Scaling Ansatz}
\label{MFSSA-SECT}
Usually, the Microcanonical FSSA takes an entropy
density scaling form~\cite{MFSS1,MFSS2,MFSS3}. In close
analogy with the canonical case, one assumes that
$s(e,\{h_{\vec x}\},N)$ can be divided into a regular part
and a singular term $s_\text{sing}(e,\{h_{\vec x}\},N)$. The
regular part is assumed to converge for large $L$ (recall
that $N=L^D$) to a smooth function of its arguments. Hence,
all critical behaviour comes from $s_\text{sing}(e,\{h_{\vec
x}\},N)$. Note as well that we write $\{h_{\vec x}\}$,
instead of $\{h_i\}$, to emphasise the spatial dependence of
the sources (supposedly very mild~\cite{VICTORAMIT}).
Hence,
\begin{equation}
s_{\text{sing}}(e,\{h_{\vec x}\},N) = L^{-D} g \left(L^{\frac{1}{\nu_\text{m}}}(e-e_\text{c}), \{ L^{y_h}h_{\vec x}\} \right)\,.\label{FSSA-OLD}
\end{equation}
Here, $g$ is a very smooth function of its arguments, while
$y_h=1+\frac{D-\eta}{2}$ is the canonical exponent, see
e.g.~\cite{VICTORAMIT}, which does not get
Fisher-renormalised. Corrections to FSS due to irrelevant scaling
fields, have not played a major role in several previous
analysis~\cite{MFSS1,MFSS2,MFSS3} (in~\cite{MFSS3} only analytical
scaling corrections were considered), but will be important for our
precision tests.  Leading order corrections were, however, explicitly
considered in Ref.~\cite{TROSTER08}.

We will propose here alternative forms of
the ansatz~(\ref{FSSA-OLD}), more suitable for a numerical work where
neither $e_\text{c}$ nor the critical exponents are known beforehand.

Our first building block is the infinite-system microcanonical
correlation length, $\xi_{\infty,e}\,$.  Indeed, ensemble equivalence implies
that, in an infinite system, the long-distance behaviour of the
microcanonical spin-spin propagator $G(\vec r;e)=\langle\sigma_{\vec
  x} \sigma_{\vec x+\vec r}\rangle_e - \langle\sigma_{\vec x}\rangle_e
\langle \sigma_{\vec x+\vec r}\rangle_e $ behaves for large $\vec r$
as in the canonical ensemble (close to a critical point
$\xi_{\infty,e}$ is large, so that rotational invariance is recovered
in our lattice systems):
\begin{equation}
G(\vec r;e) =\frac{A}{r^{D-2+\eta}} \text{e}^{-r/\xi_{\infty,e}}\,,
\label{anom_dim}
\end{equation} 
where A is a constant.
In particular, note that ensemble-equivalence implies that the
anomalous dimension $\eta$ does not get Fisher-renormalised. We expect
$\xi_{\infty,e}=\xi^\text{canonical}_{\infty,T}$ if the correspondence
between $e$ and $T$ are fixed through
$e=\langle e \rangle^\text{canonical}_{L=\infty,T}\,.$

The basic assumption underlying the FSSA is that the approach to the
$L\to\infty$ limit is governed by the dimensionless ratio
$L/\xi_{\infty,e}$. Hence, our first form of the microcanonical FSSA for the observable $O$
whose critical behaviour was discussed in referring to Eq.~(\ref{EXPO-MICRO-CANONICO})
is
\begin{equation}
\langle O\rangle_{L,e}= L^{\frac{x_{O,\text{m}}}{\nu_\text{m}}} f_O(L/\xi_{\infty,e})+\cdots\,.\label{FSSA1}
\end{equation}
In the above, the ellipsis stands for scaling-corrections, while the
function $f_O$ is expected to be very smooth (i.e., differentiable to a
large degree or even analytical). A second form of the microcanonical FSSA is
obtained by substituting the scaling behaviour $\xi_{\infty,e}\propto
|e-e_\text{c}|^{-\nu_\text{m}}$:
\begin{equation}
\langle O\rangle_{L,e}= L^{\frac{x_{O,\text{m}}}{\nu_\text{m}}} \tilde f_O\left (L^{1/\nu_\text{m}} (e-e_\text{c})\right)+\cdots\,.\label{FSSA2}
\end{equation}
Again, $\tilde f_O$ is expected to be an extremely smooth function of
its argument~\footnote{Note that the microcanonical weight (\ref{MICROPESO}) is {\em not} analytical at each energy level of the spin Hamiltonian.}. In particular, this is the form of the ansatz that
follows from Eq.~(\ref{FSSA-OLD}) by differentiating with respect to $e$ or
from the source terms.

However, the most useful form of the microcanonical FSSA is obtained by applying
Eq.~(\ref{FSSA1}) to the finite-lattice correlation length $\xi_{L,e}$,
obtained in a standard way (see Ref.~\cite{VICTORAMIT}) from the
finite-lattice microcanonical propagator.  We expect $\xi_{L,e}/L$ to
be a smooth, one-to-one function of $L/\xi_{\infty,e}$, that can be
inverted to yield $L/\xi_{\infty,e}$ as a function of
$\xi_{L,e}/L$. Hence, our preferred form of the FSSA is
\begin{equation}
\langle O\rangle_{L,e}= L^{\frac{x_{O,\text{m}}}{\nu_\text{m}}}\left[F_O\left(\frac{\xi_{L,e}}{L}\right)+L^{-\omega} G_O\left(\frac{\xi_{L,e}}{L}\right)+\cdots\right]\label{FSSA3}\,.
\end{equation}
Here, $F_O$ and $G_O$ are smooth functions of their arguments and
$\omega$ is the first universal scaling corrections exponent.

It is important to note that exponent $\omega$ does not get
Fisher-renormalised. Indeed,  let us
consider an observable $O$ with critical exponent $x_O$ at a
temperature $T$ such $e=\langle
e\rangle_{L=\infty,T}^\text{canonical}$. Now, ensemble equivalence
tells us that $O_{L=\infty,T}^\text{canonical}=O_{L=\infty,e}$ and
that $\xi_{L=\infty,T}^\text{canonical}=\xi_{L=\infty,e}$. Eliminating
$T$ in favour of $\xi_{L=\infty,T}^\text{canonical}$, see e.g.~\cite{VICTORAMIT},
we have
\begin{equation}
O_{L=\infty,T}^\text{canonical} = \xi_{L=\infty,e}^{x_O/\nu}[A_0 + B_0\xi_{L=\infty,e}^{-\omega}+\cdots]\,, 
\end{equation}
where $A_0$ and $B_0$ are scaling amplitudes. It follows that
$\omega_\text{m}=\omega$, and that
$x_O/\nu=x_{O,\text{m}}/\nu_\text{m}$.

\subsubsection{The Quotient Method}\label{subsec:quotients}

Once we have Eq.~(\ref{FSSA3}), it is straightforward to
generalise the quotient method~\cite{QUOTIENTS}. In
Appendix~\ref{Appendix_Quotient} we also describe how it should be modified in the
presence of (multiplicative) logarithmic corrections to scaling.

Let us compare data obtained {\em at the same} value of $e$ for a pair of
lattices $L_1=L$ and $L_2=sL$ with $s>1$. We expect that a single
$e_{\text{c},L_1,L_2}$ exists such that the correlation-length in units of the
lattice size coincides for both systems:
\begin{equation}
\frac{\xi_{L,e_{\text{c},L_1,L_2}}}{L}=\frac{\xi_{sL,e_{\text{c},L_1,L_2}}}{sL}\,.\label{QUO1}
\end{equation}
Hence, if we compare now in the two lattices the observable $O$ in~(\ref{FSSA3}), precisely at $e_{\text{c},L,sL}$, we have
\begin{equation}
\frac{\langle O\rangle_{sL,e_{\text{c},L_1,L_2}}}{\langle O\rangle_{L,e_{\text{c},L_1,L_2}}}=
s^{\frac{x_{O,\text{m}}}{\nu_\text{m}}}\left[1 + A_{O,s}L^{-\omega} +\cdots\right],\label{QUO2}
\end{equation}
where $A_{O,s}$ is a non-universal scaling amplitude. One
considers this equation for fixed $s$ (typically $s=2$), and uses it to
extrapolate to $L=\infty$ the $L$-dependent estimate of the critical
exponents ratio $x_{O,\text{m}}/\nu_\text{m}\,$. At the purely
numerical level, it needs to be noted as well that there are strong statistical
correlations between the quotients in \eqref{QUO1} and in
\eqref{QUO2}, that reduces the statistical errors in the estimate of 
critical exponents. These errors can be computed 
via a jack-knife method, see e.g.~\cite{VICTORAMIT}.

In this chapter, we shall compute the critical exponents from the following operators
($\chi$ is the susceptibility, while $\xi$ is the correlation length, see Sec.~\ref{FSSmicro:themodel}
for definitions):
\begin{eqnarray}
\chi & \rightarrow & x_O= \nu_\text{m} (2-\eta)\,,\\
\partial_e \xi & \rightarrow & x_O= \nu_\text{m} + 1\,.
\end{eqnarray}

The $L$ dependence of $e_{\text{c},L,s}$ follows from Eq.~(\ref{FSSA2}) as applied to $\xi_L/L$ for the two lattice sizes $L$ and $sL$~\cite{BINDER,VICTORAMIT}:
\begin{equation}
e_{\text{c},L,s}=e_\text{c}+ B \frac{1-s^{-\omega}}{s^{1/\nu_\text{m}}-1}L^{-(\omega+\frac{1}{\nu_\text{m}})}+\cdots\,,\label{SHIFT-ec-LEADING}
\end{equation}
where $B$ is again a non-universal scaling amplitude. In particular, if
one works at fixed $s$, $e_{\text{c},L,sL}$ tends to $e_\text{c}$ for
large $L$ as $L^{-(\omega+\frac{1}{\nu_\text{m}})}$~\footnote{Note
  that, Eq.~(\ref{FSSA2}) tells us that, if the energy histogram is
  double-peaked, see Sec.~\ref{SECT-DOS-PICOS}, the histogram maxima
  will tend to $e_\text{c}$ only as $L^{-1/\nu_\text{m}}\,$.}. 

\section{The Model}
\label{FSSmicro:themodel}

We will define here the model and observables of a generic $D$-dimensional
$Q$-state Potts model. The numerical study was done for two instances of
this model: the three-dimensional Ising ($Q\!=\!2$) model, and the
two-dimensional $Q\!=\!4$ Potts model.

We place the spins $\sigma_i=1,\ldots,Q$ at the nodes of a hypercubic
$D$-dimensional lattice with linear size $L$ and periodic boundary
conditions.

The Hamiltonian is
\begin{equation}
{\cal U}=-\sum_{\langle i,j \rangle}
\delta_{\sigma_i\sigma_j}\ ,\label{SPIN-HAMILTONIAN}
\end{equation}
where $\langle i,j \rangle$ denotes first nearest neighbours and $\delta_{ij}$ is the Kronecker delta. 
For a given spin, $\sigma$, we define the normalised $Q$-vector $\vec s$, 
whose $q$-th component is
\begin{equation}
s_q=\sqrt{\frac{Q}{Q-1}}\Big(\delta_{\sigma q}-\frac1Q\Big)\;.
\end{equation}
A $Q$ components order parameter for the ferromagnetic transition is
\begin{equation}
\vec{\cal M}=\frac{1}{L^D}\sum_{i}\vec s_i\;, 
\end{equation}
where $i$ runs over all the lattice sites. We will now consider microcanonical averages.
The spatial correlation function is
\begin{equation}
\begin{array}{rcl}
C({\bm r}'-{\bm r})&=&
\Big\langle \vec s({\bm r})\cdot\vec s({\bm r}')\Big\rangle_{\!e}\\
&=&\displaystyle\frac{Q}{Q-1}\Big\langle \delta_{\sigma({\bm
    r})\sigma({\bm r}')} - \frac1Q\Big\rangle_e \;.
\end{array}
\end{equation}
Our definition for the correlation length at a given internal energy density
$e$, is computed from the Fourier transform of $C$ 
\begin{equation}
\hat C({\bm k})=
\sum_{\bm r} C({\bm r})\,{\text e}^{{\text i}{\bm k}\cdot{\bm r}}\;,\label{hatC-def}
\end{equation}
at zero and minimal ($\Vert {\bm k}_\text{min}\Vert=2\pi/L$) momentum~\cite{COOPER,VICTORAMIT}:
\begin{equation}
\xi(e,L)=\frac{\sqrt{\hat C(0)/\hat C({\bm
      k}_{\text{min}})-1}}{2\sin(\pi/L)} \;.
\label{xi-def}
\end{equation}
Note that  $\hat C$ can be easily computed
in terms of the Fourier transform of the spin field, $\hat s({\bm k})$, as
\begin{equation}
\hat C({\bm k})=L^D \big\langle \hat s({\bm k})\cdot\hat s({-\bm k})\big\rangle_e\;,
\end{equation}
and that the microcanonical magnetic susceptibility is
\begin{equation}
\chi=L^D {\langle \vec{\cal M }^2 \rangle_e}=\hat C(0)\,.\label{chi-micro-def}
\end{equation}

For the specific case of the Ising model, the traditional definitions, using
$S_i=\pm1$ (recall that $s_i=\pm1/\sqrt{2}$), are related with those of the general model through:
\begin{eqnarray}
{\cal U}^\text{Ising}&=&-\sum_{<i,j>}S_i S_j=2{\cal U}-3L^D\;,\nonumber\\
\beta^\text{Ising}&=&\beta/2\;,\label{conversions}\\
\chi^\text{Ising}&=&2\chi\;.\nonumber
\end{eqnarray}

Notice that for $D=2$ this model undergoes a phase
transition at $\beta_c=\log(1+\sqrt{Q})$ which is of second
order for $Q\le4$ and first order for $Q>4$~\cite{Wu}.

\section{Numerical Results}
\label{FSSmicro:results}

\subsection{Methods}
\label{FSSmicro:methodology}

We have simulated systems of several sizes in a suitable range of energies
(see Table~\ref{SIMU}).
To update the spins we used a Swendsen-Wang (SW) version of the microcanonical cluster
method~\cite{VICTORMICRO}. This algorithm depends on a tunable
parameter, $\kappa$, which should be as close as possible to
$\langle \hat\beta \rangle_e$ in order to maximise the acceptance of the
SW attempt (SWA). This requires a start-up using a much slower Metropolis
algorithm for the determination of $\kappa$. In practice, we performed cycles 
consisting of $2\times10^3$ Metropolis steps, a $\kappa$ refresh,
$2\times10^3$ SWA, and a further $\kappa$ refresh.  We require an acceptance
exceeding $60\%$ to finish these pre-thermalization cycles fixing $\kappa$ for
the following main simulation, where only the cluster method is used.

In both cases studied, we observed a very small
autocorrelation time for all energy values at every lattice
size. In the largest lattice for the four-state Potts model
we also considered different starting configurations: hot,
cold, and mixed (strips). Although the autocorrelation time
is much smaller, for safety we decided to discard the first
10\% of the Monte Carlo history, using the last 90\% for
taking measurements.

\begin{table}[!ht]
\begin{center}
\begin{tabular}{|c|r|c|r|c|}\hline
Model 
&\multicolumn{1}{r|}{$L$} 
&\multicolumn{1}{r|}{$N_\text{m}(\times 10^6)$}
&\multicolumn{1}{c|}{$N_\text{e}$} 
&\multicolumn{1}{c|}{Energy range}\\
\hline\hline
$Q=2$, $D=3 $&8    & 20 & 42 &  $[-0.8, -0.9]$ \\
&12   & 20 & 42 &  $[-0.8, -0.9]$ \\
&16   & 20 & 49 &  $[-0.8, -0.9]$ \\
&24   & 20 & 25 &  $[-0.845, -0.875]$ \\
&32   & 20 & 16 &  $[-0.87, -0.860625]$ \\
&48   & 20 & 10 &  $[-0.87, -0.860625]$ \\
&64   & 5  & 10 &  $[-0.870625, -0.865]$ \\
&96   & 5  & 10 &  $[-0.870625, -0.865]$ \\
&128  & 5  & 7  &  $[-0.869375, -0.865625]$ \\
\hline\hline
$Q=4$, $D=2$ &32   & 1024 & 61 &  $[-1.2, -0.9]$ \\
&64   & 128  & 61 &  $[-1.2, -0.9]$ \\
&128  & 32   & 41 &  $[-1.08, -0.98]$ \\
&256  & 32   & 24 &  $[-1.08, -1.005]$ \\
&512  & 25.6 & 32 &  $[-1.07, -1.01]$ \\
&1024 & 6.4  & 30 &  $[-1.06, -1.02]$ \\
\hline
\end{tabular}
\caption{Simulation details for the two models considered. For each lattice
  size $L$ we show the number of measurements $N_\text{m}$ at each energy
  and the total number of simulated energies uniformly distributed over the
  displayed energy range $N_\text{e}$. For the $Q\!=\!4$, $D\!=\!2$ model, the values of
  $N_\text{m}$ reported were reached only at specific energies
  near the peaks of the Maxwell construction, where additional energy values were simulated.
}
\label{SIMU}
\end{center}
\end{table}

\subsection[$D=3$ Pure Ising Model]{$\bm{D=3}$ Pure Ising Model}
\label{FSSmicro:ising3D}

\begin{figure}[!ht]
\begin{center}
\includegraphics[height=0.75\columnwidth,angle=270,trim=43 50 25 25]{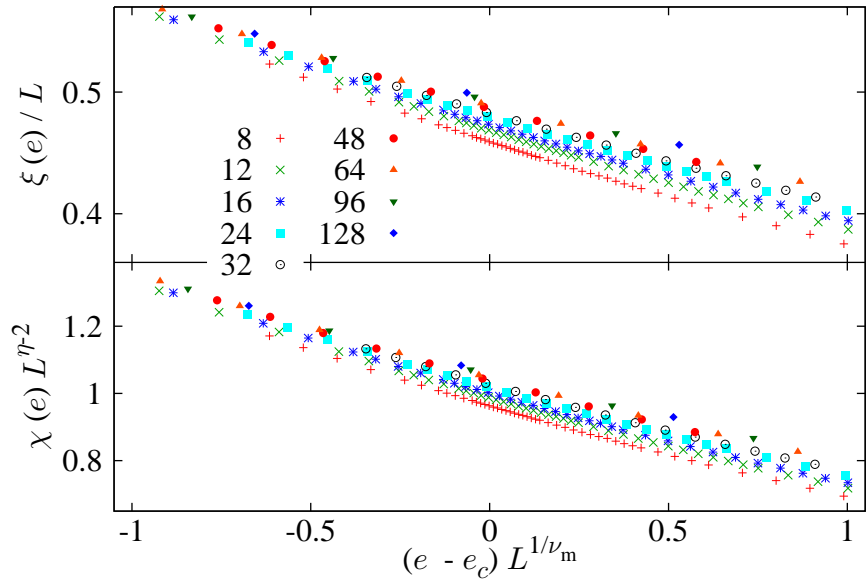}
\caption{Scaling plot of the correlation length (in
  lattice size units) and the scaled susceptibility for the
  three-dimensional Ising model. We used the critical values,
  $e_\text{c}=-0.867433$ and $\nu_\text{m}=0.7077$. Notice the strong
  scaling corrections for small systems as well as the data
  collapse for the largest ones.  }
\label{scaling_ising}
\end{center}
\end{figure}

In Fig.~\ref{scaling_ising} (upper panel) we show a scaling plot of the correlation length
(in lattice size units) against $(e-e_\text{c}) L^{1/\nu_\text{m}}$. For the
susceptibility we plot $\chi\sim L^{2-\eta}$ (lower panel). If the data
followed the expected asymptotic critical behaviour with microcanonical
critical exponents they should collapse into a single curve.
In Fig.~\ref{scaling_ising} we have used the canonical critical quantities
from Refs.~\cite{HASEN, Ising3D} transformed to the microcanonical
counterparts using Eq.~(\ref{EXPO-MICRO-CANONICO}).
From the plot it is clear that important scaling corrections exist in both
cases for the smallest lattices, although they are mainly eliminated in the largest systems.

\begin{figure}[ht!]
\begin{center}
\includegraphics[height=0.75\columnwidth,angle=270,trim=28 75 14 25]{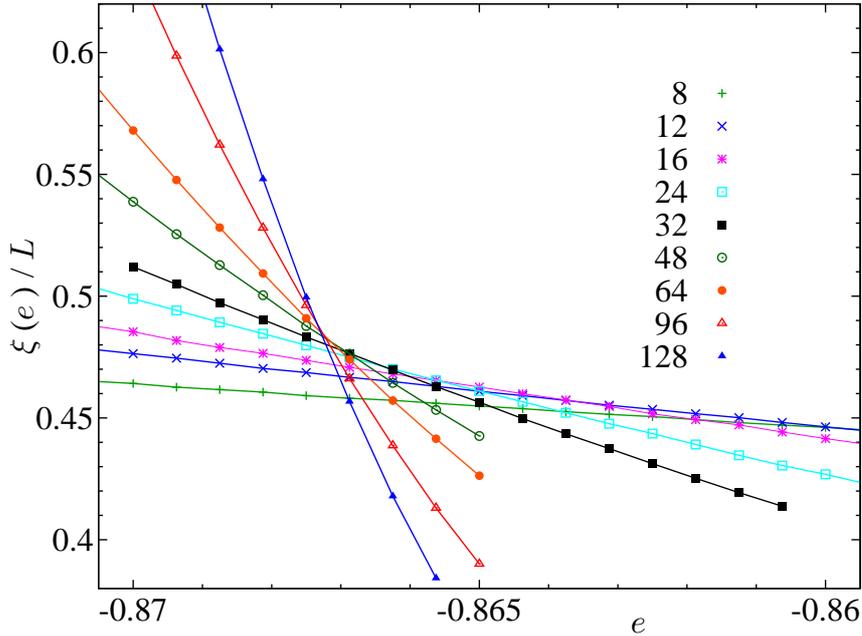}
\caption{Crossing points of the correlation length in lattice size units
for the three-dimensional Ising model. The error bars are in every case smaller than the point sizes.
The values of the different quantities at the crossing as well as the critical exponents are
given in Table~\ref{table_exp_ising}.} 
\label{fig_xi_ising}
\end{center}
\end{figure}

To obtain the microcanonical critical exponents we used the
quotient method, see Sec.~\ref{subsec:quotients}.  The
clear crossing points of the correlation length for
different lattice sizes can be seen in
Fig.~\ref{fig_xi_ising}.  The determination of the different
quantities at the crossings, and the position of the
crossing itself, requires one to interpolate the data
between consecutive simulated energies.  We found that the method of
choice, given the high number of energy values available, is
to fit, using the least squares method, a selected number of
points near the crossing to a polynomial of appropriate
degree. Straight lines do not provide good enough
fits. However, second and third order polynomials give
compatible results. In practice, we fitted a second-order
polynomial using the nine points nearest to the crossing,
also comparing the results with those using the seven
nearest points that turn out to be fully compatible. For
error determination we always used a jack-knife procedure, see Appendix~\ref{Appendix_dataanalysis}.

\renewcommand{\arraystretch}{1.1}
\begin{table}[!ht]
\begin{center}
\begin{tabular}{|r|l|l|l|l|}
\hline
$L$ 
& \multicolumn{1}{c|}{$e_{\text{c},L,2L}$}
& \multicolumn{1}{c|}{$\xi_{L,e_{\text{c},L,2L}}/L$} 
& \multicolumn{1}{c|}{$\nu_\text{m}$} 
& \multicolumn{1}{c|}{$\eta_\text{m}$} \\
\hline\hline
8   & $-0.861831(12)$ &    0.44922(3)    &   0.8033(42)  &  0.0564(2)  \\
12  & $-0.865010(10)$ &    0.46106(5)    &   0.7968(31)  &  0.0492(4)  \\
16  & $-0.866020(6)$  &    0.46710(5)    &   0.7717(22)  &  0.0469(4)  \\
24  & $-0.866767(3)$  &    0.47411(4)    &   0.7665(11)  &  0.0437(3)  \\
32  & $-0.867034(4)$  &    0.47813(6)    &   0.7594(13)  &  0.0425(5)  \\
48  & $-0.867228(2)$  &    0.48278(5)    &   0.7492(5)   &  0.0412(3)  \\
64  & $-0.867302(2)$  &    0.48555(11)   &   0.7457(16)  &  0.0397(8)  \\\cline{1-5}
\hline
\end{tabular}
\caption{Lattice size dependent estimates of critical quantities for
  the microcanonical $D=3$ Ising model. The displayed quantities are:
  crossing points $e_{\text{c},L,2L}$ for the correlation length in units of the lattice
  size, $\xi/L$ itself at those crossing points, and the estimates of the
  correlation length exponent $\nu_\text{m}$ and the anomalous
  dimension $\eta_\text{m}$. All quantities were obtained
  using parabolic interpolations.}
\label{table_exp_ising}
\end{center}
\end{table}
\renewcommand{\arraystretch}{1}

The numerical estimates for $e_\text{c}$, $\xi_{L,e_\text{c}}/L$ and
the critical exponents $\nu_\text{m}$ and $\eta_\text{m}$, obtained using the
quotient method for lattice pairs $(L,2L)$ are given in
Table~\ref{table_exp_ising}. Our small statistical errors allow one to
detect a tiny $L$ evolution. An extrapolation to infinite volume is
clearly needed.

Before continuing, let us recall our expectations as
obtained by applying Fisher renormalization to the most
accurate determination of {\em canonical} critical exponents
known to us [$\nu_\text{m}=\nu/(1-\alpha)=\nu/(D\nu-1)$]:
\begin{eqnarray}
\nu_\text{m}&=& 0.7077(5)\ \text{(from $\nu=0.6301(4)$~\cite{PELI-REP})}\,,\label{NU-FETEN}\\
\eta_\text{m}&=&\eta\ =0.03639(15)\text{~\cite{CAMPOSTRINI}}\,,\label{ETA-FETEN}\\
\omega&=&0.84(4)\text{~\cite{PELI-REP}}\,.\label{OMEGA-FETEN}
\end{eqnarray}
Besides, although non-universal, let us take $e_\text{c}=-0.867433(12)$
\footnote{For the $3D$ Ising model at criticality,
  $u_\text{c}^\text{Ising}\!=\!-0.990627(24)$~\cite{HASEN},
  and $\beta_\text{c}^\text{Ising}\!=\!0.2216546(2)$~\cite{Ising3D},
  we obtain for our Potts representation of the Ising model
  $e_\text{c}=(u_\text{c}^\text{Ising}-D)/2+1/(4\beta_\text{c}^\text{Ising})$.
}.

The results obtained from an extrapolation using only leading order
scaling corrections were:
\begin{itemize}
\item $e_\text{c}=-0.867397(6)$, $\omega+1/\nu_\text{m}=1.918(26)$\\ (we
obtained a good fit for $L\geq L_\text{min}=12$, 
with $\chi^2/\text{d.o.f.}=0.39/3$,
C.L.=94\%, where ``d.o.f.'' stands for \emph{degrees of freedom} and ``C.L.'' for
{\em confidence level}~\footnote{The confidence level is the probability that 
$\chi^2$ would be larger than the
observed value, supposing that the statistical model is correct. As
a rule, we consider a fit not good-enough whenever C.L.$<10$\%.}).
\item $\xi_{e_\text{c},L}/L=0.5003(12)$, $\omega=0.581(27)$\\ ($L_\text{min}=12$, $\chi^2/\text{d.o.f.}=0.12/3$, C.L.=99\%).
\item $\nu_\text{m}=0.714(28)$, $\omega=0.53(30)$\\ ($L_\text{min}=8$, $\chi^2/\text{d.o.f.}=3.16/4$, C.L.=53\%).
\item $\eta_\text{m}=0.0391(15)$, $\omega=1.21(24)$\\ ($L_\text{min}=8$, $\chi^2/\text{d.o.f.}=0.96/4$, C.L.=92\%).
\end{itemize}
The main conclusions that we draw from these fits are: (i) the
exponents are compatible with our expectations from
Fisher renormalization, (ii) sub-leading scaling corrections are
important given the tendency of the fits to produce a too low estimate
for $\omega$ (see below), and (iii) the estimates from canonical
exponents (themselves obtained by applying the high-temperature expansion
to improved Hamiltonians~\cite{PELI-REP,CAMPOSTRINI}) are more accurate
than our direct computation in the microcanonical ensemble.

We can, instead, take an opposite point of view.  If we take the
central values in
Eqs.~(\ref{NU-FETEN}, \ref{ETA-FETEN}, \ref{OMEGA-FETEN}) as if they
were exact, we can obtain quite detailed information on the amplitudes
for scaling corrections:
\begin{itemize}
\item We find an excellent fit to $\nu_\text{m}(L,2L)=\nu_\text{m} +
  A_1 L^{-\omega} + A_2 L^{-2\omega}$, for
  $L_\text{min}=16$: $\chi^2/\text{d.o.f.}=1.53/3$, C.L.=68\%,
  with $A_1=1.38(7)$ and $A_2=-7.6(1.1)$. This
  confirms our suspected strong sub-leading corrections. Indeed,
  according to these amplitudes $A_1$ and $A_2$, only for $L\approx
  130$ the contribution of the (sub-leading) quadratic term becomes
  $10\%$ of that of the leading one.
\item In the case of $\eta_\text{m}(L,2L)=\eta_\text{m} + B_1 L^{-\omega}+ B_2
  L^{-2\omega}\,$, for $L_\text{min}=8$: $\chi^2/\text{d.o.f.}=2.4/5$, C.L.=79\%,
  we have $B_1=0.101(10)$ and $B_2=0.07(7)$. Sub-leading scaling corrections
  are so small that, within our errors, it is not clear whether or not $B_2=0$.
\end{itemize}

The quite strong scaling corrections found for $\nu_\text{m}$ may cast
some doubt on the extrapolation for $\xi_{L,e_\text{c}}/L$, the only
quantity that we cannot double check with a canonical computation. To control
this, we proceed to a fit including terms linear and quadratic in
$L^{-\omega}$ with $\omega=0.84(4)$. We get
$$\frac{\xi_{L,e_\text{c}}}{L}=0.4952(5)(7),$$
with $L_\text{min}=12$, $\chi^2/\text{d.o.f.}=2.17/3$, C.L.=54\%.
Here, the second error is due to the quite small uncertainty in
$\omega$. It is remarkable that the contribution to the error stemming
from the error in $\omega$ is {\em larger} than the purely statistical
one.

\subsubsection{The canonical specific-heat}
\label{SUBSECT-ESPECIFICO-ISING}

Previous numerical studies of microcanonical
FSS~\cite{MFSS1,MFSS2,MFSS3} focussed on the specific heat. Although
we show all across this paper that a complete microcanonical FSS
analysis can be based only on the spin propagator, the specific heat
can be certainly studied within the present formalism.

As discussed in Sect.~\ref{SUBSECT-ENSEMBLE-EQUIVALENCE} (see
also~\cite{VICTORMICRO}), the canonical specific heat can be
estimated from the microcanonical estimator $C_\text{m}(L,e)$ defined in
Eq.~(\ref{LINK7}). The expected FSS behaviour for $C_\text{m}(L,e_{\text{c},L,2L})$
is
\begin{equation}
C_\text{m}(L,e_{\text{c},L,2L})= L^{\alpha/\nu}[A_0+ A_1 L^{-\omega}+\cdots] + B\,.\label{Cme-SCALING}
\end{equation}
Here, $A_0$ and $A_1$ are scaling amplitudes, while $B$ is a constant
background usually termed {\em analytical correction to scaling},
stemming from the non-singular part of the
free-energy~\cite{VICTORAMIT}. It is usually disregarded as it plays
the role of a subleading scaling-correction term. Yet, a peculiarity
of the $D=3$ Ising model is that $B$ is anomalously large (see
e.g.~\cite{MFSS3}) and needs to be considered.

\begin{figure}[ht!]
\begin{center}
\includegraphics[height=0.75\columnwidth,angle=270,trim=28 75 14 25]{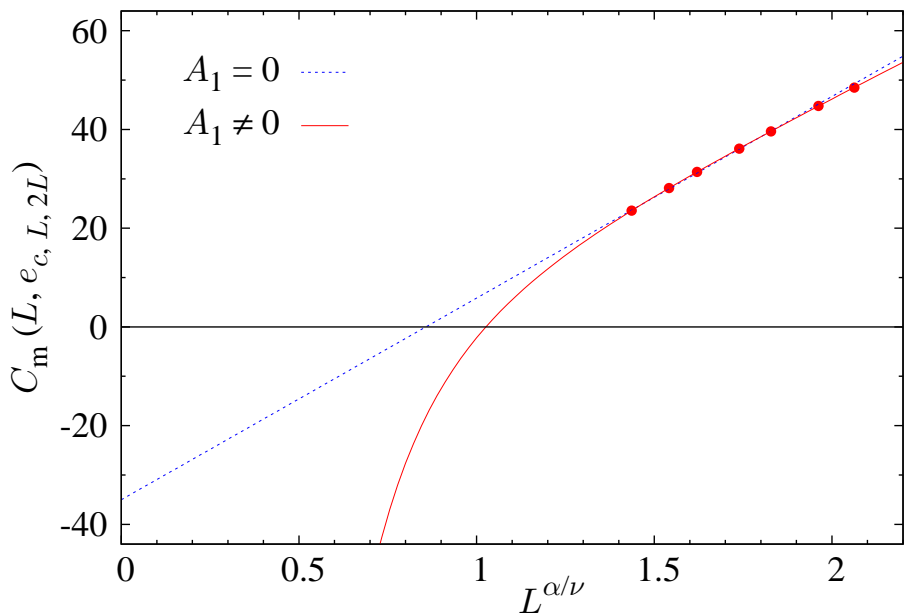}
\caption{Microcanonical estimate of the specific heat,
  $C_\text{m}(L,e)$, at $e_{\text{c},L,2L}$ for the $D=3$ Ising model,
  as a function of the system size.  The numerical estimates of
  exponents $\alpha/\nu$ and $\omega$ were taken from
  Ref.~\cite{PELI-REP}. The error bars are in every case smaller than
  the point sizes. The solid line is a fit to Eq.~(\ref{Cme-SCALING})
  (fitting parameters: $A_0,A_1$ and $B$), the dashed one is obtained
  by constraining the fit to $A_1=0$.}
\label{fig_cesp_ising}
\end{center}
\end{figure}

In Fig.~\ref{fig_cesp_ising}, we reproduce the analysis of Bruce and
Wilding~\cite{MFSS3}, where the amplitude $A_1$ in
Eq.~(\ref{Cme-SCALING}) was fixed to zero by hand. In this way, if we
consider the range of lattice sizes $8\le L \le 64$ (in~\cite{MFSS3}
only $L\le 32$ was considered), we obtain $B=-35.01(11)$ but with an
untenable $\chi^2/\text{d.o.f}=227/5$. Our value of $B$ is,
nevertheless, quite close to the result $B=-34.4(4)$ reported
in~\cite{MFSS3} (unfortunately, these authors provided no information
on fit-quality).

Once the arbitrary constraint $A_1=0$ is removed, we do obtain an
acceptable fit, $\chi^2/\text{d.o.f}=0.68/4$. Perhaps unsurprisingly,
the estimate of $B$ is largely changed, once a nonvanishing $A_1$ is
allowed: $B=-24.4(7)$.

\subsection[$D=2$\,, $Q=4$ Pure Potts Model]{$\bm{D=2\,,\ Q=4}$ Pure Potts Model}
\label{FSSmicro:potts2Dmicro}

The $Q=4$, $D=2$ Potts model involves two peculiarities that will be
explored here.  First, it suffers from quite strong logarithmic
scaling corrections. And second, it displays pseudo-metastability~\cite{Fukugita}, an ideal playground for a microcanonical
study.

The study of the FSS for the $Q=4$, $D=2$ Potts
model~\cite{Salas}, based on the analysis of the Renormalization Group (RG)
equations~\cite{Cardy}, reveals the presence of multiplicative logarithmic scaling
corrections. This is one of the possible forms that scaling
corrections can take in the limit $\omega\to 0$, and is a major
nuisance for numerical studies. A very detailed theoretical input is
mandatory to safely perform the data analysis. We shall make here an
educated guess for the {\em microcanonical} form of the scaling corrections,
based purely on ensemble equivalence and the {\em canonical} results.

From ensemble equivalence we expect
\begin{equation}	
e -e_\text{c} \sim C(L,\beta_\text{c}) \Delta \beta_L \ ,
\end{equation}
where $C(L,\beta_\text{c})$ is the finite-lattice canonical specific
heat at $\beta_\text{c}$, and $\Delta \beta=
\beta_\text{c}^{(L)}-\beta_\text{c}$ is the inverse-temperature
distance to the critical point of any $L$-dependent feature (such as
the temperature maximum of the specific heat, etc.). We borrow from
Ref.~\cite{Salas} the leading FSS behaviour for these quantities:
\begin{equation}	
C(L,\beta_\text{c}) \sim \frac{L}{(\log L)^{3/2}} \quad , \quad \Delta \beta_L \sim \frac{(\log L)^{3/4}}{L^{3/2}}\,. 
\end{equation}
Thus, we have:
\begin{equation}	
e(L) -e_c(\infty) \sim L^{-1/2} (\log L)^{-3/4} \ .
\label{potts_energy_form}
\end{equation}
This result can be derived as well by considering only the leading
terms of the first derivative of the singular part of free energy
with respect to the thermal field, $\phi\ (\propto
\beta-\beta_\mathrm{c})$~\cite{Salas}:
\begin{equation}
\frac{\partial f_\text{sing}(\phi,h,\psi)}{\partial \phi} \approx 
\frac{4}{3}D_\pm |\phi|^{1/3} (-\log |\phi|)^{-1}+
D_\pm |\phi|^{4/3} (-\log |\phi|)^{-2} \frac{1}{\phi} \,.
\end{equation}
The above equation describes the energy of the
system, and its leading term is
\begin{equation}	
e - e_\text{c} \sim \frac{4}{3}D_\pm \frac{|\phi|^{1/3}}{\log |\phi|} \ ,
\end{equation}
but 
\begin{equation}	
\phi \approx C'_\pm L^{-3/2} (\log L)^{3/4} \ ,
\end{equation}
so it is direct to obtain again Eq.~(\ref{potts_energy_form}).  Hence, 
we are compelled to recast Eq.~(\ref{FSSA2}) as
\begin{equation}
\langle O\rangle_{L,e}= L^{\frac{x_{O,\text{m}}}{\nu_\text{m}}} \tilde f_O\left (L^{1/2} (\log L)^{3/4} (e-e_\text{c})\right)+\cdots\,.\label{FSSA-POTTS}
\end{equation}
Furthermore, from the canonical analysis~\cite{Salas}, we expect
multiplicative logarithmic corrections to the susceptibility (that
do not get Fisher renormalised). 
Furthermore, the ellipsis in~(\ref{FSSA-POTTS}) stands for 
corrections of order $\log \log L / \log L $ and $1/\log L$~\cite{Salas}.

We first address in the next subsection the direct
verification of Eq.~(\ref{FSSA-POTTS}) using the quotient method. We
then consider the pseudo-metastability features.

\subsubsection{Scaling plots and critical exponents}
\label{subsec:exponents}

\begin{figure}[!ht]
\begin{center}
\includegraphics[height=0.75\columnwidth,angle=270,trim=68 58 26 25]{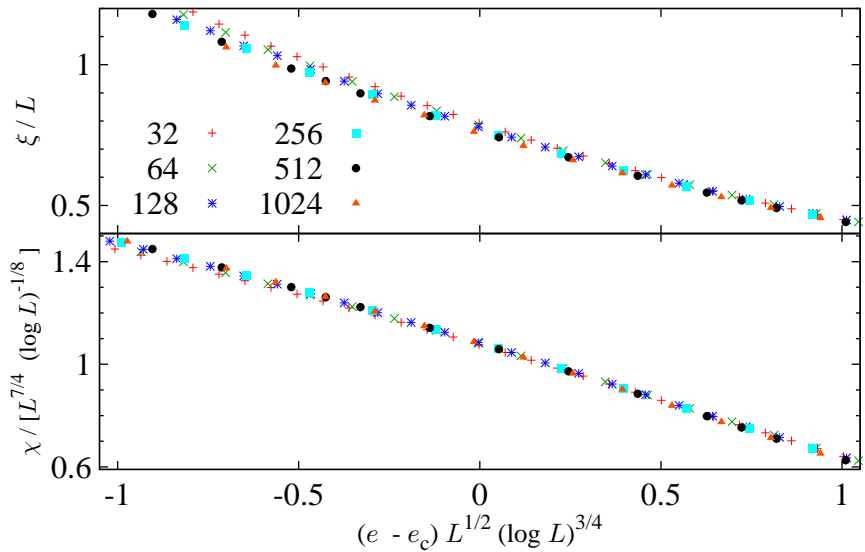}
\caption{Graphical demonstration of
  Eq.~(\ref{FSSA-POTTS}) as applied to the microcanonical $D=2$, $Q=4$
  Potts model: both the correlation length in units of the lattice
  size ({\bf top}) and the scaled susceptibility, $\overline{\chi}$ in
  Eq.~(\ref{inv_chi}) ({\bf bottom}), are functions of the scaling
  variable $(e-e_\text{c})L^{1/2}(\log L)^{3/4}$.}
\label{scaling_xiandsus_pottsmicro}
\end{center}
\end{figure}

We start with a graphical demonstration of
Eq.~(\ref{FSSA-POTTS}): $\xi/L$ as a function of
$(e-e_c)L^{1/2}(\log L)^{3/4}$ should collapse onto a single
curve (the deviation will be larger for small $L$ values due
to neglected scaling corrections of order $\log \log L /
\log L $ and $1/\log L$)~\footnote{ We obtain the exact
$e_\text{c}$ in the thermodynamic limit from
$\beta_\text{c}=\log (1+\sqrt{Q})$~\cite{Baxter}, and
$u_\text{c}=-(1+Q^{-1/2})$~\cite{Wu} by applying
$e_\text{c}=u_\text{c}+1/(2\beta_\text{c})$.}.  Similar
behaviour is expected for the scaled
susceptibility~\cite{Salas}:
\begin{equation}
\overline{\chi}=\frac{\chi}{L^{7/4}(\log L)^{-1/8}} \,.
\label{inv_chi}
\end{equation}
Note that $\xi/L$ does not need an additional logarithmic factor.
These expectations are confirmed in
Fig.~\ref{scaling_xiandsus_pottsmicro}, especially for the larger
system sizes (that are subject to smaller scaling corrections).

\begin{figure}[!ht]
\begin{center}
\includegraphics[height=0.75\columnwidth,angle=270,trim=13 58 6 25]{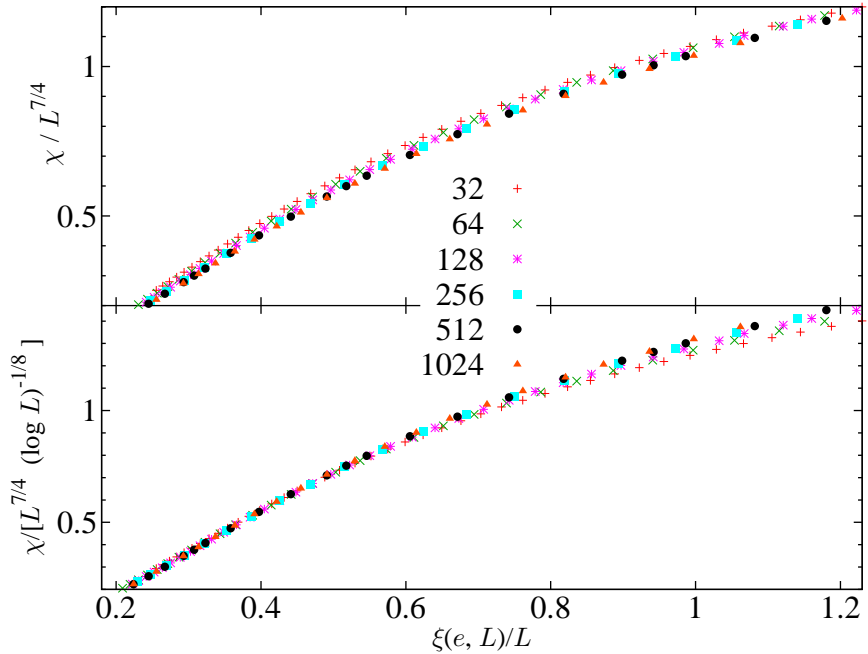}
\caption{Comparison of the scaling for the naively
  scaled susceptibility $\chi L^{-7/4}$ ({\bf top}) and for
  $\overline{\chi}$, defined in Eq.~(\ref{inv_chi}), ({\bf bottom}), as a function of the correlation
  length in units of the lattice size, for the microcanonical $D=2$,
  $Q=4$ Potts model.  }
\label{scaling_susVSxi_pottsmicro}
\end{center}
\end{figure}

We can check directly the importance of the multiplicative logarithmic
corrections for the susceptibility by comparing $\chi$ and $\overline{\chi}$
as a function of $\xi/L$, see Fig.~\ref{scaling_susVSxi_pottsmicro}. The improved scaling of $\overline{\chi}$ is apparent. We observe as well  
that the largest corrections to scaling are found at and
below the critical point (around $\xi/L \approx 1.0$).

\begin{figure}[!ht]
\begin{center}
\includegraphics[height=0.75\columnwidth,angle=270,trim=13 58 6 25]{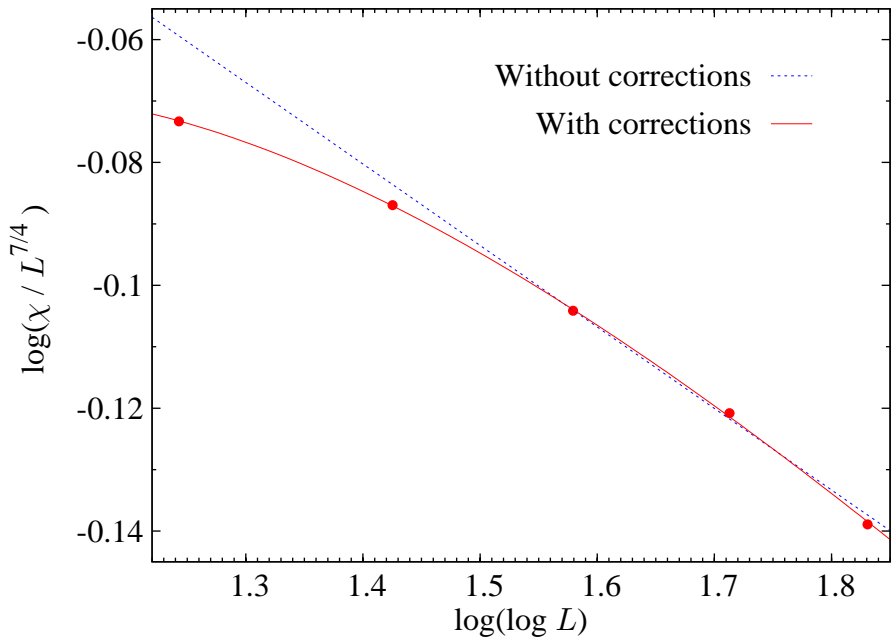}
\caption{Logarithmic scaling behaviour of the susceptibility at the
critical point. The error bars are in every case smaller than the point sizes.
The dashed line do not include the subleading additive terms of Eq.~(\ref{suscept_whole})
while the solid line do.}
\end{center}
\label{susceps_log_pottsmicro2D}
\end{figure}

The scaling proposed for the susceptibility in Ref.~\cite{Salas} can also be checked from our values
at $e_{\text{c},L,2L}$. Considering $\chi\sim L^{7/4}$ (our data is fully supportive of this point)
we can plot $\log (\chi/L^{7/4})$ versus $\log\log L$.
We obtain a linear fit for the data with $L>64$ with a slope $-0.132(3)$ ($\chi^2/\text{n.d.f.}=7.5/1$),
see the dashed line in Fig.~\ref{susceps_log_pottsmicro2D}, which can be compared with
the expected value $-1/8$~\cite{Salas}. The large value of $\chi^2/\text{d.o.f.}$ can be ascribed to the
presence of higher order correction terms. In fact the whole scaling behaviour
for the susceptibility is~\cite{Salas}
\begin{equation}
\chi \sim L^{7/4}(\log L)^{-1/8}\left(1+A\frac{\log\log L}{\log L}+B\frac{1}{\log L}+\cdots \right)\,,
\label{suscept_whole}
\end{equation}
and we can use this form for a least-square fit. Fixing both the leading and the logarithmic exponents we estimate
$A=0.80(7)$ and $B=-0.48(3)$ using all the lattice sizes with $\chi^2/\text{d.o.f.}=2.9/2$, see
the solid line in Fig.~\ref{susceps_log_pottsmicro2D}.
Therefore our data set is fully supportive of the behaviour proposed in Ref.~\cite{Salas},
including the subleading additive logarithmic corrections.

We now proceed to the numerical computation of critical exponents. We
shall use the quotient method, modified as described in
Appendix~\ref{Appendix_Quotient}.  From Fig.~\ref{fig_cortes},
we can see that the crossing points can be well obtained 
using parabolic interpolations of the nine points around the
estimated crossing energies, as done in Sec.~\ref{FSSmicro:ising3D}.  We
checked that the results do not depend on the interpolating polynomial
degree by comparing with interpolations using cubic curves. We also
compared with the results obtained using only seven points around the
crossing obtaining again full agreement.

The critical exponents  obtained are listed in Table~\ref{table_exp}. They
may be compared with the exact ones~\cite{Wu} ($\nu=2/3$, $\alpha=2/3$ and
$\eta=1/4$):
\begin{equation}
\nu_\text{m}=2 \quad \quad ;\quad \quad \eta=\eta_\text{m}=\frac{1}{4}\, .
\end{equation}
Comparing with our computed exponents, we obtain an
acceptable agreement.  In the case of the microcanonical
$\nu$ exponent, $\nu_\text{m}$, after adding the correction
for the quotient method in the presence of logarithms, the
agreement is fairly good. We can see a clear trend towards
the exact value for all the lattice sizes except the largest
(2.5 standard deviations away), which is probably due to a
bad estimate of the huge temperature derivatives of the
correlation length.  In the case of the microcanonical
$\eta$ exponent, $\eta_\text{m}$, which must be the same
as the canonical one, we can see clearly the tendency to
the analytical value $\eta_\text{m}=0.25$.  We must stress
the importance of adding the corrections described in
Appendix~\ref{Appendix_Quotient} to the quotient method.

\begin{figure}[!ht]
\begin{center}
\includegraphics[height=0.75\columnwidth,angle=270,trim=28 67 13 25]{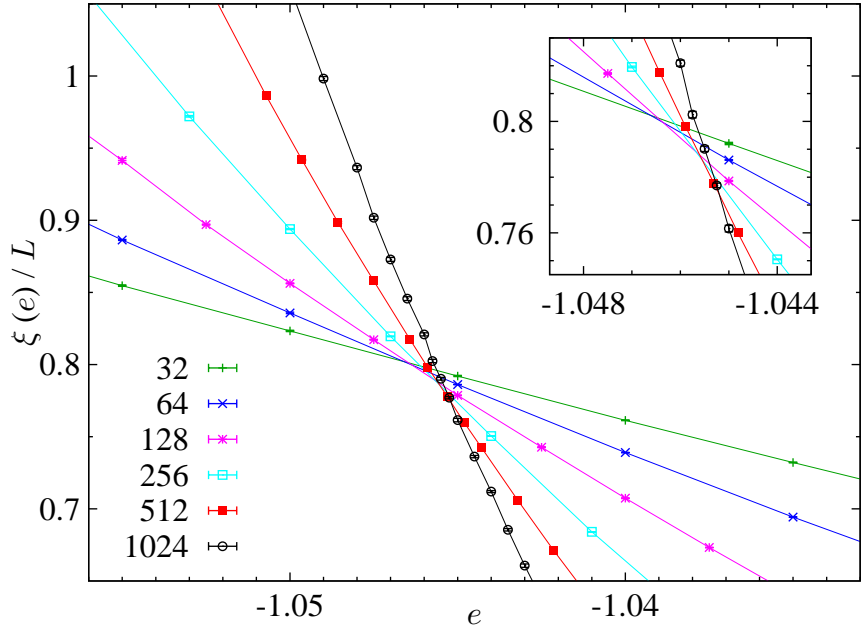}
\caption{Correlation length in lattice size units for
  the $D=2$, $Q=4$ Potts model.  The values of the different
  quantities at the crossings for lattices $L$ and $2L$, as well as
  the corresponding estimates for critical exponents, are given in
  Table~\ref{table_exp}. The inset is a magnification of the critical
  region.  }
\label{fig_cortes}
\end{center}
\end{figure}

\renewcommand{\arraystretch}{1.1}
\begin{table}[t]
\begin{center}
\begin{tabular}{|r|l|l|l|l|l|l|}
\hline
\multicolumn{1}{|c|}{$L$} 
&\multicolumn{1}{c|}{$e_{\text{c},L,2L}$} 
&\multicolumn{1}{c|}{$\xi_{L,e_{\text{c},L,2L}}/L$} 
&\multicolumn{1}{c|}{$\nu_\text{m}$} 
&\multicolumn{1}{c|}{$\nu_\text{m}'$} 
&\multicolumn{1}{c|}{$\eta_\text{m}$} 
&\multicolumn{1}{c|}{$\eta_\text{m}'$} \\
\hline\hline
32  & $-1.04659(5)$ &  0.8016(5)   &  1.534(6)  &    1.998(10)  &    0.2663(9)   &   0.2334(9)    \\
64  & $-1.04633(2)$ &  0.7990(3)   &  1.554(8)  &    1.957(12)  &    0.2638(6)   &   0.2360(6)    \\
128 & $-1.04579(1)$ &  0.7909(3)   &  1.578(5)  &    1.938(7)   &    0.2639(5)   &   0.2398(5)    \\
256 & $-1.04548(2)$ &  0.7836(5)   &  1.643(12) &    1.987(17)  &   0.2615(11)   &   0.2402(11)   \\
512 & $-1.04519(2)$ &  0.7734(9)   &  1.602(31) &    1.895(42)  &   0.2617(21)   &   0.2427(21)	\\
\hline
\end{tabular}
\caption{Crossing points of the correlation length in lattice size units as a function of the energy
for pairs of lattices \hbox{($L$, $2L$)}.  Using the
original quotient method~\cite{VICTORAMIT} we obtain the
microcanonical critical exponents, listed in Columns 4
and 6, while the corrected ones (Columns 5 and 7) are
labelled with primed symbols, see
Appendix~\ref{Appendix_Quotient}.}
\label{table_exp}
\end{center}
\end{table}
\renewcommand{\arraystretch}{1}

\subsubsection{Critical point, latent heat, and surface tension}
\label{subsec:temperature}

\begin{figure}[!ht]
\begin{center}
\includegraphics[height=0.75\columnwidth,angle=270,trim=28 76 14 25]{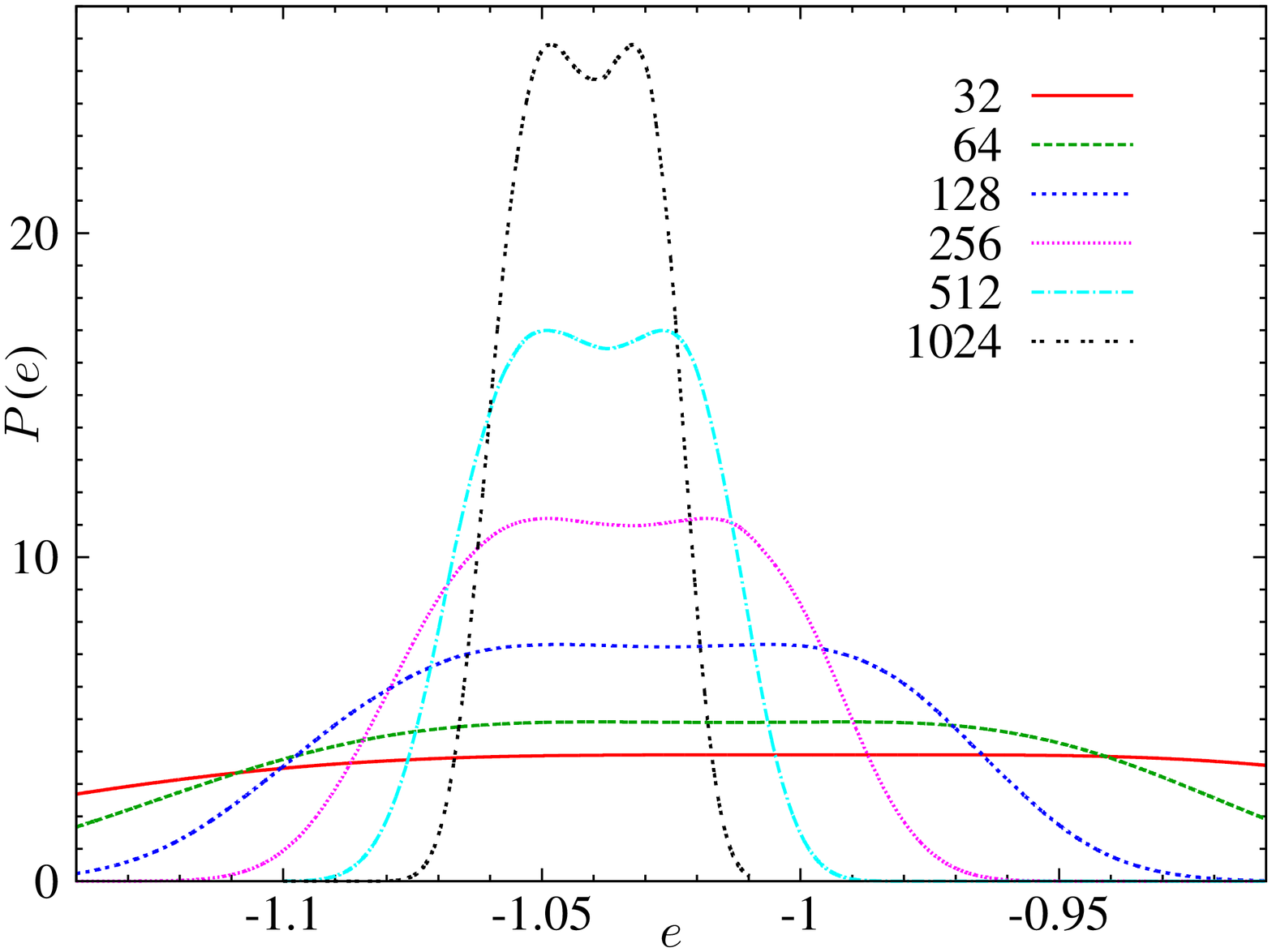}
\caption{{\em Canonical} probability distribution
  function for the energy density, $P_{\beta}^{(L)}(e)$, as
  reconstructed from microcanonical simulations of the $D=2$, $Q=4$
  Potts model for different system sizes. The $L$-dependent critical
  point $\beta_{\text{c},L}$ is computed using the Maxwell rule,
  Sec.~\ref{SECT-DOS-PICOS} (note the equal height of the two peaks
  enforced by Maxwell construction). The system displays an apparent
  latent heat that becomes smaller with increasing $L$, and vanishes in the
  large $L$ limit.
} 
\label{HISTOGRAMAS}
\end{center}
\end{figure}

It has been known for quite a long time that the $D=2$,
$Q=4$ Potts model on finite lattices shows features typical
of first-order phase transitions~\cite{Fukugita}. For
instance, see Fig.~\ref{HISTOGRAMAS}, the probability
distribution function for the internal energy, $P_\beta(e)$,
displays two peaks at energies $e_\text{d}$ (the coexisting
{\em disordered} phase) and $e_\text{o}$ (the energy of the
{\em ordered} phase) separated by a minimum at $e^*$. Of
course, since the transition is of second order,
$e_\text{c}$ is the common large $L$ limit of $e_\text{d}$,
$e_\text{o}$ and $e^*$.

\begin{figure}[!ht]
\begin{center}
\includegraphics[height=0.75\columnwidth,angle=270,trim=23 100 0 25]{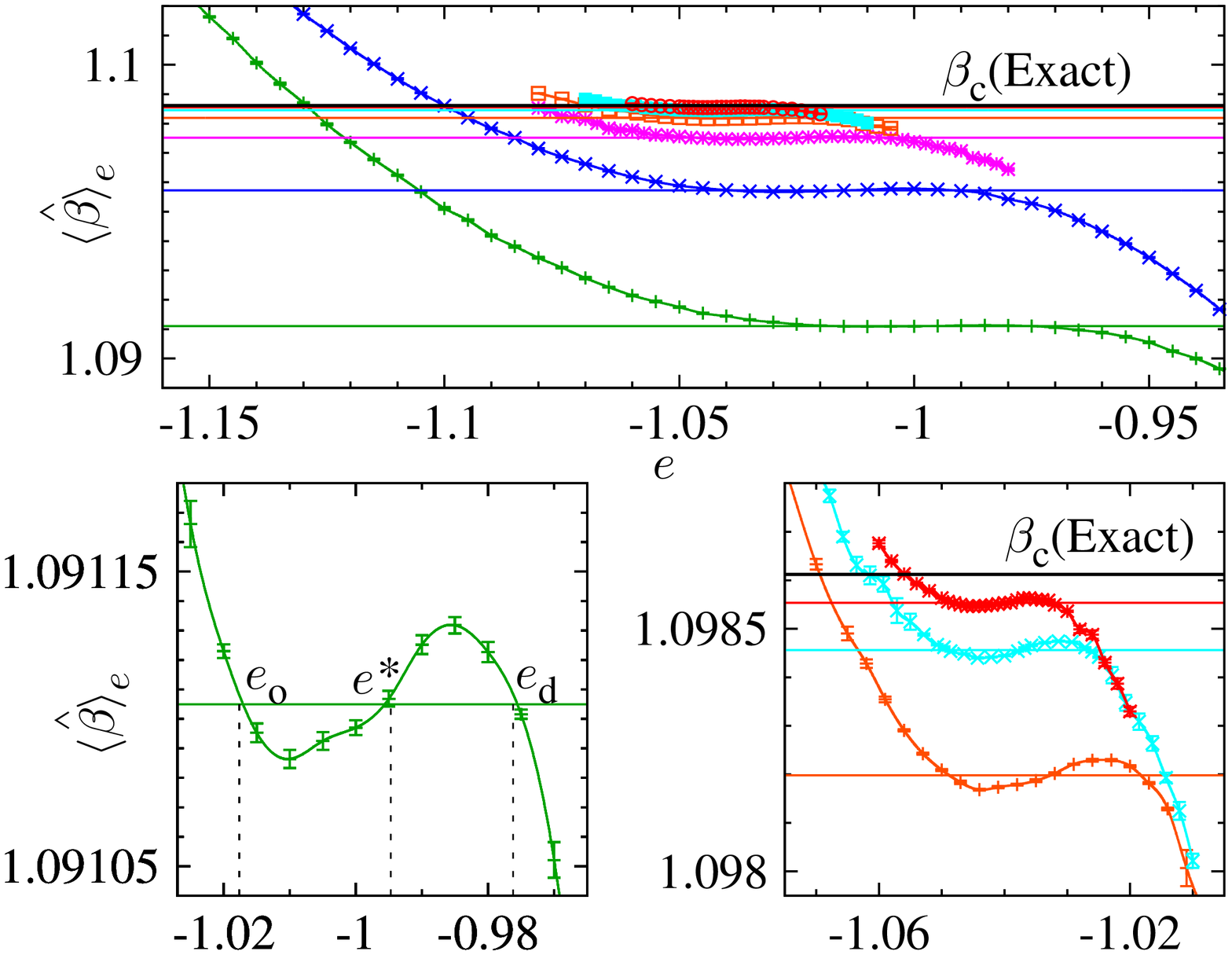}
\caption{{\bf Top:} From the microcanonical mean values
  $\langle \hat\beta\rangle_{e,L}$ for the $D=2$, $Q=4$ Potts model, we
  estimate the size dependent {\em canonical} inverse critical
  temperature $\beta_{\text{c},L}$ (horizontal lines) for all the
  simulated lattice sizes, ranging from $L=32$ (lower) to $L=1024$
  (upper). We show as well the analytical prediction (uppermost
  horizontal line).  {\bf Bottom-left:} Example of Maxwell construction
  for our $L=32$ data. The $e$-integral of $\langle
  \hat\beta\rangle_{e,L}-\beta_{\text{c},L}$ from $e_\text{o}$ to
$e_\text{d}$ vanishes. {\bf Bottom-right}: Zoom of upper panel showing
only data for lattice sizes $L=256$ (lower curve), $L=512$ (middle curve), and
$L=1024$ (upper curve). }
\label{beta_e_L}
\end{center}
\end{figure}

We discussed in Sec.~\ref{SECT-DOS-PICOS} how the Maxwell construction is
used to estimate the canonical critical point $\beta_{\text{c},L}$, as
well as $e_\text{d}$, $e_\text{o}$ and the associated surface tension.
This procedure is outlined in Fig.~\ref{beta_e_L}. The numerical results
are listed in Table~\ref{numerical_beta}, where we can see that 
$\beta_{\text{c},L}$ is a monotonically increasing
function of $L$ continuously approaching the analytical value
$\beta_\text{c}= \log(1+\sqrt{Q})=1.0986122\ldots$~\cite{Baxter}. A jack-knife
method~\cite{VICTORAMIT} was used to compute the error bars for all the quantities
in Table~\ref{numerical_beta}.

\begin{table}[!ht]
\begin{center}
\begin{tabular}{|r|l|l|l|l|}
\hline
\multicolumn{1}{|c|}{$L$} 
& \multicolumn{1}{c|}{$\beta_{\text{c},L}$} 
& \multicolumn{1}{c|}{$e_\text{o}$} 
& \multicolumn{1}{c|}{$e_\text{d}$} 
& \multicolumn{1}{c|}{$\varSigma \times 10^5$} \\\hline\hline
32    & 1.0911070(20) &  -1.0175(4) & -0.9760(2)     &  0.47(2) \\
64    & 1.0957256(14) &  -1.0392(3) & -0.9915(2)     &  2.77(7) \\
128   & 1.0975150(10) &  -1.0463(3) & -1.0062(5)     &  4.10(15)\\
256   & 1.0981989(5)  &  -1.0489(2) & -1.0183(3)     &  3.92(8) \\
512   & 1.0984570(3)  &  -1.0490(1) & -1.0266(2)     &  3.28(11)\\
1024  & 1.0985539(3)  &  -1.0483(3) & -1.0325(1)     &  2.09(17)\\
\hline
\end{tabular}
\caption{Using the Maxwell construction, we compute for the $D=2$, $Q=4$ Potts
model the $L$-dependent estimates of the (inverse) critical temperature
$\beta_{\text{c},L}$, the energies of the coexisting ordered $e_\text{o}$,
 and disordered $e_\text{d}$ phases, as well as the surface tension $\varSigma$.}
\label{numerical_beta}
\end{center}
\end{table}

To perform a first check of our data, we observe that
$\beta_{\text{c},L}$ is a typical {\em canonical} estimator of the inverse
critical temperature. As such, it is subject to standard canonical FSS,
where the main scaling corrections come from two additive logarithmic
terms~\cite{Salas}:
\begin{equation}
\beta_{\text{c},L}-\beta_\text{c}=a_1 \frac{(\log L)^{3/4}}{L^{3/2}} 
\left(1+a_2\frac{\log\log L}{\log L}+a_3 \frac{1}{\log L} \right) \;.
\end{equation}
From our data in Table~\ref{numerical_beta}, we obtain
$a_1=-0.44(7)$, $a_2=-1.15(72)$, and $a_3=2.28(26)$, 
and a good fit ($L_\text{min}=128$, $\chi^2/\text{d.o.f.}=0.28/1$, C.L.=60\%). 

\begin{figure}[!ht]
\begin{center}
\includegraphics[height=0.75\columnwidth,angle=270,trim=28 83 22 25]{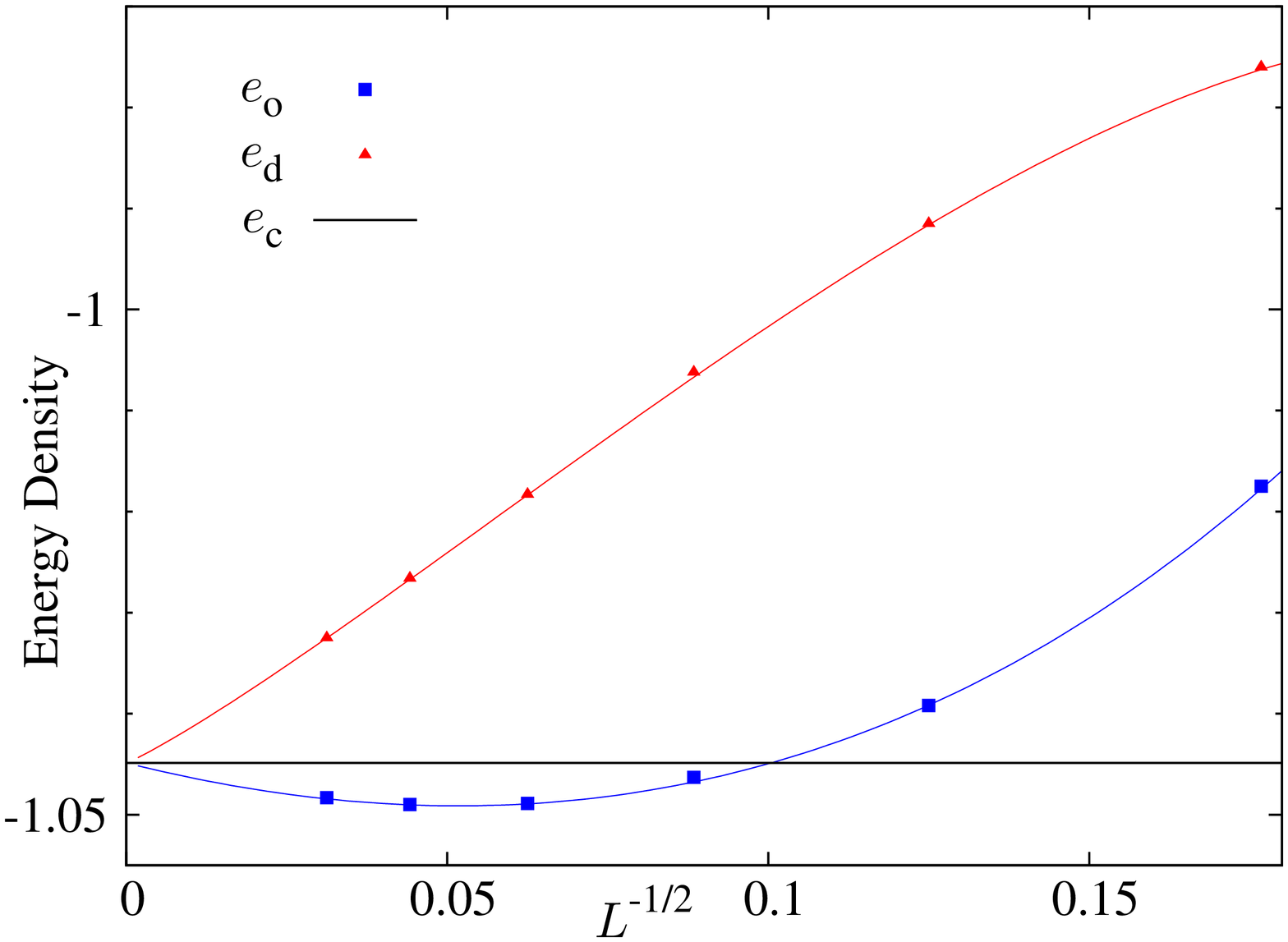}
\caption{System size dependent estimates of the
  energies of the ``coexisting'' ordered ($e_\text{o}$, blue squares)
  and disordered ($e_\text{d}$, red circles) phases of the $D=2$, $Q=4$
  Potts model, as a function of $L^{-1/2}$. The lines are fits to the
  expected analytical behaviour Eq.~(\ref{scaling_corr}). The
  horizontal line corresponds to the asymptotic value, $e_\text{c}$.  }
\label{extrap_ener_sokal}
\end{center}
\end{figure}

As for the $L$ dependence of $e_\text{d}$ and $e_\text{o}$, we try a
fit that considers the expected scaling correction terms~\cite{Salas}: 
\begin{equation}
e_{\text{c,o},L} -e_c = a_1 L^{-1/2} (\log L)^{-3/4} 
\left(1+  a_2 \frac{\log \log L}{\log L}+{a_3}\frac{1}{\log L}\right) .\label{scaling_corr}
\end{equation}
Our results for $e_\text{o}$ are: $a_\text{1o}=-2.03(20)$, $a_\text{2o}=-1.65(27)$,
and $a_\text{3o}=-2.08(41)$, with a fair fit quality
($L_\text{min}\!=\!32$, $\chi^2/\text{d.o.f.}=2/3$, C.L.=57\%).
We
obtain for $e_\text{d}$: $a_\text{1d}=2.02(14)$,
$a_\text{2d}=0.93(37)$, and $a_\text{3d}=-2.93(34)$ , with a fair fit
as well
($L_\text{min}\!=\!32$, $\chi^2/\text{d.o.f.}=0.84/3$, C.L.=84\%). 
These two fits are shown in  Fig.~\ref{extrap_ener_sokal}.

For the surface tension, one notes in
Table~\ref{numerical_beta} a non-monotonic
behaviour. Furthermore, we lack any theoretical input with
which to attempt a fit. We thus turn to a variant of the
quotient method. Were $\varSigma$ to follow pure power-law
scaling, $\varSigma\propto L^b$, the exponent $b$ would be
obtained as:
\begin{equation}
\frac{\varSigma(L_1)}{\varSigma(L_2)}= \left( \frac{L_1}{L_2} \right) ^b
\Longrightarrow
b= \frac{\log [\varSigma(L_1)/\varSigma(L_2)] }{\log (L_1/L_2)  }\,.
\label{eff_exp}
\end{equation}
The effective exponent $b$ obtained from our data is given
in Table~\ref{table_eff_exp}. We observe that it is clearly
negative (as it should be since $\varSigma$ vanishes for a
second-order phase transition). An asymptotic estimate,
however, seems to require the simulation of larger systems.
\begin{table}[!ht]
\begin{center}
\begin{tabular}{|c|l|}
\hline
$(L_1,L_2)$ 
& \multicolumn{1}{c|}{$b_\text{eff}(\varSigma)$}\\
\hline\hline
(32,64)    &  $\hphantom{-}2.56(7)$ \\
(64,128)   &  $\hphantom{-1}0.56(6)$ \\
(128,256)  &  $-0.065(60)$ \\
(256,512)  &  $-0.257(57)$ \\
(512,1024) &  $-0.650(127)$ \\
\hline
\end{tabular}
\caption{Effective exponent obtained using Eq.~(\ref{eff_exp}) for the  surface tension.}
\label{table_eff_exp}
\end{center}
\end{table}

\begin{table}[!ht]
\begin{center}
\begin{tabular}{|r|l|l|l|l|l|l|}
\hline
\multicolumn{1}{|c|}{$L$} 
& \multicolumn{1}{c|}{$\xi(e_\text{o})/L$} 
& \multicolumn{1}{c|}{$\xi(e_\text{d})/L$} 
& \multicolumn{1}{c|}{\raisebox{2pt}{$\overline{\chi}(e_\text{o})$}} 
& \multicolumn{1}{c|}{\raisebox{2pt}{$\overline{\chi}(e_\text{d})$}}
& \multicolumn{1}{c|}{$\xi^\text{canonical}/L$} 
& \multicolumn{1}{c|}{\raisebox{2pt}{$\overline{\chi}^\text{canonical}$}} \\\hline\hline
32    & 0.637(2)  & 0.453(1)  & 0.907(2) & 0.647(1)  & 0.990(3)   & 1.287(3)  \\
64    & 0.732(3)  & 0.396(1)  & 1.025(3) & 0.545(2)  & 0.995(2)   & 1.310(2)  \\
128   & 0.799(5)  & 0.357(4)  & 1.106(5) & 0.472(7)  & 1.001(3)   & 1.331(3)  \\
256   & 0.866(6)  & 0.335(3)  & 1.182(6) & 0.429(5)  & 1.001(5)   & 1.343(5)  \\
512   & 0.915(4)  & 0.315(2)  & 1.238(4) & 0.392(4)  & 1.014(8)   & 1.366(8)  \\
1024  & 0.953(15) & 0.302(2)  & 1.279(13)& 0.367(3)) & 0.997(21)  & 1.353(22) \\
\hline
\end{tabular}
\caption{Correlation length in units of the lattice size and the
RG invariant {$\overline{\chi}$} defined in
Eq.~(\ref{inv_chi}), for several $L$ values, as computed in
the microcanonical $D=2$, $Q=4$ Potts model. The values of
the energy density correspond to the ordered ($e_\text{o}$)
and disordered ($e_\text{d}$) phases. For comparison we also
display in the last two columns the {\em canonical} results at $\beta_\text{c}$
obtained in Ref.~\cite{Salas}.}
\label{numerical_corrlenght}
\end{center}
\end{table}

We have just seen that, up to scaling corrections, $e_\text{d}^{(L)}$
and $e_\text{o}^{(L)}$ correspond to (different) $L$-independent
values of the argument of the scaling function $\tilde f_\xi$ in
Eq.~(\ref{FSSA-POTTS}). Hence we expect that $\xi(e_\text{d})/L$ and
$\xi(e_\text{o})/L$, see Table~\ref{numerical_corrlenght}, approach
non-vanishing, different values in the large $L$ limit. The
FSS corrections are expected to be additive
logarithms~\cite{Salas}
\begin{equation}
\frac{\xi}{L} = a+\frac{b}{\log L}\,.
\end{equation}
The results are:
\begin{equation}
\frac{\xi(e_\text{o})}{L}=1.28(1)-\frac{2.28(5)}{\log L}\,,
\end{equation}
($L_\text{min}=32$, $\chi^2/\text{d.o.f.}=4.2/3$, C.L.=22\%), and
\begin{equation}
\frac{\xi(e_\text{d})}{L}=0.159(4)-\frac{0.98(2)}{\log L}\,,
\end{equation}
($L_\text{min}=32$, $\chi^2/\text{d.o.f.}=3.3/3$, C.L.=37\%).

A very similar analysis can be performed for the scaled
susceptibility, Eq.~(\ref{chi-micro-def}), at $e_\text{d}$ and
$e_\text{o}$.  In order to deal with the multiplicative logarithms of
the susceptibility, we used $\overline{\chi}$ defined in
Eq.~(\ref{inv_chi}).

Fitting our data set to the logarithmic form
\begin{equation}
\overline{\chi}=A+B \frac{\log \log L}{\log L}\ ,
\end{equation}
 obtained in Ref.~\cite{Salas}, we obtain a good fit in the ordered phase energy, $e_\text{o}$:
\begin{equation}
\overline{\chi}(e_\text{o})=2.41(5)-4.00(15) \frac{\log \log L}{\log L}\,,
\end{equation}
with $L_\text{min}= 128$, $\chi^2/\text{d.o.f.}=3.10/2$,
C.L.=21\%. However, the extrapolation for the
susceptibility defined in the disordered phase energy,
$e_\text{d}$, is a nonsensical negative value.

We can also fit the data to the logarithmic form also used in Ref.~\cite{Salas}:
\begin{equation}
\overline{\chi}=A+\frac{B}{\log L}\,,
\end{equation}
finding:
\begin{equation}
\overline{\chi}(e_\text{o})=1.643(5)-\frac{2.55(2)}{\log L}\,,
\end{equation}
($L_\text{min}=32$, $\chi^2/\text{d.o.f}=7.44/4$, C.L.=11\%),  and
\begin{equation}
\overline{\chi}(e_\text{d})=0.094(7)+\frac{1.87(37)}{\log L}\,,
\end{equation}
($L_\text{min}=64$, $\chi^2/\text{d.o.f}=2.94/3$, C.L.=37\%)\,.
For comparison, we recall that
Ref.~\cite{Salas} reports two different fits for $\overline{\chi}$, depending
on the logarithmic corrections they used:
\begin{eqnarray}
\overline{\chi}^{\text {canonical}}&=&1.673(33) -1.056(98) \frac{\log \log
    L}{\log L}\,,\\
\overline{\chi}^{\text{canonical}}&=&1.454(13) -\frac{0.600(55)}{\log L} \,.
\end{eqnarray}

\section{Conclusions}
\label{FSSmicro:conclusions}
\setcounter{equation}{0}

We have formulated the Finite Size Scaling Ansatz (FSSA) for microcanonical
systems in terms of quantities accessible in a finite lattice. This
form allows to extend the phenomenological renormalization approach
(the so-called quotient method) to the microcanonical framework.

Our FSSA was subjected to strong numerical testing. We
performed extensive microcanonical numerical simulations in two
archetypal systems in Statistical Mechanics: the three-dimensional
Ising model and the two-dimensional four-state Potts model. 
The two models present a power-law singularity in their
canonical specific heat, implying  non-trivial Fisher
renormalization when passing to the microcanonical ensemble.
A microcanonical cluster method works for both models, hence allowing us to
study very large system sizes ($L=128$ in $D=3$ and $L=1024$ in
$D=2$). 

In the case of the Ising model, we obtained precise
determinations of the critical exponents that provide
strong evidence for our extended microcanonical FSS ansatz.

For the Potts model, very strong logarithmic corrections (both
multiplicative and additive) plague our data. Fortunately, we have a
relatively strong command over these corrections from canonical
studies~\cite{Salas}. Our data can be fully rationalised using the scaling 
corrections suggested by the theoretical analysis~\cite{Salas}.

\clearpage{\thispagestyle{empty}\cleardoublepage}

\clearpage{\thispagestyle{empty}\cleardoublepage}

\normalfont

\vspace{4cm}

\chapter{Quenched Disorder Effect on a First-Order Phase Transition}
\label{chap:potts3D}


\section{Introduction}
\label{potts3D:intro}

Although first-order phase transitions are by far the more frequent in nature, not much is known
about the consequences of adding impurities to systems that in the pure case undergo 
this type of transition. This is due to the fact that there exist inherent difficulties
for their study.

One of the intrinsic problems in simulating first-order
phase transitions is that in this case two or more phases
coexist at the critical temperature. The system changes from
the high temperature phase to the low temperature one by
building an interface of size $L$, where $L$ is the lattice
size.  The energy cost of such a mixed configuration is
$\Sigma L^{D-1}$ (with $\Sigma$ being the surface tension and $D$
the spatial dimension). Therefore, when doing simulations
using the canonical ensemble (at fixed temperatures), the
probability of reaching such mixed configurations is attenuated
by a factor $\mathrm{exp}[-\Sigma L^{D-1}]$, and as a result
the natural time scale of the simulation grows with the
system size $L$ as $\mathrm{exp}[\Sigma L^{D-1}]$. This
huge obstacle to simulating large systems is called 
{\em Exponential Critical Slowing Down} (ECSD).

Up to now, no solution for ECSD has been found in canonical
simulations. This has motivated the popularity of
simulations within the microcanonical ensemble (at fixed
energy), see Sec.~\ref{FSSmicro:MFSSA}.  Some simulation
methods within this ensemble consider the canonical
probability density function (pdf) of the energy as a
constant within the energy interval $e^\mathrm{o}< e <
e^\mathrm{d}$ ($e^\mathrm{o}$ and $e^\mathrm{d}$ being the
energy densities of the coexisting ordered and disordered
phases respectively). This led to these methods being called
\emph{flat-histogram
methods}~\cite{MULTICANONICAL,WANGLANDAU,DEPABLO,MASFLAT}.
The canonical probability minimum in the energy gap
($\propto\mathrm{exp}[-\Sigma L^{D-1}]$) is achieved by
means of an iterative parameter optimisation. In
flat-histogram methods the system performs an energy random
walk in the energy gap.  The elementary step being of order
$L^{-D}$ (a single spin-flip), one naively expects a
tunnelling time from $e^\mathrm{o}$ to $e^\mathrm{d}$ of
order $L^{2D}$ spin-flips.  But the (one-dimensional) energy
random walk is not Markovian, and these methods still suffer
ECSD~\cite{NEUSHAGER}. In fact, for the standard benchmark
(the $Q\!=\!10$ Potts model~\cite{Wu} in $D\!=\!2$), the
barrier of $10^4$ spins was reached in
1992~\cite{MULTICANONICAL}, while the largest simulated
system (to the best of our knowledge) had $4\times 10^4$
spins~\cite{WANGLANDAU}.

ECSD in flat-histogram simulations is probably
understood~\cite{NEUSHAGER}: on its way from $e^\mathrm{d}$ to
$e^\mathrm{o}$, the system undergoes several (four in $D\!=\!2$)
``transitions''. First comes the condensation
transition~\cite{NEUSHAGER,CONDENSATION}, at a distance of order
$L^{-D/(D+1)}$ from $e^\mathrm{d}$, where a macroscopic droplet of the
ordered phase is nucleated. Decreasing $e$, the droplet grows to the
point that, for periodic boundary conditions, it reduces its surface
energy by becoming a strip~\cite{DROPLETSTRIP}, see the figures in~\cite{VICTORMICRO}
(in $D\!=\!3$, the droplet becomes a cylinder, then a slab~\cite{LUIS2}).
At lower $e$ the strip becomes a droplet of {\em
disordered} phase.  Finally, at the condensation transition close to
$e^\mathrm{o}\,,$ we encounter the homogeneous ordered phase.

In this work we will study a prototypical model of a strong first-order phase transition,
the three-dimensional Potts model with $Q>3$ states. There are numerous experimental
systems which can be mapped by this model. For instance, the $Q=4$ pure case in two dimensions describes the 
adsorption of N$_2$ molecules on Kr in graphite layers~\cite{ABSOR_DOMANY}; 
in three dimensions it describes the behaviour of FCC antiferromagnetic lattices (NdSb, NdAs, and CeAs,
for example) with the magnetic field pointing in the $\langle1, 1, 1\rangle$ direction~\cite{FCC_DOMANY}.
The site-diluted $Q=4$ case in two dimensions models the effect of oxygen impurities on a sample of nickel where 
hydrogen molecules are adsorbed~\cite{POTTS2D_EXP}. In the dilute
three-dimensional case we are not aware of any experimental realization.

It is known~\cite{VICTORMICRO,Chat05} that the pure
three-dimensional Potts model undergoes a first-order phase
transition in the pure case for $Q\ge3$. On the contrary, it
has been been found~\cite{Ball00} that for strong dilution
the system performs a second-order phase transition. A
direct question is the following: what is the exact dilution
that causes the order of the transition to change? What is
more, are we absolutely sure that first-order phase
transitions exist in the presence of dilution?  This is still an
important open problem in Statistical Mechanics, and also one
with implications in very technical fields such as highly
correlated electron systems (e.g., high temperature
superconductors or colossal magnetoresistance oxides) where
phase coexistence and chemical disorder play crucial
roles~\cite{MANGA}.

The question in the previous paragraph can be considered
exactly solved in two dimensions~\cite{Aize89}: even the
most insignificant amount of impurities is enough to switch
the phase transition from first-order to second-order (for
the Universality Classes see~\cite{Card97}).  In $D\!=\!3$
the most useful physical picture is provided by the
Cardy-Jacobsen conjecture~\cite{Card97}: considering a
ferromagnetic system undergoing a first-order phase
transition for a pure sample, with $T$ being the temperature
and $p$ the concentration of magnetic sites, a critical
line, $T_\mathrm{c}(p)$, separates the ferromagnetic and the
paramagnetic phases in the $(T,p)$ plane.  In $D\!=\!3$ a
critical concentration is expected to exist,
$1>p_\mathrm{c}>0$, such that the phase transition is of
first order for $p>p_\mathrm{c}$ and of second order for
$p<p_\mathrm{c}$ (at $p_\mathrm{c}$ one has a {\em
tricritical point}). When $p$ approaches $p_\mathrm{c}$ from
above, the latent heat must vanish with the critical
exponent of the magnetisation in the Random Field Ising
Model (RFIM). Also the surface tension, $\Sigma$, vanishes
at $p_\mathrm{c}$, while the correlation length
$\xi(T_\mathrm{c}(p))$ diverges, with critical exponents
related to those of the RFIM\footnote{ The expected
exponents $\beta$ and $\nu$ of the tricritical
point~\cite{Card97} are: $\beta=\beta_\mathrm{RFIM}$ and
$1/\nu=D-\theta_\mathrm{RFIM}-\beta_\mathrm{RFIM}/\nu_\mathrm{RFIM}$
or (modified hyperscaling relation of the RFIM)
$\nu=\nu_\mathrm{RFIM}/(2-\alpha_\mathrm{RFIM}-\beta_\mathrm{RFIM})$.
The surface tension goes to zero with an exponent
$\mu=(D-1)\nu$~\cite{Widom}.  Taking the critical exponents
of the Gaussian RFIM: $\beta_\mathrm{RFIM}=0.00(5)$,
$\nu_\mathrm{RFIM}=1.1(2)$ and
$\theta_\mathrm{RFIM}=1.53(1)$\cite{Rieger}, the exponents
for the tricritical point should be: $\nu\simeq 1.5$,
$\beta\simeq 0$ and $\mu\simeq 3$.}.  The main objection to
this argument is that the Cardy-Jacobsen conjecture relies
on a mapping from the (large $Q$) disordered Potts
model~\cite{Wu} onto the RFIM (two unsolved models in
$D=3$). As a result, if the $D\!=\!3$ RFIM phase transition
turned out to be of first order~\cite{FO-RFIM}, the
conjecture would not be valid.

The $D\!=\!3$ problem has already been numerically
studied~\cite{Ball00,Chat01,Chat05}; large regions of the
critical line $T_\mathrm{c}(p)$ were found to be second
order.  Unfortunately, the study of the tricritical point as
well as that of the first-order part of the critical line
seemed beyond hope, mainly due to two factors.  Firstly, an
important difficulty arises from the long-tailed pdf's
encountered when comparing the specific heat or the magnetic
susceptibility of different samples at
$T_\mathrm{c}(p)$~\cite{Chat05}. Note that
diverging-variance pdf's arise from the common practice of
defining the quenched free energy at temperature $T$ as the
average of the sample's free energy at $T$~\cite{GIORGIO},
which is dominated by rare events\footnote{ Equilibrium
phase-coexistence in a sample of $N$ spins occurs for a
temperature interval of width $\sim N^{-1}$~\cite{FSSFO},
where the specific heat is $C\sim N$.  Yet the {\em
sample-averaged} $C$ scales at most as
$N^{1/2}$~\cite{ESCALA-DES} because the sample dispersion
of the critical temperatures leads to the critical region
having a width $\sim N^{-1/2}$ around $T_\mathrm{c}(p)$.  For
any fixed temperature within the critical region, only a
fraction $\sim N^{-1/2}$ of the samples displays $C\sim N$.}.
Secondly, the other factor has been described above --
the simulation of a sample of linear size $L$ with previous
methods is intrinsically difficult: the required simulation time
grows exponentially with $L^{D-1}$~\cite{NEUSHAGER} due to
the ECSD. These two factors have limited previous
work~\cite{Chat01,Chat05} to $L\leq 25$.

To overcome these two difficulties, on the one hand we
propose two alternative methods of performing the sample
average, both of which reproduce the correct Thermodynamic
Limit, avoiding the diverging-variance pdf's, and providing
complementary information, and on the other we exploit a
novel microcanonic Monte Carlo method~\cite{VICTORMICRO},
which allows one to study the system entropy directly.  This
method, combined with a slightly modified typical cluster
algorithm~\cite{SW,VICTORMICRO}, permits accurate studies of systems
with more than $10^6$ spins (when the previous methods can only
handle $10^4$). In our case the method will allow us to
simulate systems of size up to $L\!=\!128$ in the case
$Q=4$, $D=3$, also making it possible to perform a Finite-Size Scaling (FSS)
study of the {\em elusive} tricritical point as well as the
associated critical behaviour.

The highly accurate numerical study presented in this
chapter has only been possible due to our capability of
using different supercomputing facilities simultaneously:
\begin{itemize}
 \item For the $Q=4$ case: on the Mare-Nostrum machine of BSC (Barcelona Supercomputing Centre) we used
160\thinspace000 computation hours (PowerPC 2.3 GHz processors); on the BIFI (Instituto de Biocomputaci\'on
y F\'{\i}sica de Sistemas Complejos de Zaragoza) cluster we used 250\thinspace000 hours
(Xeon Dual Core 3.40 GHz processors); and on computers (mostly Pentium 2.6 GHz) located 
in the UEX (Universidad de Extremadura) and UCM (Universidad Complutense de Madrid) we used
65\thinspace000 and 160\thinspace000 hours respectively. As a result we estimate that the computational
resources used for this part are equivalent to 60 years of a single last generation (Pentium 2.5 GHz) processor.

 \item For the $Q=8$ case: we used mostly the IBERCIVIS infrastructure, see Appendix~\ref{Appendix_Ibercivis},
from which we obtained the huge number of approximately 300 years of a single 
last generation (Pentium 2.5 GHz) processor. In addition we made extensive use
of the BIFI cluster, from which we  obtained around 40 years of equivalent simulation time. We also
used local resources in Badajoz but they can be disregarded compared to the aforementioned enormous numbers.
\end{itemize}


\section{Analytical Framework}
\label{potts3D:analytics}

In this section we briefly review the main analytical
results on first-order phase transitions with disorder. They have been
taken from Ref.~\cite{Aize89}, where it was demonstrated
that for $D\leq2$ even the smallest amount of impurities
(whether in the bonds or in the fields) destroys the
discontinuities of the first derivatives of the free energy
making the transition continuous (of second order type), and
from Ref.~\cite{Card97} where, after relating the dilute
Potts model with the RFIM, it was
found that for $D>2$ there must exist a region in the phase
diagram where the transition continues to be of the first order type
even in the presence of disorder. This region will end up in a
tricritical point.

\subsection{Aizenman-Wehr Theorem}
\label{potts3D:aizenman}

In Ref.~\cite{Aize89}, it was demonstrated that for $D\leq
2$ the presence of quenched random fluctuations in the
structural parameters (external field $h$, temperature $T$,
...) produces the elimination of the first-order character
of the phase transitions; in other words, it eliminates the
discontinuities in the thermodynamic expectation values of
the conjugate quantities (magnetisation if the disorder is
in the field, energy if the disorder is in the temperature, etc.).

The problem was solved for the general case of spin
variables $\sigma=\{\sigma_x\}$ located on a $D$-dimensional
lattice whose Hamiltonian is the sum of an ordered term
(translation-invariant and non-random) and a fluctuating term
with quenched randomness, represented in the following by a
collection of independent random variables $\{\eta\}$. Some
examples of this form are:
\begin{enumerate}
 \item { \underline{\itshape Random field (RF) models}
\begin{equation}
H(\sigma)= -\frac{1}{2}\sum_{x,y}J_{x-y}\sigma_x\sigma_y-\sum_x(h_x+\epsilon \eta_x)\sigma_x\,,
\end{equation}
where, in the ferromagnetic RFIM, $\sigma \in \mathbb{Z}^2$ and $J\geq 0$.
In the $O(N)$ model, $\sigma_x$ are $N$-component unit vectors with a rotation-invariant
distribution.

In RF models the spins are subjected to a fluctuating magnetic field composed of two terms:
one uniform ($h$), and the other random, with the order of magnitude of $\epsilon$. 
We assume that the random fields $\eta_x$ are independently distributed with a probability
measure $\nu(d\eta)$ (with averages denoted by ${\cal A} (f)\!\equiv\!\int\!f\nu(d\eta)$) that fulfil:
\begin{equation}
{\cal A}(\eta)=0 \quad \quad, \quad \quad {\cal A}(\eta^2)>0\,,
\label{cond_AW_1}
\end{equation}
and
\begin{equation}
{\cal A}(e^{s \eta})<\infty \quad\quad, \quad\quad \forall s<\infty \,.
\label{cond_AW_2}
\end{equation}
}

\item {\underline{\itshape Random bond (RB) models}

For example, the $Q$-state Potts model, with $\sigma\in\{1,\dots,Q\}$ and with Hamiltonian
with bond disorder
\begin{equation}
H_1(\sigma)= -\frac{1}{2}\sum_{x,y}(J_{x-y}+\epsilon_{x-y}\eta_{x,y})\delta_{\sigma_x,\sigma_y}\,,
\end{equation}
or with site disorder
\begin{equation}
H_2(\sigma)= -\frac{1}{2}\sum_{x,y}(1+\epsilon\eta_x+\epsilon\eta_y)J_{x-y}\delta_{\sigma_x,\sigma_y}
=H_0(\sigma)-\sum_x\epsilon\eta_x\sum_y J_{x-y}\delta_{\sigma_x,\sigma_y}\,.
\end{equation}

}

\item {\underline{\itshape Spin-glass models}

For example, the Ising model with Hamiltonian
\begin{equation}
H(\sigma)= -\frac{1}{2}\sum_{|x-y|=1}\eta_{x,y}\sigma_x\sigma_y-\sum_x(h+\epsilon\eta_x)\sigma_x\,.
\end{equation}
}

\end{enumerate}

In general, all the above models can be unified in a Hamiltonian of the form
\begin{equation}
H(\sigma)=H_0(\sigma)+\sum_a\sum_x(h_a+\epsilon_a\eta_{a,x})g_a(T_x\sigma)\,,
\label{H_general}
\end{equation}
where the index $a$ may parameterise pair-interaction terms of a given range or 
other multiple-spin terms, $g_a$ are bounded functions of the spin configuration,
$T_x$ are translation operators (not to be confused with
the temperature $T\equiv1/\beta$), and $\eta_{a,x}$ are a collection of random variables
satisfying the conditions~(\ref{cond_AW_1})~and~(\ref{cond_AW_2}),
with an identical distribution within each $a$ class.

The free energy, $F$, is derived from the finite volume partition function $Z_V$. 
By standard thermodynamic arguments, for almost every configuration of the disorder parameters
$\{\eta_{a,x}\}$ the limit
\begin{equation}
\lim_{V\to\infty}\frac{T}{V}\ \mathrm{log}[ Z_V(T,\{h\},\{\epsilon\},\{\eta\})]=F(T,\{h\},\{\epsilon\})
\end{equation}
converges to a non-random function in the Thermodynamic Limit. In other words, the free energy
self-averages.

In addition, it is known that the free energy is convex in $\{h\}$, for fixed $T$ and $\{\epsilon\}$, 
and therefore their directional derivatives exist; any discontinuity of
those corresponds to a first-order phase transition. One can define the following order
parameter:
\begin{equation}
M_a(T,\{h\},\{\epsilon\})=\frac{1}{2}\left[\frac{\partial}{\partial(h_a+0)}-\frac{\partial}{\partial(h_a-0)}\right]F(T,\{h\},\{\epsilon\})\,.
\end{equation}

In the case of the ferromagnetic RFIM:
\begin{equation}
M(T,h,\epsilon)=\frac{1}{2}\left[{\cal A}(\langle\sigma_0\rangle_{+})-{\cal A}(\langle\sigma_0\rangle_{-})\right]\,,
\end{equation}
where with ``$+$'' and ``$-$'' we denote the extremal Gibbs states (``pure phases'') constructed via
choices of the boundary conditions ($+$ or $-$).

The following is the main result for the general case, see Ref.~\cite{Aize89} for its demonstration:

\newtheorem*{teor}{Theorem}
\begin{teor}
In a $D\leq2$ system with quenched disorder, described by a Hamiltonian of the general
type of Eq.~(\ref{H_general}) with nearest neighbour interaction (this can be 
extended to longer range interactions) and with a continuous (non-atomic)
probability measure $\nu(d\eta)$ is
\begin{equation}
M_a(T,\{h\},\{\epsilon\})=0 \quad\quad\quad \forall\ T\geq0,\{h\},\{\epsilon\}\ \text{and\ } a\
\text{for which}\ \epsilon_a>0\,.
\end{equation}
\end{teor}

In the disordered Potts case, where the transition is due to a change in temperature, the free energy $F$
is also convex in $T$ for $\{h\}$ and $\{\epsilon\}$ fixed and therefore its partial temperature derivative 
exists. We can define in this case the latent heat as
\begin{equation}
L=\frac{1}{2}\left(\left.\frac{\partial}{\partial T}\right|_{T_c^+}
-\left.\frac{\partial}{\partial T}\right|_{T_c^-}\right) F\,,
\end{equation}
which is the order parameter for this first-order phase transition. In an analogous way it can be
demonstrated that $L=0$ for $D\leq2\,$.

\subsection{Cardy-Jacobsen Theory}
\label{potts3D:cardy}

The following results are based on a mapping between the Random Bond (RB) model (such as the 
dilute Potts model) and the Random Field (RF) model (such as the RFIM), see~\cite{Card97}. Firstly we 
will summarise the main properties of the latter model.

\subsubsection{Random Field Ising Model (RFIM)}
\label{subsubsec:RFIM}

It is defined by the Hamiltonian
\begin{equation}
H= -J\sum_{i,j}s_i s_j + \sum_i h_i s_i + h\sum_i s_i \,,
\label{H_RFIM}
\end{equation}
$h_i$ being quenched random variables satisfying
$\overline{h_i}=0$ and $\overline{h_i^2}=\Delta^2$.  The pdf
of $h_i$, $p(h_i)$, can be chosen\footnote{Some controversy
exists about the influence of this choice on the
universality class of the model, see~\cite{OUR_DAFF} and
references therein.} Gaussian or bimodal ($\pm\Delta$).

\begin{figure}[!ht]
\begin{center}
\includegraphics[width=0.45\columnwidth]{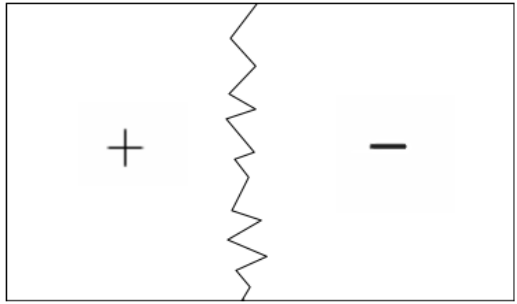}
\caption{Rough interface for a two-dimensional Ising model.}
\label{FIG_BINDER_INTER}
\end{center}
\end{figure}

There are some important general theoretical results concerning this model:
\begin{enumerate}
\item \underline{Dimensional reduction}

A $D$-dimensional system with a random field is equivalent to a system with $D-2$ dimensions
without the random field~\cite{PARI-SOUR}. Therefore the lowest critical dimension
is $D_c^\mathrm{inf}=3$ because for the pure Ising model it is $D_c^\mathrm{inf}=1$.
This result can be obtained by using supersymmetry arguments or through perturbation
theories. Nevertheless, dimensional reduction seems to fail specifically for 
this model~\cite{BRAY-MOORE}, although it is a valid result for many other models.

\item \underline{Imry-Ma argument}

Starting from $T\backsimeq0$ in the RFIM and considering that the fundamental low temperature 
ferromagnetic state is $s_i=+1$, we can analyse what happens if we form a ``droplet'' with
radius $R$ with $s_i=-1$~\cite{IMRY-MA}, see Fig.~\ref{FIG_DROPLET}.
\begin{figure}[!ht]
\begin{center}
\includegraphics[width=0.3\columnwidth]{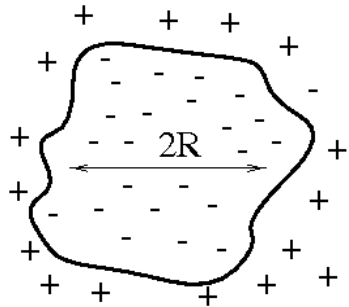}
\caption{``Droplet'' with different sign within an almost fully ordered  Ising model. We study what is the effect
of this perturbation depending on the dimension of the
space.}
\label{FIG_DROPLET}
\end{center}
\end{figure}

This ``droplet'' will present an interface with an energy cost
\begin{equation}
\Delta E=JR^{D-1}\sim R^{D-1}\,.
\end{equation}
There is also an energy variation due to the random field within the ``droplet''
\begin{equation}
\Delta E_H=\sum_{i\in R}h_i \,,
\end{equation}
which, by the definition of the random field, will fulfil
\begin{equation}
\delta\equiv\sqrt{\overline{\Delta E_H^2}}=\pm(R^Dh_{RF}^2)^{1/2}\sim R^\frac{D}{2}\quad,  \quad\quad h_{RF}^2\equiv\Delta^2 \,.
\end{equation}

We can always choose a point in the lattice where $\delta<0$, so that the energy balance
between $\Delta E$ and $\Delta E_H$ produces the following results depending on the dimensionality of
the space:
\begin{itemize}
\item For $D>2$ the fundamental low temperature state is stable. Consequently the low
temperature ferromagnetic phase exists and a phase transition at finite temperature
can be found.

\item For $D<2$ the fundamental state is not stable and there will exist no phase transition.

\item For $D=2$ we are in the marginal case. In Ref.~\cite{BINDER_INTER} it was demonstrated that
the rugosity of the interface (which is obviously not flat)
destabilises the ferromagnetic state, see
Fig.~\ref{FIG_BINDER_INTER}. This result is included in the Aizenman-Wehr theorem.

\end{itemize}
\end{enumerate}

The rugosity of the interface in $D=2$ produces the energy~\cite{BINDER_INTER}
\begin{equation}
E=-C\frac{h_{RF}^2}{J}R\log R\,,
\end{equation}
with $C>0$, which makes the ferromagnetic state unstable. By defining (for $D=2$)
\begin{equation}
J(L)=JL-C\frac{h_{RF}^2}{J}L\log L=JL(1-Cw_{RF}^2\mathrm{log}L)\quad
\text {with:  } w_{RF}^2\equiv\frac{h_{RF}^2}{J^2}\,,
\label{w_RF_def}
\end{equation}
with $L$ being the linear lattice size, we can obtain~\cite{BRAY-MOORE}
\begin{equation}
\frac{d J(L)}{d\log L }=J(L)(1-Cw_{RF}^2)+O(w_{RF}^4)\,,
\end{equation}
which is the Renormalization Group (RG) equation for the
coupling $J$. We can easily obtain the remaining RG
equations by taking into account that $h$ has dimensions of
$R^D$, $h_{RF}^2$ has dimensions of $R^D$, and $J$ has
dimensions of $R^{D-1}$. Therefore it will be to leading
order

\begin{numcases}{}
&\hspace{-0.7cm}$\genfrac{}{}{}{0}{d h_{RF}}{d l}=\frac{D}{2}h_{RF} \quad \quad , \quad\quad l\equiv \log(b)\,,$\notag\\[0.3cm]
&\hspace{-0.7cm}$\genfrac{}{}{}{0}{d h}{d l}\ = D h\,,$\label{RG_RFIM}\\[0.3cm]
&\hspace{-0.7cm}$\genfrac{}{}{}{0}{d J}{d l}\ = J[(D-1)-Cw_{RF}^2] \longleftarrow \text{(generalising to dimension\ } D)\,.$\notag
\end{numcases}

In addition, by definition, $w_{RF}\equiv h_{RF}/J$ and therefore
\begin{equation}
\frac{d w_{RF}}{d l}=-\frac{\epsilon}{2}w_{RF}+Cw_{RF}^3\ ,\quad \text{with:}\quad \epsilon \equiv D-2\,.
\end{equation}

\begin{figure}[!ht]
\begin{center}
\includegraphics[width=0.75\columnwidth]{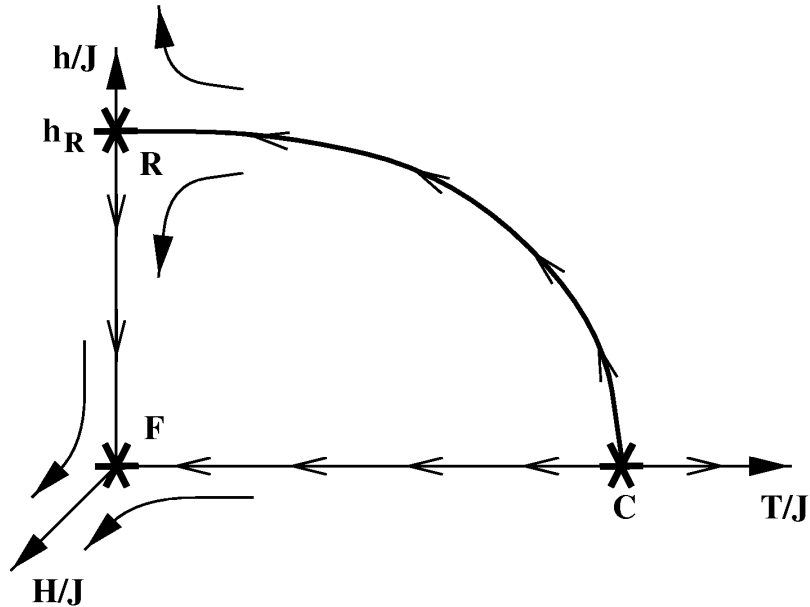}
\caption{RG flows for the RFIM for $D>2\,$.}
\label{FIG_RFIM_FLOW}
\end{center}
\end{figure}

If $D>2$ then $\epsilon>0$ and the RG flow has a non-trivial
\emph{random} fixed point (with
$w\sim\epsilon^\frac{1}{2}\neq 0$) in agreement
with~\cite{IMRY-MA}. Using these results, the phase diagram
for the RFIM can be obtained, see Fig.~\ref{FIG_RFIM_FLOW}.

\subsubsection{Cardy-Jacobsen mapping}
\label{potts3D:CJmap}

In the case of a pure system undergoing a first-order phase transition there
will be coexistence of a (generally unique) disordered phase and the (generally non-unique)
ordered phases. The internal energies $U_1$ and $U_2$ of these two phases differ by the latent heat.
Consider, see Ref.~\cite{Card97}, a large (say horizontal) interface between the disordered
phase and one of the ordered ones,
with surface tension $\Sigma$. If $\Sigma\gg1$, there will be a really small number of isolated bubbles of the
opposite phase above or below the main interface. The free energy of these bubbles is proportional
to their areas multiplied by the surface tension.

On the other hand, let us consider an Ising model and build
an interface between the two possible ordered phases (all
spins taking the values $\pm1$). At a very low temperature,
there will basically exist no ``bubbles'' above or below the
interface and the surface tension will be $\sim2J$, where
$J$ is the reduced coupling of the Hamiltonian.  In the
limit $\Sigma\sim 2J\gg1$, these two interface models will
be {\itshape identical}.

We will analyse the effect of disorder in these two
models. In the first one, we introduce random bonds (disorder
coupled with the energy) while in the second we
introduce a random field (disorder coupled with the
magnetisation). The changes in the energy due to the
introduction of the disorder are:
\begin{enumerate}
\item Random Field:
\begin{equation}
\sum_{r^{>}}h(r)-\sum_{r^{<}}h(r)\,,
\label{deltaE_RF}
\end{equation}
where the sums are defined over all the points above ($>$) or below ($<$) the interface.

\item Random bonds
\begin{equation}
U_1\sum_{r^{>}}\delta x(r)+U_2\sum_{r^{<}}\delta x(r)=
\frac{1}{2} L \left( \sum_{r^{>}}\delta x(r) - \sum_{r^{<}}\delta x(r) \right)+\text{ const.} \,,
\label{deltaE_RB}
\end{equation}
with $\delta x(r)$ being the local impurities density, $L=U_1-U_2$ is the latent heat, and
the final constant is independent of the interface location\footnote{
The energy can be split in the following way, $\Delta$ being the difference between the
two sides of Eq.~(\ref{deltaE_RB}):
\begin{equation}
\Delta=\frac{1}{2}U_2\sum_{r^{<}}\delta x(r)+\frac{1}{2}U_1\sum_{r^{>}}\delta x(r)+
\frac{1}{2}U_1\sum_{r^{<}}\delta x(r)+\frac{1}{2}U_2\sum_{r^{>}}\delta x(r)=\nonumber
\end{equation}
\begin{equation}
\frac{1}{2}U_1\left[ \sum_{r^{<}}\delta x(r)+\sum_{r^{>}}\delta x(r)\right] +
\frac{1}{2}U_2\left[ \sum_{r^{<}}\delta x(r)+\sum_{r^{>}}\delta x(r)\right]=\nonumber
\end{equation}
\begin{equation}
\frac{1}{2}U_1\sum_{\forall i}\delta x_i+\frac{1}{2}U_2\sum_{\forall i}\delta x_i\nonumber
\end{equation}
which is independent of the interface location.
}. The latter equation has the same form as that corresponding to the random
field.
\end{enumerate}

Therefore the thermal variables of the random bond system are related to the magnetic variables of the RFIM
by the following mapping:
\begin{eqnarray}
\text{\underline{Random Bond}} & \quad \quad & \text{\underline{Random Field}}\notag\\
\Sigma/kT_c &  \quad\Longleftrightarrow\quad & J/kT \notag\\
(L/kT_c)x   &  \quad\Longleftrightarrow\quad & h_{RF}/kT\\
(T-T_c)L    &  \quad\Longleftrightarrow\quad & H M \notag
\end{eqnarray}
%

The last relationship is between the ``field'' $(T-T_c)$ and
a uniform external field $H$, which helps to distinguish
between the two phases (in the same way as $(T-T_c)$). One
of the possible problems of this mapping is its use of the
local energy density as a kind of order parameter. However it
can be made completely explicit, for example for the
$Q$-state Potts model, through the mapping to the random
cluster model where $\Sigma\sim L\sim \log Q,\ Q\to\infty$,
see Ref.~\cite{Card97}.

\subsubsection{Explicit relationship with the Potts model}
\label{potts3D:RelPotts}

We can now derive~\cite{Card97} the specific relationship of the previous section with 
the $Q$-state Potts model with quenched disorder, with Hamiltonian
\begin{equation}
{\cal H}=-\sum_{<i,j>}K_{ij} \delta_{s_i s_j} \,,
\end{equation}
where the sum extends only over nearest neighbours. The
ferromagnetic couplings $K_{ij}$ are quenched random
variables, taking the values $K_1$ and $K_2$, each with
probability $\frac{1}{2}$; in other words, their pdf is
\begin{equation}
p(K_{ij})=\frac{1}{2}\delta(K_{ij}-K_1)-\frac{1}{2}\delta(K_{ij}-K_2)\,.
\end{equation}

When $(e^{K_1}-1)(e^{K_2}-1)=1$ this model is, on average, self-dual, and, if the transition
is unique, is at its critical point~\cite{Kinzel}. It is useful to parameterise the
model through
\begin{equation}
u_{ij}=(e^{K_{ij}}-1)=Q^{\frac{1}{2}+w_{ij}}\,,
\end{equation}
with $w_{ij}=\pm w$. $w\geq0$ measures the strength of the
randomness, with $w=0$ being the case without disorder. We can
solve for $K_{ij}$
\begin{equation}
K_{ij}=\log(1+Q^{\frac{1}{2}+w_{ij}})\,,
\end{equation}
and consider the limit $Q\to\infty$ to approximate
\begin{equation}
K_{ij}=\log (1+Q^{\frac{1}{2}+w_{ij}})\simeq\log (Q^{\frac{1}{2}+w_{ij}})=
\left(\frac{1}{2}+w_{ij}\right)\log Q \,.
\end{equation}
By substituting in the Hamiltonian we obtain
\begin{equation}
{\cal H}=-\frac{1}{2}\log Q\sum_{<i,j>} \delta_{s_i s_j} - \log Q\sum_{<i,j>}w_{ij} \delta_{s_i s_j}\,,
\end{equation}
where the first term corresponds to an ordered model while the second term corresponds
to a disordered one. Therefore the term added to the pure model is
\begin{equation}
{\cal H}_\mathrm{added}\sim\sum_{<i,j>}w_{ij} \log Q\ \delta_{s_i s_j}\,.
\end{equation}
We will work in the following in two dimensions; i.e. the
label $i$ identifying the lattice sites means
$i\equiv(x,y)$, with $(x,y)$ being Cartesian
coordinates. Using the same arguments as in the previous
section, if an interface divides the space into two parts,
denoted by ``$>$'' and ``$<$'', then
\begin{equation}
{\cal H}_\mathrm{added}\sim\sum_{>}w\log Q+\sum_{<}w\log Q\,,
\label{H_addde_tworegions}
\end{equation}
noting that within each of these two homogeneous regions $\delta_{s_i s_j}=1$. Each
of the two terms of Eq.~(\ref{H_addde_tworegions}) is
\begin{equation}
\sum_{<}w\log Q=\frac{1}{2}\left(\sum_{>} + \sum_{<}\right)+\frac{1}{2}\left(\sum_{<} - \sum_{>}\right)\,,
\end{equation}
but, as in the previous section, the term $(\sum_{>} + \sum_{<})$ does not depend on the interface position
and as a consequence
\begin{equation}
\frac{1}{2}\left(\sum_{>} - \sum_{<}\right)w\log Q
\end{equation}
is an analogue of the random field term of Eq.~(\ref{deltaE_RF}). Therefore we obtain
the relationship
\begin{equation}
h_{RF}\longleftrightarrow\frac{1}{2}w\log Q\,.
\label{w_RB}
\end{equation}

In addition, following Ref.~\cite{Wu}, when $Q\to\infty$, the surface tension is:
\begin{equation}
\Sigma\sim\frac{1}{4}\log Q\,.
\end{equation}
But as was seen in Sec.~\ref{potts3D:CJmap} for the RFIM $\Sigma\sim2J$, and therefore
we have the relation
\begin{equation}
J\longleftrightarrow\frac{1}{8}\log Q\,.
\label{jtoq}
\end{equation}

Finally, while a uniform field in the RF model distinguishes
between the two phases, in the RB model this is the task of
the reduced temperature $t\equiv\frac{T-T_c}{T}$; provided
that $t$ is coupled to the energy density we can make the
identification
\begin{equation}
h\longleftrightarrow\frac{1}{4}t\log Q\,.
\label{htot}
\end{equation}

To summarise, the mapping between the Potts model and the RFIM is:
\renewcommand{\arraystretch}{1.2}
\begin{eqnarray}
\text{\underline{RFIM}} & \quad \quad & \text{\underline{$Q$-state Potts model}} \notag\\
J     &  \quad\Longleftrightarrow\quad & \tfrac{1}{8}\log Q \notag\\
h     &  \quad\Longleftrightarrow\quad & \tfrac{1}{4} t \log Q\\
h_{RF}&  \quad\Longleftrightarrow\quad & \tfrac{1}{2}w\log Q \notag
\end{eqnarray}
%

%

We can use this mapping to derive the RG equations for the
Potts model starting from those for the
RFIM~\cite{BRAY-MOORE}, see Eq.~(\ref{RG_RFIM}). We will
denote for the sake of clarity the $w$ used in
Eq.~(\ref{w_RB}) as $w_{RB}\equiv w$, to
distinguish between this $w_{RB}$ and the $w_{RF}$ defined
in Eq.~(\ref{w_RF_def}). Therefore from
Eqs.~(\ref{w_RB})~and~(\ref{jtoq}) one obtains
\begin{equation}
w_{RB}=\frac{2h_{RF}}{\log Q}=\frac{h_{RF}}{4J}\,.
\end{equation}
But $w_{RF}=h_{RF}/J$, hence
\begin{equation}
w_{RB}=\frac{1}{4}w_{RF}\,,
\end{equation}
and therefore the RG equation for $w_{RF}$ is also valid for $w_{RB}$, i.e.,
\begin{equation}
\frac{d w_{RB}}{d l}=-\frac{\epsilon}{2}w_{RB}+Aw_{RB}^3\,\ ,\quad \text{with:}\quad \epsilon \equiv D-2\quad, \quad l\equiv \log(b)\,.
\end{equation}
From the last RG equation of Eq.~(\ref{RG_RFIM}) and from Eq.~(\ref{jtoq}), one
easily derives the RG relation
\begin{equation}
\frac{d (\log Q)^{-1}}{d l}=-(\log Q)^{-1}[(D-1)-Aw_{RB}^2]\,.
\end{equation}
Finally, using again Eq.~(\ref{RG_RFIM}) and Eq.~(\ref{htot}), one obtains:
\begin{equation}
\frac{d t}{d l}=t(1+Aw_{RB}^2)\,.
\end{equation}

Summarising, the set of RG equations for the $Q$-state Potts model with quenched disorder is:

\begin{numcases}{}
&\hspace{-0.7cm}$\genfrac{}{}{}{0}{d w_{RB}}{d l}=-\frac{\epsilon}{2}w_{RB}+Aw_{RB}^3\,\ ,\quad \text{with}\quad \epsilon \equiv D-2 \quad,\quad l\equiv \log(b)\,,$\notag\\[0.3cm]
&\hspace{-0.7cm}$\genfrac{}{}{}{0}{d t}{d l}=t(1+Aw_{RB}^2)\,,$\label{RG_POTTS}\\[0.3cm]
&\hspace{-0.7cm}$\genfrac{}{}{}{0}{d (\log Q)^{-1}}{d l}=-(\log Q)^{-1}[\epsilon+1-Aw_{RB}^2]\,.$\notag
\end{numcases}

We can now analyse the stability of the fixed points of the RG transformations to
trace the phase diagram of the model. To simplify we make the change
\begin{equation}
 \hat{Q}\equiv\frac{1}{\log Q}\,.
\end{equation}
Then the fixed points of the RG transformation are the Gaussian one:
\begin{equation}
 \hat{Q}=t=w_{RB}=0\,,
\end{equation}
and the tricritical one:
\begin{equation}
 \hat{Q}=t=0\quad ,\quad w_{RB}=\sqrt{\frac{\epsilon}{2A}}\,.
\end{equation}

The Jacobian matrix for the transformation is
\[
J= \begin{pmatrix}
-\frac{\epsilon}{2}+3Aw_{RB}^2& 0 &0\\[0.3cm]
2Atw_{RB}&1+Aw_{RB}^2&0\\[0.3cm]
2A\hat{Q}w_{RB}&0&-(\epsilon+1-Aw_{RB}^2)\\
\end{pmatrix}
\]
which evaluated at the Gaussian point is
\[
J|_\mathrm{Gauss}= \begin{pmatrix}
-\frac{\epsilon}{2}& 0 &0\\[0.3cm]
0 & 1 &0\\[0.3cm]
0 & 0&-(\epsilon+1)\\
\end{pmatrix}
\]
resulting in that at the Gaussian fixed point the ``fields''
$w_{RB}$ and $Q$ are irrelevant while $t$ is relevant.  At
the tricritical (TC) point, the Jacobian will be
\[
J|_{TC}= \begin{pmatrix}
\epsilon& 0 &0\\[0.3cm]
0 & 1+\frac{\epsilon}{2} &0\\[0.3cm]
0 & 0&-(1+\frac{\epsilon}{2})\\
\end{pmatrix}
\]
and therefore $w_{RB}$ and $t$ are relevant ``fields'' while $Q$ is irrelevant. For $D>2$ the
eigenvalues are
\begin{eqnarray}
y_w=\epsilon>0 \quad\quad\quad &\Longrightarrow& \quad \text{Relevant} \nonumber\\
y_t=1 +\dfrac{\epsilon}{2}\quad\quad\quad &\Longrightarrow&\quad \text{Relevant}\nonumber\\
y_Q=-\left(1 +\dfrac{\epsilon}{2}\right)\quad &\Longrightarrow&\quad \text{Irrelevant}\nonumber
\end{eqnarray}

\begin{figure}[!ht]
\begin{center}
\includegraphics[width=0.65\columnwidth]{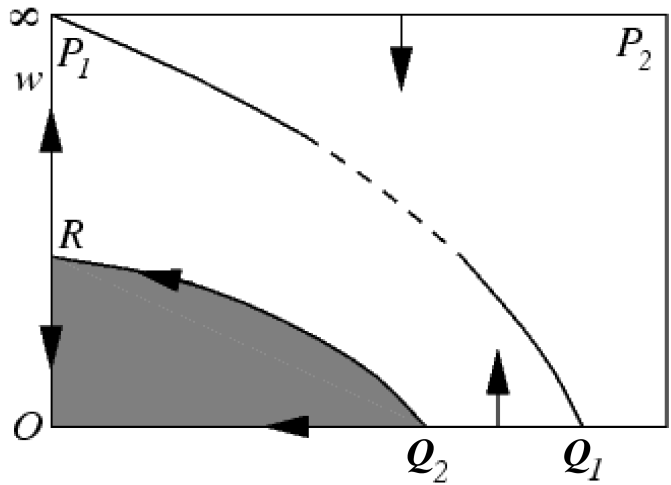}
\caption{Phase diagram of the dilute Potts model with $Q$ states for $D>2$, obtained 
in Ref.~\cite{Card97}. $Q$ grows towards the left; $w$ is a measure of the disorder strength, $w=0$
being the pure case. The latent heat do not vanishes within the shaded region while outside
the transition is continuous. $P_1$ and $P_2$ are the percolation thresholds.}
\label{FIG_DIG_CARDY}
\end{center}
\end{figure}

The $(Q,w)$ plane of the phase diagram is depicted in
Fig.~(\ref{FIG_DIG_CARDY}). In the pure system, for $Q>Q_2$
there is a phase transition with non-vanishing latent heat
controlled by a fixed point at infinite $Q$. For $D>2$ it
will continue into the shaded region bounded by a line of
tricritical points whose exponents are related to those
of the RFIM~\cite{Card97}. It also may be shown that the
latent heat vanishes as $(w_c-w)^{\beta_\mathrm{RF}}$ as the
line $RQ_2$ is approached from below. Let us note also that
$Q=1$ corresponds to the percolation model, where the
disorder is irrelevant ($\alpha\approx -0.64$), while $Q=2$
corresponds to an Ising model, where the phase transition is
always of second order but the disorder is relevant
($\alpha\approx 0.1118$). Therefore there exists a $Q_1$,
with $1<Q_1<2$, at which the sign of $\alpha$ changes.  In
addition, for $Q>Q_2=2+\epsilon$ the transition becomes first order.

Above the line $RQ_2$, as $w$ grows the RG equations lose
validity. In addition the surface tension goes to zero and
the mapping between the two models disappears. Nevertheless,
for $Q\to\infty$ the mapping remains exact and the flow goes
to infinite $w$. But this can not happen for finite $Q$
because this is the percolation limit $K_1/K_2=0$, at which
the disorder is relevant~\cite{Stauff}. Therefore there
should exist~\cite{Card97} another line of stable fixed
points emerging from $P_1$, which control the universal
continuous transition for large, but finite, values of $w$
and $Q$.

\section{The Model}
\label{potts3D:themodel}

In the three-dimensional site-diluted $Q$-state Potts model~\cite{Wu} the spins, $\sigma_i$, take the
values $\sigma_i\!=\!1,\ldots,Q$ and are defined at the nodes of a cubic lattice
with probability $p$. We consider only nearest neighbour interaction and periodic 
boundary conditions. Therefore the Hamiltonian takes the form:
\begin{equation}
{\cal H}^\mathrm{spin}=-\sum_{\langle i,j \rangle}\epsilon_i \epsilon_j
\delta_{\sigma_i \sigma_j}\,,
\end{equation}
with $\epsilon_i$ being quenched occupation variables
($\epsilon_i\!=\!0$ or 1 with probability $1-p$ and $p$
respectively)\footnote{To reduce statistical fluctuations, we kept
  only the spins in the percolating cluster~\cite{Stauffer} that
  control the critical behaviour.  However, in the most interesting
  region ($p \gtrsim 0.9$) this correction is quite small.},
and $\langle i,j \rangle$ denoting
nearest neighbours. Each one of the specific disorder
realizations ($\{\epsilon\}$ spatial distribution) is called a \emph{sample}. The
pure system is recovered for $p\!=\!1$, and is known to undergo a first-order phase
transition for $Q\geq3$~\cite{Chat05,VICTORMICRO} generally regarded as {\em very strong}.

A valid order parameter for the model is the magnetisation density (a $Q$-dimensional vector)
defined as
\begin{equation}
M_q=\frac{1}{V}\sum_{i} \epsilon_i \left[ \frac{Q\delta_{\sigma_i,q}-1}{\sqrt{Q(Q-1)}} \right]\,,
\label{magnet}
\end{equation}
with $V=L^3$ being the volume and $L$ the linear size of the system. We can define the magnetic susceptibility as
\begin{equation}
\chi=V {\vert M \vert}^2 \,.
\label{suscept}
\end{equation}

A well-behaved definition for the correlation length in a finite system is obtained from the correlation
function as~\cite{VICTORAMIT}
\begin{equation}
\xi \equiv \left(\frac{\overline{\chi} /
\overline{F}-1}{4\sin^2(\pi/L)}\right)^\frac{1}{2}\,,
\label{xi}
\end{equation}
where
\begin{equation}
\overline{F} \equiv \frac{V}{3}\overline{\left\langle |F(2\pi/L,0,0)|^2+|F(0,2\pi/L,0)|^2+|F(0,0,2\pi/L)|^2\right\rangle}\,,
\end{equation}
and where we denote the thermal averages with brackets while the sample average is overlined. In addition
\begin{equation}
F({\bm{k}}) \equiv \frac{1}{V}\sum_{\bm{r}}e^{\mathrm i
\bm{k}\cdot\bm{r}} \epsilon_{\bm r}\sigma_{\bm r}\,.
\end{equation}

In this work we use the microcanonical simulation method
defined in Ref.~\cite{VICTORMICRO}, see also
Sec.~\ref{FSSmicro:MFSSA}, so that, by using the Maxwell
construction, see Appendix~\ref{Appendix_Max}, we can
directly obtain some quantities characteristic of the phase
transition. Firstly the critical temperature is fixed by the
definition of the Maxwell construction: the $e$-integral
of $\beta_{\{\epsilon\}}(e) -1/T_\mathrm{c}$ from
$e_\mathrm{d}$ to $e_\mathrm{o}$ must vanish, where $e$ is
the energy density and $\beta_{\{\epsilon\}}(e)$ was defined in Eq.~(\ref{beta_micro}).
This fact also implies that
\begin{equation}
s_\mathrm{d}-s_\mathrm{o}\!=\!(e_\mathrm{d}-e_\mathrm{o})/T_\mathrm{c}\,,
\end{equation}
$s$ being the entropy density. The latent heat is defined directly as
\begin{equation}
\upDelta e\!=\!e_\mathrm{d}-e_\mathrm{o}\,. 
\label{latent}
\end{equation}
Finally the surface-tension, $\Sigma$, is $L^{2}/2$ times
the integral of the positive part of
$\beta_{\{\epsilon\}}(e) -1/T_\mathrm{c}\,$, see
Ref.~\cite{VICTORMICRO}.


\section{Numerical Results}
\label{potts3D:results}

We have studied numerically two cases of the three-dimensional site-diluted Potts model:
the four-state ($Q\!=\!4$) and the eight-state ($Q\!=\!8$) cases. Both cases undergo
a well-known~\cite{Wu} strong first-order phase transition in the pure case; the strength of the
first-order character of the phase transition will grow with $Q$. In 
both cases a softening of the discontinuities of the first derivatives of the free energy
is expected to appear with increasing dilution. The critical concentration, $p_c$,
at which the character of the phase transition switches to second order will depend on $Q$, being
smaller for increasing $Q$.

\subsection{Methods}
\label{potts3D:methodology}

\subsubsection{Simulation method}
\label{potts3D:simumethod}

To update the Potts spins of our systems, we used a
microcanonical version~\cite{VICTORMICRO} of the Swendsen-Wang (SW)~\cite{SW}
cluster method.  For disordered systems,
SW updates loosely connected regions properly~\cite{ISDIL4D}
and does not require tedious parameter tuning.  The
microcanonical cluster method, which is not rejection-free,
depends on a tunable parameter, $\kappa$.  In order to
maximise the acceptance of the SW attempt (SWA), $\kappa$
should be chosen as close as possible to
$\beta_{\{\epsilon\}}(e)$.  After every $e$ change, we
performed cycles consisting of $10^3$ Metropolis steps,
a $\kappa$ refresh, then $10^3$ SWAs, and a new $\kappa$
refresh.  The cycling was stopped, and $\kappa$ fixed,
when the SWA acceptance exceeded $60\%$.   We then
performed a number of SWAs depending on the lattice size,
taking measurements every 2 SWAs.

For $Q=4$ we performed thermalization checks that included comparisons of hot and cold
starts or even mixed configurations (bands and strips~\cite{VICTORMICRO}). We checked 
that the Maxwell construction obtained for the pure case of the largest
system ($L=128$) does not depend on the initial configuration, after discarding
a part of the initial Monte Carlo history.

In the $Q=8$ case, reaching the thermodynamic equilibrium is a far more complicated task, especially
on the first-order side of the phase diagram. For a first-order phase transition, it is known
that metastable states do exist. These states can have a very long life even when they are not the
true equilibrated states. This is most dramatic for large systems in which the simulation times are
intrinsically longer. Therefore the thermalization issue in this case deserves a special
treatment that will be described in Sec.~\ref{potts3D:eight-state}.

\subsubsection{Sample averaging methods}
\label{potts3D:sampleaverages}

For a disordered system, one has to analyse a set of functions
$\beta_{\{\epsilon\}}(e)$ corresponding to a large enough number of
samples.  There are two natural possibilities. On the one hand, one can use the Maxwell
construction for each sample extracting $T_\mathrm{c}$,
$e_\mathrm{d}$, $e_\mathrm{o}$ and $\Sigma$, and then considering
their sample average, median, or even their pdf, see Fig.~\ref{HISTOGRAMS_Q4}. This is the
most usual approach.

On the other hand, one can compute the sample average of the (inverse) temperature
defined at each simulation energy, $e$,
$\beta(e)=\overline{\beta_{\{\epsilon\}}(e)}$, and then perform the
Maxwell construction  on it  (i.e., take the sample average of $s(e)$, rather
than the sample average of the free energy at fixed $T$).

We found empirically that the two sample averagings are equivalent
in the first-order piece of the critical line. This is hardly
surprising, because the internal energy as a function of $T$ is a
self-averaging quantity, for all temperatures but the critical one.
Therefore, also $e_\mathrm{d}$, $e_\mathrm{o}$, and $T_\mathrm{c}$ are
self-averaging properties in the first-order part of the critical
line. 

While the first method offers more information, it is
computationally more demanding (it requires high accuracy for each
sample). The method featuring $\beta(e)$ can be used as well in the
second-order part of the critical line, but its merit in that region
is yet to be investigated.

\subsection[$D=3$\,, $Q=4$ Site-Diluted Potts Model]{$\bm{D=3\,,\ Q=4}$ Site-Diluted Potts Model}
\label{potts3D:four-state}

For $Q=4$ we  investigated the phase transition for
several $p$ values in the range $ 0.75 \leq p\leq 1$.  As a
rule, we found that at fixed $p$ the latent heat is a
monotonically decreasing function of $L$, see
Fig.~\ref{LATENT-SIGMA_Q4}. For each $p$ value, we simulated
$L=16$, $32$, $64$, and $128$ (for a given $p$, we did not
consider larger lattices once the latent heat vanished). For
all pairs ($p$, $L$) we simulated 128 samples. Also, some
intermediate $p$ values were added for the FSS study,
see Fig.~\ref{cortes_xi_ed_Q4}, and we
raised to 512 the number of samples for ($L=16$ and $32$,\,
$p=0.86$ and $0.875$).

\subsubsection{General behaviour}
\label{potts3D:four-state:general}

\begin{figure}[h!]
\begin{center}
\includegraphics[height=0.60\textheight,angle=270]{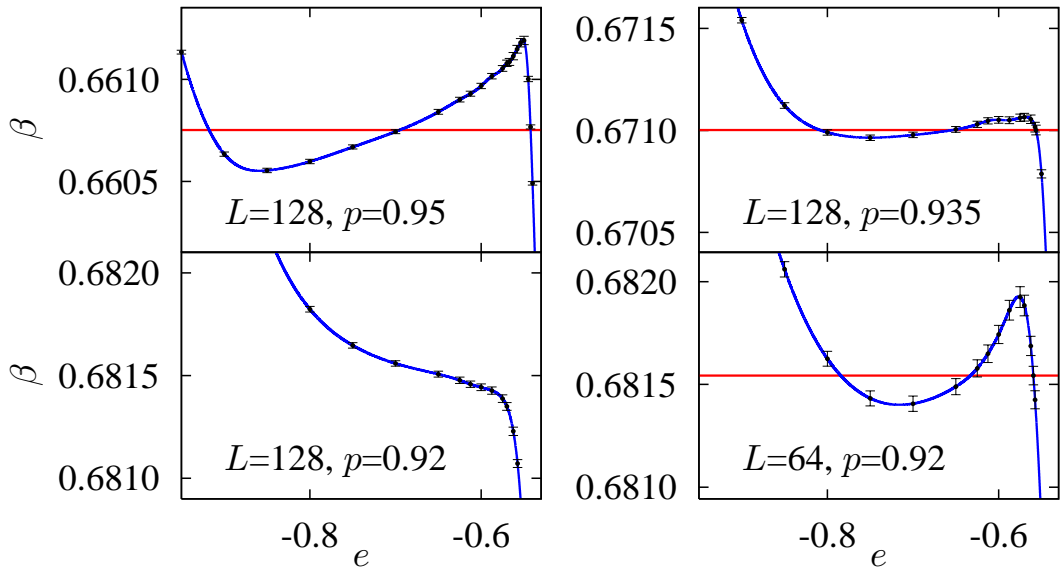}
\caption{Sample-averaged $e$-derivative of the
  entropy, $\beta(e)$, for several lattice sizes, $L$, and spin
  concentrations, $p$.  Metastability requires a non-decreasing
  $\beta(e)$. The horizontal line marks the critical (inverse)
  temperature $1/T_\mathrm{c}$, obtained through Maxwell
  construction. At fixed $L$ the surface tension increases for increasing
  $p$. Note that, for a fixed dilution, a seemingly first-order
  transition ($L=64$, bottom-right), may actually be of second
  order if studied on larger lattices ($L=128$, bottom-left).}
\label{BETAPROMEDIO_Q4}
\end{center}
\end{figure}

Following the procedure of Sec.~\ref{potts3D:sampleaverages}
we performed sample averages of the Maxwell constructions
for each $L$ and $p$. In Fig.~\ref{BETAPROMEDIO_Q4} one can
see the general behaviour of the Maxwell constructions as
the spin concentration, $p$, varies: while for large $p$
($p\approx0.95$) we can form the Maxwell construction for
every system size, it softens with decreasing $p$ up to a
point, $p_c$, at which both latent heat and surface tension
vanish. This $p_c$ depends on the system size.

\begin{figure}
\begin{center}
\includegraphics[height=0.55\textheight, angle=270]{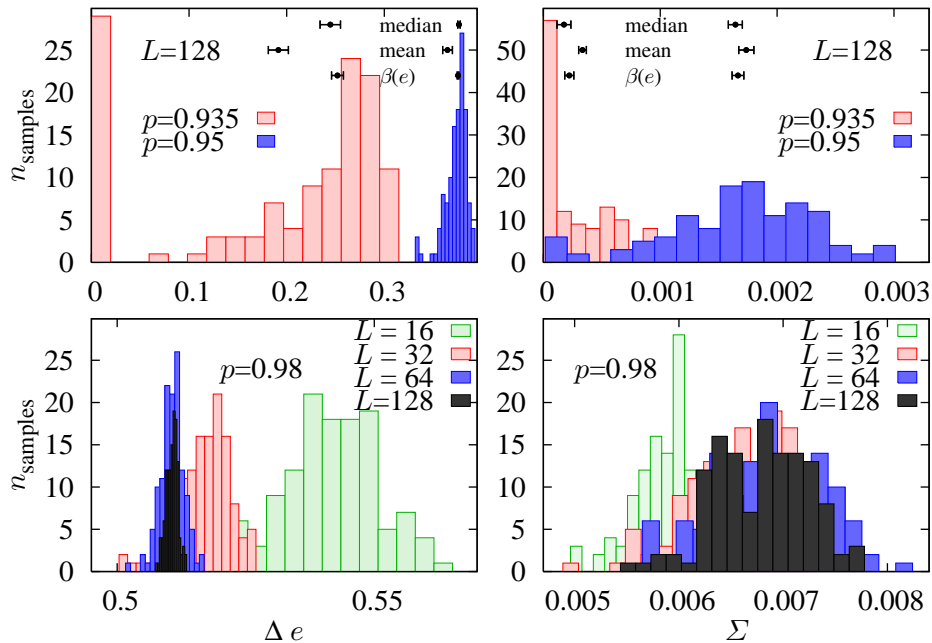}
\caption{Histograms for the sample-dependent
latent heat $\upDelta_{\{\epsilon\}} e= e_\mathrm{d}-e_\mathrm{o}$
({\bf left}) and surface tension, $\Sigma$({\bf right}). In the top panels we
show results in the largest lattice, where two very close spin
concentrations behave very differently.  The three types of
horizontal lines  drawn (indicating central value and statistical error)
correspond, from top to bottom, to the median, the mean, and the value
obtained from $\beta(e)$. In the lower panels we show the histograms
for $p=0.98$ and different $L$'s (note the difference in the
horizontal scales with the upper part). As can be seen, the latent heat is
self-averaging while the surface tension is not.}
\label{HISTOGRAMS_Q4}
\end{center}
\end{figure}

As was said before, for each sample we can define the
different thermal-averaged quantities, and then determine
their mean, median, or pdf.  We can also compute the
sample-average
$\beta(e)=\overline{\beta_{\{\epsilon\}}(e)}$, and then
perform the Maxwell construction on it. We  compared the
two approaches for this model both for the latent heat and
for the surface tension, see Fig.~\ref{HISTOGRAMS_Q4}. We found
 that the two approaches are equivalent, although the
second one requires less statistical accuracy for each
sample and is therefore less numerically demanding. Also
from Fig.~\ref{HISTOGRAMS_Q4} (top row) we can see that as
the dilution is slightly decreased (the tricritical point is
reached), a great number of samples present vanishing latent
heat and surface tension; the transition has become
continuous. Finally we found that within the first-order
part of the phase diagram ($p=0.98$) the width of the
histograms of the latent heat decreases as the lattice size
increases; this is the definition of a self-averaging
quantity. On the contrary, we can not see this behaviour for
the surface tension and therefore we can state that it is
not self-averaging.

\subsubsection{Latent heat and surface tension}
\label{potts3D:four-state:lat&tens}

\begin{figure}
\begin{center}
\includegraphics[height=0.55\textheight,angle=270]{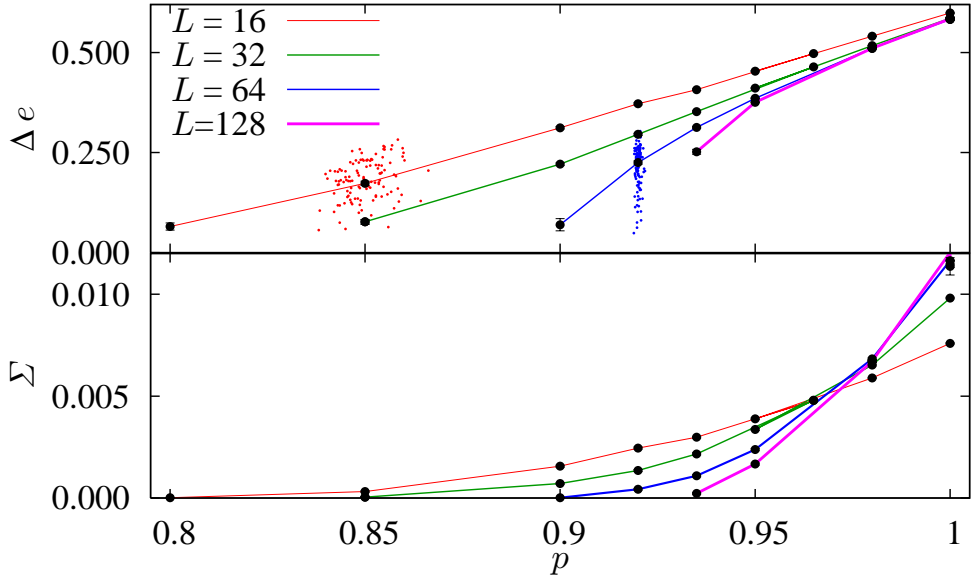}
\caption{{\bf Top:} Latent heat as obtained from
$\beta(e)$ as a function of spin concentration for several lattice sizes
(lines are linear interpolations). Data for $p=1$ and
$L\!=\!128$ were taken from Ref.~\cite{VICTORMICRO}.  To
illustrate the sample dispersion, we also show the
scatter-plot of ($N/L^D$, $\upDelta_{\{\epsilon\}}e$) for
the 128 samples at $L\!=\!16$ $p\!=\!0.85$ and $L\!=\!64$
$p\!=\!0.92$. {\bf Bottom:} As the top panel, but for the surface
tension.}
\label{LATENT-SIGMA_Q4}
\end{center}
\end{figure}

Our results for the behaviour of the latent heat and the surface tension
obtained from $\beta(e)$ as the dilution changes are shown in
Fig.~\ref{LATENT-SIGMA_Q4}. The apparent location of the tricritical
point (i.e., the $p$ where both $\upDelta e$ and $\Sigma$ vanish)
shifts to higher $p$ for increasing $L$ rather fast. For lattice sizes
comparable with those of previous work~\cite{Chat05}, $L=16$, we obtain a sizeable
value $p_\mathrm{c}^{L=16}\approx 0.75$, but the estimate of $p_\mathrm{c}$
increases very rapidly with $L$. An extrapolation to $L\to\infty$ is called for.

The pdf's for $\upDelta e$ and $\Sigma$, Fig.~\ref{HISTOGRAMS_Q4},
display an interesting $L$ evolution.  When the $\beta(e)$ changes
behaviour from non-monotonic ($L=64$, Fig.~\ref{BETAPROMEDIO_Q4},
bottom-right) to monotonic ($L=128$, Fig.~\ref{BETAPROMEDIO_Q4},
bottom-left), the two pdf's becomes enormously wide\footnote{The
estimates for $\upDelta e$ and $\Sigma$ are consistent with the {\em
median} of their (non-Gaussian) pdf's.}, see the top panels in
Fig~\ref{HISTOGRAMS_Q4}. This arises because for many $L=128$ samples
the curve $\beta_{\{\epsilon\}}(e)$ is becoming flat, or even
monotonically decreasing (i.e., $\upDelta e\!=\!\Sigma=0$), while no such
behaviour was seen for $L\!=\!64$.  Only for $p\!=\!0.98$ does the width of the
pdf's for $\upDelta e$ scale as $L^{-D/2}$, as expected for a
self-averaging quantity, see Fig.~\ref{HISTOGRAMS_Q4} -- bottom-left.  The
surface-tension is {\em not} self-averaging,
see Fig.~\ref{HISTOGRAMS_Q4} -- bottom-right.

\subsubsection{Finite Size Scaling study}
\label{potts3D:four-state:FSS}

From Figs.~\ref{BETAPROMEDIO_Q4},~\ref{HISTOGRAMS_Q4}, and~\ref{LATENT-SIGMA_Q4}
one cannot rule out that $p_c\neq1$: a
disordered first-order transition would not exist.  Fortunately we can
solve this dilemma by considering the correlation length, obtained
from the {\em sample-averaged} correlation function, Eq.~(\ref{xi}).

\begin{figure}
\begin{center}
  \includegraphics[height=0.55\textheight,angle=270]{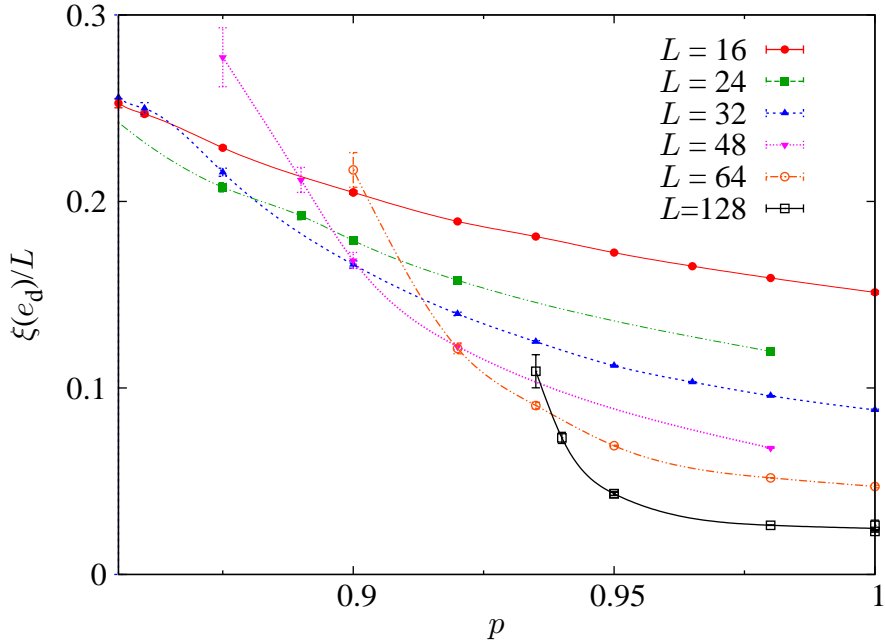}
  \caption{Correlation length in units of
    the lattice size in the $Q=4$ case, at phase coexistence for the paramagnetic
    phases, $e_\mathrm{d}$, as a function of concentration, $p$, for several system sizes, $L$ (lines are cubic spline
    interpolations for data at fixed $L$).}
\label{cortes_xi_ed_Q4}
\end{center}
\end{figure}

\begin{figure}
\begin{center}

  \includegraphics[height=0.55\textheight,angle=270]{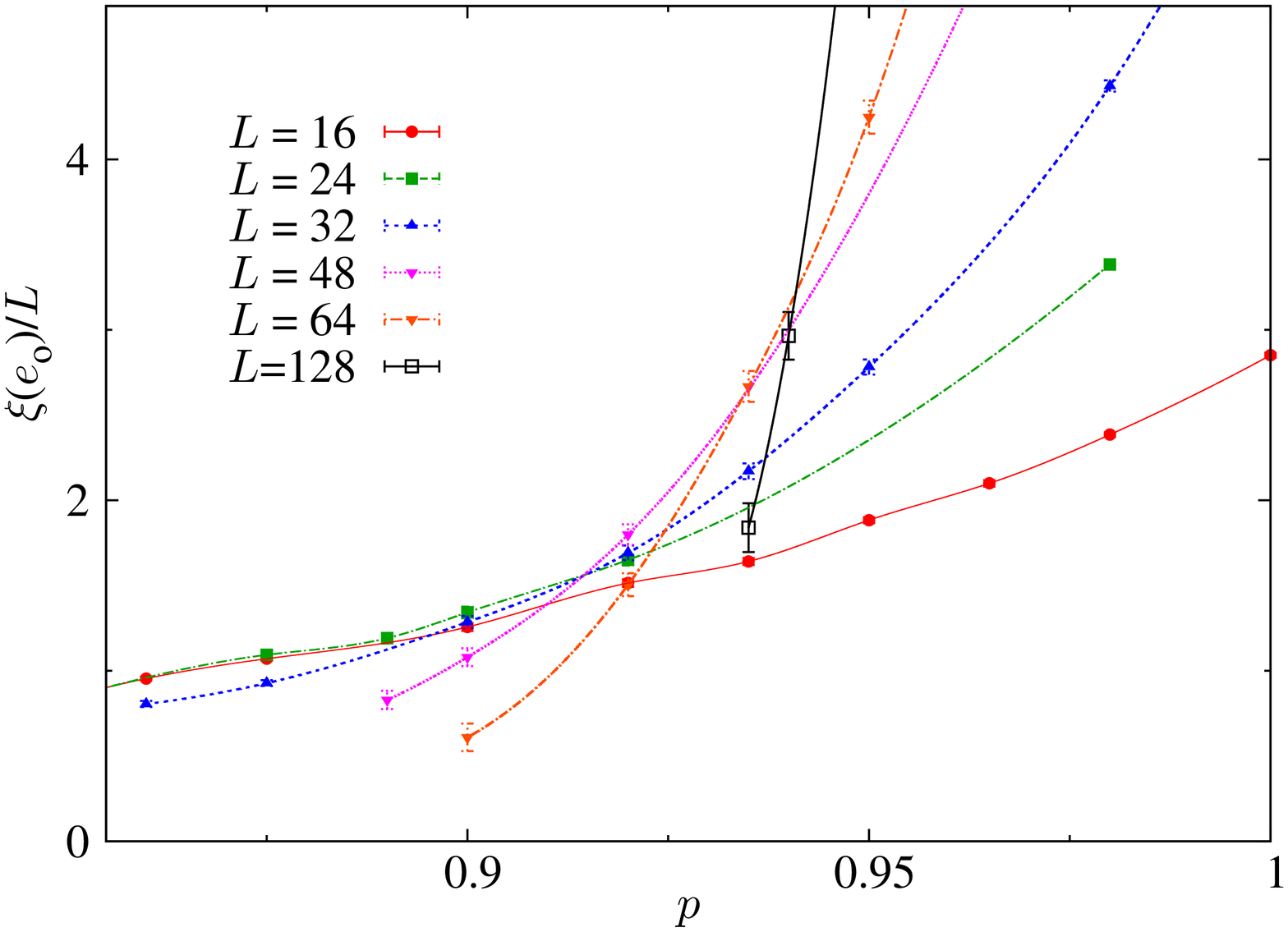}
  \caption{Correlation length in units of the lattice size in the $Q=4$ case, at phase coexistence for the ferromagnetic
    phases, $e_\mathrm{o}$, as a function of concentration, $p$, for several system sizes, $L$.}
\label{cortes_xi_eo_Q4}
\end{center}
\end{figure}

We take the correlation length in units of the lattice size at
$e_\mathrm{d}$ (see Fig.~\ref{cortes_xi_ed_Q4}), and $e_\mathrm{o}$
(see Fig.~\ref{cortes_xi_eo_Q4}), as obtained from $\beta(e)$ (a jack-knife
method~\cite{VICTORAMIT} takes care of the statistical correlations). For all
$p<p_\mathrm{c}$, one expects that both $\xi(e_\mathrm{d})/L$ and
$\xi(e_\mathrm{o})/L$ tend to non-vanishing and different limits for large
$L$\footnote{We have checked numerically that this is indeed the case
  for the $D\!=\!2$, $Q\!=\!4$, pure Potts model (a prototypical
  example of a second-order phase transition with a double-peaked
  canonical pdf for $e$ at $T_\mathrm{c}$), see Sec.~\ref{subsec:temperature}.}.
For
$p>p_\mathrm{c}$, $\xi(e_\mathrm{d})/L$ is of order $1/L$, while
$\xi(e_\mathrm{o})/L\sim L^{D/2}$. For a fixed $L$, with increasing
$p$, the behaviour goes from a second-order type to first-order (see
Fig~\ref{BETAPROMEDIO_Q4}). Hence, a FSS approach~\cite{VICTORAMIT} is needed.

\begin{figure}
\begin{center}
  \includegraphics[height=0.55\textheight,angle=270]{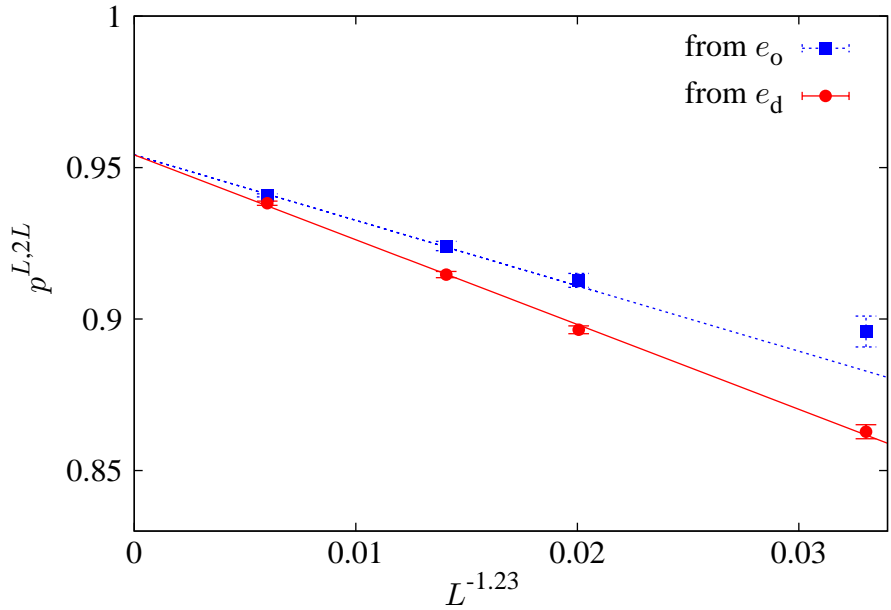}
  \caption{Spin
    concentration where the values of $\xi/L$ (data from Figs.~\ref{cortes_xi_ed_Q4} and 
    \ref{cortes_xi_eo_Q4}) coincide for
    lattices $L$ and $2L$ versus $1/L^{x}$, see
    Eqs.~(\ref{SCALING}) and~(\ref{RESULTADO-GORDO}). Lines are a combined fit
    for $x$, $p_\mathrm{c}$, $A_\mathrm{d}$, and $A_\mathrm{o}$.}
\label{ajuste_conjunto_Q4}
\end{center}
\end{figure}

Consider the curves of $\xi(e_\mathrm{d})/L$ versus $p$ for different
$L$, see Fig.~\ref{cortes_xi_ed_Q4}. There is a unique concentration,
$p^{L,2L}$, where the correlation lengths in units of the lattice size
coincide for pairs of lattices of sizes $L$ and $2L$. One has~\footnote{The tricritical point has no basin of
  attraction for the RG flow in the $(T,p)$ plane.
  Although two relevant scaling fields are to be expected, the Maxwell
  construction allows us to eliminate one of them and hence we use
  the formulae for a standard critical point.}
\begin{equation}
  p^{L,2L}\approx p_\mathrm{c} + A_\mathrm{d} L^{-x}\,.\label{SCALING}
\end{equation}
An exactly analogous result holds for $\xi(e_\mathrm{o})/L$, see Fig.~\ref{cortes_xi_eo_Q4}. Since
$A_\mathrm{d}$ and $A_\mathrm{o}$ are quite different, see
Fig.~\ref{ajuste_conjunto_Q4}, a combined fit of all the data yields an accurate
estimate of the location of the tricritical point:
\begin{equation}
  p_\mathrm{c}=0.954(3),\quad x= 1.23(9),\quad \frac{\chi^2}{\mathrm{d.o.f.}}=\frac{4.23}{3},\quad \text{C.L.}=24\%\, .\label{RESULTADO-GORDO}
\end{equation}
Of course, due to higher-order scaling corrections, Eq.~(\ref{SCALING})
should be used only for lattices larger than some
$L^\mathrm{min}$~\cite{QUOTIENTS}. The fit $\chi^2$ was acceptable taking
$L^\mathrm{min}_\mathrm{d}=12$ and $L^\mathrm{min}_\mathrm{o}=16$ (for
the sake of clarity we do not display data for $L=12$ in the figures).
Therefore we can conclude that $p=0.98$ is
definitively in the first-order part of the critical line.

\begin{figure}
\begin{center}
  \includegraphics[height=0.5\textheight,angle=270]{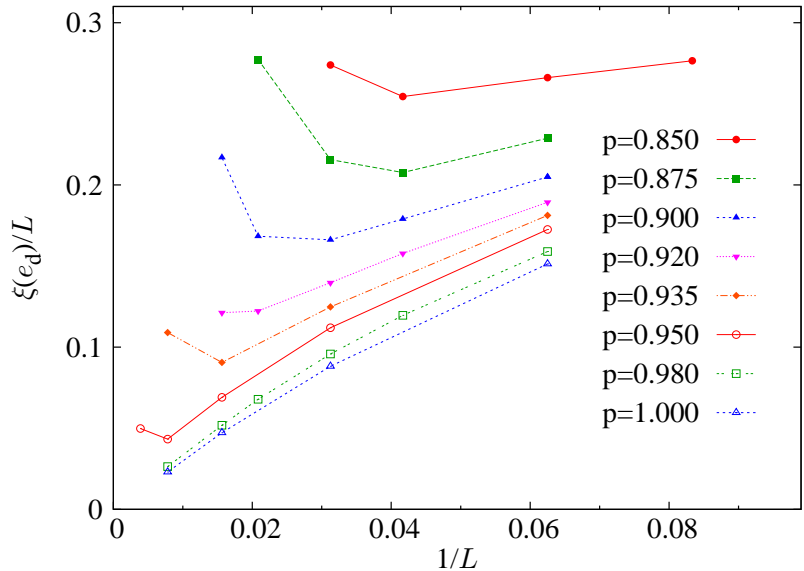}
  \caption{Correlation length at $e_d$ in units of
    the lattice size, for fixed dilutions as a function of the inverse
    lattice size. For fixed $p<p_\mathrm{c}$,
    $\xi(e_\mathrm{d})/L$ has a minimum at a crossover length scale,
    $L_\mathrm{cr}(p)$, that separates the first-order type of behaviour 
    from the second-order type.}
\label{XI2L_Q4}
\end{center}
\end{figure}

We now look at $\xi/L$ at $p^{L,2L}$, see Figs.~\ref{cortes_xi_ed_Q4} and \ref{cortes_xi_eo_Q4}.
Consider $\xi(e_\mathrm{d})/L$ as a function of $(L,p)$, see Fig.~\ref{XI2L_Q4}.
Its salient features are:
\begin{enumerate}
 \item For fixed $L$, $\xi(e_\mathrm{d})/L$ is a decreasing function of $p$
(while $\xi(e_\mathrm{o})/L$ is increasing).
\item  For fixed $p$, $\xi(e_\mathrm{d})/L$ has a minimum (while $\xi(e_\mathrm{o})/L$ has a
maximum) at a crossover length scale, $L_\mathrm{cr}(p)$, that
separates the first-order type of behaviour from the second-order type,
see Figs.~\ref{cortes_xi_ed_Q4} and \ref{cortes_xi_eo_Q4}.
\item At the crossing point $p^{L,2L}$ we have $L <
L_\mathrm{cr}(p^{L,2L}) < 2L$.
\item At least within the range of our
simulations, $L_\mathrm{cr}(p)$ is an increasing function of $p$.
\end{enumerate}
A standard scaling argument, combined with (1--4), yields that
$\xi(e_\mathrm{d})/L$ at $p^{L,2L}$ is of order
$1/L_\mathrm{cr}(p^{L,2L})$ ($\xi(e_\mathrm{o})/L\sim
L^{D/2}_\mathrm{cr}$).  If $L_\mathrm{cr}(p)$ diverges at
$p_\mathrm{c}$, $\xi(e_\mathrm{d})/L$ at $p^{L,2L}$ should tend to
zero for large $L$, which is indeed consistent with our data.

\subsection[$D=3$\,, $Q=8$ Site-Diluted Potts Model]{$\bm{D=3\,,\ Q=8}$ Site-Diluted Potts Model}
\label{potts3D:eight-state}

We also present in this chapter some of the preliminary
results of our study of the eight-state ($Q=8$) site-diluted
Potts model using basically the same methodological approach
as in the $Q=4$ case. First, however, it has to be stressed
that there are two important differences in this case:

\begin{itemize}
 \item We used chiefly another kind of computing
 platform. While the $Q=4$ case was entirely simulated on
 typical cluster facilities, i.e., the BSC and BIFI, the $Q=8$ case case
 was simulated on IBERCIVIS, a distributed computing
 platform based on BOINC, see
 Appendix~\ref{Appendix_Ibercivis}. This change in platform
 involved both advantages and disadvantages. By using
 IBERCIVIS, we were able to outperform broadly all previous
 statistical accuracies both in the number of samples (we
 were able to simulate up to 2000 samples of a system with
 $64^3$ spins) and in the number of dilution levels
 (around ten for each system size). In addition, we did more
 than $3\times10^6$ Swendsen-Wang steps at each energy of a
 system with $64^3$ spins. In particular, with
 IBERCIVIS we obtained more than 300 years of
 computation time in less than a year of wall clock time.
 This would have been hard to achieve using a traditional
 cluster facility.

On the other hand, the use of IBERCIVIS, a novel infrastructure,
led to numerous unusual problems in the adaptation and stabilisation of 
the original code to the new computing paradigm, see again Appendix~\ref{Appendix_Ibercivis}.
The huge output of the computations has to be carefully analysed, and a major
effort must be made to identify all the possible error sources.
For example, the connection of the individual
parts of each BOINC job is an extremely delicate issue, and the development of a secure mechanism
for the detection of corrupted outputs is fundamental.

\item It is known that the pure Potts model undergoes a first-order phase transition for $Q>2$ in three
dimensions~\cite{Wu}, with the strength of the first-order
character being larger as $Q$ grows. Therefore the $Q=8$ case will
show more evidently the features of this kind of transition
(i.e., latent heat, metastabilities, phase coexistence,
etc.).  This fact has pros and cons. The main benefit is
that if we want to see a first-order phase transition in the
presence of disorder, the first-order region in the phase
diagram is expected to be larger for $Q=8$ than for $Q=4$,
in other words $p_\text{c}$ will be smaller. This will
mean stronger evidence for the main result of this
chapter -- first-order phase transitions do exist in the presence
of disorder in three-dimensional systems.

On the contrary, the stronger first-order character of the
transition produces much more palpable metastability
effects. This is a huge problem in Monte Carlo simulations,
because exponential autocorrelation times will grow
substantially, see Appendix~\ref{Appendix_autocorr} and
Ref.~\cite{sokal96}, making thermalization really hard to
achieve for large systems at given values of their internal
energy.  This fact restricted us to thermalizing
systems with ``only'' $64^3$ spins on the first-order
side of the phase diagram, compared with the $Q=4$ case for which we
were able to thermalize systems with up to $128^3$
spins. Anyway we plan to study the first-order side of
the transition by making estimates of the errors due to not
having reached the asymptotic states of the system.
\end{itemize}

For $Q=8$ we simulated the model for $p$ values in the
range $ 0.65 \leq p\leq 1$.  For each $p$ value, we
simulated $L=12$, $16$, $24$, $32$, $48$, $64$, and $96$ (for
a given $p$, we did not consider larger lattices once the
latent heat vanished). Finally, we disregarded our
simulations for $L=96$ because of the impossibility of
thermalization in a reasonable time. For all pairs ($p$,
$L$) we simulated at least 500 samples.

\begin{figure}
\begin{center}
  \includegraphics[height=0.55\textheight,angle=270]{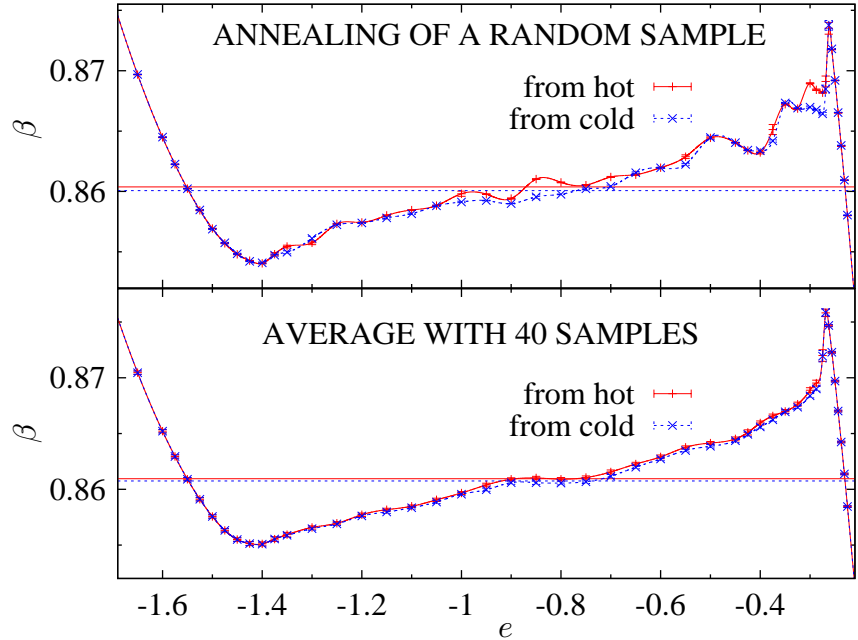}
  \caption{Comparison of the different ``energy walks'' for the system with $L=48$ and $p=0.95$.
  The upper part correspond to the Maxwell construction of a random sample. Note
  the clear difference between the walk starting from hot (red solid line) and cold (blue dashed line).
The lower part shows the softening of the difference as the sample average (with 40 samples)
  is performed. Horizontal lines mark the corresponding estimates of $\beta_\text{c}$.
}
\label{terma_Q8_L48p95}
\end{center}
\end{figure}

To check the thermalization of the systems we compared
simulations of the same samples (distribution of the
vacancies), performing annealings starting from both random
configurations at high temperatures and cold (all the spins in the same state)
configurations at low temperatures. As we performed our
``energy walk'' we found that for energies corresponding to
pure states (with no ``islands'' of the other phase) both
annealings will agree fully. However, between $e_\text{o}$
and $e_\text{d}$ there will exist some energies where the
two annealings will produce different estimates of the
observables (especially for the largest lattices).
These energies are precisely those at which
the system switches between the different configurations of
the ``islands'', for example from a ``droplet'' to a
``strip''.  In the case that the thermal averages of the
different observables from the two annealings were similar,
we would be fairly confident of the equilibration of the
system. If they were not, we could at least estimate the
error due to the lack of thermalization from their
difference.

In Fig.~\ref{terma_Q8_L48p95} we plot the comparison of the
annealings of the system with $48^3$ spins and
$p=0.95$. The system is clearly undergoing a first-order
phase transition. The simulation of \emph{each sample} used
in Fig.~\ref{terma_Q8_L48p95} took around three days of a
last generation Pentium $I7$ 3.0 GHz core and are clearly
not thermalized! With this in mind, it is clear that on the
first-order side of the phase diagram the simulations must
be really long in time to reach equilibrated states.  As was
said before, this fact will critically restrict us in
simulating large systems.

Due to the lack of mixed phases, thermalization is quite easy
to achieve on the second-order side of the phase diagram. No metastability will
exist and the cluster update method will work very well.

Therefore the approach to the problem must be very
different depending on the dilution of the system. If the
dilution is weak, the systems will undergo first-order phase
transitions and we will not be able to simulate large
systems. Nevertheless we can estimate the latent heat and
the surface tension to obtain the exponents of their
scaling, always taking into account the possibility of
unequilibrated systems. On the contrary, if the dilution is
important, since the systems will undergo second-order phase
transitions, we will be able to equilibrate large systems
(with $64^3$ spins) and to obtain accurate results
for the tricritical point location. In this work we
present only the latter study, i.e., the study of the exact
location of the tricritical point. The study of the
 first-order side will be left for further research.

\subsubsection{General behaviour}
\label{potts3D:eight-state:general}

\begin{figure}
\begin{center}
  \includegraphics[height=0.52\textheight,angle=270]{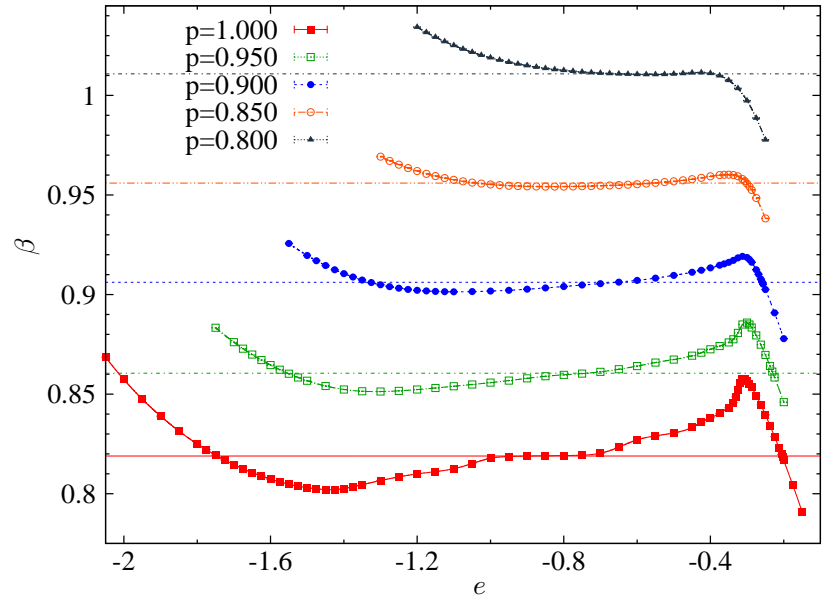}
  \caption{Maxwell constructions for the system with $L=24$ as a function of
the spin concentration, $p$. In the dilute cases, each point represents the average of the temperature
for 500 samples. Note that for this size we can form the construction even for $p=0.800$.
}
\label{Maxwell_Q8_L24}
\end{center}
\end{figure}

First we outline the behaviour of the model, which is very similar to
that of the $Q=4$ case although the first-order character is stronger. 
Firstly, by taking the sample average of the Maxwell constructions, see Sec.~\ref{potts3D:sampleaverages},
we can obtain for each system size the behaviour of the model
as the dilution changes, see Fig.~\ref{Maxwell_Q8_L24} for the system with $L=24$. Note that 
in this case we obtain 
clean Maxwell constructions up to $p=0.800$. The system is undergoing (on average)
a first-order phase transition even with $20\%$ of vacancies! It is also remarkable
that in the pure case, $p=1$, the Maxwell construction is smooth,
without flat parts or strong steps between consecutive energies.

\begin{figure}
\begin{center}
  \includegraphics[height=0.52\textheight,angle=270]{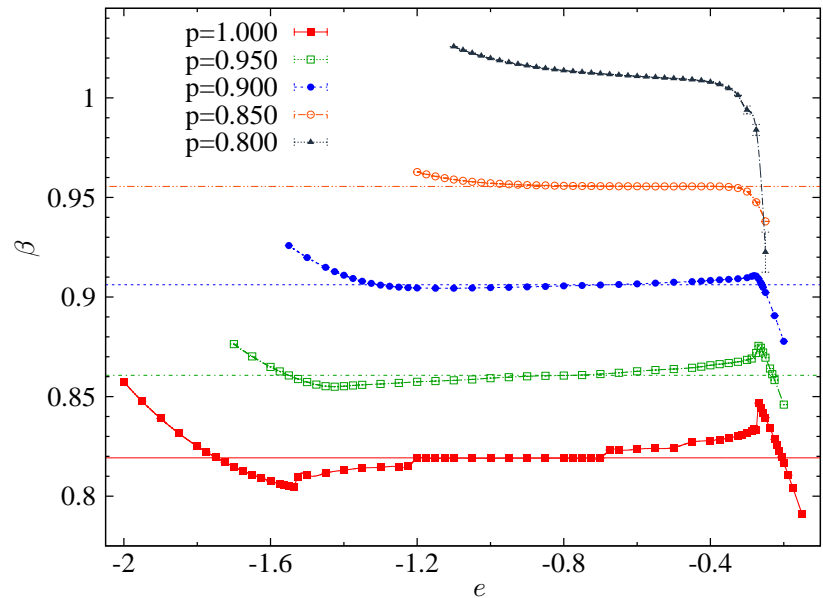}
  \caption{Maxwell constructions for the system with $L=48$
  as a function of the spin concentration, $p$. In the
  dilute cases, each point represents the average of the
  temperature for 500 samples. Note that we can form the
  construction only up to $p=0.850$, the transition has
  become of first order between $p=0.850$ and $p=0.800$.  }
\label{Maxwell_Q8_L48}
\end{center}
\end{figure}

By representing the same plot for a larger system, see
Fig.~\ref{Maxwell_Q8_L24} for the $L=48$ case, we find that
Maxwell constructions can not be formed for dilutions
less than $p=0.850$.  As was said in
Sec.~\ref{potts3D:four-state:general}, the apparent tricritical point
depends on the system size (a FSS study is again called
for). We can also see that the Maxwell construction in the
pure case presents flat parts and clear steps in the
temperature. They are due to the existence of clear mixed
regimes with droplet or strip-like configurations in which
the internal energy  basically does not change over a range of
energy densities. This is also manifest in the case of the
mildly dilute samples prior to the sample-averaging
process. The location of the flat parts and the steps is by
far the part of the ``energy walk'' that is most sensitive to the
metastability effects; it is really difficult for any Monte Carlo
spin update method to perform properly with
configurations of this kind.

\begin{figure}
\begin{center}
\includegraphics[height=0.6\textheight,angle=270]{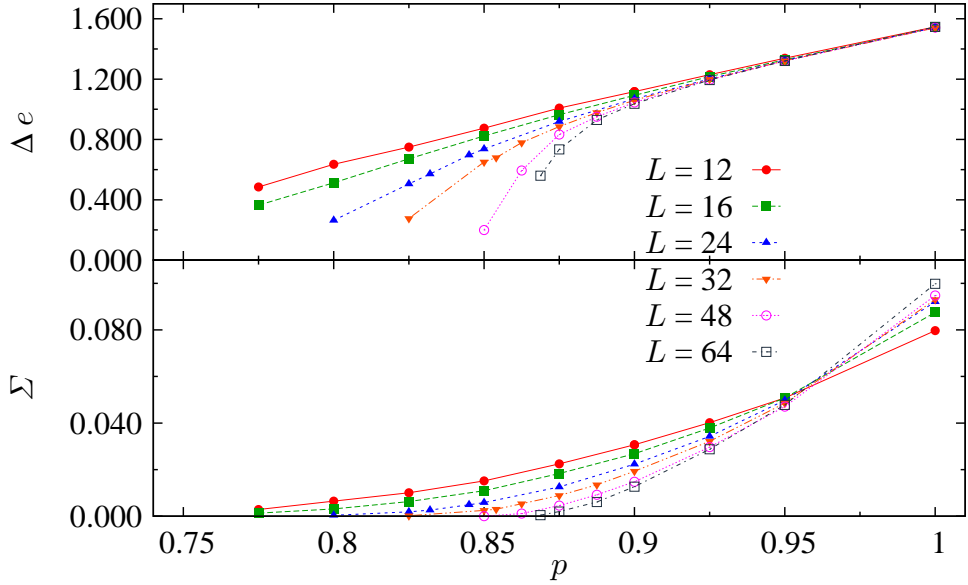}
\caption{{\bf Top:} Latent heat as obtained from the averaged Maxwell construction as a function
of $p$ for each simulated lattice size (lines are linear
interpolations) for the $Q=8$ case. {\bf Bottom:} As in
the top panel, but for the surface tension.}
\label{LATENT-SIGMA_Q8}
\end{center}
\end{figure}

Again we can plot the entire behaviour of the latent heat
and the surface tension as a function of the system size and
the dilution, see Fig.~\ref{LATENT-SIGMA_Q8}. This
figure must be compared with that corresponding to the $Q=4$
case, Fig.~\ref{LATENT-SIGMA_Q4}. While in the $Q=4$
case the tricritical dilution is clearly above $p=0.95$,
this is not so for $Q=8$.  The data for both the latent
heat and the surface tension show the clear trend of the
tricritical dilution towards larger values as
the number of Potts states grows. The points in
Fig.~\ref{LATENT-SIGMA_Q8} corresponding to large values of
the dilution and the lattice size ($p>0.925,\ L>32$) are
possibly not fully equilibrated, so special treatment of
the data is needed to obtain accurate information of the
scaling in this part of the phase diagram. This will be done
in future work.


\subsubsection{Behaviour of the model on the second-order side}
\label{potts3D:eight:secondorderside}

\begin{figure}
\begin{center}
  \includegraphics[height=0.52\textheight,angle=270]{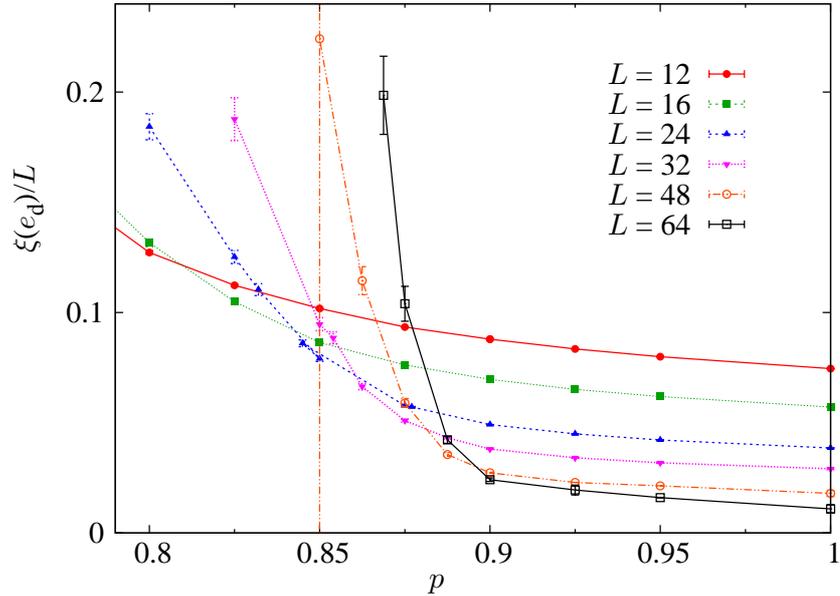}
  \caption{Correlation length for the $Q=8$ case in units of
    the lattice size, at phase coexistence for the paramagnetic
    phases, $e_\mathrm{d}$, as a function of concentration for several system sizes, $L$.
    The huge error bar for $L=48,\ p=0.85$ is due to the enormous
    sample-to-sample dispersion of this observable close to the tricritical point.}
\label{cortes_xi_ed_Q8}
\end{center}
\end{figure}

\begin{figure}
\begin{center}

  \includegraphics[height=0.52\textheight,angle=270]{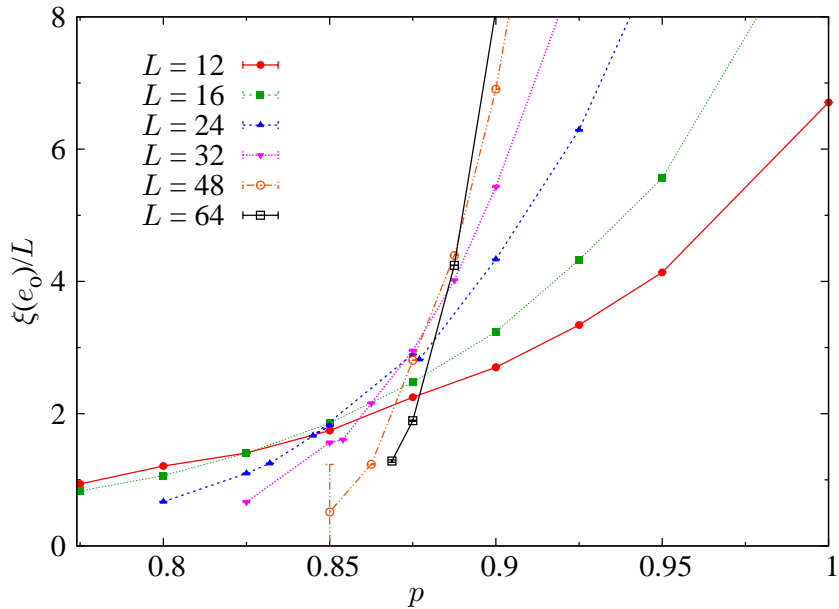}
  \caption{Correlation length for the $Q=8$ case in units of the lattice size, at phase coexistence for the ferromagnetic
    phases, $e_\mathrm{o}$, as a function of concentration for several system sizes, $L$.}
\label{cortes_xi_eo_Q8}
\end{center}
\end{figure}

From Fig.~\ref{LATENT-SIGMA_Q8}, while one can not obtain an accurate estimate of $p_c\neq1$,
one can again consider the correlation length obtained
from the {\em sample-averaged} correlation function, Eq.~(\ref{xi}). We
use then the same approach as we used in Sec.~\ref{potts3D:four-state:FSS}
computing the crossings of the correlation length in units of the lattice size at
$e_\mathrm{d}$, see Fig.~\ref{cortes_xi_ed_Q8}, and $e_\mathrm{o}$,
see Fig.~\ref{cortes_xi_eo_Q8}, as obtained from $\beta(e)$. 

\begin{figure}
\begin{center}
  \includegraphics[height=0.52\textheight,angle=270]{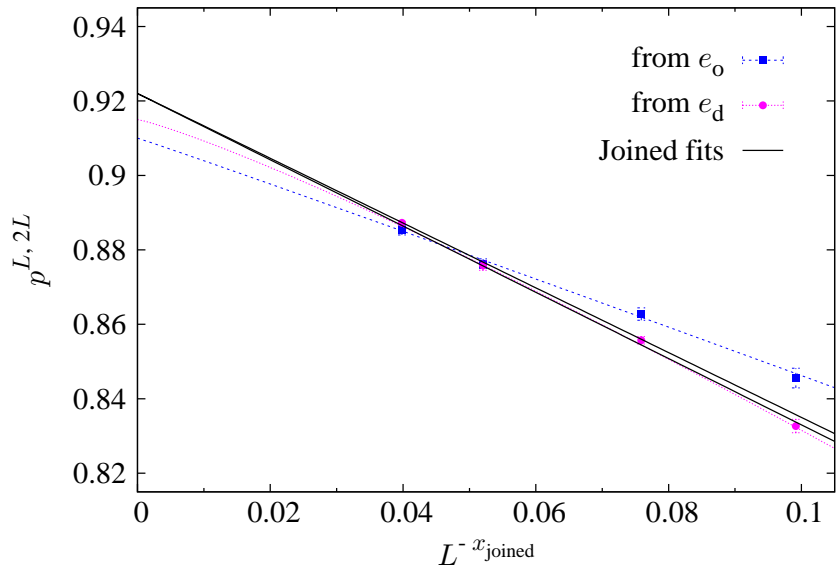}
  \caption{Spin concentration where values of $\xi/L$ (data
  from Figs.~\ref{cortes_xi_ed_Q8} and
  \ref{cortes_xi_eo_Q8}) coincide for lattices $L$ and $2L$
  versus $1/L^{x_\mathrm{joined}}$, see
  Eq.~(\ref{RESULTADO-joined-Q8}).  Solid lines are the
  combined fit, while the dashed and dotted curves correspond
  to the individual fits of the crossing of
  $\xi(e_\mathrm{d})$ and $\xi(e_\mathrm{o})$ respectively, Eqs.~(\ref{RESULTADO_ed_Q8}) and~(\ref{RESULTADO_eo_Q8}).}
\label{ajuste_conjunto_Q8}
\end{center}
\end{figure}

As was done in Sec.~\ref{potts3D:four-state:FSS}, we define
$p^{L,2L}$ as the crossing points of the correlation length
$\xi(e_\mathrm{d})$ (in lattice size units) for pairs of
lattices with $L$ and $2L$. We can fit these points again to the
form of Eq.~(\ref{SCALING}) to obtain the value of the tricritical dilution $p_\mathrm{c}$. Fitting our
data set in $e_\mathrm{d}$, see Fig.~\ref{cortes_xi_ed_Q8}, we obtain:
\begin{equation}
p_{\mathrm{c},\, e_\mathrm{d}}=0.915(1),\quad x= 1.14(28),\quad \frac{\chi^2}{\mathrm{d.o.f.}}=\frac{1.5}{2}, \quad \text{C.L.}=47\% \,,
\label{RESULTADO_ed_Q8}
\end{equation}
which is a perfectly valid fit producing a value for $p_\mathrm{c}$ clearly less than unity.
Using the same approach for the crossings of $\xi(e_\mathrm{o})$, see Fig.~\ref{cortes_xi_eo_Q8}, we obtain
a valid fit with parameters
\begin{equation}
 p_{\mathrm{c},\, e_\mathrm{o}}=0.910(2),\quad x= 0.95(49),\quad \frac{\chi^2}{\mathrm{d.o.f.}}=\frac{1.1}{2}, \quad \text{C.L.}=58\% \,.
\label{RESULTADO_eo_Q8}
\end{equation}
Finally we can fit both data series to the form~(\ref{SCALING}) sharing the same coefficients $p_\mathrm{c}$
and $x$. To get an acceptable value for the $\chi^2$ of the fit we had to disregard
the data with $L<16$ for $\xi(e_\mathrm{d})$ and $L<24$ for $\xi(e_\mathrm{o})$ obtaining the
fitting parameters:
\begin{equation}
p_\mathrm{c,\, joined}=0.922(1),\quad x_\mathrm{joined}= 0.93(47),\quad \frac{\chi^2}{\mathrm{d.o.f.}}=\frac{1.50}{2}, \quad \text{C.L.}=47\% \,.
\label{RESULTADO-joined-Q8}
\end{equation}
A plot of all the above fits is shown in
Fig.~\ref{ajuste_conjunto_Q8}. Therefore we can firmly
conclude that $p=0.925$ is in the first-order part of the
critical line for the three-dimensional Potts model with
$Q=8$. This is another result that reinforces our
main conclusion of the previous section, i.e., first-order phase transitions do exist in $D=3$.

\section{Conclusions}
\label{potts3D:conclusions}

In this chapter we have performed a detailed study of the
effects of quenched disorder on a three-dimensional system
undergoing a first-order transition in the pure case.  We
studied the site-diluted version of both the $Q\!=\!4$ and
the $Q\!=\!8$ Potts model, a model undergoing a
prototypically strong first-order transition, with the
strength being proportional to the value of $Q$.  A small
degree of dilution smooths the transition up to the point of
becoming second order at a tricritical point,
$p_\mathrm{c}$. We observed strong finite-size effects
in both the location of the tricritical point and
the behaviour of the most relevant quantities (latent heat,
surface tension, correlation length, etc.).  A delicate
FSS analysis  allowed us to firmly
conclude that $p_\mathrm{c}<1$, with
$p_\mathrm{c}=0.954(3)$ and $p_\mathrm{c}=0.922(1)$ in the
$Q=4$ and $Q=8$ cases respectively. We are then able to
claim that (quenched) disordered first-order transitions do
exist in three dimensions, although quenched disorder is
unreasonably effective in smoothing the transition (we
speculate that the percolation mechanism for colossal
magnetoresistance proposed in~\cite{MANGA} could be fairly
common in $D\!=\!3$).

We also observed that, for a given $p<p_\mathrm{c}$, a
crossover length scale $L_\mathrm{cr}(p)$ exists such that for
$L<L_\mathrm{cr}(p)$ the behaviour is of first-order type.  The
asymptotic second-order behaviour appears only for
$L>L_\mathrm{cr}(p)$. 

In the $Q=4$ case, we also verified that the latent heat is a
self-averaging quantity for random first-order phase
transitions while the surface tension is not. We will try to
verify this point for the $Q=8$ case in future work.

All these results were made possible first by a
new definition of the quenched average that avoids long-tailed
pdf's~\cite{Chat05}, and second by the use of a recently introduced
microcanonical Monte Carlo method that features the entropy density
rather than the free energy~\cite{VICTORMICRO}.

As further research, we will obtain novel information on
the scaling of some quantities on the first-order part of
the critical line in the $Q=8$ case. To perform this
analysis, we will have to deal with systems that are not fully
equilibrated. The characterisation of the effects due to the
metastable states will be done by comparing pairs of
simulations performing annealings from hot and cold states.

\clearpage{\thispagestyle{empty}\cleardoublepage}

\clearpage{\thispagestyle{empty}\cleardoublepage}

\normalfont

\vspace{4cm}

\selectlanguage{british}

\chapter{The Site-Diluted Heisenberg Model in Three Dimensions}
\label{chap:O3}

\section{Introduction}
\label{O3:intro}

The three-dimensional Heisenberg model is the most general
representation of the interaction of the spins within an
isotropic magnetic material, where isotropic means that the
magnetisation does not have any preferential direction to
point to.  Besides, other popular models such as the Ising or
$XY$ models describe materials with a plane or axis
of easy magnetisation, as is the case for instance of
hexagonal lattices where the magnetisation usually chooses
as preferential orientation either the $c$ axis (correctly
described then by the Ising model) or its orthogonal plane
(an $XY$ model is then correct).

The three-dimensional site-diluted Heisenberg model
correctly describes the experimental behaviour of a large
number of real dilute magnetic materials, see
Table~\ref{experiments_O3}, so we will be able to compare
our numerical results with some experimental estimates.
\begin{table}[ht]
\begin{center}
\begin{tabular}{|c|c|c|c|c|}\hline
Ref. & Material & $\gamma$ & $\beta$& $\delta$\\\hline\hline
\cite{kaul}$_{1994}$          & Fe$_{10}$Ni$_{70}$Bi$_{19}$Si      & 1.387(12) & 0.378(15) & 4.50(5)\\\cline{1-5}
\cite{kaul}$_{1994}$          & Fe$_{13}$Ni$_{67}$Bi$_{19}$Si      & 1.386(12) & 0.367(15) & 4.50(5)\\\cline{1-5}
\cite{kaul}$_{1994}$          & Fe$_{16}$Ni$_{64}$Bi$_{19}$Si      & 1.386(14) & 0.360(15) & 4.86(4)\\\cline{1-5}
\cite{samba1,samba2}$_{1995}$ & Fe$_{20}$Ni$_{60}$P$_{14}$B$_{6}$  & 1.386(10) & 0.367(10) & 4.77(5)\\\cline{1-5}
\cite{samba1,samba2}$_{1995}$ & Fe$_{40}$Ni$_{40}$P$_{14}$B$_{6}$  & 1.385(10) & 0.364(5)  & 4.79(5)\\\cline{1-5}
\cite{babu}$_{1997}$          & Fe$_{91}$Zr$_{9}$                  & 1.383(4)  & 0.366(4)  & 4.75(5)\\\cline{1-5}
\cite{babu}$_{1997}$          & Fe$_{89}$CoZr$_{10}$               & 1.385(5)  & 0.368(6)  & 4.80(4)\\\cline{1-5}
\cite{babu}$_{1997}$          & Fe$_{88}$Co$_{2}$Zr$_{10}$         & 1.389(6)  & 0.363(5)  & 4.81(5)\\\cline{1-5}
\cite{babu}$_{1997}$          & Fe$_{84}$Co$_{6}$Zr$_{10}$         & 1.386(6)  & 0.370(5)  & 4.84(5)\\\cline{1-5}
\cite{said}$_{1999}$          & Fe$_{1.85}$Mn$_{1.15}$Si           & 1.543(20) & 0.408(60) & 4.74(7)\\\cline{1-5}
\cite{said}$_{1999}$          & Fe$_{1.50}$Mn$_{1.50}$Si           & 1.274(60) & 0.383(10) & 4.45(19)\\\cline{1-5}
\cite{baran}$_{1999}$         & MnCr$_{1.9}$In$_{0.1}$S$_4$        & 1.39(1)   & 0.36(1)   & 4.814(14)\\\cline{1-5}
\cite{baran}$_{1999}$         & MnCr$_{1.8}$In$_{0.2}$S$_4$        & 1.39(1)   & 0.36(1)   & 4.795(10)\\\cline{1-5}
\cite{peru1}$_{2000}$         & Fe$_{86}$Mn$_{4}$Zr$_{10}$         & 1.381(12) & 0.361     &   \\\cline{1-5}
\cite{peru1}$_{2000}$         & Fe$_{82}$Mn$_{8}$Zr$_{10}$         & 1.367(12) & 0.363     &   \\\cline{1-5}
\cite{peru2}$_{2001}$         & Fe$_{84}$Mn$_{6}$Zr$_{10}$         & 1.37(3)   & 0.359     & 4.81(4)\\\cline{1-5}
\cite{peru2}$_{2001}$         & Fe$_{74}$Mn$_{16}$Zr$_{10}$        & 1.39(5)   & 0.361     & 4.86(3)\\\hline
\end{tabular}
\caption{Experimentally-obtained critical exponents of materials which are expected to be
described by the three-dimensional site-diluted Heisenberg model with quenched disorder.
Table from Ref.~\cite{PELI-REP}. The results we obtain in this work are:
$\gamma=1.398(6)$, $\beta=0.370(2)$, and $\delta=4.775(5)$.}
\label{experiments_O3}
\end{center}
\end{table}

In this model, according to the Harris
criterion~\cite{critharris}, see
Appendix~\ref{Appendix_Harris}, the disorder is
irrelevant. We want to check this point through numerical simulation
by measuring critical exponents and different cumulants for different
values of the dilution. If they do not depend on the
dilution and agree with the pure case values, they will all
belong to the same Universality Class (UC) and the Harris
criterion will be re-verified.

In addition we will study the self-averaging properties of the model
computing at criticality the quantity $R_\chi$, which will be
defined below and is a measure of the self-averageness of
the susceptibility. We will show results strongly supporting
that this $R_\chi$ cumulant is zero at the critical point,
but only taking into account the scaling corrections. This
runs against some theoretical predictions~\cite{korut}
but supports others~\cite{harris,harris2}.

We will obtain high-precision measurements of the observables for
each lattice size near the critical point, so it will be
necessary to take into account their finite-size effects in
order to obtain asymptotic results. This implies estimating
the correction to scaling exponents, whose leading term is
denoted $\omega$, related to irrelevant operators in the
Renormalization Group (RG) language. To this end, we will use the shift of the
crossing points both for the Binder cumulant and for the
correlation length for  lattice pairs of different sizes
near the critical point. This study will also provide
estimates of the asymptotic critical temperature value.  We will also
check that including the correction to scaling terms is
crucial for the comparison of the values we obtain for the critical
exponents with those of other workers.

The simulations of this chapter were done mainly on the
BIFI cluster.  This consists of Xeon Dual Core 64-bit
3.40 GHz processors, with 2 GB of shared RAM.  We used
around fifty nodes for nine months making a total of around
17 years of computation time.

\section{Analytical Framework}
\label{O3:analytics}

The self-averaging (SA) of the susceptibility is defined in
terms of:
\begin{equation}
R_\chi \equiv \frac{\overline{ \langle {\cal M}^2 \rangle^2 - 
\overline{\langle {\cal M}^2 \rangle}^2
}}
{\overline{ \langle {\cal M}^2 \rangle}^2 }\quad\,,
\label{RsubChi}
\end{equation}
with $\cal M$ being the total magnetisation. The
susceptibility is self-averaging if $R_\chi \to 0$ as $L\to
\infty$.

In Ref.~\cite{harris}, the following picture was found:
\begin{enumerate}
\item Away from the critical temperature: $R_\chi=0$.
 On the basis of the RG or using general statistical
 arguments, one can find that $R_\chi \propto (\xi/L)^d$ in a finite geometry,
 $L$ being the system size and $\xi$ the correlation length which is
 finite for $T \neq T_c$. Then $R_\chi\to 0$ as $L\to \infty$. This is
 called \emph{Strong SA}.
\item At the critical temperature, a RG analysis opens up two
  possible scenarios:
\begin{itemize}

\item Models in which according to the Harris criterion the disorder
  is relevant ($\alpha_\mathrm{pure}>0$): $R_\chi\neq 0$. The
  susceptibility at the critical point is not self-averaging. In
  particular, Ref.~\cite{harris} shows that under these conditions
  $R_\chi$ is proportional to the fixed-point value of the coupling
  which induces the disorder in the Hamiltonian, which controls the
  new UC. This is called \emph{No SA}.
\item Models in which  according to the Harris criterion the disorder
  is not relevant ($\alpha_\mathrm{pure}<0$):
  $R_\chi=0$. The susceptibility at the critical point is
  self-averaging. In a finite geometry $R_\chi$ scales as
  $L^{\alpha/\nu}\to 0$, where $\alpha$ and $\nu$ are the
  critical exponents of the pure system, which are the same
  in the disordered one. This is called \emph{Weak SA}.
\end{itemize}
\end{enumerate}

The observable $R_\chi$ has been measured in other dilute models, for
example in the four-dimensional dilute Ising model, see
Ref.~\cite{ISDIL4D}. In this model a Mean Field computation and a
numerical one found a non-zero value for $R_\chi$ although the dilute
model was shown to belong to the same UC as the pure
model, contradicting the conclusions of Ref.~\cite{harris}.  One can
claim that the logarithms involved in the upper critical dimension 
make the numerical analysis difficult. In particular it was found
analytically in the mean field that $R_\chi=0.31024$ and numerically
that $R_\chi\in[0.15,0.32]$. Because of the logarithms, it was impossible to
make an infinite volume extrapolation for the numerical values of
$R_\chi$.  Notice that in this model the only fixed point is the
Gaussian one (all the values of the couplings are zero) and, following
Ref.~\cite{harris}, $R_\chi$ should be zero.

In addition a two-loop field theory calculation done in
Ref.~\cite{korut} predicts a non-zero value for $R_\chi$ for the
dilute Heisenberg model (in which the disorder is irrelevant,
$\alpha_\mathrm{pure}=-0.134$, see Ref.~\cite{peli}). The two-loop
field theoretical prediction for $\alpha$ in the pure case was
$\alpha_\mathrm{pure}>0$, so that apparently this work is consistent with
the findings of Ref.~\cite{harris}. The starting point in
Ref.~\cite{korut} was the mean field computation done in
Ref.~\cite{ISDIL4D}, modifying it to take into account the vector
degrees of freedom, introducing the fluctuations using the
Brezin-Zinn-Justin (BZJ) method, Ref.~\cite{BZJ}. They found
analytically $R_\chi=0.022688$ for the vector channel and universally
(independent of the dilution for all $p<1$). It is important to remark
that in the BZJ method one fixes from the beginning the temperature of
the system to the infinite volume critical value, working in a finite
geometry, so in order to compute $R_\chi$ in this scheme the following
sequence of limits is used:
\begin{equation}
R_\chi^*=\lim_{L\to \infty} \lim_{T\to T_c} R_\chi(L,T)\,,
\end{equation}
where $R_\chi^*$ is the infinite volume extrapolation at
criticality of $R_\chi(L,T)$, and $T_c$ is the infinite
volume critical temperature of the system.  The other
possible limit sequence that can be computed is:
\begin{equation}
\lim_{T\to T_c} \lim_{L\to \infty}  R_\chi(L,T)\,,
\end{equation}
which is zero even when the disorder is relevant since $R_\chi \propto
L^{-d}$ as $T \neq T_c$.

Hence, in order to test these discrepancies we simulated
numerically the site-diluted three-dimensional Heisenberg
model computing $R_\chi^*$ in the vector and tensor
channels.  To perform this programme, in particular in doing
the infinite volume extrapolations of cumulants and
exponents, a proper use of the corrections to scaling is
really important.

\section{The Model}
\label{O3:themodel}

The Heisenberg site-diluted model in three dimensions is defined in terms
of O($3$) spin variables placed at the nodes of a cubic three-dimensional lattice, with
Hamiltonian
\begin{equation}
H=-\beta\sum_{<i,j>} \epsilon_i \epsilon_j \boldsymbol{\mathit{S}_i}\cdot\boldsymbol{\mathit{S}_j}\, ,
\label{heismodel}
\end{equation}
where the $\boldsymbol{\mathit{S}_i}$ are three-dimensional vectors of
unit modulus, and the sum is extended only over nearest
neighbours. The disorder is introduced by the random variables
$\epsilon_i$ which take value unity with probability $p$ and zero with
probability $1-p$. An actual $\lbrace\epsilon_i\rbrace$ configuration
will be called a \emph{sample}.

In addition, as done in Ref.~\cite{UCMOND3}, we define a tensorial
channel associated with the vector $\boldsymbol{\mathit{S}}$ through
the traceless tensor
\begin{equation}
\tau_i^{\alpha\beta}=S_i^\alpha S_i^\beta-\frac{1}{3}\delta^{\alpha\beta}\ ,\qquad
\alpha,\beta=1,2,3\ .
\label{tensorialspin}
\end{equation}

We define the
total nearest-neighbour energy as
\begin{equation}
{\cal E} =\sum_{\langle
i,j\rangle}\epsilon_i \epsilon_j \boldsymbol{\mathit{S}_i} \cdot \boldsymbol{\mathit{S}_j}\ ,
\label{totalenergy}
\end{equation}
and the normalised magnetisation for both channels as
\begin{equation}
{\cal M}=\frac{1}{V}\sum_i \epsilon_i\boldsymbol{\mathit{S}_i}\ ,
\label{vectorialmag}
\end{equation}
\begin{equation}
{\mathcal M}^{\alpha \beta}_T =\frac{1}{V}\sum_i \epsilon_i(
\mathit{S}^\alpha_i \mathit{S}^\beta_i-\frac{1}{3}\delta^{\alpha\beta}) \,,
\label{tensorialmag}
\end{equation}
with $V=L^3$ and $L$ is the linear lattice size.
Because of the finite probability of reaching every minimal value
for the free energy, the thermal average of
Eqs.~(\ref{vectorialmag})~and~(\ref{tensorialmag}) is zero in a
finite lattice. Therefore, we have to define the order parameters as
the O(3) invariant scalars
\begin{equation}
M=\overline{\left\langle \sqrt{{\cal M}^2}\right\rangle} \quad ,
\quad M_T=\overline{\left\langle \sqrt{\mathrm{tr}
{\mathcal M}_T^2}\right\rangle}\,.
\label{orderparameters}
\end{equation}
Notice that the mean value of a non-invariant O(3) observable is automatically zero.

We also define the two susceptibilities as:
\begin{equation}
\chi=V\overline{\left\langle {\cal M}^2 \right\rangle}\quad ,\quad
\chi_T=V \overline{\left\langle 
                \mathrm{tr}{\mathcal M}_T^2\right\rangle}\,.
\label{susceptibilities}
\end{equation}

A very useful quantity is the Binder parameter, defined as
\begin{equation}
g^V_4=1-\frac{1}{3}\frac{\overline{\langle {\cal M}^4\rangle}}
           {\overline{\langle {\cal M}^2 \rangle}^2} \quad ,\quad
g^T_{4}=1-\frac{\overline{\langle {(\mathrm{tr}
{\mathcal M}_T^2)}^2\rangle}} 
{3 \overline{{\langle {\mathrm{tr} {\mathcal M}_T^2}\rangle}}^2}\,.
\label{g4cumulants}
\end{equation}

Another kind of Binder parameter, meaningless for the pure system, can be
defined as
\begin{equation}
g^V_2=\frac{\overline{ \langle {\cal M}^2 \rangle^2 - 
\overline{\langle {\cal M}^2 \rangle}^2
}}
{\overline{ \langle {\cal M}^2 \rangle}^2 }\quad ,\quad 
g^T_{2}=\frac{\overline{ \langle \mathrm{tr}
{\mathcal M}_T^2 \rangle^2 - 
\overline{\langle \mathrm{tr}{\mathcal M}_T^2 \rangle}^2
}}
{\overline{ \langle \mathrm{tr}{\mathcal M}_T^2 \rangle}^2 }\,,
\label{g2cumulants}
\end{equation}
and these are the quantities we use to estimate the self-averaging properties
of the susceptibility ($R_\chi$) in both channels.

A very convenient definition of the correlation length in a finite lattice 
is, see Ref.~\cite{COOPER},
\begin{equation}
\xi=\left(\frac{\chi/F-1}{4\sin^2(\pi/L)}\right)^\frac{1}{2},
\label{XI}
\end{equation}
where $F$ is defined in terms of the Fourier transform of the
magnetisation
\begin{equation}
\boldsymbol{{\cal F}}(\boldsymbol{\mathit{k}})=\frac{1}{V}\sum_{\boldsymbol{\mathit{r}}}e^{\mathrm i
\boldsymbol{\mathit{k}\cdot\mathit{r}}} \epsilon_{\boldsymbol{\mathit r}}\boldsymbol{\mathit{S}_r}
\end{equation}
as
\begin{equation}
F=\frac{V}{3}\overline{\left\langle |{\cal F}(2\pi/L,0,0)|^2+|{\cal F}(0,2\pi/L,0)|^2+|{\cal F}(0,0,2\pi/L)|^2\right\rangle}\ .
\end{equation}
The same definition is also valid in the tensorial case. This
definition is very well behaved for the FSS
method we have employed, see Ref.~\cite{UCMOND3}. Finally, we measure
the specific heat as
\begin{equation}
C= V^{-1} \overline{\langle {\cal E}^2 \rangle  -\langle{\cal E} \rangle ^2} \ .
\label{cesp}
\end{equation}

\section{Numerical Results}
\label{O3:simulations}

\subsection{Methods}
\label{O3:methodology}

The lattice sizes $L$ we have studied are $8,12,16,$ $24,32,48,64$,
and, only in the pure model, $L=96$. We have simulated five values
of the dilution apart from the pure case, $p=1$. These values are
$p=0.97,\ 0.95,\ 0.9,\ 0.7$, and $0.5$. 

Between each measurement of the observable described in
Sec.~\ref{O3:themodel}, firstly, we update the spin variables using a
Metropolis method over 10\% of the individuals spins, chosen at
random, then we perform a number (increasing with $L$) of cluster updates
using a Wolff method -- see Ref.~\cite{VICTORAMIT}. This is our elementary
Monte Carlo step (EMCS). The number of clusters traced (or Wolff
updates) between measurements was chosen to yield a good value of
the self-correlation time, see Ref.~\cite{VICTORAMIT}, in our case always 
$1<\tau<2$ ($\tau$ being the integrated autocorrelation time of the energy, see Appendix~\ref{Appendix_dataanalysis}).

In order to work in thermally equilibrated systems, we
perform a great number of EMCSs before starting
measurements. We start the simulation always from random
(hot) distributions of the spin variables, although we have
checked that the averages do not change if we begin from
cold configurations (i.e., all spins pointing in the same
direction). In particular, we took $4\times10^6$
measurements for the pure model, discarding about $10^5$ of
the first measurements for $L=8$ and increasing this number
with the lattice size.  For every lattice size, we
performed $2\times 10^4$ quenched disorder realizations in
the dilute models (except for $p=0.97$ and $p=0.95$ with
only $10^3$ realizations) taking 100 measurements per sample
after equilibration, in accordance with Ballesteros et
al.~\cite{ISDIL4D} who demonstrated that the best approach
to minimising the statistical error is to simulate a great
number of samples with just a few measurements in each one.

To measure the critical exponents, we use the so-called quotient
method~\cite{QUOTIENTS}, which allows great statistical
accuracy, see Appendix~\ref{Appendix_Quotient}. Therefore, firstly we needed to estimate by
successive simulations the $\beta$ point where
\begin{equation}
\frac{\xi(2L,\beta,p)}{2L}=\frac{\xi(L,\beta,p)}{L}\ ,
\label{BETANAIVE}
\end{equation}
for each pair of lattices $(L,2L)$. Then we used re-weighting techniques 
to fine-tune this condition. These re-weighting techniques are used to
$\beta$ extrapolate the observables and calculate their $\beta$ derivatives, always before the
sample averaging is performed. The equations used are, see Appendix~\ref{Appendix_Extrapolations},
\begin{equation}
\langle {\cal O} \rangle (\beta+\Delta\beta)=
\langle {\cal O} e^{\Delta\beta{\cal E}}\rangle / \langle e^{\Delta\beta{\cal E}}\rangle\ ,
\label{FS_O3}
\end{equation}
\begin{equation}
\partial_\beta \overline{\langle {\cal O}\rangle}=
 \overline{\partial_\beta\langle {\cal O}\rangle}=
\overline{\left\langle_{\vphantom{|}} {\cal{OE}} - \langle{\cal O}
 \rangle \langle {\cal E} \rangle\right\rangle}.
\label{DERIVADA_O3}
\end{equation}

\begin{figure}[!ht]
\begin{center}
\includegraphics[height=0.55\textheight,trim=8 10 18 10, angle=270]{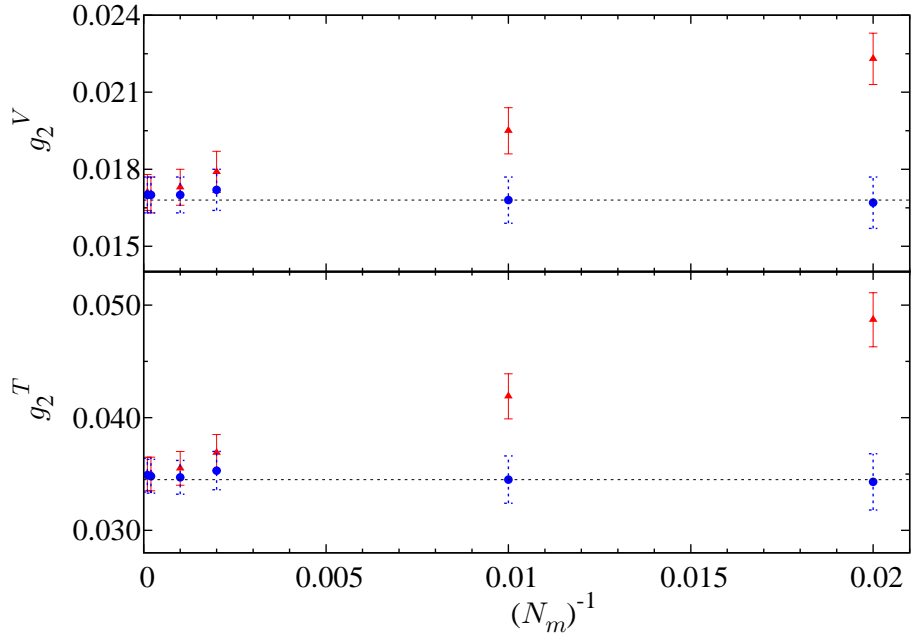}
\caption{The $g_2$ cumulant in both channels for $L=64$, $p=0.9$ with
1000 samples, $\beta_\mathrm{simulation}=0.79112$, re-weighted at
$\beta=0.79082$ as a function of $1/N_m$, with $N_m$ being
the number of measurements in each sample. We report data with
$N_m=50,100,500,1000,5000,$ and 10000. The data without the bias correction
proposed in Ref.~\cite{ISDIL4D} are marked with triangles while the
corrected ones are marked with circles. We also mark with the dotted
lines the selection used in this work (which corresponds to
$N_m=100$). Notice the importance of the correction of the bias if one
performs re-weighting with the data.
\label{fig_extrap_Nsamples}}
\end{center}
\end{figure}

These extrapolations are biased. For instance, the expectation value of equation (\ref{DERIVADA_O3}),
when the averages are calculated with $N_m$ measurements is
\begin{equation}
\overline{\left(1-\frac{2\tau}{N_m}\right)\partial_\beta \langle\cal O
\rangle}\ .
\label{BIASDER_O3}
\end{equation}
Hence, we have to correct this bias, see again Appendix~\ref{Appendix_Extrapolations}.
An example of the effect of this correction is
found in Fig.~\ref{fig_extrap_Nsamples}: a major bias affects the
uncorrected numerical data, and the importance of taking
this effect into account is clear. In addition, it is clear that the recipe of
Ref.~\cite{ISDIL4D} is working perfectly for $N_m=100$,
which is the number of measurements per sample we have taken in this
work. Therefore, we are very confident that all the data presented in this
work are not biased due to the re-weighting technique.

Also, we tried to use the solution for the bias obtained in
Ref.~\cite{Hasen}, where each sample is split into four parts, but
the results were poor. This was due to the small number of
measurements we take in each sample ($10^2$), which leads to large differences
between the averages in each quarter.

To compute errors in the averages we used a jack-knife method, see
Appendix~\ref{Appendix_dataanalysis}. We defined twenty jack-knife blocks for the pure
model in a \emph{single} sample and one block for \emph{each} sample
in the dilute ($p<1$) models.

The calculated observables and critical exponents sometimes
present, instead of a stable value, a monotonically
decreasing one. For $\eta$, there is found this type of
evolution with increasing $L$, but it is clearly weaker than
for $\nu$, see Tables~\ref{expmagpuro}--\ref{expter05}. In these cases an infinite volume extrapolation
is called for.  If hyperscaling holds, we expect
finite-volume scaling corrections proportional to
$L^{-\omega}$. This issue will be addressed in the next
subsection.

\subsection[The Scaling Exponent $\omega$]{The Scaling Exponent \boldmath$\omega$}
\label{O3:omega}

As will be seen in Tables~\ref{expmagpuro}~to~\ref{expter05}
in Sec.~\ref{O3:exponents}, there are evident finite volume
effects, especially for the thermal exponents and the
cumulants ($g_4$ and $g_2$).  So we have to use the equation
\begin{equation}
\left.\frac{x_O}{\nu}\right|_{\infty} -\left.\frac{x_O}{\nu}
\right|_{(L,2L)}\propto L^{-\omega}\,,
\label{Xw}
\end{equation}
which is a consequence of the scale hypothesis first derived in
Ref.~\cite{HIPERSCALA}. Consequently, choosing a good value for
$\omega$ is a crucial question.

Exact results and RG calculations tell us that the disorder, being
irrelevant in this model, induces scaling corrections with an exponent
$\alpha/\nu\simeq -0.188$ (in $L$)~\cite{PELI-REP}. In addition to
this new scaling correction one must have that of the pure model,
which is related to the coupling of the $(\phi^2)^2$ term in the
Ginzburg-Landau theory. This exponent is assumed to be
$0.8$~\cite{guida,hasen2} (for the pure model). Hence, the leading correction
is the exponent induced by the disorder. We will try to check this
scenario by computing the `leading' correction to the scaling
exponent from the numerical data.

First of all, we  tried to estimate $\omega$ just by considering
it as another tunable parameter in Eq.~(\ref{Xw}) applied to some
physical quantities. In these fits, as a first approximation, we
disregarded the possible correlations between the data for different
$L$ values. The results are presented in
Table~\ref{omega_weightmean}. If we perform a weighted averaging with
these results we obtain $\omega=1.07(9)$ for the pure model and
$\omega=0.92(9)$, $\omega=0.81(7)$, and $\omega=0.88(4)$ for the
dilute model with $p=0.9, 0.7$, and $0.5$ respectively, in very good
agreement with the value of the scaling correction exponent of the pure
model. However, we
think this method is not very consistent because of the variability of
the results from one quantity to another as seen in
Table~\ref{omega_weightmean}.

\renewcommand{\arraystretch}{1.2}

\begin{table}[ht]
\begin{center}
\begin{tabular}{|c|c|c|c|c|}\hline
$\cal O$ & $\omega_{p=1.0}$ & $\omega_{p=0.9}$ & $\omega_{p=0.7}$& $\omega_{p=0.5}$\\\hline\hline
$\eta_{\chi^V}$               & 1.45(52)  & ---      & ---       & ---\\\cline{1-5}
$\eta_{M^V}$                  & 1.62(80)  & ---      & ---       & ---\\\cline{1-5}
$\eta_{\chi^T}$               & ---       & ---      & 1.2 (1.1) & 0.68(46) \\\cline{1-5}
$\eta_{M^T}$                  & ---       & ---      & ---       & 0.73(46) \\\cline{1-5}
$\nu_{\partial_\beta g_4^V}$  & ---       & ---      & ---       & ---\\\cline{1-5}
$\nu_{\partial_\beta\xi^V}$   & 2.30(61)  & ---      & ---       & 0.62(47)\\\cline{1-5}
$\nu_{\partial_\beta g_4^T}$  & ---       & ---      & ---       & --- \\\cline{1-5}
$\nu_{\partial_\beta\xi^T}$   & 2.12(52)  & 1.76(60) & 1.09(40)  & 1.34 (27)\\\cline{1-5}
$\xi^V/L$                     & 1.08(21)  & 1.21(31) & 0.61(12)  & 0.45(10)\\\cline{1-5}
$\xi^T/L$                     & ---       & ---      & 1.55(76)  & 1.64(17)\\\cline{1-5}
$g_4^V$                       & 0.85(14)  & 2.00(61) & 1.21(15)  & 1.19(13)\\\cline{1-5}
$g_4^T$                       & 1.06(14)  & ---      & 1.35(33)  & 1.41(42)\\\cline{1-5}
$g_2^V$                       & ---       & 0.81(16) & 0.89(9)   & 0.94(7)\\\cline{1-5}
$g_2^T$                       & ---       & ---      & 0.63(12)  & 0.72(10)\\\hline\hline
$\bar{\omega}_\mathrm{weighted}$     & 1.07(9)   & 0.92(9) & 0.81(7)   & 0.88(4)\\\hline
\end{tabular}
\caption{$\omega$ values from the $L\rightarrow\infty$
extrapolations of some quantities.  The last row gives the
weighted average of each column. We disregarded data with error bars
larger than  100\% of the values themselves. Those disregarded data
are shown in the table as ---.}
\label{omega_weightmean}
\end{center}
\end{table}

\renewcommand{\arraystretch}{1.2}

Another approach, following Ref.~\cite{ISDIL3D}, is to study the
crossing points of scaling functions (such as $\xi/L$ and $g_4$) measured
in pairs of lattices with sizes $L$ and $sL$. The deviation of these
crossing points from the infinite volume critical coupling will behave
as
\begin{equation} 
\Delta\beta(L,sL)\equiv \\
\beta(L,sL)-\beta_{\mathrm{c}}(\infty) \propto
\frac{1-s^{-\omega}}{s^{\frac{1}{\nu}}-1}L^{-\omega-\frac{1}{\nu}}\,.
\label{BETADESVIATION}
\end{equation} 
With this method we need an additional estimate for the thermal
exponent $\nu$. We used,
following~\cite{peli}, the value
$\nu=0.7113(11)$ for the pure model (notice the really small error in
$\nu$, so that we will discard it in the following), which is also a valid value for the dilute
models because of the validity of the Harris criterion, and as can be
checked with the data below. We fixed $s=2$. In this
approach, we only use the crossing points in the vectorial channel
because they are cleaner.

\begin{figure}[!ht]
\begin{center}
\includegraphics[height=0.55\textheight,trim=18 10 28 10, angle=270]{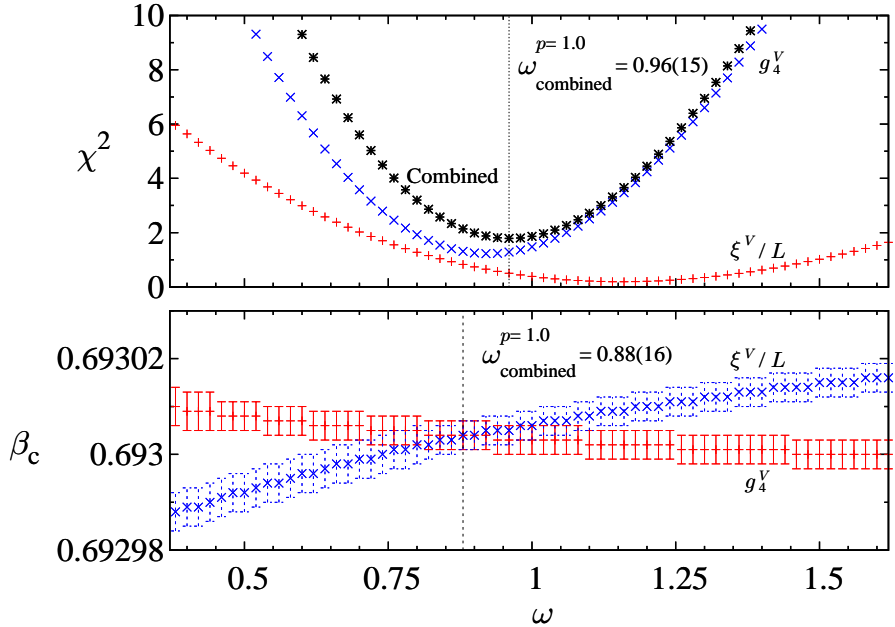}
\caption{ {\bf Top:} $\chi^2$ as a function of $\omega$
deduced from the fits to $L\rightarrow\infty$, Eq.~(\ref{BETADESVIATION}),
for the crossing point of $\xi/L$ and $g_4$ for the ($L$,\ $2L$) pair for
the pure model. Also shown is the combined $\chi^2$, whose minimum is marked
with the dotted line. {\bf Bottom:}
Extrapolated $\beta_c(\infty)$ as a function of $\omega$. The point where
the two observables give the same extrapolated value is marked with the
dashed line.
\label{calculoomega_puro}}
\end{center}
\end{figure}

\begin{figure}[!ht]
\begin{center}
\includegraphics[height=0.55\textheight,trim=18 10 0 10, angle=270]{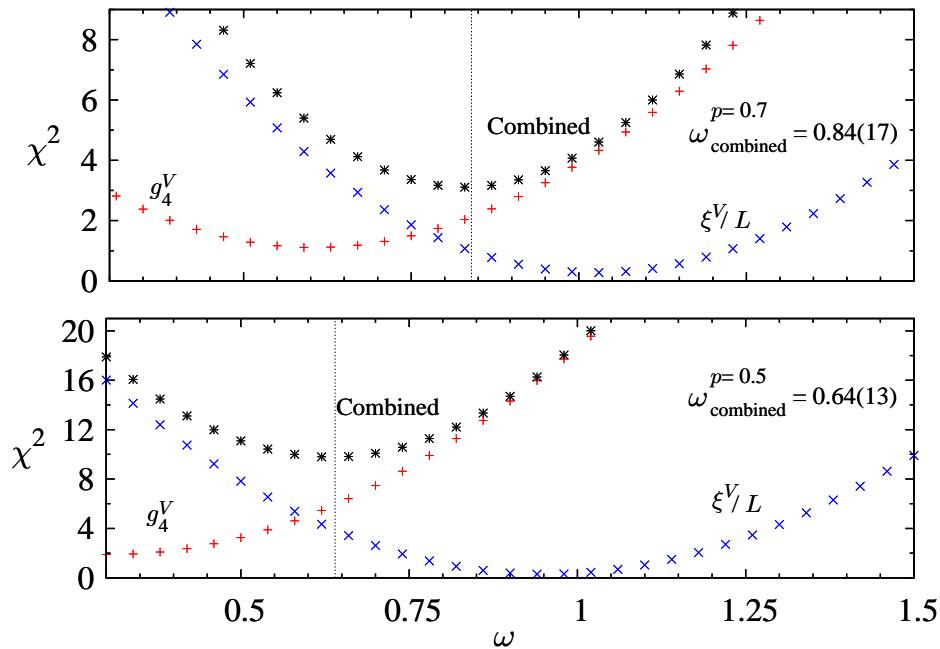}
\caption{ {\bf Top:} $\chi^2$ as a function of $\omega$ for
the dilute ($p=0.7$) model. Also shown is the combined
$\chi^2$. {\bf Bottom:} $\chi^2$ as a function of $\omega$ for
the dilute ($p=0.5$) model.
\label{calculoomega_diluidos}}
\end{center}
\end{figure}

Extrapolating these crossing points using Eq.(\ref{BETADESVIATION}), we
can plot the minimum of the $\chi^2$ of the fit as a function of
$\omega$ obtaining the upper part of Fig.~\ref{calculoomega_puro}~and
the whole of Fig~\ref{calculoomega_diluidos}. To carry out these
extrapolations we must take into account that the measurements of the crossing
points are correlated in pairs, so that we have to use the $\chi^2$
definition that includes the whole self-covariance matrix
\begin{equation}
\chi_x^2\ =\ \sum_{l=1}^{N} \sum_{m=1}^{N} (x_l-\mathrm{fit})(\mathrm{cov}^{-1})_{l,m}(x_m-\mathrm{fit})\,,
\label{CHISQU} 
\end{equation} 
with $N$ being the number of crossing points, that is to say, the number of
simulated $L$ values minus two; $x_{l}$ is the value obtained for the
observable $x$ (in our case the coupling) at the crossing point for $L_l$
and $2L_l$, and ``$\mathrm{fit}$'' is the value fitted to the form of
Eq.~(\ref{BETADESVIATION}) (or to another scaling form) for $L_l$. In addition
\begin{equation} 
\mathrm{(cov)}_{l,m}=\langle x_m x_l \rangle - \langle x_m \rangle
\langle x_l \rangle
\label{COV} 
\end{equation} 
can also be defined in terms of jack-knife blocks, see Ref.~\cite{VICTORAMIT}, as
\begin{equation} 
\mathrm{(cov)}_{l,m}=\frac{N_{b}-1}{N_{b}} \sum_{i=1}^{N_{b}}
(x_{l,i}^{\mathrm{J-K}}- \langle x_l \rangle)(x_{m,i}^{\mathrm{J-K}}
- \langle x_m \rangle)\,,
\label{COV_JK} 
\end{equation} 
where $N_b$ is the number of jack-knife blocks,
$x_{l,i}^{\mathrm{J-K}}$ are block variables, where the first subindex
runs over $L$ values while the second one runs over jack-knife blocks, and
$\langle x_l \rangle$ is the average of all block variables given
$L=L_l$.

Also, following Ref.~\cite{ISDIL3D}, we can do a combined
fit in $\omega$ of the crossing points of $\xi^V/L$ and $g^V_4$ by defining
\begin{equation} 
\chi^2_\mathrm{combined}=\chi^2_{\xi^V/L}+\chi^2_{g^{V}_4}\,,
\label{CHICUAD_JOINT} 
\end{equation} 
using Eq.~(\ref{CHISQU}) to calculate each of the right-hand-side terms and
searching for the minimum of $\chi^2_\mathrm{combined}$. We can obtain
the error in $\omega$ by searching for the point
$\omega_1$ at which
$\chi^2_\mathrm{combined}(\omega_1)=
\chi^2_\mathrm{combined}(\omega_\mathrm{min})+1$,
so that the error  is
$\Delta\omega=|\omega_\mathrm{min}-\omega_1|$. The results for these
combined fits are shown in the upper part of
Fig.~\ref{calculoomega_puro} and in the whole of
Fig.~\ref{calculoomega_diluidos}.  With this method we find the values
\begin{equation} 
\omega=0.96(15)\,,
\end{equation} 
for the pure model and
\begin{equation} 
\omega=2.29(70)\ ,\ \ 0.84(17)\ ,\ \ 0.64(13)\,, 
\end{equation} 
for the dilute models with $p=0.9,\ 0.7,\ \mathrm{and}\ 0.5$
respectively, in agreement with the value obtained in the
pure model~\cite{peli,UCMOND3,guida,hasen2}, except in the
$p=0.9$ case for which the value is two standard
deviations away from $\omega=0.8$~\cite{guida,hasen2}.  One
possibility is that we are computing the leading correction
to the scaling exponent but with a large error. Another
possibility is that in the $p=0.9$ model the coefficient of
the leading correction to the scaling vanishes or is very
small.  This result and the change in the slope of the $g_4$
data for $p<1$ with respect to the $p=1$ ones, as can be seen in
Table~\ref{cumulantes_todos}, constitute evidence for the possible
\emph{improved action} found for $p=0.9$, see
Ref.~\cite{Hasen}. Therefore the $\omega$ exponent that we
are measuring in this case could correspond to the third
irrelevant operator, instead of the second one (remember
that following RG the first one is $\alpha/\nu\simeq
-0.188$).

In addition, as also was done in Ref.~\cite{ISDIL3D}, we
were able to estimate the correct value for $\omega$ as that
producing the same $\beta_c(\infty)$ value for both the
crossings of $\xi/L$ and
$g_4$, as can be seen in the lower part of
Fig.~\ref{calculoomega_puro} marked with the dotted line at
$\omega=0.88$. This approach only works for the pure model
in which such a point is found. With another $p$ value the
$\beta_c(\infty)$ estimates from $\xi/L$ and $g_4$ do not
cross each other.

In conclusion, we have shown that our data (for both the pure and
the dilute models) are fully compatible with the value $\omega=0.80(1)$
obtained previously both numerically and analytically for the pure
model~\footnote{Field theoretical approaches (both
fixed dimension and $\epsilon$-expansion) provide very accurate values
for $\omega$: 0.782(13) and 0.794(18)
(respectively)~\cite{guida}. Recent numerical simulations provide the
values 0.775(13) and 0.799(13)~\cite{hasen2} and 0.64(13) and
0.71(15)~\cite{UCMOND3}.}. In addition, since the error bars in
$\omega$ are really small (1\% of error) we have discarded
the uncertainty in $\omega$ in the analysis presented in this work.
Since the error bars in the extrapolated quantities are much larger than the
uncertainty caused by the error bars in $\omega$, we fixed
$\omega=0.80$.  The extrapolations obtained in the rest of the chapter
are all obtained using this value.

Finally, it is interesting to note that in the analysis
presented in this subsection we have seen no traces of the
leading correction to the scaling exponent even for the
strongest dilution we have simulated, which should be
$\alpha/\nu \simeq -0.188$. One can explain this fact by
assuming that the amplitudes of this scaling correction
exponent are really small, so that we are seeing only the
next-to-leading scaling correction.

\subsection{Self-Averaging of the Susceptibility}
\label{O3:selfaveraging}

Having checked that the value $\omega=0.80$ describes the
corrections to the scaling for both the pure and the dilute models, we can try to
extrapolate the values of $g_2$ to infinite volume.

Numerical results for $g_2$ and $g_4$ in both channels are presented in
Table~\ref{cumulantes_todos} for both pure (only $g_4$) and dilute
models.

\renewcommand{\arraystretch}{1.1}

\begin{table}[ht]
\begin{center}
\begin{tabular}{|c|c|c|c|c|c|}\cline{1-6}
$p$& $L$ & $g_2^V$ & $g_2^T$ & $g_4^V$ & $g_4^T$ \\\hline\hline
1.0 & 8 & 0 & 0 & 0.62243(4) & 0.5216(1) \\
   & 12 & 0 & 0 & 0.62172(5) & 0.5189(2) \\
   & 16 & 0 & 0 & 0.62152(6) & 0.5181(2) \\
   & 24 & 0 & 0 & 0.62100(5) & 0.5166(2) \\
   & 32 & 0 & 0 & 0.62092(3) & 0.5162(1) \\
   & 48 & 0 & 0 & 0.62066(5) & 0.5156(2) \\\cline{1-6}

0.9 & 8 & 0.0327(4) & 0.0576(7)  & 0.6151(2) & 0.5102(3)\\
   & 12 & 0.0273(3) & 0.0518(6)  & 0.6163(1) & 0.5104(3) \\
   & 16 & 0.0253(3) & 0.0499(6)  & 0.6166(1) & 0.5100(3) \\
   & 24 & 0.0226(3) & 0.0453(6)  & 0.6168(1) & 0.5098(3) \\
   & 32 & 0.0208(2) & 0.0421(5)  & 0.6171(1) & 0.5100(3) \\\cline{1-6}

0.7 & 8 & 0.0780(8) & 0.1406(16)  & 0.6061(3) & 0.4994(6)\\
   & 12 & 0.0610(6) & 0.1177(13)  & 0.6108(2) & 0.5039(5) \\
   & 16 & 0.0512(5) & 0.1009(11)  & 0.6131(2) & 0.5064(4) \\
   & 24 & 0.0423(4) & 0.0868(10)  & 0.6150(2) & 0.5077(4) \\
   & 32 & 0.0371(4) & 0.0770(9)   & 0.6160(2) & 0.5089(4) \\\cline{1-6}

0.5 & 8 & 0.1130(11) & 0.2061(24) & 0.6006(4) & 0.4999(8)\\
   & 12 & 0.0834(8)  & 0.1600(18) & 0.6072(3) & 0.5047(6) \\
   & 16 & 0.0702(7)  & 0.1395(16) & 0.6107(3) & 0.5070(6) \\
   & 24 & 0.0553(6)  & 0.1138(13) & 0.6138(2) & 0.5085(5) \\
   & 32 & 0.0474(5)  & 0.0980(11) & 0.6151(2) & 0.5095(4) \\\cline{1-6}

\end{tabular}
\caption{Cumulants for the O($3$) model.  The first column is the spin density $p$. All the cumulants are calculated at
the crossing points of $\xi/L$ for $L$ and $2L$. The
averages were computed using $10^4$ samples (except in the
$p=1$ case).}
\label{cumulantes_todos}
\end{center}
\end{table}

\begin{table}[ht]
\begin{center}
\begin{tabular}{|c|c|c|c|c|c|}\cline{1-6}
$p$& $L$ & $g_2^V$ & $g_2^T$ & $g_4^V$ & $g_4^T$ \\\hline\hline
0.97 & 8  & 0.0108(6) & 0.0181(13)  & 0.6201(4) & 0.5187(10) \\
     & 12 & 0.0102(6) & 0.0189(14)  & 0.6195(4) & 0.5164(10) \\
     & 16 & 0.0084(6) & 0.0158(12)  & 0.6201(4) & 0.5159(10) \\
     & 24 & 0.0072(5) & 0.0146(11)  & 0.6206(4) & 0.5162(9)  \\
     & 32 & 0.0074(5) & 0.0152(12)  & 0.6206(4) & 0.5152(10) \\\cline{1-6}
0.95 & 8  & 0.0179(10) & 0.0290(18) & 0.6180(5) & 0.5158(11)\\
     & 12 & 0.0167(9)  & 0.0329(20) & 0.6182(5) & 0.5116(12) \\
     & 16 & 0.0150(9)  & 0.0286(18) & 0.6181(5) & 0.5129(11) \\
     & 24 & 0.0117(7)  & 0.0228(14  & 0.6186(4) & 0.5135(11) \\
     & 32 & 0.0118(7)  & 0.0251(17) & 0.6193(4) & 0.5140(10) \\\cline{1-6}
\end{tabular}
\caption{Cumulants for the O($3$) model with high $p$ values (very soft dilution). In this case the cumulants are computed 
averaging $10^3$ samples.}
\label{cumulantes_pgrande}
\end{center}
\end{table}

\begin{figure}[htbp]
\begin{center}
\includegraphics[height=0.55\textheight,trim=18 10 0 10, angle=270]{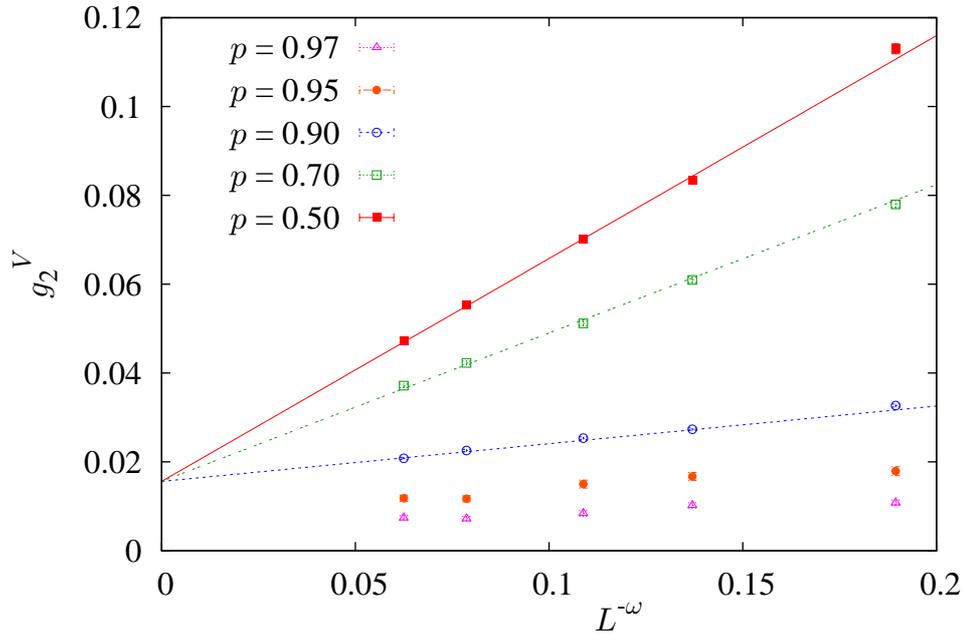}
\caption{ Combined extrapolation to $L\rightarrow\infty$ for the $g_2$
cumulant of the vectorial susceptibility. Extrapolations were carried
out by choosing a common value for the first term of Eq.~(\ref{Xw})
for all dilutions and by minimising the combined $\chi{^2}$. We
disregarded the data with $L=8$ to obtain a good value for the $\chi^2$.}
\label{fig:extrap_g2V}
\end{center}
\end{figure}

\begin{figure}[htbp]
\begin{center}
\includegraphics[height=0.55\textheight,trim=18 10 0 10, angle=270]{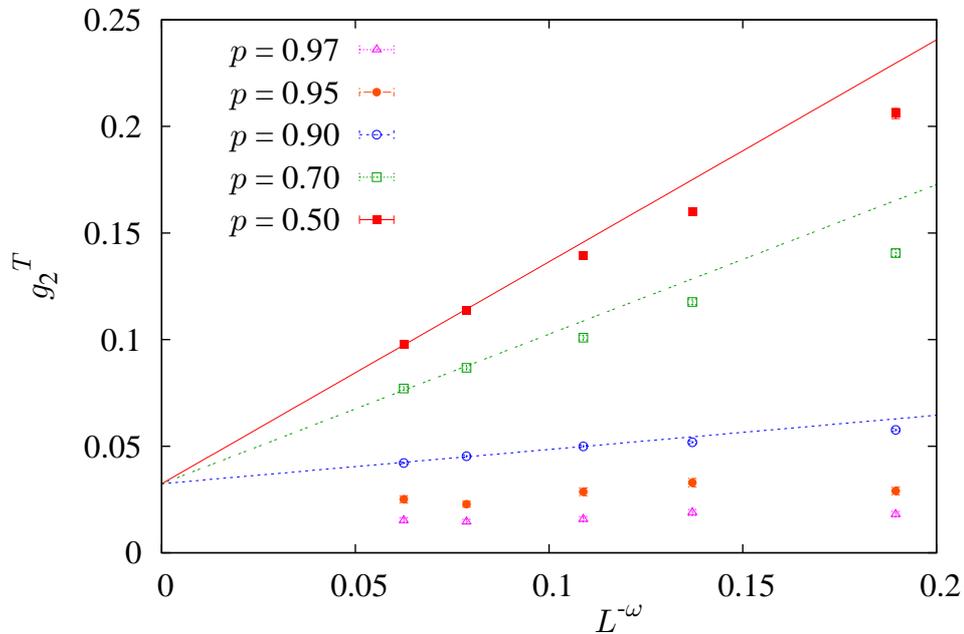}
\caption{ Combined extrapolation to $L\rightarrow\infty$ for the $g_2$
cumulant of the tensorial susceptibility, for the form of Eq.~(\ref{Xw}).}
\label{fig:extrap_g2T}
\end{center}
\end{figure}

First of all, we will try to check the non-zero $g_2$ scenario with
the correction to the scaling exponent fixed to that obtained in the
previous section. We found that it is possible, using the form of
Eq.~(\ref{Xw}) (performing a combined 
fit)
to extrapolate the values of $g_2$ to a value
(depending only on the channel) which is independent of the dilution,
and near the analytical prediction of reference \cite{korut}. However,
simulations at dilutions $p=0.95$
and $p=0.97$ do not follow the scaling found for $p\le 0.90$ (see
Table~\ref{cumulantes_pgrande}). Hence, as a whole our numerical data do not
support the scenario $g_2\neq 0$, see Figs.~\ref{fig:extrap_g2V}
and~\ref{fig:extrap_g2T} for the two channels. Notice, see also
Table~\ref{cumulantes_pgrande}, that all the values for these two
lowest dilutions are smaller than the extrapolated point and they are
decreasing (for both channels and taking into account the error bars).

\begin{figure}[htbp]
\begin{center}
\includegraphics[height=0.55\textheight,trim=18 10 0 10, angle=270]{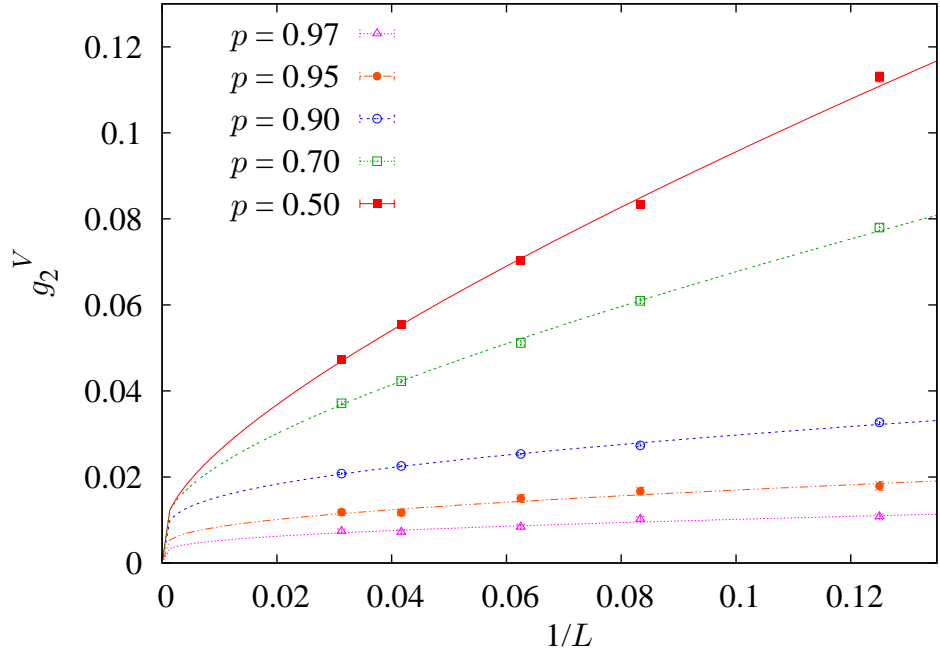}
\caption{Extrapolation to $L\rightarrow\infty$ for the $g_2$
cumulant of the vectorial susceptibility. The fitting function is in this case
given in Eq.~(\ref{g2_zero}).}
\label{fig:extrap_g2V_pelissetto}
\end{center}
\end{figure}

\begin{figure}[htbp]
\begin{center}
\includegraphics[height=0.55\textheight,trim=18 10 0 10, angle=270]{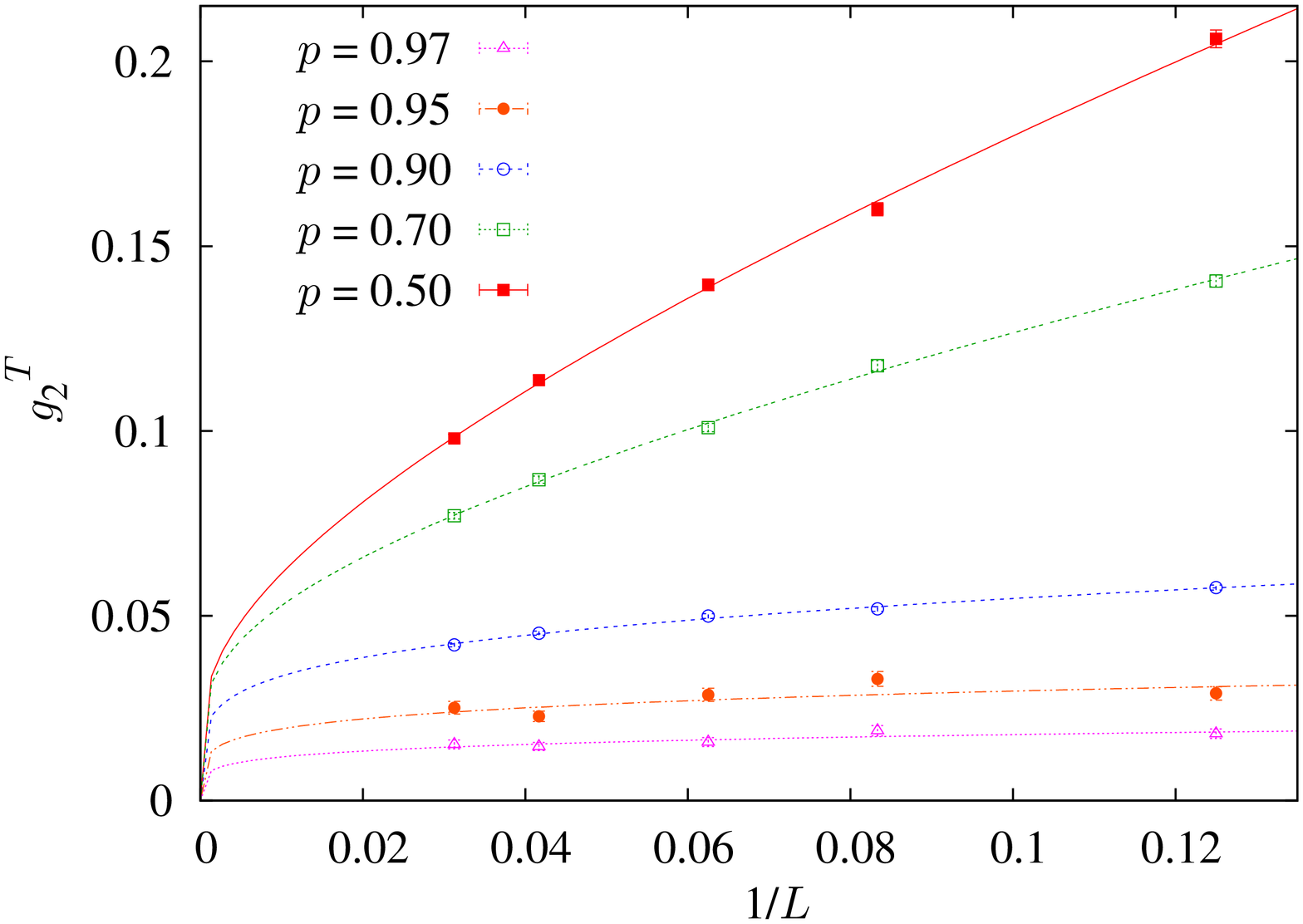}
\caption{Extrapolation to $L\rightarrow\infty$ for the $g_2$
cumulant of the tensorial susceptibility for the fitting form $g_2 = aL^{\alpha/\nu}+bL^{-\omega}$.}
\label{fig:extrap_g2T_pelissetto}
\end{center}
\end{figure}

Secondly, we will check the $g_2=0$ scenario. To do this, we
extrapolate $g_2$ using the form proposed in Ref.~\cite{harris} ($g_2
\backsim L^{\alpha/\nu}$) {\em but} also including the term
$L^{-\omega}$, i.e. we fit to:
\begin{equation}
g_2=a L^{\alpha/\nu} + b L^{-\omega}\,.
\label{g2_zero}
\end{equation}
We obtain the fits shown in Figs.~\ref{fig:extrap_g2V_pelissetto}
and~\ref{fig:extrap_g2T_pelissetto} for the two channels.  The $\chi^2$ of
these fits are really good. Hence, we have obtained strong evidence
supporting this $g_2=0$ scenario. Notice that the introduction of the two
scaling correction exponents has been of paramount importance for obtaining a
very good $\chi^2$ for all the fits. The numerical data, for the simulated lattice
size, do not follow the one-term dependence $g_2 \propto L
^{\alpha/\nu}$.

\subsection{Critical Exponents and Cumulants}
\label{O3:exponents}

In this subsection we will check the consistency of the $\omega$
exponent obtained in the text by means of the computation of critical
exponents and cumulants. In addition, we will check whether or not these sets of
exponents are universal by comparing different dilutions
with the pure model.  In this analysis we will use the data for
$p=0.9, 0.7$, and 0.5.

\begin{figure}[htbp]
\begin{center}
\includegraphics[height=0.55\textheight,trim=18 10 0 10, angle=270]{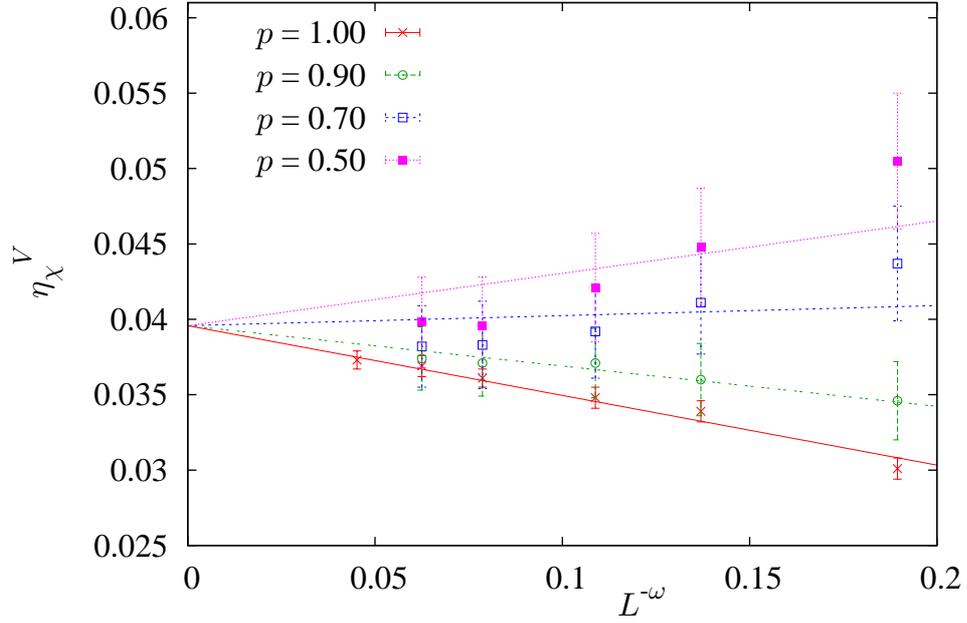}
\caption{Combined extrapolation to $L\rightarrow\infty$ for the $\eta$ exponent deduced from the vectorial susceptibility ($\chi^V$). Extrapolations were carried out by choosing a common value for the first term of Eq.~(\ref{Xw}) for all dilutions, and by minimising the combined $\chi^2$.}
\label{fig:extrap_eta_susV}
\end{center}
\end{figure}

\begin{figure}[htbp]
\begin{center}
\includegraphics[height=0.55\textheight,trim=18 10 0 10, angle=270]{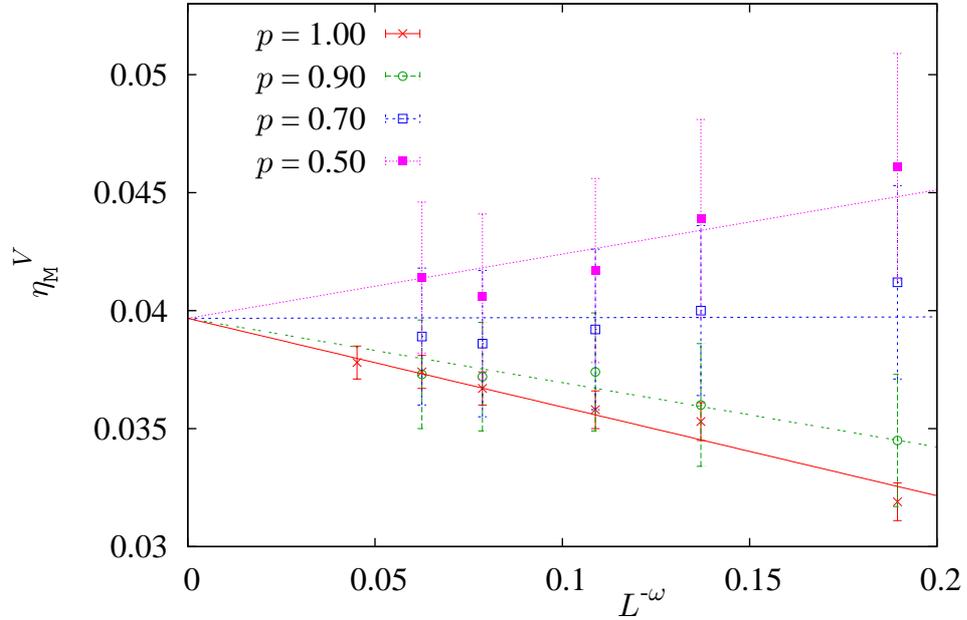}
\caption{ Combined extrapolation with all $p$ values to $L\rightarrow\infty$ for the $\eta$ exponent deduced from  the vectorial magnetisation ($M^V$). }
\label{fig:extrap_eta_magV}
\end{center}
\end{figure}

\begin{figure}[htbp]
\begin{center}
\includegraphics[height=0.55\textheight,trim=18 10 0 10, angle=270]{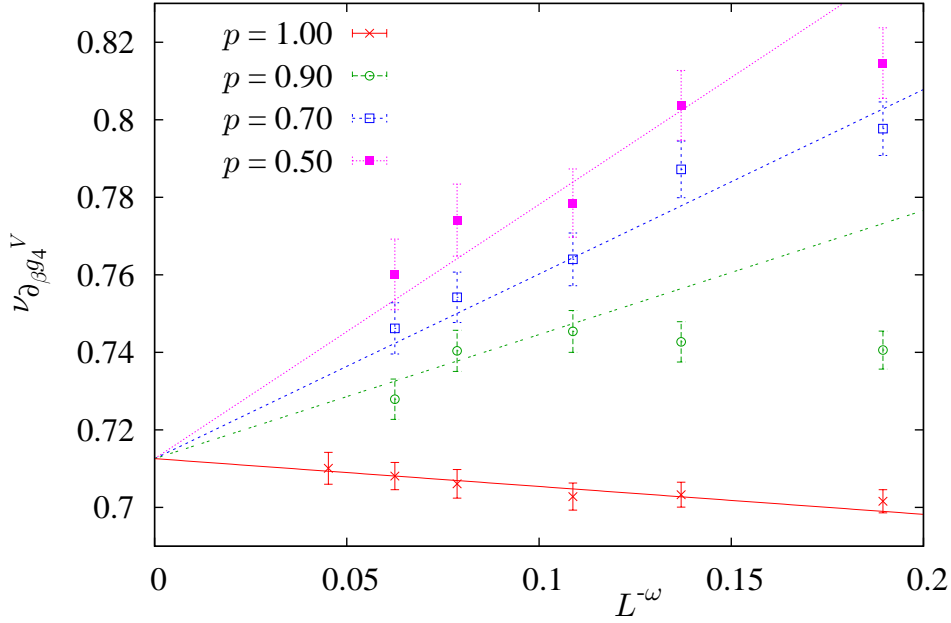}
\caption{ Combined extrapolation with all $p$ values to $L\rightarrow\infty$ for the $\nu$ exponent deduced from $\partial_\beta g_4^V$.}
\label{fig:extrap_nu_dg4V}
\end{center}
\end{figure}

\begin{figure}[!ht]
\begin{center}
\includegraphics[height=0.55\textheight,trim=18 10 0 10, angle=270]{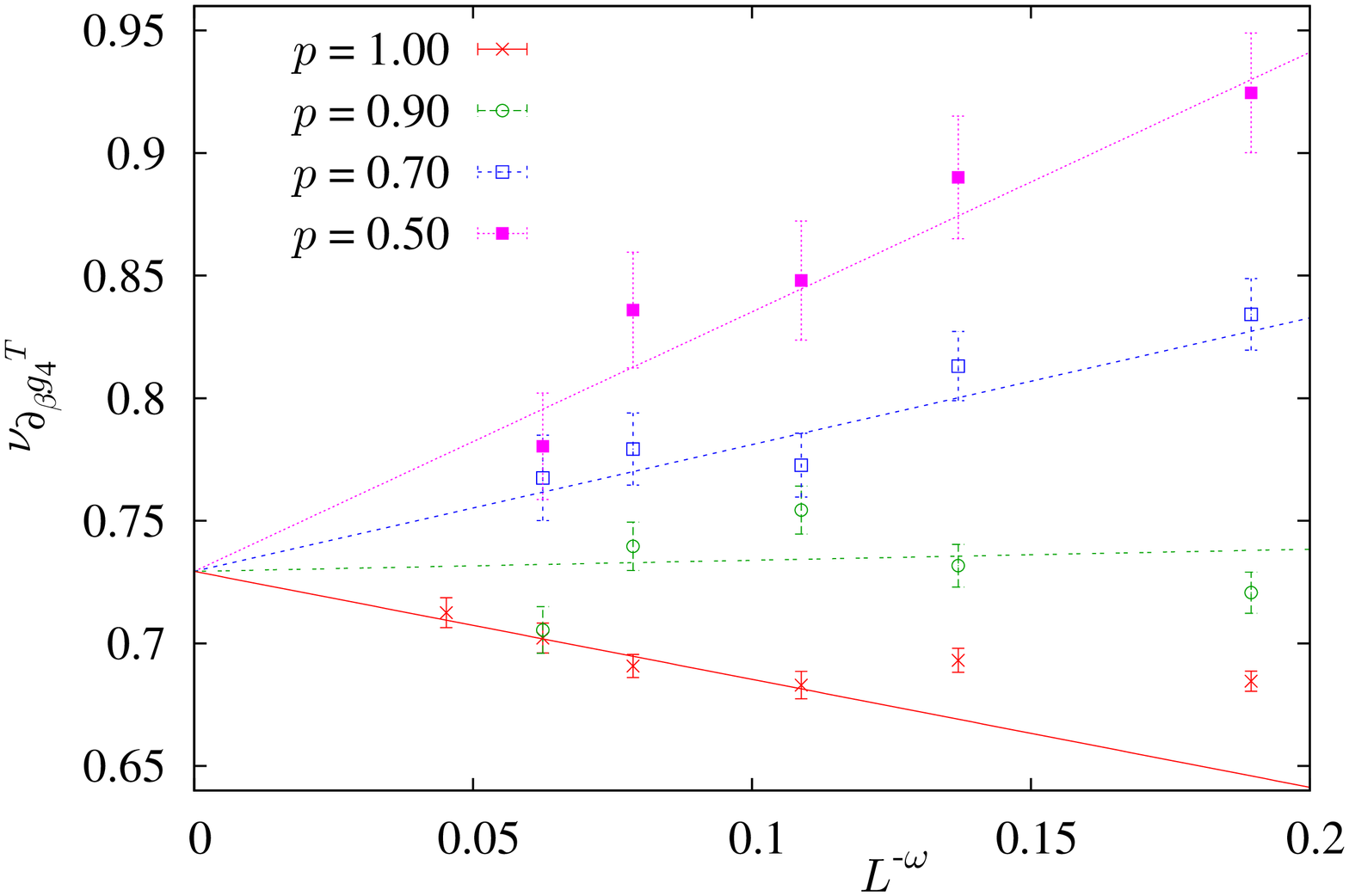}
\caption{Combined extrapolation to $L\rightarrow\infty$ for the $\nu$ exponent deduced from $\partial_\beta g_4^T$.}
\label{fig:extrap_nu_dg4T}
\end{center}
\end{figure}

Equation~(\ref{naive_quotient_method}) applied to the quantities $\partial_\beta \xi$,
$\partial_\beta g_4$, $M$, and $\chi$ yields respectively the critical
exponents $1+1/\nu$, $1/\nu$, $(D-2+\eta)/2$, and $2-\eta$.  Their
numerical results are given in Tables~\ref{expmagpuro}
and~\ref{expterpuro} for the pure model, Tables~\ref{expmag09}
and~\ref{expter09} for the $p$~=~0.9 case, Tables~\ref{expmag07}
and~\ref{expter07} for~$p$~=~0.7, and Tables~\ref{expmag05}
and~\ref{expter05} for~$p$~=~0.5.  We also carried out combined
extrapolations for all $p$ values by fixing the same value of the
extrapolated exponents for every $p$ value. Some of these fits are shown in
Figs.~\ref{fig:extrap_eta_susV} to~\ref{fig:extrap_nu_dg4T}, and
the compared results are presented in Tables~\ref{comparationmag}
and~\ref{comparationtherm}.

The combined extrapolation of the Binder cumulant $g_4$ is given in
Table~\ref{comparationbind}.  The agreement of our
results with those obtained in Refs.~\cite{peli} (numerical for the
pure model) and \cite{korut} (analytical) is really very good.
We also obtain complete agreement with previous numerical estimates of the pure
model critical exponents, see Ref.~\cite{peli}.

We obtain non-universal critical exponents and cumulants if
instead of $\omega=0.8$ we use $\omega=-\alpha/\nu$ as the
correction to scaling exponent. In addition, the dilution
dependent exponents and cumulants are clearly different from
the pure ones. Furthermore, this scenario does not change if
we fit the data using both $-\alpha/\nu$ and
$\omega=0.8$.

\begin{table}[htbp]
\begin{center}
\begin{tabular}{c|c|c|c|c|}\cline{2-5}
& \multicolumn{2}{c|}{$\eta$} & \multicolumn{2}{c|}{$\eta_T$}\\
\cline{1-5}
\multicolumn{1}{|c|}{$L$}& \multicolumn{1}{c|}{$\chi$}      
    & \multicolumn{1}{c|}{\lower2pt\hbox{$M$}}
    & \multicolumn{1}{c|}{$\chi_T$}        
    & \multicolumn{1}{c|}{\lower2pt\hbox{$M_T$}}\\\hline\hline
\multicolumn{1}{|c|}{8}  & 0.0301(7) & 0.0319(8) & 1.4301(12) & 1.4343(13)\\\cline{1-5}
\multicolumn{1}{|c|}{12} & 0.0339(7) & 0.0353(8) & 1.4324(11) & 1.4352(12)\\\cline{1-5}
\multicolumn{1}{|c|}{16} & 0.0348(7) & 0.0358(8) & 1.4310(11) & 1.4335(12)\\\cline{1-5}
\multicolumn{1}{|c|}{24} & 0.0361(6) & 0.0367(7) & 1.4293(9)  & 1.4307(10)\\\cline{1-5}
\multicolumn{1}{|c|}{32} & 0.0369(7) & 0.0374(7) & 1.4289(11) & 1.4300(12)\\\cline{1-5}
\multicolumn{1}{|c|}{48} & 0.0373(6) & 0.0378(7) & 1.4271(9)  & 1.4280(10)\\\hline\hline
\multicolumn{1}{|c|}{$L\rightarrow\infty$}   & 0.0391(9)& 0.0390(10)& 1.4250(13)& 1.4249(15)\\\cline{1-5}
\multicolumn{1}{|c|}{$\chi^2/\mathrm{d.o.f}$}& 0.138/3  & 0.354/3   & 1.047/3   & 1.952/3   \\\cline{1-5}
\multicolumn{1}{|c|}{$\mathrm{C.L.}$}        & 0.987    & 0.950     & 0.790     & 0.582     \\\hline
\end{tabular}
\caption{Magnetic exponents for the pure O($3$) model. The last three rows correspond to the $L\rightarrow\infty$ extrapolation (disregarding data with $L=8$).}
\label{expmagpuro}
\end{center}
\end{table}

\begin{table}[htbp]
\begin{center}
\begin{tabular}{c|c|c|c|c|}\cline{2-5}
& \multicolumn{4}{c|}{$\nu$} \\
\cline{1-5}
\multicolumn{1}{|c|}{$L$} & \multicolumn{1}{c|}{$\partial_\beta g_4^V$}      
    & \multicolumn{1}{c|}{$\partial_\beta \xi^V$} 
    & \multicolumn{1}{c|}{$\partial_\beta g_4^T$}        
    & \multicolumn{1}{c|}{$\partial_\beta \xi^T$}\\\hline\hline
\multicolumn{1}{|c|}{8}  & 0.7016(30) & 0.7217(13) & 0.6846(41) & 0.7306(14)\\\cline{1-5} 
\multicolumn{1}{|c|}{12} & 0.7033(32) & 0.7162(14) & 0.6931(49) & 0.7188(13)\\\cline{1-5} 
\multicolumn{1}{|c|}{16} & 0.7028(35) & 0.7123(16) & 0.6830(56) & 0.7118(17)\\\cline{1-5} 
\multicolumn{1}{|c|}{24} & 0.7061(37) & 0.7123(17) & 0.6908(47) & 0.7112(18)\\\cline{1-5} 
\multicolumn{1}{|c|}{32} & 0.7081(35) & 0.7121(19) & 0.7022(61) & 0.7116(23)\\\cline{1-5} 
\multicolumn{1}{|c|}{48} & 0.7101(41) & 0.7118(19) & 0.7125(61) & 0.7085(21)\\\hline\hline

\multicolumn{1}{|c|}{$L\rightarrow\infty$}    & 0.7109(38)& 0.7071(19)& 0.7082(51) & 0.7071(35)\\\cline{1-5}
\multicolumn{1}{|c|}{$\chi^2/\mathrm{d.o.f}$} & 0.667/4   & 4.104/4   & 7.039/4    & 0.565/2 \\\cline{1-5}
\multicolumn{1}{|c|}{$\mathrm{C.L.}$}         & 0.954     & 0.392     & 0.134      & 0.754 \\\hline  

\end{tabular}
\caption{Thermal critical exponents for the pure O($3$) model. In the last column we have disregarded data with $L<16$.}
\label{expterpuro}
\end{center}
\end{table}

\begin{table}[htbp]
\begin{center}
\begin{tabular}{c|c|c|c|c|}\cline{2-5}
& \multicolumn{2}{c|}{$\eta$} & \multicolumn{2}{c|}{$\eta_T$}\\\cline{1-5}
\multicolumn{1}{|c|}{$L$}& \multicolumn{1}{c|}{$\chi$}
    & \multicolumn{1}{c|}{\lower2pt\hbox{$M$}}
    & \multicolumn{1}{c|}{$\chi_T$}
    & \multicolumn{1}{c|}{\lower2pt\hbox{$M_T$}}\\\hline\hline
\multicolumn{1}{|c|}{8}  & 0.0346(26) & 0.0345(28) & 1.4154(36) & 1.4176(37) \\\cline{1-5}
\multicolumn{1}{|c|}{12} & 0.0360(24) & 0.0360(26) & 1.4195(34) & 1.4207(36) \\\cline{1-5}
\multicolumn{1}{|c|}{16} & 0.0371(23) & 0.0374(25) & 1.4207(34) & 1.4218(35) \\\cline{1-5}
\multicolumn{1}{|c|}{24} & 0.0373(22) & 0.0375(24) & 1.4204(32) & 1.4221(34) \\\cline{1-5}
\multicolumn{1}{|c|}{32} & 0.0383(21) & 0.0383(23) & 1.4219(31) & 1.4227(33) \\\hline\hline

\multicolumn{1}{|c|}{$L\rightarrow\infty$}   & 0.0397(29) & 0.0399(31) & 1.4245(41) & 1.4252(43)\\\cline{1-5}
\multicolumn{1}{|c|}{$\chi^2/\mathrm{d.o.f}$}& 0.292/3    & 0.124/3    & 0.544/3    & 0.137/3 \\\cline{1-5}
\multicolumn{1}{|c|}{$\mathrm{C.L.}$}        & 0.962      & 0.989      & 0.909      & 0.987 \\\hline
\end{tabular}
\caption{Magnetic exponents for the dilute O($3$) model with
$p=0.9\,$. Extrapolations were carried out without disregarding data.}
\label{expmag09}
\end{center}
\end{table}

\begin{table}[htbp]
\begin{center}
\begin{tabular}{c|c|c|c|c|}\cline{2-5}
& \multicolumn{4}{c|}{$\nu$} \\
\cline{1-5}
\multicolumn{1}{|c|}{$L$} & \multicolumn{1}{c|}{$\partial_\beta g_4^V$}      
    & \multicolumn{1}{c|}{$\partial_\beta \xi^V$} 
    & \multicolumn{1}{c|}{$\partial_\beta g_4^T$}        
    & \multicolumn{1}{c|}{$\partial_\beta \xi^T$}\\\hline\hline
\multicolumn{1}{|c|}{8}  & 0.7319(49) & 0.7443(24) & 0.7128(83) & 0.7709(29)\\\cline{1-5}
\multicolumn{1}{|c|}{12} & 0.7381(53) & 0.7411(25) & 0.7267(86) & 0.7514(29)\\\cline{1-5}
\multicolumn{1}{|c|}{16} & 0.7430(55) & 0.7381(26) & 0.7536(99) & 0.7426(31)\\\cline{1-5}
\multicolumn{1}{|c|}{24} & 0.7384(57) & 0.7368(28) & 0.7337(95) & 0.7395(32)\\\cline{1-5}
\multicolumn{1}{|c|}{32} & 0.7398(54) & 0.7365(29) & 0.7241(97) & 0.7345(33)\\\hline\hline
                                                
\multicolumn{1}{|c|}{$L\rightarrow\infty$ }  & 0.734(15) & 0.7318(33) & 0.728(17) & 0.7152(39)\\\cline{1-5}
\multicolumn{1}{|c|}{$\chi^2/\mathrm{d.o.f}$}& 0.134/1   & 0.168/3    & 5.468/2   & 3.156/3 \\\cline{1-5}
\multicolumn{1}{|c|}{$\mathrm{C.L.}$}        & 0.714     & 0.983      & 0.065     & 0.368\\\hline
    
\end{tabular}
\caption{Thermal exponents for the dilute O($3$) model with $p=0.9\,$. In the second and fourth columns we obtain poor results because the series are not monotonically decreasing.}
\label{expter09}
\end{center}
\end{table}

\begin{table}[htbp]
\begin{center}
\begin{tabular}{c|c|c|c|c|}\cline{2-5}
& \multicolumn{2}{c|}{$\eta$} &
\multicolumn{2}{c|}{$\eta_T$}\\
\cline{1-5}
\multicolumn{1}{|c|}{$L$}& \multicolumn{1}{c|}{$\chi$}      
    & \multicolumn{1}{c|}{\lower2pt\hbox{$M$}}
    & \multicolumn{1}{c|}{$\chi_T$}        
    & \multicolumn{1}{c|}{\lower2pt\hbox{$M_T$}}\\\hline\hline
\multicolumn{1}{|c|}{8 } & 0.0436(38) & 0.0412(41) & 1.3882(52) & 1.3879(53)\\\cline{1-5}
\multicolumn{1}{|c|}{12} & 0.0411(34) & 0.0401(36) & 1.4005(48) & 1.4007(49)\\\cline{1-5}
\multicolumn{1}{|c|}{16} & 0.0392(31) & 0.0392(34) & 1.4061(45) & 1.4073(46)\\\cline{1-5}
\multicolumn{1}{|c|}{24} & 0.0383(29) & 0.0386(31) & 1.4131(41) & 1.4136(43)\\\cline{1-5}
\multicolumn{1}{|c|}{32} & 0.0382(27) & 0.0389(29) & 1.4142(40) & 1.4149(41)\\\hline\hline

\multicolumn{1}{|c|}{$L\rightarrow\infty$}   & 0.0343(57) & 0.0370(58) & 1.4299(72) & 1.4318(76)\\\cline{1-5}
\multicolumn{1}{|c|}{$\chi^2/\mathrm{d.o.f}$}& 0.232/3    & 0.059/3    & 0.472/3    & 0.567/3   \\\cline{1-5}
\multicolumn{1}{|c|}{$\mathrm{C.L.}$}        & 0.972      & 0.996      & 0.925      & 0.904     \\\hline
\end{tabular}
\caption{Magnetic exponents for the dilute O($3$) model with $p=0.7\,$. Extrapolations were carried out without disregarding data.}
\label{expmag07}
\end{center}
\end{table}

\begin{table}[htbp]
\begin{center}
\begin{tabular}{c|c|c|c|c|}\cline{2-5}
& \multicolumn{4}{c|}{$\nu$}\\
\cline{1-5}
\multicolumn{1}{|c|}{$L$} & \multicolumn{1}{c|}{$\partial_\beta g_4^V$}      
    & \multicolumn{1}{c|}{$\partial_\beta \xi^V$} 
    & \multicolumn{1}{c|}{$\partial_\beta g_4^T$}        
    & \multicolumn{1}{c|}{$\partial_\beta \xi^T$}\\\hline\hline
\multicolumn{1}{|c|}{8}  & 0.7888(69)  & 0.7881(31) & 0.8256(143) & 0.8422(42)\\\cline{1-5}
\multicolumn{1}{|c|}{12} & 0.7810(74)  & 0.7806(33) & 0.8078(140) & 0.8067(41)\\\cline{1-5}
\multicolumn{1}{|c|}{16} & 0.7633(70)  & 0.7760(35) & 0.7739(131) & 0.7897(43)\\\cline{1-5}
\multicolumn{1}{|c|}{24} & 0.7491(66)  & 0.7628(37) & 0.7719(146) & 0.7792(47)\\\cline{1-5}
\multicolumn{1}{|c|}{32} & 0.7400(67)  & 0.7521(42) & 0.7656(178) & 0.7627(56)\\\hline\hline

\multicolumn{1}{|c|}{$L\rightarrow\infty$ }  &  0.7206(88)& 0.723(10)& 0.729(19)& 0.7255(61)\\\cline{1-5}
\multicolumn{1}{|c|}{$\chi^2/\mathrm{d.o.f}$}&  2.313/3   & 0.281/1  & 1.314/3  & 1.965/3 \\\cline{1-5}
\multicolumn{1}{|c|}{$\mathrm{C.L.}$}        &  0.510     & 0.596    & 0.726    & 0.580 \\\hline
\end{tabular}
\caption{Thermal exponents for the dilute O($3$) model with $p=0.7\,$. In the third column the fit was obtained disregarding data with $L<16$.}
\label{expter07}
\end{center}
\end{table}

\begin{table}[htbp]
\begin{center}
\begin{tabular}{c|c|c|c|c|}\cline{2-5}
& \multicolumn{2}{c|}{$\eta$} & \multicolumn{2}{c|}{$\eta_T$}\\
\cline{1-5}
\multicolumn{1}{|c|}{$L$}& \multicolumn{1}{c|}{$\chi$}      
    & \multicolumn{1}{c|}{\lower2pt\hbox{$M$}}
    & \multicolumn{1}{c|}{$\chi_T$}        
    & \multicolumn{1}{c|}{\lower2pt\hbox{$M_T$}}\\\hline\hline
\multicolumn{1}{|c|}{8 } & 0.0505(45) & 0.0461(48) & 1.3435(61) & 1.3431(62) \\\cline{1-5}
\multicolumn{1}{|c|}{12} & 0.0448(39) & 0.0439(42) & 1.3684(54) & 1.3702(56) \\\cline{1-5}
\multicolumn{1}{|c|}{16} & 0.0421(36) & 0.0417(39) & 1.3877(51) & 1.3896(52) \\\cline{1-5}
\multicolumn{1}{|c|}{24} & 0.0396(32) & 0.0406(35) & 1.4033(46) & 1.4053(48) \\\cline{1-5}
\multicolumn{1}{|c|}{32} & 0.0399(30) & 0.0414(32) & 1.4126(43) & 1.4152(45)\\\hline\hline

\multicolumn{1}{|c|}{$L\rightarrow\infty$}   & 0.0346(60) & 0.0378(46) & 1.446(12) & 1.449(12)\\\cline{1-5}
\multicolumn{1}{|c|}{$\chi^2/\mathrm{d.o.f}$}& 2.225/2    & 2.191/3    & 0.119/1   & 0.327/1  \\\cline{1-5}
\multicolumn{1}{|c|}{$\mathrm{C.L.}$}        & 0.329      & 0.534      & 0.730     & 0.568    \\\hline
\end{tabular}
\caption{Magnetic exponents for the dilute O($3$) model with $p=0.5\,$. In the fourth and fifth columns we 
disregarded data with $L<16$.}
\label{expmag05}
\end{center}
\end{table}

\clearpage

\begin{table}[htbp]
\begin{center}
\begin{tabular}{c|c|c|c|c|}\cline{2-5}
& \multicolumn{4}{c|}{$\nu$}\\
\cline{1-5}
\multicolumn{1}{|c|}{$L$} & \multicolumn{1}{c|}{$\partial_\beta g_4^V$}      
    & \multicolumn{1}{c|}{$\partial_\beta \xi^V$} 
    & \multicolumn{1}{c|}{$\partial_\beta g_4^T$}        
    & \multicolumn{1}{c|}{$\partial_\beta \xi^T$}\\\hline\hline
\multicolumn{1}{|c|}{8}  & 0.8102(91) & 0.8357(46) & 0.9180(241) & 0.9540(72) \\\cline{1-5}
\multicolumn{1}{|c|}{12} & 0.8042(90) & 0.8322(50) & 0.8880(248) & 0.8866(71) \\\cline{1-5}
\multicolumn{1}{|c|}{16} & 0.7764(89) & 0.7862(48) & 0.8449(242) & 0.8136(64) \\\cline{1-5}
\multicolumn{1}{|c|}{24} & 0.7702(93) & 0.7778(52) & 0.8311(234) & 0.7952(66) \\\cline{1-5}
\multicolumn{1}{|c|}{32} & 0.7562(91) & 0.7779(56) & 0.7812(220) & 0.7833(70)\\\hline\hline

\multicolumn{1}{|c|}{$L\rightarrow\infty$ }  & 0.720(16) & 0.764(14)  & 0.735(28) & 0.744(17)\\\cline{1-5}
\multicolumn{1}{|c|}{$\chi^2/\mathrm{d.o.f}$}& 1.149/2   & 0.208/1    & 1.565/3   & 0.025/1 \\\cline{1-5}
\multicolumn{1}{|c|}{$\mathrm{C.L.}$}        & 0.563     & 0.649      & 0.667     & 0.874 \\\hline
\end{tabular}
\caption{Thermal exponents for the dilute O($3$) model with $p=0.5\,$. In the second column we only used data with $L>8$ while
in the third and fifth columns we only used data with $L>12$.}
\label{expter05}
\end{center}
\end{table}

\begin{table}[!ht]
\begin{center}
\begin{tabular}{c|c|c|c|c|c|}\cline{2-5}
& \multicolumn{2}{c|}{$\eta$} & \multicolumn{2}{c|}{$\eta_T$}     \\

\cline{2-5}
    & \multicolumn{1}{c|}{$\chi$}
    & \multicolumn{1}{c|}{\lower2pt\hbox{$M$}} 
    & \multicolumn{1}{c|}{$\chi_T$}
    & \multicolumn{1}{c|}{\lower2pt\hbox{$M_T$}}
    \\\cline{2-5}\hline
     
\multicolumn{1}{|c|}{$\mathrm{Our\ results}$}    & 0.0390(9)& 0.0389(10)& 1.4251(13)& 1.4251(14) \\\cline{1-5}
\multicolumn{1}{|c|}{$\chi^2/\mathrm{d.o.f}$}    & 6.675/12 & 5.104/15  & 9.151/10  & 13.931/11  \\\cline{1-5}
\multicolumn{1}{|c|}{$\mathrm{C.L.}$}            & 0.878    & 0.991     & 0.518     & 0.237  \\\hline\hline     
\multicolumn{1}{|c|}{$\mathrm{Ref.}$~\cite{peli}}& 0.0378(6)& ---       & ---       & --- \\\cline{1-5}

\end{tabular}
\caption{Combined extrapolation with \emph{all} $p$ values for the magnetic exponent $\eta$ compared with the results from Ref.~\cite{peli}. The first three rows correspond to our $L\rightarrow\infty$ extrapolation.
}
\label{comparationmag}
\end{center}
\end{table}

\begin{table}[htbp]
\begin{center}
\begin{tabular}{c|c|c|c|c|c|}\cline{2-5}
&  \multicolumn{4}{c|}{$\nu$} \\

\cline{2-5}
    & \multicolumn{1}{c|}{$\partial_\beta g_4^V$}      
    & \multicolumn{1}{c|}{$\partial_\beta \xi^V$} 
    & \multicolumn{1}{c|}{$\partial_\beta g_4^T$}        
    & \multicolumn{1}{c|}{$\partial_\beta \xi^T$}
    \\\cline{2-5}\hline
     
\multicolumn{1}{|c|}{$\mathrm{Our\ results}$}    & 0.7126(46)& 0.7129(31)& 0.7294(81)& 0.7089(32)\\\cline{1-5}
\multicolumn{1}{|c|}{$\chi^2/\mathrm{d.o.f}$}    & 4.831/11  & 6.606/6   & 9.009/13  & 9.609/7   \\\cline{1-5}
\multicolumn{1}{|c|}{$\mathrm{C.L.}$}            & 0.939     & 0.359     & 0.772     & 0.212\\\hline\hline
\multicolumn{1}{|c|}{$\mathrm{Ref.}$~\cite{peli}}& 0.7113(11)& ---       & ---       & --- \\\cline{1-5}

\end{tabular}
\caption{Combined extrapolation with \emph{all} $p$ values for the thermal exponent $\nu$ compared with the results from Ref.~\cite{peli}.}
\label{comparationtherm}
\end{center}
\end{table}

\begin{table}[htbp]
\begin{center}
\begin{tabular}{c|c|c|}\cline{2-3}
    & \multicolumn{1}{c|}{$g_4^V$}        
    & \multicolumn{1}{c|}{$g_4^T$}\\\cline{2-3}\hline
     
\multicolumn{1}{|c|}{$\mathrm{Our\ results}$}     & 0.62018(6) & 0.51366(19)\\\cline{1-3}
\multicolumn{1}{|c|}{$\chi^2/\mathrm{d.o.f}$}     & 10.324/9   & 5.980/10   \\\cline{1-3}
\multicolumn{1}{|c|}{$\mathrm{C.L.}$}             & 0.325      & 0.817      \\\hline\hline
\multicolumn{1}{|c|}{$\mathrm{Ref.}$ \cite{korut}}& 0.625783   & ---        \\\cline{1-3}
\multicolumn{1}{|c|}{$\mathrm{Ref.}$ \cite{peli}} & 0.6202(1)  & ---        \\\cline{1-3}

\end{tabular}
\caption{Combined extrapolation to $L\rightarrow\infty$ with \emph{all} $p$ values for the Binder cumulant $g_4$ defined in Eq.~(\ref{g4cumulants}), compared with results from Refs.\cite{korut}~and~\cite{peli}.
}
\label{comparationbind}
\end{center}
\end{table}

\newpage
\section{Conclusions}
\label{O3:conclusions}

We have studied the critical properties of the site-dilute Heisenberg model
for different values of the dilution. Our main aims were both to re-verify 
the Harris criterion and to check the self-averaging properties of the
susceptibility.

We studied in great detail the corrections to the scaling in
the model, finding that the numerical data follow the
next-to-leading correction to the scaling exponent instead
of the leading one. We obtained all the critical exponents
and cumulants using this next-to-leading exponent.  Also,
the result of this analysis was found to be fully compatible
with the RG predictions and the Harris criterion: our
exponents and cumulants are compatible with those of the
pure model and independent of the dilution to a high
degree of precision.

In addition, we showed that we obtain non-universal
quantities if we assume $\alpha/\nu$ to be the main scaling
correction even if we add the $\omega$ correction to the
scaling exponent, using two correction-to-scaling exponents
in the analysis.

Finally, we showed strong evidence for a zero $g_2$
cumulant, in both the vector and the tensor channels, in the
thermodynamic limit at criticality, contrasting with some
analytical predictions~\cite{korut}, but in agreement with
others~\cite{harris}.  The introduction of scaling
corrections in the analysis was crucial to obtain the
$g_2=0$ scenario. In addition, simulations of samples with
very soft dilution ($p>0.9$) helped us to discard the $g_2 \neq 0$
scenario.

\renewcommand{\arraystretch}{1}

\clearpage{\thispagestyle{empty}\cleardoublepage}

\clearpage{\thispagestyle{empty}\cleardoublepage}

\normalfont

\vspace{4cm}

\selectlanguage{british}

\chapter{The Site-Diluted Ising Model in Four Dimensions}
\label{chap:RSIM4D}

\section{Introduction}
\label{RSIM4D:intro}

One of the major achievements of statistical physics is the fundamental  explanation 
of critical behaviour at continuous phase transitions through Wilson's Renormalization Group (RG) approach. 
While this has mostly provided a satisfying picture 
for over thirty years, certain types of phase transitions
have resisted full treatment. Such stubborn cases, which have been the
subject of conflicting proposals and analyses,  
include systems with in-built disorder.

The Ising model with uncorrelated, quenched random-site or random-bond 
disorder is a classic example 
of such systems and has been controversial in both two and four dimensions. 
In these dimensions,
the leading exponent $\alpha$ which characterises the specific heat 
critical behaviour vanishes and 
no Harris prediction for the consequences of quenched disorder can be 
made~\cite{critharris}, see Appendix~\ref{Appendix_Harris}.
In the two-dimensional case, the controversy concerns the strong 
universality hypothesis
which maintains that the leading critical exponents remain the same as in the pure case, and the weak universality
hypothesis, which favours dilution-dependent leading critical exponents
(see~\cite{KeRu08} and references therein). 

Since $D=4$ marks the upper critical dimensionality of the model, 
the leading critical exponents there must be given by mean field theory
and there is no weak universality hypothesis.
However, unusual corrections to scaling characterise this model, 
and the precise nature of these corrections has been debated.
This debate motivates the work presented in this chapter: methods similar to 
those employed in ~\cite{KeRu08},
namely a high-statistics Monte Carlo (MC) approach coupled with 
finite-size scaling (FSS), are used 
to advance our understanding of the  four-dimensional version of the 
random-site Ising model (RSIM).

While not directly experimentally accessable, the four-dimensional 
RSIM  is of interest for the
following reasons: (i) it is closely related to the experimentally 
important dipolar Ising systems in three dimensions,
(ii) it is an important testing ground for the widespread 
applicability of the RG, (iii) it presents unusual corrections
to scaling, (iv) in high energy physics, the establishment 
of a non-trivial Higgs sector~\cite{FeFro92} for the standard model requires
a non-Gaussian fixed point and a new universality class which 
may, in principle, result from site dilution, and (v) it is 
the subject of at least {\emph{five}}  analytical papers which 
differ in the detail of the scaling behaviour at the phase transition.


\section{Analytical Framework}
\label{RSIM4D:analytics}

\subsection{Scaling in the RSIM in Four Dimensions}
\label{RSIM4D:scaling}

The consensus in the literature is that the following structure characterises the scaling behaviour of the
specific heat, the susceptibility, and the correlation length at the second-order phase transition in the RSIM in four dimensions 
(up to higher-order corrections to scaling terms)
\cite{Ah76,Boris,Jug,GeDe93,ISDIL4D,HeJa06}:
\begin{eqnarray} 
C_\infty(t)    & \approx & A - B|t|^{-\alpha} \exp{ \left( {-2  \sqrt{\frac{6}{53}|\log{|t|}|} }\right) } |\log{|t|}|^{\hat{\alpha}} \,,  \label{C1} \\
\chi_\infty(t) & \sim & |t|^{-\gamma}  \exp{\left({\sqrt{\frac{6}{53}|\log{|t|}|} }\right)}    |\log{|t|}|^{\hat{\gamma}} \,, \label{chiinfty} \\
\xi_\infty(t)  & \sim & |t|^{-\nu}    \exp{\left(\frac{1}{2}{\sqrt{\frac{6}{53}|\log{|t|}|} }\right)}     |\log{|t|}|^{\hat{\nu}} \,. \label{xiinfty}
\end{eqnarray}
Here, the subscript indicates the size of 
the system,  the reduced temperature $t=(T-T_\text{c})/T_\text{c}$
marks the distance of the temperature 
$T$ from its critical value $T_\text{c}$, and  $A$ and $B>0$ are  constants. 
The correlation function at criticality decays as~\cite{Boris,GeDe93}
\begin{equation}
{\cal{G}}_\infty(x)     =  x^{-(D-2+\eta)}       |\log{x}|^{\hat{\eta}} \,,
\label{G}
\end{equation}
where $x$ measures distance across the lattice, the 
dimensionality of which is $D$. The correlation length 
for a system of finite linear extent $L$ also exhibits a logarithmic correction and is of the form
\begin{equation}
 \xi_L(t=0) \sim L  (\log{L})^{\hat{q}} \,.
\label{xiq}
\end{equation}

The leading power-law behaviour is believed to be
mean field because the fixed point is expected to be Gaussian and therefore
\begin{equation}
\alpha = 0\,, \quad \beta = \frac{1}{2}\,, \quad \gamma = 1\,, \quad \delta = 3\,, \quad \nu = \frac{1}{2}\,, \quad \eta = 0\,, \quad \Delta = \frac{3}{2}\,.
\label{MF}
\end{equation}
Here, $\beta$ and $\delta$ are, in standard notation, the critical
exponents for the magnetisation out of field and in field respectively
while $\Delta$ is the gap exponent characterising the Yang-Lee 
edge.
There is no dispute in the literature regarding
these leading exponents, some of which will be re-verified in this chapter. 
Neither is there any dispute regarding the details of the unusual exponential 
correction terms in (\ref{C1})--(\ref{xiinfty}).
However there are at least {\emph{five}} different sets of 
predictions for the exponents of the logarithmic terms,
which differ from their counterparts in the pure model, and 
a principle aim of this work is to investigate these predictions numerically.

Aharony used a two-loop RG analysis to derive the unusual exponential terms in (\ref{C1})--(\ref{xiinfty}), and also found~\cite{Ah76} 
\begin{equation}
 \hat{\alpha}= \frac{1}{2}\,, \quad \hat{\gamma} = 0\,, \quad \hat{\nu}=0\,.
\label{Ah76}
\end{equation}
In~\cite{Boris}, Shalaev pointed out that Aharony's results needed 
to be refined and, by determining the beta function 
to three loops,  gave predictions for the specific heat and the susceptibility which differ from those in~\cite{Ah76}
in the slowly varying multiplicative logarithmic factors: 
\begin{equation}
\hat{\alpha}= 1.2368\,, \quad \hat{\gamma} = -0.3684\,, \quad \hat{\eta}=0.0094\,.
\label{Borisalpha}
\end{equation}
Jug studied the $\alpha = 0$ line  of $n$-component spin models 
in $(n,D)$ space where $D$ is the system's dimensionality, 
and thereby worked out the logarithmic corrections
for the $D=4$ $n$-vector model~\cite{Jug}. 
For the case at hand ($n=1$), he obtained
\begin{equation}
\hat{\alpha}= 1/2\,, \quad \hat{\gamma} = 1/212 \approx 0.0047\,.
\label{Jugalpha}
\end{equation}
In~\cite{GeDe93}, Geldart and De'Bell confirmed that to obtain the correct powers of $|\log{|t|}|$ the beta function has 
to be calculated to three loops, but the results of~\cite{GeDe93}
differ from those of 
\cite{Boris} in the powers of the logarithms
which appear in the specific heat and in the correlation function:
\begin{equation}
 \hat{\alpha}\approx 1.2463\,, \quad \hat{\gamma} \approx -0.3684\,, \quad \hat{\eta}=\frac{1}{212}=0.0047\,.
\label{GaDe93}
\end{equation}
Finally Ballesteros et al.~\cite{ISDIL4D} extended and corrected Aharony's computation to give the correction exponents:
\begin{equation}
 \hat{\alpha}= \frac{1}{2}\,, \quad \hat{\gamma} = \frac{1}{106} \approx 0.0094\,, \quad \hat{\nu}=0\,, \quad \hat{q}=\frac{1}{8}\,.
\label{BaFe98}
\end{equation}

So the detailed analytic scaling predictions of at least five groups of workers clash, and a number of questions arise:
(i) Is each set of predictions self-consistent?
(ii) What is the full set of predictions (i.e., extended to all 
observables) originated by each set?
(iii) Can a simulation approach provide numerical support for the shift in the
correction terms from their counterparts in the pure model?
(iv) And can such a computational approach lend  support to one or other of these five
different sets of analytic predictions?
Here the scaling relations for logarithmic corrections developed in~\cite{KeJo06a,KeJo06b} are used
to answer (ii), and it is shown that the answers to questions (i) and (iii) 
and to some extent (iv) are affirmative. 
In particular, numerical support is presented for the broad scenarios presented in~\cite{Ah76,Jug,ISDIL4D}.

Modification of the self-consistent scaling theory
for logarithmic corrections of~\cite{KeJo06a,KeJo06b}
to incorporate the exponential terms leads to the following forms 
for the  behaviour of the magnetisation in the 4D RSIM:
\begin{eqnarray}
m_\infty(t) & = & t^{\beta} \exp{\left({-\frac{1}{2}\sqrt{\frac{6}{53}|\log{|t|}|} }\right)}  |\log{t}|^{\hat{\beta}} \,, \label{m1} \\
m_\infty(h) & = & h^{\frac{1}{\delta}}  |\log{h}|^{\hat{\delta}} \,.\label{m2}
\end{eqnarray}

The Lee-Yang edge, denoted by $r_{LY}(t)$, is related to the locus of the Lee-Yang zeros along
the imaginary $h$-axis, see Appendix~\ref{Appendix_LY}, and marks the end of their
distribution. From Eq.(15) of~\cite{KeJo06a}, we also write for its scaling in the
paramagnetic phase
\begin{equation}
r_{\rm{YL}}(t) \sim t^{\Delta} \exp{\left({-\frac{3}{2}\sqrt{\frac{6}{53}|\log{|t|}|}}\right)} |\log{t}|^{\hat{\Delta}} \,.
\label{r}
\end{equation}

Besides the scaling behaviour of the Yang-Lee edge, defined in Eq.~(\ref{r}),
we also consider the \emph{density} of zeros which, for an
infinitely large system, we write as $g_\infty(r)$, where $r$ parameterises their locus along
the imaginary $h$-axis (assuming the Lee-Yang theorem holds). In fact it is more convenient to consider
the integrated, or cumulative, distribution function of zeros, defined as
\begin{equation}
G_\infty(r,t)=\int_{r_{YL}(t)}^r g_\infty (s,t) ds \ .
\label{dens_integr_LY}
\end{equation}
Following the approach outlined in~\cite{KeJo06a}, its
critical behaviour can be determined as
\begin{equation}
G_\infty(r) \sim r^{\frac{2-\alpha}{\varDelta}} \exp \left( \left(1 -\frac{3\gamma}{2\varDelta} \right) \sqrt{\frac{6}{53}|\log r|} \right) |\log r |^{\hat{\alpha}-(2-\alpha)\frac{\hat{\varDelta}}{\varDelta}} \ ,
\label{dens_integr_beha_LY}
\end{equation}
where the exponential term drops out by using the mean-field values $\gamma=1$ and $\varDelta=3/2$.

The scaling relations for logarithmic corrections in this 4D model are~\cite{KeJo06a,KeJo06b}\footnote{
The relation (\ref{J1}) is modified to read $ \hat{\alpha}  = 1+  d \hat{q} -  d \hat{\nu}$ when $\alpha = 0$
and when the impact angle of Fisher zeros onto the real axis is any value other than $\pi/4$, which is not expected to be the case in this 4D model~\cite{KeJo06b}.}
\begin{eqnarray}
 \hat{\alpha}                 & = &   D \hat{q} -  D \hat{\nu}  \,,          \label{J1} \\
 2 \hat{\beta} - \hat{\gamma} & = &   D \hat{q} -  D \hat{\nu}  \,,          \label{R1}\\
 \hat{\beta} (\delta - 1)     & = &  \delta \hat{\delta} - \hat{\gamma} \,,  \label{G1}\\
\hat{\eta}                    & = & \hat{\gamma} - \hat{\nu} (2 - \eta )\,,  \label{F1} \\
 \hat{\Delta}                 & = & \hat{\beta} - \hat{\gamma}\,.            \label{A1}
\end{eqnarray}

These scaling relations are now used to generate a complete scaling picture from the 
fragments available in the literature~\cite{Ah76,Boris,Jug,GeDe93,ISDIL4D}. 
This complete picture is given in Table~\ref{exp_analytic}, where the exponents of the logarithmic correction terms are listed.
Values for the exponents in boldface come directly from the reference concerned and the remaining values
are consequences of the scaling relations  (\ref{J1})--(\ref{A1}) .
\begin{table}
\begin{center}
\begin{tabular}{|r|l|l|l|l|l|l|} \hline
\footnotesize Log  &  \footnotesize Pure   model       &\footnotesize   Aharony ~\cite{Ah76}  &\footnotesize  Shalaev~\cite{Boris} &\footnotesize  Jug~\cite{Jug} &\footnotesize  Geldart          &\footnotesize  Ballesteros   \\
\footnotesize exp        &\footnotesize  \cite{ISDIL4D,KeLa94}&\footnotesize                       &\footnotesize                       &\footnotesize                 &\footnotesize   \& De'Bell~\cite{GeDe93}  &\footnotesize   et al ~\cite{ISDIL4D} \\\hline\hline
$\hat{\alpha}$ & 1/3            & {\bf{0.5}}            &   \,{\bf{1.237}}     &   {\bf{0.5}}   &  \, {\bf{1.246}}       &  {\bf{0.5}}      \\
$\hat{\beta}$  & 1/3            &  0.25                 &   \,0.434            &      0.252     &  \, 0.439              &  0.255           \\
$\hat{\gamma}$ & 1/3            & {\bf{0}}              &  {\bf{-0.368}}       &  {\bf{0.005}}  &   {\bf{-0.368}}        &  {\bf{0.009}}  \\
$\hat{\delta}$ & 1/3            &  0.167                &  \,0.167             &     0.170      &  \,  0.170             & 0.173            \\
$\hat{\nu}$    & 1/6            &  {\bf{0}}             &  -0.189              &                &   -0.187               & {\bf{0}}          \\
$\hat{\eta}$   & 0              & 0                     &  \,{\bf{0.009}}      &                &  \, {\bf{0.005}}       & 0.009            \\
$\hat{q}$      & 1/4            &  0.125                &   \,0.120            &                &   \,  0.125            & {\bf{0.125}}   \\
$\hat{\Delta}$ & 0              &  0.25                 &   \,0.803            &    0.248       &  \, 0.807              & 0.245            \\
\hline
\end{tabular}
\caption{Theoretical predictions for the exponents of the logarithmic corrections to scaling for the pure Ising model in 
four dimensions and for its random-site counterpart. 
The latter exponents are listed in boldface 
if they come directly from the cited literature. 
The remaining values are extended from those of the literature using the
scaling relations (\ref{J1})--(\ref{A1}). 
}  
\label{exp_analytic}
\end{center}
\end{table}
Each of the five papers~\cite{Ah76,Boris,Jug,GeDe93,ISDIL4D} is self-consistent in that the exponents
given within them do not violate logarithmic scaling relations. 
However, there are clear discrepancies {\emph{between}} 
each of the five papers.

The presence of the special exponential corrections has recently 
been verified by Hellmund and Janke in the case of 
the susceptibility~\cite{HeJa06}. 
These exponential terms mask the purely logarithmic corrections, 
so in order to detect and measure the latter
one needs to cancel the former. 
Certain combinations of thermodynamic functions achieve this,
but it turns out that FSS does this also.
FSS therefore offers an ideal method to determine the exponents of the logarithmic corrections numerically~\cite{KeRu08}.

\subsection{Finite-Size Scaling}
\label{RSIM4D:FSS}

Fixing the ratio of  $\xi_\infty(t)$ in (\ref{xiinfty}) and $\xi_L(0)$ in (\ref{xiq}) to $x$, one has
\begin{equation}
 t^{-\nu}  \exp{\left(\frac{1}{2}{\sqrt{\frac{6}{53}|\log{|t|}|} }\right)}  |\log{|t|}|^{\hat{\nu}}  =  x L (\log{L})^{\hat{q}}
\,.
\label{star}
\end{equation}
Taking logarithms of both sides, one obtains 
\begin{equation}
| \log{|t|}|  \approx \frac{1}{\nu} {\log{L}}\,,
\label{lnLt}
\end{equation}
which re-inserted into (\ref{star}) gives
\begin{eqnarray}
 t & \sim & 
 L^{-\frac{1}{\nu}} 
 \left({ \log{L} }\right)^{ \frac{ \hat{\nu}-\hat{q} }{\nu} }
 \exp{\left({ \frac{1}{2\nu} \sqrt{\frac{6}{53} \frac{1}{\nu} \log{L}} }\right)}
 \left\{{
    1 + {\cal{O}} 
                     \left({ 
                              \frac{1}{\sqrt{\log{L}}} 
                       }\right)
 }\right\}
\\
 & \sim & 
L^{-2} 
 \left({ \log{L} }\right)^{ -\frac{\hat{\alpha}}{2} }
 \exp{\left({ \sqrt{\frac{12}{53} \log{L}} }\right)}
 \left\{{
    1 + {\cal{O}} 
                     \left({ 
                              \frac{1}{\sqrt{\log{L}}} 
                       }\right)
  }\right\}\,,
\label{FSSpres}
\end{eqnarray}
having used the mean-field value (\ref{MF}) for the leading exponent 
$\nu$ and the logarithmic scaling relation (\ref{J1}).
If $\hat{\alpha}= 1/2$, this recovers a result in~\cite{ISDIL4D} for the FSS of the pseudo-critical point.

Inserting (\ref{FSSpres}) into (\ref{xiinfty}) recovers (\ref{xiq}), as it should.
The FSS's of the remaining functions are determined by inserting (\ref{FSSpres}) into  (\ref{C1}) to (\ref{xiinfty}) and (\ref{m1}) to (\ref{r}).
One finds 
\begin{equation}
C_L(0)    \approx A           - B^\prime   L^{\frac{\alpha}{\nu}}
   \exp{\left({ -\left({2+\frac{\alpha}{2\nu}}\right)\sqrt{\frac{6}{53\nu} \log{L}} }\right)}    
(\log{L})^{\hat{\alpha} + \frac{\alpha }{\nu}(\hat{q}-\hat{\nu})} \,, 
\label{C3long}
\end{equation}
where $B^\prime \propto B$ is a positive constant ~\cite{Ah76,Boris,Jug,GeDe93,ISDIL4D}.
Inserting the mean-field values $\alpha = 0$, $\nu = 1/2$, one obtains the
simpler form
\begin{equation}
C_L(0)    \approx A           - B^\prime   
\exp{\left({ -2\sqrt{\frac{12}{53} \log{L}} }\right)}    (\log{L})^{\hat{\alpha}} \,.
\label{C3}
\end{equation}
Similarly,  the FSS for the susceptibility is
\begin{equation}
\chi_L(0)  \sim  L^{\frac{\gamma}{\nu}} 
                        |\log{L}|^{\hat{\gamma}-\frac{\gamma}{\nu}(\hat{\nu} - \hat{q})}
 = L^{2} 
                        |\log{L}|^{\hat{\zeta} }
\,,
\label{chi3}
\end{equation}
where 
\begin{equation}
 \hat{\zeta}= \hat{\gamma}-2(\hat{\nu} - \hat{q})
 = \frac{1}{2} \hat{\alpha}+\hat{\gamma}\,.
\label{zeta}
\end{equation}
The FSS for the Yang-Lee edge is 
\begin{equation}
r_1(L)  \sim   L^{-\frac{\Delta}{\nu}} 
                        |\log{L}|^{\hat{\Delta}+\frac{\Delta}{\nu}(\hat{\nu} - \hat{q})}
=
 L^{-3} 
                        |\log{L}|^{\hat{\rho} }
\,,
\label{r2}
\end{equation}
where 
\begin{equation}
 \hat{\rho}=\hat{\Delta}+3(\hat{\nu} - \hat{q})
 = -\frac{1}{4}\hat{\alpha}- \frac{1}{2}\hat{\gamma} \,.
\label{rho}
\end{equation}
Each of these also has sub-leading scaling corrections
of strength $\mathcal{O}(1/\sqrt{\log{L}})$ times the leading behaviour.
One notes, however, that the unusual exponential terms, which swamp the 
logarithmic corrections in the
thermal scaling formulae (\ref{chiinfty})  and (\ref{r}), drop out 
of their FSS counterparts (\ref{chi3}) and (\ref{r2}).
These are therefore ideal quantities to study the logarithmic corrections. 
The theoretical analytical predictions of each of the five sources in the literature are now used to construct five possible FSS scenarios for 
the specific heat, the susceptibility, and the Lee-Yang zeros.
While Jug did not calculate the critical correlator or
correlation length in 4D, the  FSS picture corresponding to~\cite{Jug} can still be constructed through the scaling relations for logarithmic corrections.  
The FSS scenarios are listed  in Table~\ref{FSS_analytic}.
\begin{table}
\begin{center}
\begin{tabular}{|r|l|l|l|l|l|l|} \hline
Exponent   & Pure        & Aharony         & Shalaev          & Jug          &  Geldart \&              & Ballesteros   \\
           & model       &   ~\cite{Ah76}  & ~\cite{Boris}    &~\cite{Jug}   &   De'Bell~\cite{GeDe93}  &  et al ~\cite{ISDIL4D}   \\\hline\hline
Susceptibility  $\hat{\zeta}$    &  1/2   &  \,0.25  &  \,0.25  &  \, 0.255  &  \, 0.255                &  0.259           \\
Lee-Yang zeros  $\hat{\rho}$     & -1/4   &  -0.125  &  -0.125  &  -0.127    &   -0.127                 &  -0.130  \\
\hline \hline
\end{tabular}
\caption{The exponents of the multiplicative logarithmic corrections 
to FSS for 
the magnetic susceptibility and for the Lee-Yang zeros
coming from the literature and compared to their equivalents in the pure case. 
The FSS exponents are 
$\hat{\zeta}$ for the susceptibility and
$ {\hat{\rho}}$ for the Yang-Lee edge.
}
\label{FSS_analytic}
\end{center}
\end{table}

The remainder of this chapter is concerned with Tables~\ref{exp_analytic}~and~\ref{FSS_analytic}. 
The primary objective is to  verify that 
the exponents for the logarithmic-correction terms in the RSIM are indeed different from those of the pure model. 
Once this is established, one would like to determine which of the five sets of analytical predictions are supported numerically.
From Table~\ref{FSS_analytic}, it is clear that present-day numerics can not 
be sensitive enough to distinguish between all  five scenarios for
the susceptibility or individual zeros. However, there are clear differences between the predictions from 
\cite{Ah76,Jug,ISDIL4D} and from~\cite{Boris,GeDe93} for the specific heat
(Table~\ref{exp_analytic}),
and it will turn out that the numerical data is indeed sensitive enough to favour the former over the latter.

\section{The Model}
\label{RSIM4D:themodel}

The partition function of the RSIM in a reduced magnetic field $h$ is
\begin{equation}
Z_L(\beta,h)=\sum_{\{ \sigma_i \}}\text{exp}\left( \beta\sum_{\langle i, j \rangle} \epsilon_i \epsilon_j \sigma_i \sigma_j +
 h \sum_i \epsilon_i \sigma_i\right) \,,
\label{Z_RSIM}
\end{equation}
where $L$ denotes the linear extent of the lattice, the sum over configurations $\{ \sigma_i \}$ is taken over Ising spins
$\sigma_i \in \{ \pm 1\}$, $\langle i, j \rangle$ denotes nearest neighbours, and $\epsilon_i$ are independent quenched random variables 
which take the value unity with probability $p$ and zero with probability $1-p$. Below the percolation
threshold ($p_\text{c}=0.197$ in four dimensions), the phase transition is expected to
disappear, while for every $p>p_\text{c}$ there exists a critical (inverse) temperature $\beta_\text{c}(p)$ for
each given dilution.

In order to find the Lee-Yang zeros we define the energy, $E$, and the magnetisation,
$M$, of the system as
\begin{equation}
E=-\sum_{\langle i j \rangle} \epsilon_i \epsilon_j \sigma_i \sigma_j \quad , \quad
M=\sum_{i} \epsilon_i\sigma_i \,,
\label{quant_LY}
\end{equation}
and
\begin{equation}
\rho_L(\beta\,;M)=\sum_{E} \rho_L(E,M)\text{exp}(-\beta E) \,,
\label{rho_LY}
\end{equation}
where the spectral density $\rho_L(E,M)$ gives the relative weight of
configurations with given values of $E$ and $M$, the partition function
in an imaginary field $ih$ is therefore
\begin{equation}
Z_L(\beta,h) = \sum_M \rho_L(\beta\,;M)\text{exp}(ihM) =  Z_L(\beta,0) \langle \cos(hM) +i \sin(hM) \rangle \,,
\label{Z_imaginary}
\end{equation}
where the thermal average $\langle (\cdots) \rangle$ is a real measure, i.e., it is taken with
$Z(\beta,h=0)$. Assuming the Lee-Yang theorem holds~\cite{LEEYANG,JJRu97}, since
odd moments of the magnetisation vanish in the paramagnetic phase, the zeros
for a given realization of the disorder are given by the values of $h$ for which
\begin{equation}
\langle \cos (hM) \rangle = 0 \,.
\label{cos_LY}
\end{equation}
In this way we obtain the zeros of the partition function for each value of $p$ and $L$.
Then we average over realizations of the disorder (samples), and the resulting $j$th
Lee-Yang zero is denoted by $r_j(L)$,  the zero with $j=1$ being the smallest.

A robust method to determine the density of zeros, defined in Eq.~(\ref{dens_integr_LY}), from simulation
data was presented in Ref.~\cite{KENNA-DENS}. Defining the
density of zeros for a finite system of size $L$ along the singular line $r>r_{YL}(t)$ as
\begin{equation}
g_L(r)=L^{-D}\sum_j \delta [r-r_j(L)] \,,
\label{dens_delta}
\end{equation}
we can insert it into the cumulative density of zeros to obtain
\begin{equation}
G_L(r)=\int_0^r g_L (s) ds = \frac{j}{L^D} \quad \text{for: }\ r_j(L)<r<r_{j+1}(L)\,,
\label{dens_integr_LY2}
\end{equation}
so that it is given at a zero by the average
\begin{equation}
G_L[r_j(L)]= \frac{2j-1}{2L^D}\,.
\label{dens_integr_LY3}
\end{equation}

We also measure the non-connected susceptibility, $\chi_W$, defined as
\begin{equation}
\chi_L=\frac{1}{V}\langle M^2 \rangle \,,
\label{sus_W}
\end{equation}
with $V=L^4$ being the volume of the system.  This quantity
is directly related to the average size of the clusters
constructed using a Wolff algorithm~\cite{Wolff89}.  We
checked this point in this work. In all cases, the two
definitions of the non-connected susceptibility are fully
compatible.

Finally we measure the specific heat of the system, defined as
\begin{equation}
C=\frac{1}{V}(\langle E^2 \rangle - \langle E \rangle^2 )\,.
\label{C_LY}
\end{equation}

\section{Numerical Results}
\label{RSIM4D:simulations}

\subsection{Methods}
\label{RSIM4D:methodology}

\begin{figure}[!ht]
\begin{center}
\includegraphics[width=0.6\columnwidth, angle=270]{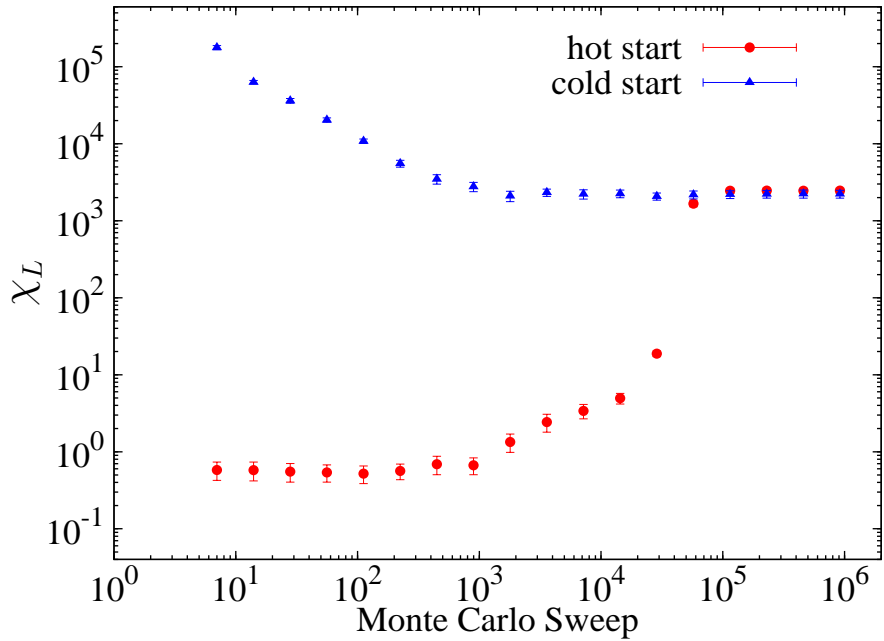}
\caption{Averaged behaviour of the susceptibility with the 
MC time for 20 samples at $L=32$ and $p=0.800$. 
Measurements were performed after every MC sweep (Wolff update). The plateau is reached more easily 
starting from a cold configuration.}
\label{termatest1}
\end{center}
\end{figure}

We performed extensive simulations of the model for linear
lattice sizes from $L=8$ to $L=48$ at dilutions $p=1$,
$p=0.8$, and $p=0.5$. In each case, we employed a Wolff
single-cluster algorithm~\cite{Wolff89} to update the spin
variables using periodic boundary conditions.
Thermalization tests including the comparison between cold
(all spins up) and hot (all spins random) starts were
carried out. We found that the plateau for the
susceptibility is quickly reached by starting from cold
configurations, see Fig.~\ref{termatest1}.  Indeed, the
results for the susceptibilities from hot and cold starts
are fully compatible (and are less than two standard
deviations away from each other, even at the level of
logarithmic corrections).
The information about the numerical details is given in
Table~\ref{tablesimu}.  We took 1000 disorder realizations
in all the cases except for $L=48$, where only 800 samples
were used. We estimate that the total simulation time was
equivalent to 20 years of a single node of a Pentium Intel
Core2 Quad 2.66 GHz processor.  Since our aim is to estimate
the scaling of quantities right at the critical point,
simulations must be performed at the critical temperature of
the model. We used the estimates for the critical
temperature given in~\cite{ISDIL4D}. In terms of $\beta =
1/kT$, where $k$ is the Boltzmann constant, these are
$\beta_\mathrm{c} = 0.149695$, $\beta_\mathrm{c} = 0.188864$,
and $\beta_\mathrm{c} = 0.317368$, for $p=1$, $p=0.8$, and
$p=0.5$, respectively.
%

\begin{figure}[!ht]
\begin{center}
\includegraphics[width=0.6\columnwidth, angle=270]{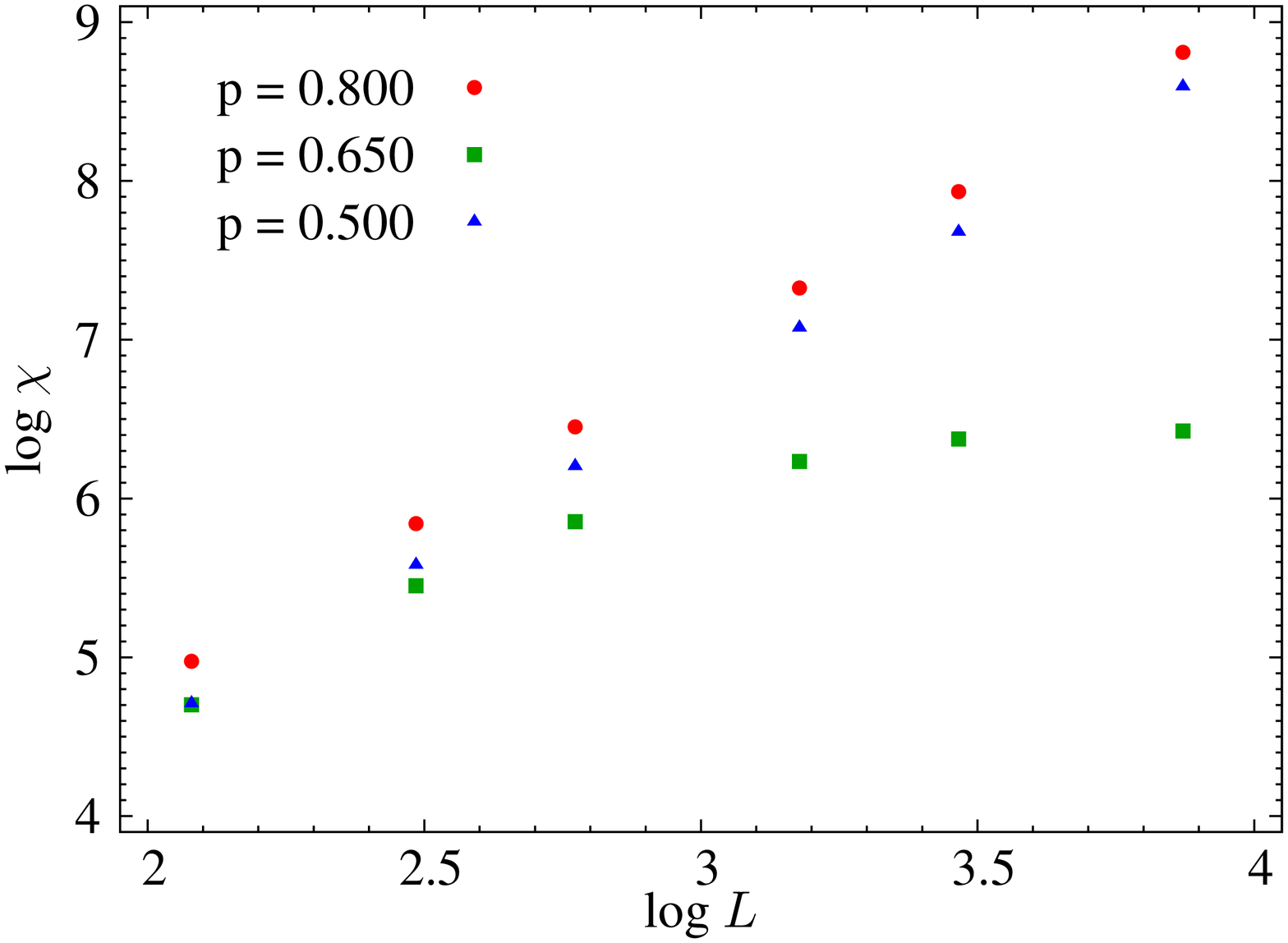}
\caption{Scaling of the susceptibility comparing the $p=0.650$ 
case at $\beta_\mathrm{c} = 0.235049$~\cite{ISDIL4D} with the other dilute cases (also
simulated at their corresponding $\beta_\mathrm{c}$'s obtained from~\cite{ISDIL4D}). There can be seen 
the strong deviation for $p=0.650$ from the expected leading behaviour $\chi\sim L^2$ .
The point size is in every case larger than the corresponding error bar.
}
\label{bias_p0650}
\end{center}
\end{figure}

In addition we simulated the dilution $p=0.650$ at
$\beta_\mathrm{c} = 0.235049$~\cite{ISDIL4D} using the same
statistics as for the other dilutions. In this case we found the
behaviour of the observables to differ from the expected. For
example,  Fig.~\ref{bias_p0650} shows the strong
deviation of the leading scaling behaviour of the
susceptibility compared with that of the other dilutions.
We re-checked this point starting from different initial
configurations and even using different random number
generators. This deviation is surely due to a biased
estimate of the critical temperature in~\cite{ISDIL4D}. For
this reason we omit $p=0.650$ from our analysis.

\begin{table}
\begin{center}
\begin{tabular}{|r|c|c|c|} \hline 
Spin Concentration & $L$ & $N_\mathrm{Wolff}$   & $N_\mathrm{d}$     \\\hline\hline
${\bm p}\mathbf{=1.000}$\,   &      8       &  200    &  2          \\
($\beta_\mathrm{c} = 0.149695$)	      &      12      &  400    &  8  \\
	      &      16      &  1600   &  32      \\
	      &      24      &  2000   &  128     \\
	      &      32      &  3000   &  400     \\
	      &      48      &  4000   &  1600    \\\hline
${\bm p}\mathbf{=0.800}$\,   &      8       &  100    &  1    \\
($\beta_\mathrm{c} =  0.188864$)	      &      12      &  200    &  4\\
	      &      16      &  800    &  16   \\
	      &      24      &  1000   &  64   \\
	      &      32      &  1500   &  200  \\
	      &      48      &  2000   &  1250 \\\hline
${\bm p}\mathbf{=0.500}$\,   &      8       &  100    &  2   \\
($\beta_\mathrm{c} =  0.317368$)	      &      12      &  200    &  8    \\
	      &      16      &  800    &  32   \\
	      &      24      &  1000   &  128  \\
	      &      32      &  1500   &  512  \\
	      &      48      &  2000   &  1250 \\
\hline 
\end{tabular}
\caption{Simulation details for each spin concentration $p$
and system size $L$.
Here, $N_\mathrm{Wolff}$ denotes the number of Wolff updates between consecutive
measurements, $N_\mathrm{d}$ is the number of dropped measurements  at the
beginning of the MC history (in units of $10^3$). We always performed $10^3$ measurements
within each sample after equilibration.
}  
\label{tablesimu}
\end{center}
\end{table}

\subsection[The Pure Case $p=1$]{The Pure Case $\bm{p=1}$}
\label{RSIM4D:purecase}

To establish confidence in the present approach, the pure system is analysed first to test whether the method employed
successfully quantitatively identifies the logarithmic corrections which are well established there.

The scaling and FSS of the pure model ($p=1$) are well understood~\cite{ISDIL4D,KeLa94}. 
The specific heat FSS behaviour is given by
\begin{equation}
  C_L(0)   \sim (\log{L})^{\hat{\alpha}} \sim  (\log{L})^{1/3}\,,
\label{C3pure}
\end{equation}
up to additive corrections. Fitting to this form for
$\hat{\alpha}$ over the full data set $ 8 \le L \le 48$, one
finds the estimate $\hat{\alpha}= 0.42(4)$ with a goodness
of fit given by a $\chi^2/\mathrm{d.o.f.}=3.9/3$,
C.L.=27\%. The estimate is two standard deviations away from
the known value $1/3$. As elsewhere in this analysis,
inclusion of sub-leading scaling correction terms in the
fits does not ameliorate this result, which is similar to
that reported in \cite{ISDIL4D}.

The FSS for the susceptibility is given in (\ref{chi3}) with $\hat{\zeta} = 1/2$. Fitting to the leading form 
\begin{equation}
 \chi_L(0) \sim  L^{\frac{\gamma}{\nu}}
\label{chifit}
\end{equation}
gives $\gamma/\nu=2.16(1)$ for $ 8 \le L \le 48$ and
 $\gamma/\nu=2.13(2)$  for $ 12 \le L \le 32$, the difference from the theoretical value $\gamma/\nu=2$ being
ascribable to the presence of the logarithmic correction term.  
Accepting this mean-field value for $\gamma/\nu$ and fitting to 
\begin{equation}
 \chi_L(0) \sim L^{2} (\log{L})^{\hat{\zeta}}
\label{chifitlog}
\end{equation}
gives the estimate
 $\hat{\zeta} = 0.48 \pm 0.02$   in the range $8 \le L \le 48$, albeit with $\chi^2/\mathrm{d.o.f.}=12.3/3$, C.L.=1\%.

The FSS for the individual Lee-Yang zeros is given in (\ref{r2}) with $\hat{\rho} = -1/4$ in the pure case. Fitting to the leading form 
\begin{equation}
 r_j(L) \sim L^{-\frac{\Delta}{\nu}}
\label{rfit}
\end{equation}
gives $\Delta / \nu= 3.074(5)  $ for $ 8 \le L \le 48$,
the difference from the theoretical mean-field value $\Delta / \nu=3$ being
due to the corrections.  Accepting this value and fitting to 
\begin{equation}
 r_j(L) \sim L^{-3} (\log{L})^{\hat{\rho}}
\label{rfitlog}
\end{equation}
gives  $\hat{\rho} = -0.22(2)$   in the range $8 \le L \le 48$.
This estimate is compatible with the known value $\hat{\rho}=-1/4$.
As one would expect,  the higher zeros yield less accurate estimates
(as they are further from the real simulation points) with 
 $\hat{\rho} = -0.18(3)$ ,
 $\hat{\rho} = -0.17(7)$ ,
and
 $\hat{\rho} = -0.10(14)$ 
from the second, third and fourth zeros respectively. 
These estimates are listed in Table~\ref{FSS_estimates}.

Having established that the numerics give reasonable agreement
with the  pure theory at the leading and the 
logarithmic levels, we now perform a similar analysis 
in the presence of disorder.

\subsection[The Dilute Cases $p=0.8$ and $p=0.5$]{The Dilute Cases $\bm{p=0.8}$ and $\bm{p=0.5}$}
\label{RSIM4D:dilutedcases}

Since the ansatz (\ref{C3}) for the specific heat in
disordered systems is somewhat more complex than that for the
pure case (\ref{C3pure}), we begin the $p \ne 1$ analyses
with the susceptibility and the Lee-Yang zeros. It will turn
out that our analyses will reinforce the analytical
predictions that scaling is governed by the Gaussian fixed
point and that the logarithmic corrections in the RSIM
differ from those in the pure model. Indeed, the results for
the zeros will be seen to be broadly compatible with the
analytic predictions contained
in~\cite{Ah76,Boris,Jug,GeDe93,ISDIL4D}.

\begin{table}
\begin{center}
\begin{tabular}{|l|l|l|l|l|l|l|} \hline 
   $p$   &          & $\hat{\zeta}$&\multicolumn{4}{c|}{$\hat{\rho}$}  \\\hline
\hline
         &          &             &        \multicolumn{4}{c|}{} \\ 
         & Theory ($p=1$) $\Rightarrow$& 1/2&\multicolumn{4}{c|}{ -1/4}\\ 
         & Theory ($p\ne 1$)$\Rightarrow$&$0.25$ to $0.26$&\multicolumn{4}{c|}{$-0.125$ to $-0.13$}\\
\hline
   &           &             &  $j=1$ & $j=2$  &$j=3$&$j=4$  \\ 
\hline
1  & $L=8-48$    & 0.48(2)  &-0.22(2)&-0.18(3)&-0.17(7) &-0.10(14)\\ 
\hline
0.8& $L=8-48$    & 0.39(3)  &-0.15(2)&-0.16(3)&-0.20(3) &-0.17(3) \\ 
0.8& $L=12-48$   & 0.42(4)  &-0.17(4)&-0.16(4)&-0.17(5) &-0.18(4)   \\ 
\hline
0.5& $L=8-48$    &  0.37(4) &-0.20(4)&-0.22(4)&-0.21(4) &-0.21(4) \\ 
0.5& $L=12-48$   &  0.40(6) &-0.16(5)&-0.20(5)&-0.18(5) &-0.19(5) \\ 
\hline 
\end{tabular}
\caption{FSS estimates for the various dilution values, using a range
of lattice sizes. For $p<1$ the susceptibility is expected to scale as $\chi_L\sim L^2(\log{L})^{\hat{\zeta}}$ and the Lee-Yang zeros 
as $r_j\sim L^{-3}(\log{L})^{\hat{\rho}}$, where $\hat{\zeta} \approx 0.25$ to $0.259$ and $\hat{\rho} \approx -0.125$ to $-0.130$. (For comparison, the pure theory with $p=1$ has $\hat{\zeta} = 1/2$ and $\hat{\rho} = -1/4$.)
}  
\label{FSS_estimates}
\end{center}
\end{table}
For the weaker dilution value $p=0.8$, a fit using all
lattice sizes to the leading form (\ref{chifit}) for the
susceptibility yields the estimate $\gamma/\nu = 2.14 \pm
0.01$. Ascribing the difference from the Gaussian value
$\gamma/\nu=2$ as being due to the correction terms and, as
in the pure case, fitting to (\ref{chifitlog}), one finds an
estimate for the correction exponent $\hat{\zeta} = 0.39(3)$
for $8 \le L \le 48$.  This value lies between the pure value
$\hat{\zeta} = 0.5$ and the theoretical estimates for the
dilute value which are $\hat{\zeta}\approx 0.25$ to
$0.26$. Thus, while the FSS for the susceptibility does not
capture the theoretical estimates for the dilute case, the
fitted values have moved away from the pure value and
towards the lower value listed in Table~\ref{FSS_analytic}.
As elsewhere in this work, the inclusion of scaling
corrections does not alter these results significantly.
\begin{figure}[!ht]
\begin{center}
\includegraphics[width=0.45\columnwidth, angle=270, trim=220 40 0 0, clip]{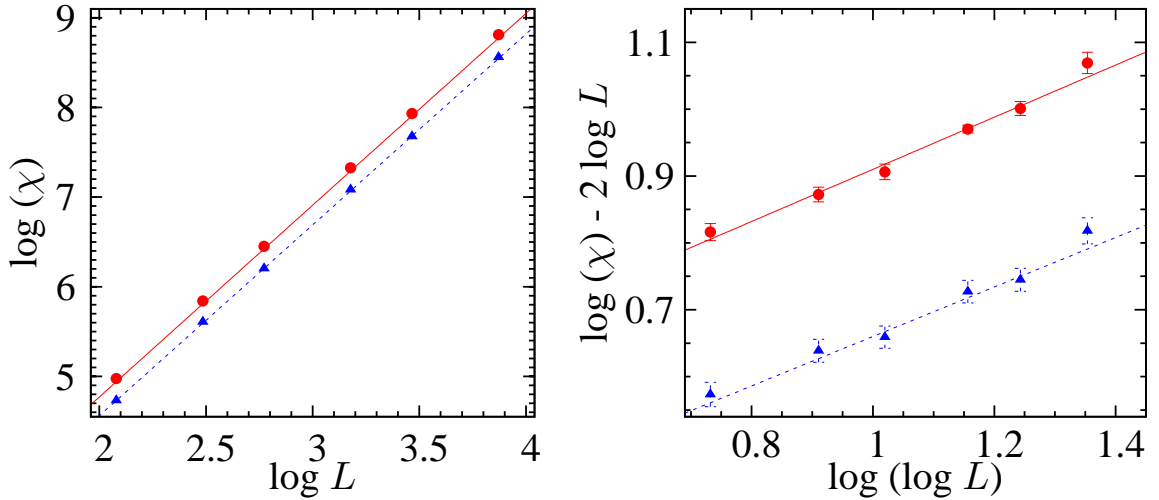}
\caption{
{\bf Left}: FSS plot for $\chi_L$ at $p = 0.8$  (circles)
and $p=0.5$ (triangles) at the critical point. The slopes of
the fitted solid and dashed lines give estimates for $\gamma
/ \nu$ of $2.14(1)$ and $2.13(2)$, respectively.
{\bf Right}: Plot of  $\log{\chi_L}-2 \log{L}$ against $\log{(\log{L})}$ 
at $p = 0.8$  (circles) and $p=0.5$ (triangles)
giving  slopes $0.39(3)$ and $0.37(4)$, respectively, 
indicating slow crossover of  multiplicative logarithmic 
corrections from the pure case (where $\hat{\zeta} = 0.5$) to the dilute
case, where the theoretical value is $\hat{\zeta} \approx 0.25$.
}
\label{fig_susceptibility}
\end{center}
\end{figure}

A similar analysis for the FSS of the susceptibility at the
stronger dilution value $p=0.5$ gives similar results: the
leading form (\ref{chifit}) yields an estimate $\gamma/\nu =
2.13 \pm 0.02$ with a goodness of fit given by
$\chi^2/\mathrm{d.o.f.}=1.2/3$, C.L.=75\%.  Ascribing the
difference from the mean-field value $\gamma/\nu = 2$ as
being due to the logarithmic corrections, and fitting to
(\ref{chifitlog}), one obtains the estimate $\hat{\zeta} =
0.37(4)$ for $8 \le L \le 48$.  Again this result is between
the theoretical predictions for the pure ($\hat{\zeta} =
0.5$) and the dilute ($\hat{\zeta} \approx 0.25$ to $0.26$)
cases. These results are summarised in
Table~\ref{FSS_estimates}, together with the results for the
same fits but with the smallest lattices removed.

Since in each of the dilute cases the susceptibility results
lie between what is expected for the pure and for the dilute
theories, we appeal to the Lee-Yang zeros since they are
expected to provide a cleaner signal.

\begin{figure}[!ht]
\begin{center}
\includegraphics[width=0.45\columnwidth, angle=270, trim=220 40 0 0, clip]{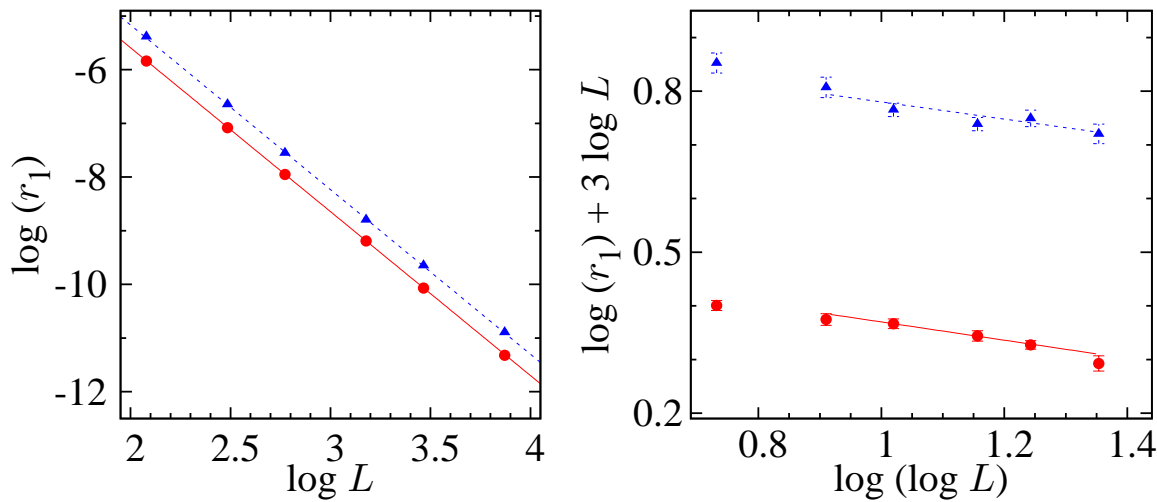}
\caption{
{\bf Left}: FSS plot for the Yang-Lee edge at $p = 0.8$  
(circles) and $p=0.5$ (triangles). 
The slopes of the fitted solid and dashed lines give estimates for 
$\Delta / \nu$ of $3.055(4)$ and $3.07(2)$, respectively.
{\bf Right}: Plot of  $\log{r_1}+3 \log{L}$ against $\log{(\log{L})}$ 
at $p = 0.8$  (circles) and $p=0.5$ (triangles).
Fits in the range $L=12$ to $L=48$ (plotted) give
 slopes $-0.17(4)$ and $-0.16(5)$, compatible with the literature
predictions that range from $\hat{\rho}\approx -0.125$ to $\hat{\rho}\approx -0.13$.
(For comparison, in the pure model, $\hat{\rho} = -1/4$.)
}
\label{fig_firstzero}
\end{center}
\end{figure}

The leading behaviour is first examined by fitting each of the 
first four Lee-Yang zeros to Eq.~(\ref{rfit}). 
For the weaker dilution given by $p=0.8$, one obtains 
$\Delta/\nu = 3.055(8)$, $3.056(9)$, $3.069(11)$, and $3.060(10)$
from fits to the first, second, third, and fourth zeros,
respectively, using all lattice sizes.
The equivalent results for the stronger dilution value
$p=0.5$ are 
$\Delta/\nu = 3.068(13)$, $3.071(15)$, $3.072(12)$, and $3.071(11)$,
respectively.
All fits are of good quality with acceptable values of $\chi^2/\mathrm{d.o.f.}$,
which we refrain from detailing.
Again, these are interpreted as being supportive of the mean-field
leading behaviour $\varDelta/\nu=3$ with logarithmic corrections.

The logarithmic-correction exponents are estimated by fitting to 
Eq.~(\ref{rfitlog}), with the various theories indicating that 
$\hat{\rho}= -0.125$ to $-0.13$. The strongest evidence supporting this
comes, as it should, from the first zero (the 
Yang-Lee edge) for $p=0.8$, which yields the estimate $\hat{\rho} = -0.15(2)$
(with $\chi^2/\mathrm{d.o.f.}=1.8/3$, C.L.=61\%).
As in the pure case, and as expected, estimates for $\hat{\rho}$
deteriorate as higher-index zeros are used. Dropping the smallest 
lattices from the analysis, however, leads to these estimates for  $\hat{\rho}$
becoming more compatible with~\cite{Ah76,Boris,Jug,GeDe93,ISDIL4D}. These results are
summarised in Table~\ref{FSS_estimates}.

The equivalent analysis for the stronger dilution value $p=0.5$ is less clear,
with an estimate $\hat{\rho}=-0.20(4)$ coming from the first zero
when all lattices are included in the fit (with $\chi^2/\mathrm{d.o.f.}=3.3/3$, C.L.=35\%).
Dropping the smallest  lattices, however,
gives $\hat{\rho}=-0.16(5)$ (with $\chi^2/\mathrm{d.o.f.}=1.8/2$, C.L.=40\%),
closer to the values coming from~\cite{Ah76,Boris,Jug,GeDe93,ISDIL4D}. 
Similar results are obtained for the higher zeros, and these are also 
summarised in Table~\ref{FSS_estimates}.

As a final check of the reliability of our results, we used
the spectral energy method~\cite{REWEIGHT1,REWEIGHT2} to
re-weight the data obtained at $\beta_\mathrm{c}$ to
$\beta_\mathrm{c}\pm\Delta\beta_\mathrm{c}$ (taken again
from~\cite{ISDIL4D}), finding that the new data sets are
fully supportive of the previous results\footnote{ We
followed the recipe given in
Appendix~\ref{Appendix_Extrapolations} to perform the
extrapolation to infinite number of measurements per sample.
}.

%
%

Having established that the leading FSS behaviour
corresponds to that originating from the Gaussian fixed
point, and that the logarithmic corrections to scaling are
different from those in the pure model and moreover are (at
least in the case of the Yang-Lee edge) broadly compatible
with the literature
predictions~\cite{Ah76,Boris,Jug,GeDe93,ISDIL4D}, we now
attempt to distinguish {\emph{between}} these broad
predictions.  To this end we turn to the specific heat.

\begin{figure}[!ht]
\begin{center}
\includegraphics[width=0.6\columnwidth, angle=270, trim=0 40 0 0, clip]{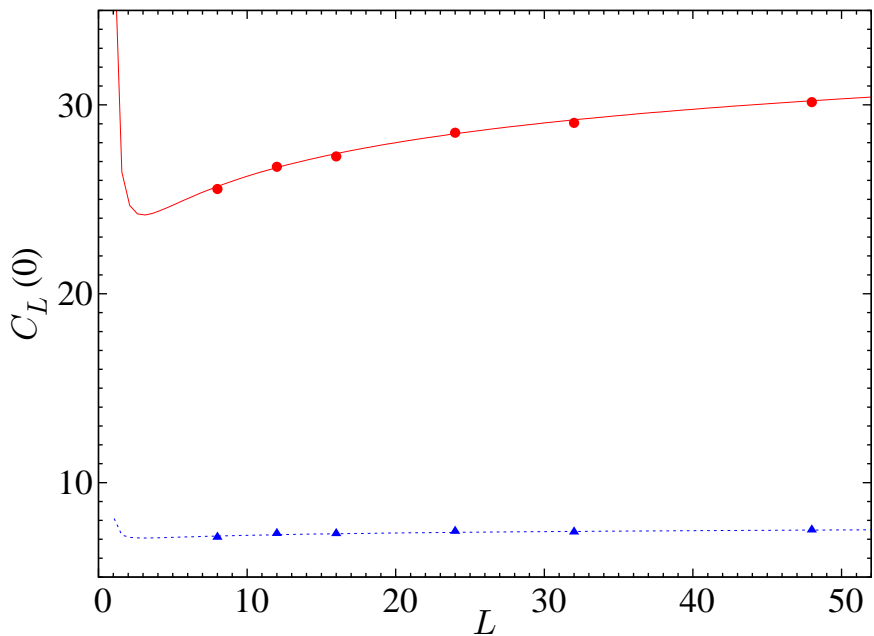}
\caption{The specific heat for $p=0.8$ (circles) and 
$p=0.5$ (triangles). The error bars are in every case smaller than the point size.
The solid and dashed curves are best fits to the
ansatz Eq.~(\ref{C3}), with $\hat{\alpha}=1/2$.
}
\label{fig_cesp}
\end{center}
\end{figure}

Having established confidence in the validity of the
mean-field values $\gamma=1$ and $\Delta=3/2$ for the 4D
RSIM, we may use the scaling relation $\alpha = 2-2\Delta
+\gamma $ to also establish the mean-field value $\alpha =
0$. The ansatz Eq.~(\ref{C3}) for the specific heat may now
be used.  This contains information which can be used to
discriminate between some of the scenarios in the
literature. In Table~\ref{exp_analytic}, one observes that
there is a striking difference between the estimates for the
specific heat logarithmic-correction exponent $\hat{\alpha}$
coming from~\cite{Boris,GeDe93} and
from~\cite{Ah76,Jug,ISDIL4D}. While the former have
relatively large values of $\hat{\alpha}$, the latter agree
on $\hat{\alpha} = 0.5$. The simulated values of the
specific heat at $p=0.8$ and $p=0.5$ are plotted in
Fig.~\ref{fig_cesp}.  The slope of the full specific heat
curve (\ref{C3}) is
\begin{equation}
 \frac{d C_L}{dL} = [A-C_L(0)]\frac{\sqrt{12/53}}{L\sqrt{\log{L}}}\left({ 1 - \frac{\hat{\alpha} \sqrt{53/12}}{\sqrt{\log{L}}} }\right)
\,.
\label{slope}
\end{equation}
This  vanishes when 
$C_L(0) = A$ and when $\sqrt{\log{L}} = \hat{\alpha} \sqrt{53/12}$.
The first of these is the asymptote 
$L \rightarrow \infty$, from which  $A$ can be determined
for each dilution. 
The second occurrence of zero slope is for quite small lattice sizes,
i.e., beneath lattice size $L=8$.
Therefore 
$ \hat{\alpha}
{\rm{\raisebox{-.75ex}{ {\small \shortstack{$<$ \\ $\sim$}} }}}
\sqrt{12/53}\sqrt{\log{8}}
\approx 0.7
$, which excludes the values $ \hat{\alpha}\approx 1.237$
and $\hat{\alpha}\approx 1.246$ given
in~\cite{Boris,GeDe93}.  In fact, a best fit to the ansatz
Eq.~(\ref{C3}) gives $A=89(34)$, $B^\prime=142(71)$, and
$\hat{\alpha} = 0.57(14)$ for $p=0.8$, and $A=60(25)$,
$B^\prime=102(54)$, and $\hat{\alpha} = 0.76(3)$ for
$p=0.5$.  Fixing $\hat{\alpha}=1/2$ in each case gives
$A=76(4)$, $B^\prime=115(13)$ for $p=0.8$, and $A=22(3)$,
$B^\prime=19(9)$ for $p=0.5$. These curves are plotted along
with the specific heat measurements in Fig.~\ref{fig_cesp}.
However, fixing the correction exponent $\hat{\alpha}$ to the value
given in~\cite{Boris,GeDe93} yields a
best-fit value of $B^\prime$ which is negative in each case,
contradictory to
\cite{Ah76,GeDe93}. Thus we can deem these values to be unlikely.

\section{Conclusions}
\label{RSIM4D:conclusions}

Numerical measurements of the leading critical exponents
in the 4D RSIM have been presented, confirming that the phase transition 
in this model is governed by the Gaussian fixed point.
We then turned to the corrections to scaling, for which there exist 
{\emph{five distinct sets of predictions}} in the literature 
\cite{Ah76,Boris,Jug,GeDe93,ISDIL4D}. 
The scaling relations for logarithmic
corrections were used to complete these sets,
and their counterparts for finite-size systems were given.

The measured values of the susceptibility FSS correction exponent, 
$\hat{\zeta}$, for the site-diluted model lie between
the known value for the pure model and the theoretical estimates
coming from~\cite{Ah76,Boris,Jug,GeDe93,ISDIL4D} for the disordered system.
While this result illustrates slow crossover of the susceptibility,
the lowest lying Lee-Yang zeros give a cleaner signal. The measured
value for their logarithmic correction exponents were
 indeed found to be compatible with the theories.

To discriminate between the five theories, the detailed
finite-size scaling behaviour of the specific heat was also
examined. The analysis was clearly in favour of the
analytical predictions of~\cite{Ah76,Jug,ISDIL4D} over those
of~\cite{Boris,GeDe93}. This was contrary to expectation
since the former involve only two loops in the perturbative
RG expansion, while the latter take the expansion to three
loops in the beta function.

\renewcommand{\arraystretch}{1}

\clearpage{\thispagestyle{empty}\cleardoublepage}

\clearpage{\thispagestyle{empty}\cleardoublepage}

\normalfont

\selectlanguage{british}

\vspace{4cm}
\chapter{Conclusions}
\label{chap:conclusions}

We have presented in this work highly accurate numerical simulations
of various models of phase transitions in the presence of dilution. We
checked the validity of some recent work, being able to outperform their
statistical accuracy.

Firstly, we checked the goodness of a recently proposed
microcanonical simulation method~\cite{VICTORMICRO} that computes entropy,
rather than free energy, to derive all the thermodynamic
information. The results, both for the pure four-state
($Q=4$) Potts model in two dimensions and for the pure Ising
model in three dimensions, were fully compatible with the
most recent canonical simulations~\cite{PELI-REP, Salas}. To achieve this, we
applied (for the first time, to the best of our knowledge)
the Nightingale phenomenological renormalization~\cite{NIGHTINGALE} (also
called the Quotient Method~\cite{QUOTIENTS}) within a microcanonical
ensemble. In addition we obtained the critical exponents in
the microcanonical ensemble, checking the validity of the
Fisher renormalization~\cite{FISHER-RENORM} for models with a constraint in the
internal energy.

Once we had set up a correct microcanonical simulation
method, we used it to study the (inherently complicated)
strong first-order phase transition of the three-dimensional
Potts model with $Q>3$ states. We confirmed that dilution
dramatically smooths the transition up to a tricritical
point, with spin concentration $p_\mathrm{c}$, where it
becomes of second order~\cite{Ball00, Chat01, Chat05}. We were able to claim that
$p_\mathrm{c}$ is definitely less than unity, having
obtained $p_\mathrm{c}=0.954(3)$ and $p_\mathrm{c}=0.922(1)$
in the $Q=4$ and $Q=8$ cases, respectively. We also obtained
that within the first order region in the $Q=4$ case the
latent heat is a self-averaging quantity while the surface
tension is not.  As a future development we will study the
scaling of latent heat, surface tension, and critical
temperature within the first-order region for the $Q=8$
case.  In this case we will have to deal with not fully
equilibrated systems, so that we will have to quantify the effects
of the lack of thermalization.

Within the canonical ensemble, we studied the critical
properties of the Heisenberg dilute model in three
dimensions for different values of the dilution.  Using the
next-to-leading scaling correction, we obtained results
fully compatible with the Renormalization Group predictions and with the Harris
criterion: our exponents and cumulants in the dilute cases
were compatibles with those of the pure model and independent
of the dilution. We also obtained strong evidence for
a zero $g_2$ cumulant, in both the vector and the tensor
channels, in the thermodynamic limit at criticality,
contrasting with some analytical predictions~\cite{korut}, but in
agreement with others~\cite{harris}. The introduction of
scaling corrections into the analysis was found to be crucial to
obtain the $g_2=0$ scenario.

We also studied the site-diluted version of the Ising model
in four dimensions, confirming that the phase transition in
this model is governed by the Gaussian fixed point.  The
logarithmic corrections to scaling were analysed to try to
discriminate between five distinct sets of
predictions~\cite{Ah76,Boris,Jug,GeDe93,ISDIL4D}. The
measured values of the susceptibility logarithmic correction
exponent in the dilute case lie between the known value for
the pure model and all the theoretical estimates for the
disordered system, indicative of a slow crossover to the
dilute universality class. We were able to discriminate
between the five theories by a detailed study of the
finite-size scaling behaviour of the specific heat.  The
analysis is clearly in favour of the analytical predictions
of~\cite{Ah76,Jug,ISDIL4D} over those
of~\cite{Boris,GeDe93}. Further theoretical effort should be
made in this field because the favoured scenerio stems from
computation up to two loops in the perturbative RG
expansion, while the rejected scenario involves expansion up
to three loops in the beta function.

The numerical results of this thesis were only made possible
by the intensive use of important supercomputing
facilities. We obtained from their resources more than the
equivalent of 400 years of computation time of a single last
generation Pentium 2.5 GHz CPU. Specifically, we used
the clusters at the ``Instituto de Biocomputaci\'on y F\'{i}sica
de Sistemas Complejos'' (BIFI) and the ``Barcelona
Supercomputing Centre'' (BSC). In addition, we exploited
the volunteer computing platform IBERCIVIS, for which we are in debt
with all its developers and volunteers.

\selectlanguage{british}

\clearpage{\thispagestyle{empty}\cleardoublepage}

\addtocounter{chapter}{-1}

\normalfont

\selectlanguage{spanish}

\vspace{4cm}
\chapter{Conclusiones}
\label{conclusiones}

En el presente trabajo hemos realizado simulaciones num\'ericas de alta precisi\'on
de varios modelos de transiciones de fase en presencia de desorden.
Con dichas simulaciones hemos logrado comprobar la validez de los trabajos
m\'as recientes y hemos mejorado la precisi\'on de sus resultados.

En primer lugar, hemos comprobado la validez de un m\'etodo de simulaci\'on
dentro del colectivo microcan\'onico propuesto recientemente~\cite{VICTORMICRO}. Dicho
m\'etodo utiliza la entrop\'{\i}a, en lugar de la energ\'{\i}a,
para obtener toda la informaci\'on termodin\'amica del sistema.
Los resultados obtenidos, tanto para el modelo de Potts puro con cuatro estados ($Q=4$)
en dos dimensiones como para el modelo de Ising puro en tres dimensiones, son completamente
compatibles con las simulaciones dentro del colectivo can\'onico m\'as recientes~\cite{PELI-REP, Salas}.
Para lograrlo hemos aplicado por primera vez, que nosotros sepamos, la
renormalizaci\'on fenomenol\'ogica de Nightingale~\cite{NIGHTINGALE} (tambi\'en llamada M\'etodo de
los Cocientes~\cite{QUOTIENTS}) dentro del colectivo microcan\'onico.
Adem\'as hemos obtenido los exponentes cr\'{\i}ticos microcan\'onicos
comprobando la validez de la renormalizaci\'on de Fisher~\cite{FISHER-RENORM} para un modelo
con una ligadura en la energ\'{\i}a.

Una vez que hemos asegurado la bondad de nuestro m\'etodo de simulaci\'on microcan\'onico,
lo hemos utilizado para el estudio de la (inherentemente complicada) transici\'on
de primer orden fuerte que tiene lugar en el modelo de Potts tridimensional
con $Q>3$ estados. Hemos comprobado que la diluci\'on suaviza dr\'asticamente la
transici\'on hasta llegar a un punto tricr\'{\i}tico, con concentraci\'on de espines
$p_\mathrm{c}$, donde la transici\'on pasa a ser de segundo orden~\cite{Ball00, Chat01, Chat05}. Podemos afirmar
que $p_\mathrm{c}$ es definitivamente m\'as peque\~na que la unidad, habiendo
obtenido $p_\mathrm{c}=0.954(3)$ y $p_\mathrm{c}=0.922(1)$ en los casos
$Q=4$ y $Q=8$ respectivamente.
Tambi\'en hemos obtenido que dentro de la regi\'on de primer orden para el caso $Q=4$ 
el calor latente es una magnitud autopromediante mientras que la tensi\'on superficial no lo es.
Como una futura investigaci\'on se pretende estudiar el escalado del
calor latente, la tensi\'on superficial y la temperatura cr\'{\i}tica dentro de la regi\'on
de primer orden para el caso con $Q=8$, para ello tendremos que tratar con estados
no completamente equilibrados por lo que tendremos que cuantificar los efectos debidos
a la falta de termalizaci\'on.

Dentro del colectivo can\'onico hemos estudiado las propiedades cr\'{\i}ticas
del modelo de Heisenberg diluido en tres dimensiones para diferentes valores
de la diluci\'on. Usando hasta segundo orden en correcciones de escala hemos
obtenido resultados completamente compatibles con las predicciones del Grupo
de Renormalizaci\'on y con el criterio de Harris: los exponentes
y cumulantes obtenidos son compatibles con los del modelo puro
e independientes de la diluci\'on. Adem\'as hemos obtenido evidencias
importantes de un cumulante $g_2$ nulo, tanto en el canal vectorial como en el tensorial,
en el l\'{\i}mite termodin\'amico y en el punto cr\'{\i}tico. \'Este \'ultimo resultado
contrasta con algunas predicciones anal\'{\i}ticas~\cite{korut} y concuerda
con otras~\cite{harris}. La introducci\'on de correcciones al escalado en
nuestro an\'alisis ha sido fundamental para obtener el escenario con $g_2=0$.

Tambi\'en hemos estudiado la versi\'on con diluci\'on por sitios
del modelo de Ising en cuatro dimensiones, confirmando que la transici\'on
de fase de este modelo est\'a gobernada por el punto fijo Gaussiano.
Las correcciones logar\'{\i}tmicas al escalado fueron analizadas para
tratar de discriminar entre cinco conjuntos distintos de 
predicciones~\cite{Ah76,Boris,Jug,GeDe93,ISDIL4D}. 
Los valores medidos del exponente de correcciones logar\'{\i}tmicas de la
susceptibilidad en el caso diluido se sit\'uan entre el valor
conocido del modelo puro y todos los valores te\'oricos estimados para el modelo diluido, esto
es un signo de la existencia un fen\'omeno de paso (\emph{crossover}) muy lento hacia la clase
de universalidad del modelo diluido.
Hemos logrado discriminar entre las cinco teor\'{\i}as en conflicto haciendo un estudio
detallado de comportamiento de escalado con el tama\~no del sistema
del calor espec\'{\i}fico. El an\'alisis claramente favorece las predicciones
anal\'{\i}ticas propuestas en~\cite{Ah76,Jug,ISDIL4D} sobre las propuestas en~\cite{Boris,GeDe93}. 
Un esfuerzo te\'orico adicional parece necesario ya que
el escenario favorecido procede de un c\'alculo hasta dos ``loops''
de la expansi\'on perturbativa del Grupo de Renormalizaci\'on mientras
que el escenario descartado procede de una expansi\'on hasta tres ``loops''.

Los resultados num\'ericos de esta tesis doctoral solo han sido posibles
debido al uso exhaustivo de importantes infraestructuras de supercomputaci\'on.
Hemos obtenido el equivalente a m\'as de 400 a\~nos de tiempo de c\'omputo de un \'unico
procesador Pentium 2.5 GHz de \'ultima generaci\'on. Espec\'{\i}ficamente
hemos usado los \emph{clusters} del ``Instituto de Biocomputaci\'on y F\'{i}sica de Sistemas
Complejos'' (BIFI) y del ``Barcelona Supercomputing Centre'' (BSC). 
Adem\'as hemos explotado la plataforma de computaci\'on voluntaria IBERCIVIS, 
por lo que estamos en una deuda profunda con los desarrolladores y voluntarios.

\selectlanguage{british}

\clearpage{\thispagestyle{empty}\cleardoublepage}

\normalfont

\selectlanguage{british}

\renewcommand{\appendixtocname}{List of Appendices}

\appendixtitletocon

\begin{appendices}
 

\appendixpage

\noappendicestocpagenum

\addappheadtotoc

\appendixheaderoff

\vspace{4cm}
\chapter{The Harris Criterion}
\label{Appendix_Harris}

Given the fact that real systems are almost always impure, it is crucial to quantify
to what extent, if any, disorder affects their critical behaviour.
A criterion, the so-called Harris criterion, makes it possible to
predict quantitatively the effect of disorder by using the critical exponents
of the pure system only~\cite{critharris}. According to this criterion,
the impurities change the critical behaviour only if the specific heat exponent $\alpha$
of the pure system is positive (the specific heat of the pure system is divergent). In the
opposite case, $\alpha<0$ (the specific heat is finite), the impurities appear to be irrelevant,
i.e. their presence does not alter the critical behaviour. We will derive the criterion following the approach
of Ref.~\cite{Dots-book}.

Let us consider a system with quenched disorder.  This can
be for example the presence of impurities at random sites in
a crystal lattice. In the pure case, this system undergoes a
continuous phase transition at a temperature $T_\text{c}$.
This critical temperature is expected to change in presence of disorder
because it introduces spatial inhomogeneities in
the coordination number\footnote{This is by definition the
number of interacting neighbours of a given spin. In a mean
field calculation, one obtains that the critical temperature
of the Ising model is $T_\text{c}=2qJ/k_B$, where $q$ is the
coordination number, $J$ the coupling, and $k_B$ the
Boltzmann constant.}.
The thermodynamics of second-order phase transitions is dominated by
large-scale fluctuations. The dominant scale, or the correlation
length, $\xi \sim |t|^{-\nu}$ goes to infinite as $ t\to 0$, with $t=(T-T_\text{c})/T_\text{c}$ being the reduced temperature.

The strength of the disorder (in our example, the impurity
concentration) is denoted by $\rho$, with $\rho=0$  being the pure case.
As $T_c$ is approached the following change of scale length takes place. First
the correlation length of the fluctuations becomes much larger
than the lattice spacing, and the system 'forgets' about the lattice.
The only relevant scale that remains in the system is $\xi(t)$. When we move closer
to the critical point, $\xi$ grows and becomes larger than the average distance
between impurities, so that the effective concentration of impurities, measured with respect to the correlation
length, becomes larger. It should be stressed that such a situation is reached for an
arbitrary small initial concentration $\rho$. If $\xi^D \rho\gg1$ there is no reason for believing that the effect of impurities
will be small.

We will discuss, for the sake of simplicity, the Harris criterion using a
particular model: the D-dimensional Ising-like system described in terms of the scalar 
field Ginzburg-Landau Hamiltonian, see for example~\cite{Dots-book}:
\begin{equation}
H= \int \text{d}^Dx\left[ \frac{1}{2}(\nabla\phi(x))^2+\frac{1}{2}(t-\delta t(x))\phi^2(x)+\frac{1}{4}g\phi^4(x) \right]\,,
\label{GL-Hamil}
\end{equation}
where the quenched disorder is described by random fluctuations of the effective transition temperature $\delta t(x)$ whose probability
distribution is taken to be symmetric and Gausssian:
\begin{equation}
P[\delta t]=p_0 \exp\left( - \frac{1}{4\rho} \int \text{d}x\, (\delta t(x))^2 \right)\,,
\label{P_delta_t}
\end{equation}
where $p_0$ is the normalisation constant. For notational simplicity,
the sign of $\delta t (x)$ in Eq.~(\ref{GL-Hamil}) is defined such that positive fluctuations lead
to locally ordered regions.

Configurations of the fields $\phi(x)$ that correspond local minima in $H$ satisfy the saddle-point equation:
\begin{equation}
-\varDelta \phi(x)+t\phi(x)+g\phi^3(x) =\delta t(x) \phi(x)\,,
\label{saddle_cond}
\end{equation}
Such localised solutions exist in regions of space where $t-\delta t(x)$ assumes negative values.
Clearly, the solutions of Eq.~(\ref{saddle_cond}) depend on a particular configuration
of the function $\delta t(x)$ being inhomogeneous. Let us estimate under which conditions the quenched
fluctuations of the effective transition temperature are the dominant factor for the local minima field
configurations.

Let us consider a large region $\Omega_L$ of linear size $L\gg 1$. The spatially
average value of the function $\delta t(x)$ in this region can be defined as follows:
\begin{equation}
\delta t(\Omega_L)= \frac{1}{L^D} \int_{x\in \Omega_L} \text{d}x\,\delta t(x)\,.
\label{spat_aver}
\end{equation}
Correspondingly, for the characteristic values of the temperature fluctuations (averaged
over realizations) in this region we get:
\begin{equation}
\delta t_L= \sqrt{\overline{\delta t^2(\Omega_L)}}= \sqrt{2\rho} L^{-D/2}\,.
\label{sampl_aver}
\end{equation}
Then, according to Eq.~(\ref{saddle_cond}) the average value of the order parameter $\phi(\Omega_L)$
in this region can be estimated from the equation:
\begin{equation}
t+g\phi^2 =\delta t(\Omega_L)\,.
\label{saddle_cond2}
\end{equation}
One can obtain that if the value of $t$ is sufficiently small, i.e. if
\begin{equation}
\delta t(\Omega_L)\gg t \,,
\label{t_small}
\end{equation}
then the solutions of Eq.~(\ref{saddle_cond2}) are defined only by the value of the random temperature fluctuation
\begin{equation}
\phi(\Omega_L) \simeq \pm \left( \frac{\delta t(\Omega_L)}{g} \right)^{1/2}\,.
\label{field_cond}
\end{equation}
Now let us estimate up to which sizes of locally ordered regions this may occur. According to Eq.~(\ref{sampl_aver})
the condition $\delta t_L\gg t$ yields:
\begin{equation}
L\ll\frac{\rho^{1/D}}{t^{2/D}} \,.
\label{L_cond}
\end{equation}
On the other hand, the estimation of the order parameter in terms of the saddle-point equation~(\ref{saddle_cond2})
can be correct only at scales much larger than the correlation length $\xi\sim t^{-\nu}$. Thus one has a lower bound
for $L$
\begin{equation}
L\gg\ t^{-\nu} \,.
\label{L_cond2}
\end{equation}
Therefore, quenched temperature fluctuations are relevant only when
\begin{equation}
t^{-\nu}\ll\frac{\rho^{1/D}}{t^{2/D}} \,,
\label{t_cond}
\end{equation}
or
\begin{equation}
t^{2-\nu D}\ll \rho \,.
\label{t_cond2}
\end{equation}
According to the Josephson scaling relation, $\alpha=2-\nu D$, see for example~\cite{GOLDEN}. Thus
one recovers the Harris criterion: if the specific heat critical exponent of the pure system
is positive, then in the critical interval:
\begin{equation}
t<t_*\equiv \rho^{1/\alpha} \,
\label{crit_int}
\end{equation}
the disorder becomes relevant. This argument identifies $1/\alpha$ as the crossover exponent associated with
randomness. In this case, the critical exponents of the disordered systems differ from those of the
pure one. In particular, the value of $\alpha$  for the disordered systems is never positive~\footnote{A rigorous
argument for $\alpha<0$ in the disordered case, applicable
to many situations, is given in~\cite{ESCALA-DES}.}.

On the other hand, if the exponent $\alpha=2-\nu D<0$, the condition~(\ref{crit_int}) can not be
satisfied near $T_\text{c}$ (at $t\ll1$), and therefore in this case a weak disorder remains irrelevant
in the critical region.

In the marginal situation, i.e. $\alpha=0$, which is the case, for instance, in the four-dimensional
Ising model (Chapter~\ref{chap:RSIM4D}) or in the two-dimensional Ising model~\cite{KeRu08}, it
can be demonstrated~\cite{Dots-book} that although the specific heat exponent in the disordered model
remains zero, the forms of the logarithmic singularities are affected by the disorder.

\clearpage{\thispagestyle{empty}\cleardoublepage}

\chapter{Finite Size Scaling and the Quotient Method}
\label{Appendix_Quotient}

When doing numerical simulations we are restricted to 
finite systems and therefore we will never obtain infinite specific heats
or susceptibilities at the critical point. Nonetheless, there exist different methods
to study the critical  behaviour of a physical system, working with a finite number of degrees
of freedom.

Probably the most popular approach is the use of Finite Size Scaling (FSS) techniques.
They are based on the study of the evolution of observables with the system
size in order to obtain information about the behaviour of the system at the Thermodynamic Limit (TL).
FSS is based on the scaling hypothesis~\cite{HIPERSCALA}, which states that
the behaviour of the system is governed by the ratio $L/\xi(\infty,t)$, where
$L$ is a characteristic scale of the system (for example its linear dimension)
and $\xi(\infty,t)$ is the correlation length of the infinite system. If this ratio
is large, the system is basically in its TL; if it is small, it will
be in the FSS regime.

One of the consequences of the above statement is 
that the evolution of the mean value of a given observable, $O$, with
the system size will follow the functional form
\begin{equation}
\frac{\langle O(L,t) \rangle}{\langle O(\infty,t) \rangle} = f_O(L/\xi(\infty,t))
+{\cal O}(\xi^{-\omega}, L^{-\omega}) \ ,
\label{FSS_ansatz1}
\end{equation}
with $t=(T-T_\text{c})/T_\text{c}$ being the reduced temperature and $f_O$ is a smooth function depending on the
observable. The leading correction term exponent, $\omega$,
is minus the eigenvalue of the first irrelevant operator of
the theory, in terms of RG language. In the following we
assume that we are in the critical region, so that $\xi \gg L$.
Then in the last term we can neglect $\xi^{-\omega}$
compared with $L^{-\omega}$.  The above equation is one of
the multiple forms of the FSS ansatz, which can
also be derived from a pure RG analysis, see for example~\cite{VICTORAMIT} or~\cite{BARBER} for detailed
explanations.

There exist more practical forms of the ansatz. The observable $O$ diverges in the TL according to:
\begin{equation}
\langle O(\infty,t) \rangle = t^{-x_O}+\cdots \,.
\label{obs_divergence}
\end{equation}
For the correlation length, $x_\xi=\nu$, so that we can make the change
\begin{equation}
t^{-x_O}= t^{-\frac{x_O}{\nu}  \nu} \propto \xi(\infty,t)^{x_O/\nu} L^{-x_O/\nu} L^{x_O/\nu}
= g(L/\xi(\infty,t)) L^{x_O/\nu} \,, 
\label{tranformation}
\end{equation}
and Eq.~(\ref{FSS_ansatz1}) can be rewritten as
\begin{equation}
\langle O(L,t)\rangle= L^{x_O/\nu}[\hat{F}_O(L/\xi(\infty,t))
+L^{-\omega} \hat{G}_O(L/\xi(\infty,t))+\cdots] \,,
\label{FSS_ansatz2}
\end{equation}
where it can be shown that the correction term is also a
function of $L/\xi(\infty,t)$. Given that
$\xi(\infty,t)\propto t^{-\nu}$, the scale variable can be
written in terms of the reduced temperature to obtain
an alternative form of the ansatz:
\begin{equation}
\langle O(L,t)\rangle= L^{x_O/\nu}[\tilde{F}_O(t L^\frac{1}{\nu})
+L^{-\omega} \tilde{G}_O(t L^\frac{1}{\nu})+\cdots] \ .
\label{FSS_ansatz3}
\end{equation}

Moreover, since we can use Eq.~(\ref{FSS_ansatz2}) for the correlation length and $\hat{F}_O$ is
smooth, we can invert it to obtain $\xi(\infty,t)$ as a function of $\xi(L,t)$,
and thus arrive at the most useful form of the ansatz:
\begin{equation}
\langle O(L,t)\rangle= L^{x_O/\nu}[F_O(L/\xi(L,t))
+L^{-\omega} G_O(L/\xi(L,t))+\cdots] \ .
\label{FSS_ansatz4}
\end{equation}
All the quantities in the above equation can be measured on
a finite lattice, so that this will be our starting point for the
explanation of the quotient method~\cite{QUOTIENTS}.

If we form the quotient, $Q_O$, of a given observable, $O$,
measured for two lattice sizes $L_1=L$ and $L_2=sL$, with
$s>1$ , at just such a temperature that $Q_\xi= s$, or
explicitly
\begin{equation}
\frac{\xi(sL,t)}{sL}=\frac{\xi(L,t)}{L}\,,
\label{XI_QUOT}
\end{equation}
the result will be the elimination of the scaling function $F_O$ in Eq.~(\ref{FSS_ansatz4}),
\begin{equation}
\left.Q_O\right|_{Q_\xi=s}=s^{x_O/\nu}+{\cal O}(L^{-\omega}) \ .
\label{naive_quotient_method}
\end{equation}
Typically one chooses $s=2$. The critical exponent $x_O$ can
easily be derived from the above equation.  The fact that
there are strong statistical correlations between the
quotients in Eqs.~(\ref{XI_QUOT})
and~(\ref{naive_quotient_method}) can be used via a
jack-knife method to decrease the statistical errors in the
numerical estimates of critical exponents, see
Appendix~\ref{Appendix_errors} or Ref.~\cite{VICTORAMIT}.

In the present work we used the quantities:
\begin{eqnarray}
\chi & \to & x_o=\nu(2-\eta)\,, \\
\cal{M} & \to & x_o=\frac{\nu}{2}(D-2-\eta)\,, \\
\partial_\beta \xi\ ,\  \partial_e \xi & \to & x_o=\nu+1\,, \\
\partial_\beta g_4 & \to & x_o=1\,.
\end{eqnarray}

In addition, the crossing points of the correlation length, i.e. the temperatures
where the condition of Eq.~(\ref{XI_QUOT}) is satisfied, provide an estimate of the critical temperature
of the transition. By applying Eq.~(\ref{FSS_ansatz3}) to the correlation length, assuming that
the scaling functions are smooth, we can obtain for the (inverse) temperature of the crossing the following
behaviour:
\begin{equation}
\beta_c^{L,s} -\beta_c \propto \frac{1-s^{-\omega}}{s^{1/\nu}-1} L^{-\omega -\frac{1}{\nu}} \ .
\label{beta_deviation}
\end{equation}

The method can be also applied in a microcanonical context if a valid FSS ansatz is
available. In this case the role of the reduced temperature $t$ is played by $e-e_\text{c}$, where
$e$ is the energy density of the system and $e_\text{c}$ is the
energy density at the critical point, see Sec.~\ref{subsec:quotients} for more details.

The quotient method can be improved to speed up convergence if logarithmic corrections are present.
In particular, if a given quantity, $O$, behaves in the TL as
\begin{equation}
O(L,z)=L^{x_O/\nu}(\log L)^{\widehat{x}_O}\left[ F_O\left(\frac{L}{\xi(L,z)}\right) + {\cal O}(L^{-\omega}) \right] \,,
\end{equation}
where $z$ can be either the reduced temperature $t$ or $e-e_\text{c}$, the critical canonical
or microcanonical exponent 
calculated using Eq.~(\ref{naive_quotient_method}) must be corrected according to:
\begin{equation}
\frac{x_O'}{\nu}=\frac{x_O}{\nu}-\frac{\widehat{x}_O}
{\log(L_2/L_1)}\log \left( \frac{\log L_2}{\log L_1} \right) \,,
\end{equation}
where we use primes to label corrected exponents.

If we have enough analytical information about the
logarithmic term exponents we can apply the correction
exactly. For example, for the two-dimensional four-state
Potts model the values of the logarithmic correction
exponents are known analytically~\cite{Salas, Cardy}. Thus we
can calculate the corrections accurately. The
susceptibility behaves as
\begin{equation}
\chi \sim L^{7/4}(\log L)^{-1/8}\,
\label{scaling_chi}
\end{equation}
and we easily arrive at
\begin{equation}
\eta'=\eta-\frac{1}{8 \log(L_2/L_1)}\log \left( \frac{\log L_2}{\log L_1} \right) \,.
\end{equation}
For the correlation length it is known that
\begin{equation}
\xi \sim |t|^{-2/3}(- \log t)^{1/2}  \quad\quad ; \quad\quad  t \sim L^{-3/2}(\log L)^{3/4}\,,
\end{equation}
and therefore its temperature derivative scales as
\begin{equation}
\partial_\beta \xi \sim L^{5/2}(\log L)^{-3/4}\,,
\end{equation}
resulting in a $\nu$ canonical exponent correction of
\begin{equation}
\nu'=\nu\left[1-\frac{3}{4}\frac{\nu}{\log(L_2/L_1)}\log \left( \frac{\log L_2}{\log L_1} \right) \right] \,.
\end{equation}

For the microcanonical $\nu$ exponent, $\nu_\text{m}$, we use that
\begin{equation}
e \sim L^{-1/2}(\log L)^{-3/4}\,,
\end{equation}
so that
\begin{equation}
\partial_e \xi \sim L^{3/2}(\log L)^{3/4}\,,
\end{equation}
and
\begin{equation}
\nu_\text{m}'=\nu_\text{m} \left[ 1+\frac{3}{4}\frac{\nu_\text{m}}{\log(L_2/L_1)}\log \left( \frac{\log L_2}{\log L_1} \right) \right] \,.
\end{equation}
\bigskip

\clearpage{\thispagestyle{empty}\cleardoublepage}


\chapter{Data Analysis: Autocorrelations and Error Estimation}
\label{Appendix_dataanalysis}

The goal of this appendix is to provide a brief resume of
the main ideas for the data analysis of the output of a
dynamic Monte Carlo (MC) simulation.  For a more detailed study
see for example Refs.~\cite{VICTORAMIT,sokal96}.  Our aim is
to describe the modern techniques that avoid the usual error
sources in this kind of numerical study.

Given that the output of a dynamic MC simulation
is a sequence of system configurations\footnote{As one does
not need to store all the configurations, but only the
values of a few functions of them (observables), what one
really has is a sequence of numbers $O_0, O_1, O_2, \ldots,
O_N$, where $O$ is a generic observable.}, $\{\phi\}_0,
\{\phi\}_1, \{\phi\}_2, \ldots, \{\phi\}_N$, we have to take
two crucial issues into account:

\begin{enumerate}
\item \emph{The initial bias:} We have to start every simulation from a physically
unrepresentative configuration (usually ``hot'', all the
spins in random configurations, or ``cold'' , every spin in
the same state). The first configurations are thus not
representative of the equilibrium distribution (the
Boltzmann weight).  There will be an initial transient
regime which must be discarded to avoid a systematic source
of error. If we discard the $n_\text{d}$ initial data in
estimating the mean value $\langle O
\rangle_\beta$ at the inverse temperature $\beta$, then:
\begin{equation}
\langle O \rangle_\beta \approx \bar{O} \equiv \frac{1}{N-n_\text{d}}\sum_{t=n_\text{d}+1}^{N}O_t \ ,
\label{mean_discard}
\end{equation}
where we have distinguished the true mean value $\langle O \rangle$ from its estimate $\bar{O}$. 

\item \emph{Error estimates in equilibrium:} The output of every MC simulation
must be a confidence interval around the estimated mean
value. The true mean value must lie within this interval at
a reasonable level of confidence if the correct procedures
have been applied. Once equilibrium is reached, correlations
between consecutive system configurations make the
statistical error a factor $2\tau_{\text{int}, O}$ larger
than that of the corresponding  independent
sampling case, where $\tau_{\text{int}, O}$ is the integrated
correlation time of the observable $O$, see below.
\end{enumerate}

Both these issues are related to the same object, namely the
\emph{autocorrelation function}.

\section*{The Autocorrelation Function}
\label{Appendix_autocorr}

By definition, the equilibrium autocorrelation function of the observable $O$ at time $t$ is~\cite{VICTORAMIT}:
\begin{eqnarray}
C_{O O}(t) & \equiv & \langle O_s O_{s+t}\rangle_\beta - \langle O\rangle^2_\beta \\
  & = & \sum_Y\sum_X O(Y)\left[ [T^t]_{YX}-\frac{e^{-\beta H(Y)}}{Z} \right] \frac{e^{-\beta H(X)}}{Z}O(X)\,,
\end{eqnarray}
where:
\begin{itemize}
\item $O(Y)$ is the value of the observable $O$ for the system configuration $Y$.

\item $[T]^t_{YX}$ is the probability of reaching the configuration $Y$ starting from the configuration
$X$ in $t$ steps; i.e., it is a sum over all possible paths connecting $X$ and $Y$ in $t$ steps.

\item $\exp(-\beta H(Y))/Z$ is the Boltzmann weight of the configuration $Y$, with $\beta$ being the inverse
temperature, $H$ the Hamiltonian, and $Z$ the partition
function.

\end{itemize}
A normalised form is often used, defined as:
\begin{equation}
\rho_{O O}(t)\equiv\frac{C_{O O}(t)}{C_{O O}(0)} \,.
\label{autocorr_norm}
\end{equation}
Typically, for long times, $C_{OO}(t)$ decays exponentially with time.

The \emph{exponential autocorrelation time} is defined by
\begin{equation}
\tau_{\text{exp}, O}= \lim_{t\to\infty} \sup \frac{t}{-\log \rho_{O O}(t)} \,.
\label{exp_autocorr}
\end{equation}
It is useful to define the maximum over all the measured observables, 
\begin{equation}
\tau_{\text{exp}}= \sup_O \tau_{\text{exp}, O} \,.
\label{exp_autocorr_max}
\end{equation}
In Ref.~\cite{sokal96} it is demonstrated that the rate of convergence to
equilibrium from an initial non-equilibrium distribution can be bounded
in terms of $\tau_{\text{exp}}$. In particular:
\begin{equation}
\left| \sum_Y O(Y) [T^t]_{YX}-\langle O\rangle_\beta \right| \sim e^{-t/\tau_{\text{exp}}}
 \,, 
\label{term_rate}
\end{equation}
From this, it can be said that setting
$n_\text{d}=20\tau_{\text{exp}}$ in Eq.~(\ref{mean_discard})
is enough for all practical purposes. Hence, waiting for
this time before starting to save the measurements for
averaging, we can avoid the initial bias of the MC
simulation. The problem with this approach is the difficulty
in estimating $\tau_{\text{exp}}$ in some cases.

Usually the convergence to equilibrium is determined
\emph{empirically} by plotting certain observables as a
function of time and noting when the initial transient seems
to end.  This includes the comparison between hot and cold
starts. The main objection to this is the possibility of
metastability, especially for first-order phase transitions.
In such cases the equilibrium appears to be reached but
really the system has just settled down into a long-lived
metastable region of the configuration space. One has to be
extremely careful in these cases, see
Chapter~\ref{chap:potts3D}.

Once in equilibrium, to what extent are the measurements
taken in the system representative? This issue reflects the
fact that consecutive measurements are usually close in
configuration space (and are thus \emph{correlated}) so they do
not provide the same information as if they were
independent.

We can resolve this question in terms of the
\emph{integrated autocorrelation time}, defined as:
\begin{equation}
\tau_{\text{int}, O}= \frac{1}{2}+ \sum_{t=1}^{\infty}\rho_{O O}(t) \,.
\label{int_autocorr}
\end{equation}
This time controls the error estimates in MC simulations. In particular, the sample mean
\begin{equation}
\bar{O} \equiv \frac{1}{N}\sum_{t=1}^{N}O_t \,, 
\label{sample_mean}
\end{equation}
assuming for brevity that the data at $t=1$ are already good, has a variance
\begin{eqnarray}
\langle (\bar{O}- \langle O \rangle_\beta)^2 \rangle_\beta & = & \frac{1}{N^2}\sum_{r,s=1}^{N}
\langle (O_r - \langle O \rangle_\beta)(O_s - \langle O \rangle_\beta) \rangle_\beta \\
& = & \frac{1}{N^2}\sum_{r,s=1}^{N} C_{O O}(r-s)\\
& = & \frac{1}{N}\sum_{t=-(N-1)}^{N-1} \left(1-\frac{|t|}{N} \right) C_{O O}(t)\\
& \approx & \frac{2\tau_{\text{int}, O}}{N} (\langle O^2 \rangle_\beta - \langle O \rangle_\beta^2)\,.
\end{eqnarray}
To derive these last relationships, we made use of
$C_{O O}(t)=C_{O O}(-t)$ in equilibrium and assumed that $N$ is large
enough to neglect $C_{O O}(t)$ for $|t|\sim N$ -- recall that
$C_{O O}(t)\to0$ exponentially for increasing $t$.

Therefore the variance of $\bar{O}$ is a factor
$2\tau_{\text{int}, O}$ larger that it would be if all the
configurations $\{\phi\}_i$ were statistically
independent. In other words, the number of effective
measurements in a MC simulation of length $N$ is
reduced to $N/(2\tau_{\text{int}, O})$. Roughly speaking,
the error bars will be of order $(\tau_{\text{int},
O}/N)^{1/2}$, so that if we want an error bar for our simulation of
around $1\%$ precision we will need a run of length $\approx
10000\tau_{\text{int}}$.

Now we will define a more practical estimate of the
correlation times~\cite{sokal96}. The direct estimate from a
run of length $N$ (supposing again that at $t=1$ the data are
already good) is:
\begin{equation}
\overline{C_{O O}(t)} =\frac{1}{N-|t|}\sum_{s=1}^{N -|t|}(O_s-\bar{O})(O_{s+|t|}-\bar{O})\,, 
\label{autocorr_estimate_nonorm}
\end{equation}
\begin{equation}
\overline{\rho_{O O}(t)} =\overline{C_{O O}(t)}/\overline{C_{O O}(0)}\,. 
\label{autocorr_estimate_norm}
\end{equation}
At first sight, one would estimate the integrated autocorrelation time as 
\begin{equation}
\overline{\tau_{\text{int},O}} = \frac{1}{2} + \sum_{t=1}^{N-1}\overline{\rho_{O O}(t)}\,.
\label{int_estimate_bad}
\end{equation}
But this is wrong because this estimator has a variance that
does not go to zero for large $N$. Each of the terms
$\rho_{O O}(t)$ decreases exponentially with $t$ but is a
random variable obtained by averaging $N-t$ data.  For
$t>\tau_{\text{int},O}$ each $\rho_{O O}(t)$ is very small,
but its error is not null. Therefore most of the terms in
the definition Eq.~(\ref{int_estimate_bad}) carry little
information but much noise for large $t$, and there are very
many of them. The cure is the truncation of
Eq.~(\ref{int_estimate_bad}) to estimate
$\tau_{\text{int},O}$ self-consistently:
\begin{equation}
\overline{\tau_{\text{int},O}} = \frac{1}{2} + \sum_{t=1}^{\alpha \overline{\tau_{\text{int},O}}}\overline{\rho_{O O}(t)}\,,
\label{int_estimate_good}
\end{equation}
where $\alpha$ is a small fixed number, around 6. Clearly $\tau_{\text{int},O}$ is biased,
but the contribution of the neglected terms should be negligible (at $6\tau_{\text{int},O}$,
it is expected that $\overline{\rho_{O O}(t)}\ll1$). See~\cite{sokal96} for more details 
about this ``automatic windowing'' algorithm.

For $\tau_{\text{exp},O}$, one can use two different estimates, see~\cite{VICTORAMIT}:
\begin{equation}
\overline{\tau_{\text{exp},O_{t,A}}} = \frac{t}{|\log \overline{\rho_{O O}(t)}|}\,,
\label{exp_estimate_A}
\end{equation}
\begin{equation}
\overline{\tau_{\text{exp},O_{t,B}}} = \left[ \log \left( \frac{ \overline{\rho_{O O}(t)}}{\overline{\rho_{O O}(t+1)}} \right) \right]^{-1}\,.
\label{exp_estimate_B}
\end{equation}
The autocorrelation functions decay as pure exponentials, $e^{-t/\tau_{\text{exp},O}}$, for large $t$.
Then both $\overline{\tau_{\text{exp},O_{t,A}}}$ and $\overline{\tau_{\text{exp},O_{t,B}}}$ 
become equal to $\tau_{\text{exp},O}$. Nevertheless as was found before, the information carried
by the signal decreases rapidly with $t$, and it can not be said which definition
will first reach the $t$-independent region, see~\cite{VICTORAMIT}.

\section*{Error Estimation}
\label{Appendix_errors}

Having demonstrated that the effective number of
measurements is $N/(2\tau_{\text{int},O})$, where
$\tau_{\text{int},O}$ is not known accurately, we will now
describe the most usual method of obtaining independent quantities:
\emph{making data blocks}. This will also allow the
estimate of functions of observables, and we will see that
the time correlations between different observables can even be
beneficial.

Starting from a set of $N$ data, let us form $N/b$ data
blocks of size $b$:
\begin{equation}
O_{b,i} = \frac{1}{b} \sum_{t=(i-1)b+1}^{bi}O_t\,.
\label{data_block}
\end{equation}
The autocorrelation times for the blocked data are divided by a factor $b$.
The mean value of the block data, $\hat{O}$, is $b$-invariant, while
the error can be estimated (if $b$ is large enough) as for statistically independent data:
\begin{equation}
\overline{\varDelta_O^2} = \frac{1}{\frac{N}{b}-1} \left[ \frac{b}{N}\sum_{i=1}^{N/b}(O_{b,i})^2 
- \left( \frac{b}{N}\sum_{i=1}^{N/b}O_{b,i} \right)^2\right]\,.
\label{block_error}
\end{equation}
The error estimate first grows with $b$ until the data
blocks become effectively independent ($b\approx
20\tau_{\text{int},O}$). Then $\overline{\varDelta_O^2}$
becomes $b$-independent (the expression in brackets in
Eq~(\ref{block_error}) decreases as $1/b$ but the prefactor
grows as $b$). As can be seen in Ref.~\cite{VICTORAMIT}, an
assumption that the data are independent would underestimate
the errors by a factor $\sqrt{2\tau_{\text{int},O}}$. Once
$b$ is large enough, the error estimate reaches a plateau,
but from then on the fluctuations grow as the block size is
increased. Therefore $N/b$ should be kept large enough.
Typically, a ratio of between 10 and 50 is chosen.

Now we have to consider the crucial issue of the error
estimation of \emph{functions} of observables.  Let $f(\langle
O^{(1)} \rangle_\beta, \langle O^{(2)} \rangle_\beta,
\ldots, \langle O^{(R)} \rangle_\beta)$ be a function of $R$
observables (R could be one). The best thing we can do is
estimate $f(\{\langle O^{(R)} \rangle_\beta\})$ by $f(
\{\overline{O^{(R)}}\})$, although we know that this estimate
is biased unless $f(x)$ is a linear function.  The estimate
of the error could be done by linear error propagation, but
when $f$ depends on several observables this may
overestimate the size of the errors due to correlations
between the observables. For some clear examples of this last
point, see Ref.~\cite{VICTORAMIT}.

The fact that correlations can be beneficial is exploited by
the \emph{jack-knife} method. This allows one to
estimate the error bars of derivative quantities easily and
coherently. The procedure is the
following~\cite{VICTORAMIT}:
\begin{enumerate}
\item Estimate $f(\langle O^{(1)} \rangle_\beta, \langle O^{(2)} \rangle_\beta, \ldots, \langle
O^{(R)} \rangle_\beta)$ by $f(\overline{O^{(1)}}, \overline{O^{(2)}}, \ldots, \overline{O^{(R)}})\,$.

\item For each observable, form the corresponding block data, as in Eq.~(\ref{data_block}), with
large enough $b$ (the same for every observable).

\item Make jack-knife blocks from the blocked data. This means that the $i$-th jack-knife block
is formed by averaging all the blocks formed in the previous step except the $i$-th. I.e.:

\begin{equation}
O_{J K, b,i}^{(r)} \equiv \frac{1}{\frac{N}{b}-1} \sum_{j\neq i} O_{b,j}^{(r)}\quad , \quad r=1,2,\ldots, R\,.
\label{jk_block}
\end{equation}

\item Estimate the function value for the jack-knife blocked observables as
\begin{equation}
f_{J K, b,i} \equiv f(O_{J K, b,i}^{(1)}, O_{J K, b,i}^{(1)}, \ldots, O_{J K, b,i}^{(R)}) \,.
\label{f_jk_block}
\end{equation}

\item Estimate the variance of the function as
\begin{equation}
\overline{\varDelta^2_f} = \left( \frac{N}{b} -1 \right) \left[ \frac{b}{N} \sum_{i=1}^{N/b} f_{J K, b,i}^2 -
 \left( \frac{b}{N} \sum_{i=1}^{N/b} f_{J K, b,i} \right)^2 \right] \,.
\label{var_jk}
\end{equation}

Since the expression in brackets is an average of blocked
data, it is smaller than usual.  This is the reason for the
multiplication (instead of division) by the number of blocks
minus one. In the case of correlations between observables,
their jack-knife blocks will fluctuate simultaneously, thus
reproducing the possible positive effect on the error.

\end{enumerate}

\clearpage{\thispagestyle{empty}\cleardoublepage}

\chapter{Temperature Extrapolations}
\label{Appendix_Extrapolations}

Within a canonical Monte Carlo method, the temperature of
the system is kept fixed, and all the information about the
observable corresponds to the simulation temperature. It is
often very desirable to obtain an accurate estimate of a
given quantity at a temperature different from the simulated
one.  This may be the case for example when one tries to
obtain the absolute maxima in temperature of some quantities
to estimate critical exponents, or when one has to fine-tune
some condition as in the quotient method, see
Eq.~(\ref{XI_QUOT}).

Using the energy histogram of the system at a given
temperature, one can obtain accurate information at nearby
temperatures. The method was first proposed
in~\cite{REWEIGHT1}, and was recovered in~\cite{REWEIGHT2}.
If we are working with disordered systems, the temperature
extrapolation must be performed before averaging over the
different disorder realizations.

The following formulae allow one to calculate the
thermal derivative of an observable, $O$, and its value at a
temperature close to the simulated one:
\begin{equation}
\partial_\beta \overline{\langle {O}\rangle}=
 \overline{\partial_\beta\langle {O}\rangle}=
\overline{\left\langle_{\vphantom{|}} {OE} - \langle{O}
 \rangle \langle {E} \rangle\right\rangle}\,,
\label{DERIVADA}
\end{equation}
\begin{equation}
\langle { O} \rangle (\beta+ \Delta \beta)= 
\frac{\langle { O} {\mathrm e}^{-\Delta \beta{\cal E}}\rangle } 
{\langle {\mathrm e}^{-\Delta \beta{\cal E}} \rangle\ } \,,
\label{FS}
\end{equation}
where $\beta=1/T$ is the inverse simulation temperature.

Nevertheless, it must be borne in mind that the two above
expressions involve a systematic bias whose correction can
become critical.  We shall follow the approach proposed
in~\cite{TESIS-BALLES}.  Using Eq.~(\ref{DERIVADA}) with $N$
different measurements, we are really obtaining
\begin{equation}
\overline{\left(1-\frac{2\tau}{N}\right)\partial_\beta \langle O
\rangle}\,,
\label{BIASDER}
\end{equation}
where $\tau$ represents the integrated autocorrelation time,
see Appendix~\ref{Appendix_dataanalysis}, between the
observable $O$ and the energy. Since the $\tau$ value is
different for different samples, we shall have to
obtain the disorder average.

Let us demonstrate the validity of Eq.~(\ref{BIASDER}).
Consider two observables, ${O}^a$ and ${O}^b$, and, using
$N$ measurements, calculate their connected correlation
\begin{equation}
\langle {O}^a {O}^b \rangle - \langle {O}^a \rangle \langle
{O}^b \rangle \,,
\label{DER}
\end{equation}
Since the mean value of ${O}^a$ is by definition
\begin{equation}
\langle {O}^a \rangle=\frac{1}{N}\sum_{i=1}^N {O}^a_i\,,
\end{equation}
Eq.~(\ref{DER}) can be written as
\begin{equation}
\frac{1}{N}\sum_{i=1}^N {O}^a_i \, {O}^b_i - 
\frac{1}{N^2}\sum_{i=1}^N \sum_{j=1}^N {O}^a_i \, {O}^b_j\,,
\end{equation}
where the latter term is more complex given that the measurements of the two observable may be correlated.
We can rewrite it as
\begin{equation}
\sum_{i=1}^N\sum_{j=1}^N {O}^a_i \,{\cal  O}^b_j=
\sum_{i=1}^N\sum_{t=1-i}^{N-i} {O}^a_i \, {O}^b_{i+t}\,.
\end{equation}

Recall that the behaviour of the correlation between two observables is for
large $|t|$:
\begin{equation}
\langle {O}^a_i \, {O}^b_{i+t}\rangle \equiv 
C_t= \langle {O}^a \rangle \langle {O}^b \rangle + 
C_0 \exp \left[-{\frac{|t|}{\tau}}\right]\,, 
\label{CORR}
\end{equation}
where
\begin{equation}
C_0=\langle {O}^a \, {O}^b\rangle 
-\langle {O}^a \rangle \langle {O}^b \rangle\,.
\label{COO}
\end{equation}
As we are summing over $i$, we can replace ${O}^a_i \,
{O}^b_{i+t}$ by its mean value, and in order to perform the
sum we shall take $N\gg |t|$. From the exponential decay
with $|t|$ in Eq.~(\ref{CORR}) we can extend the limit
of the sum to infinity, and simplify things by replacing the sum
with an integral. Therefore
\begin{equation}
\sum_{i=1}^N\sum_{j=1}^N {O}^a_i {O}^b_j=
N^2 \left( \langle {O}^a \rangle \langle {O}^b \rangle \right) +
N  C_0
\int_{-\infty}^{+\infty} \mathrm{d} t \ 
\exp \left[-{\frac{|t|}{\tau}}\right]  \, .
\end{equation}
But the integral is $2\tau$, see again Appendix~\ref{Appendix_dataanalysis},
and replacing $C_0$ by the expression~(\ref{COO}) one recovers
Eq.~(\ref{BIASDER}).

Thus it has been shown that the derivative of the mean value
of an observable is subject to a systematic bias of order
$2\tau/N$, although corrections of higher order can be
expected. This systematic error is not considered in general
in other MC studies mainly because it is usually
masked by the statistical error, of order $1/\sqrt{N}$, that
is  far larger than the bias.

Nevertheless, in a dilute model the situation is more
complicated because there are two parameters involved: the
number of measurements that we perform within each realization
of the disorder (Ising average), denoted by $N_I$, and the
number of realizations of the disorder (sample average),
$N_M$.

We can associate with every observable $\sigma_M$,
representing the deviation of this observable between
different samples, and $\sigma_I$, representing the average
of the deviations within each sample. Assuming statistical
independence between different samples (this is the case if
there are no correlations in the random numbers) and
independence between different measurements within a sample
(possible using a cluster method), the variance of the mean
value of an observable can be written as
\begin{equation}
\sigma^2_T=\frac{1}{N_M}\left(\sigma^2_M+\frac{\sigma^2_I}{N_I}\right)
\, .
\label{ERROR}
\end{equation}

From this equation the optimal choice of $N_I$ and $N_M$ can
be deduced if we want to optimise the simulation time to
obtain a given error bar.  Note that the simulation time is
proportional to $N_M N_I$, because the simulation spends an
approximately fixed portion of time taking measurements. For
this reason the optimal value of $N_I$ can not be much
larger than $\sigma^2_I/\sigma^2_M$, i.e., $N_I$ is limited
by this latter expression. Therefore the best procedure to
improve the statistical errors is to increase the number of
samples, performing a not too large number of measurements
within each one.  Numerous studies on dilute models have
avoided tackling this question by working in general in the
regime $N_I\gg N_M$. Besides, a value of $N_I$ that is large does not
improve the result for the deviation $\sigma^2_M$ because it
is only suppressed by a factor $N_M$.

However, if we need to obtain the temperature derivatives
there is a term in the statistical error proportional to
$1/\sqrt{N_M}$. If this number is of order $1/N_I$
(systematic error in the extrapolation) a detectable bias
appears. Unfortunately this is indeed the scenario in some
cases.  For example for the Heisenberg model, see
Sec.~\ref{O3:methodology}, we used $N_M=10^4$ and $N_I=100$
measurements per sample.

As a result it is necessary to find some algorithm to obtain
correct results from the simulation data. The first
possibility is to use fully independent measurements (by
assuring $2\tau=1$) for each disorder realization. In such a
case, recalling Eq.~(\ref{BIASDER}), we could build a
correct unbiased estimator by multiplying by the factor
$1/(1-1/N_I)$ in the case of the derivative and by the
corresponding quantity in the case of the temperature
extrapolations. However this is too expensive in simulation
effort, due both to the need for too wide a time interval
between measurements and to the fluctuations of $\tau$
between different samples, which forces one to disregard too
many measurements in samples with $\tau$ smaller than the
maximum. This makes the use of this approach inefficient.

Another possibility, see Ref.~\cite{ISDIL4D}, is to correct
the systematic bias by splitting the measurements into
statistically independent groups. In this way, by
multiplying estimators from the different groups, correct
values of $\langle {O}^a \rangle \langle {O}^b \rangle$ can
be obtained.  Nevertheless, in a real situation independence
is complicated because, in general, consecutive measurements
in MC time are used to form these groups.

In this work the approach that will be used for the dilute
models, see again Ref.~\cite{ISDIL4D}, is to
extrapolate to $N_I\to\infty$ by using estimators coming
from different values of $N_I$. We will work with
$2\tau\gtrsim 1$, but the equality is not essential, as we
will demonstrate in the following.

\begin{figure}[!ht]
\begin{center}
\includegraphics[width=0.85\columnwidth]{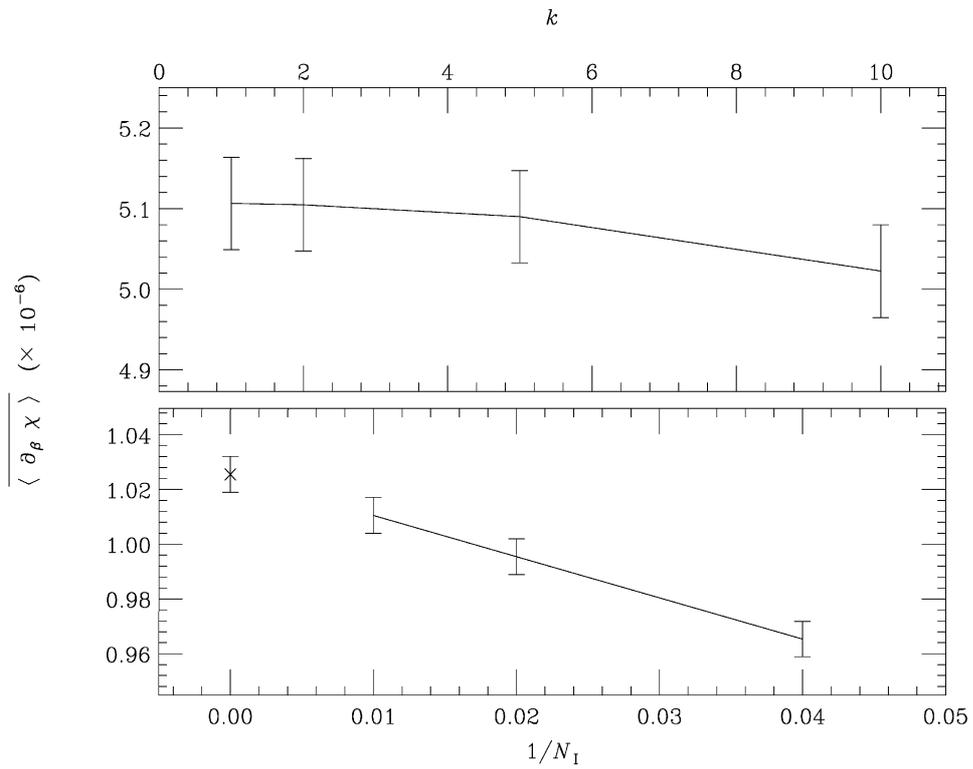}
\caption{
\emph{Upper part:} Sample average of the estimator $y_1$ corresponding
to $\partial_\beta\chi$ considering just one of each $k$ measurements,
for a four-dimensional Ising model on an $L=48$ lattice, with $p=0.5$ and $\tau\approx 8$. 
\emph{Lower part:} Sample average of $y_1$, $y_2$, and $y_3$ as a function of the inverse of $N_I$, 
on a lattice, with $L=32$, $p=0.5\,$, and $\tau\approx
0.8$. Figure taken from Ref.~\cite{ISDIL4D}.}
\label{DOSTAU}
\end{center}
\end{figure}

In particular, for each disorder realization we can
calculate the derivative with the entire MC history
($N_I$ measurements), obtaining the estimator, $y_1$, whose
systematic error is proportional to $2\tau/N_I$. If we
consider the two contiguous halves of the total history, we
obtain two estimates for the derivative that, once averaged,
produce the estimator, $y_2$, whose systematic error is
proportional to $4\tau/N_I$. If the history is divided into
four, the derivative in each part is obtained and the
average of the four parts is calculated, the estimator,
$y_3$, will have a bias of $8\tau/N_I$. 
In Fig.~\ref{DOSTAU}, obtained in Ref.~\cite{ISDIL4D}, is
shown the observable $\partial_\beta\chi$ for a
four-dimensional site-diluted Ising model in a system with
$L=48$, $p=0.5$. The estimator $y_1$ is plotted, but using
only one of every $k$ measurements of the history and
averaging the results. It is obvious that doing so reduces
the $\tau$ by a factor $k$, as also is the case for the
number of measurements, $N_I$. Therefore the bias,
proportional to $2\tau/N_I$, remains constant. This effect is
maintained up to $k\approx 10$. The correlation time between
energy and $\chi$, $\tau$, was about eight measurements.

Using the $y_i$ values, linear and quadratic extrapolations
to $N_I\to\infty$ in the variable $1/N_I$ can be obtained
with the form
\begin{equation}
y_i = y_\infty + \frac{A}{N_I}+ \frac{B}{N_I^2}\,.
\label{CUAD}
\end{equation}
For the linear case ($B=0$) we obtain
\begin{equation}
y_L= y_\infty = 2 y_1 - y_2\,,
\end{equation}
and for the quadratic case
\begin{equation}
y_Q = y_\infty = \frac{8}{3} y_1 - 2 y_2 + \frac{1}{3} y_3\,. 
\end{equation}

The procedure is repeated for each sample and averaged over
the disorder. The value $y_Q$ is used to verify that the
higher-order effects are negligible compared with the
statistical error of the extrapolation in temperature. If this
were not the case, to obtain measurements of the observable at
the shifted temperature it would be advisable to make another
simulation closer to that point.

The estimate for the temperature extrapolations of the $O(3)$ model 
was obtained in this work in a similar way, see
Chap.~\ref{chap:O3}, Fig.~\ref{fig_extrap_Nsamples}.

%

\clearpage{\thispagestyle{empty}\cleardoublepage}

\vspace{4cm}
\chapter{The Maxwell Construction}
\label{Appendix_Max}

Let us consider a system with action $S[\psi]$, depending on the local
variables $\psi$, coupled with a source $J$. If 
$\phi(\beta,J)$ is the Gibbs free energy\footnote{
We have included in its definition the $\beta$ term.}
and $O(x)\equiv O({\psi(x)})$ is a generic observable then
\begin{equation}
e^{-V\phi(\beta,J)}=\int \mathrm{d}[\psi] e^{-\beta S[\psi] -J\int \mathrm{d^D}x\,O(x)}\,,
\end{equation}
where $V$ is the system volume and $D$ is the spatial dimensionality. The equation of state is
\begin{equation}
\frac{\partial \phi}{\partial J}=\frac{1}{V} \int \mathrm{d^D}x\,O(x)\equiv \overline{O}(J) \,,
\end{equation}
relating the observable, $\overline{O}$, with its conjugate variable, $J$.

\begin{figure}[!ht]
\begin{center}
\includegraphics[width=0.65\columnwidth]{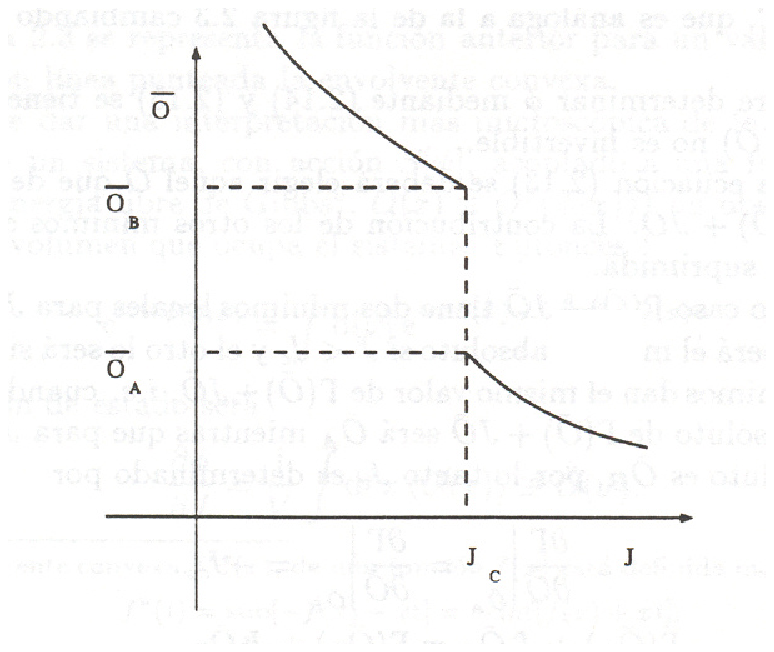}
\caption{Lack of continuity of the function $\overline{O}=\overline{O}(J)$ at a first-order phase transition.}
\label{discont_maxwell}
\end{center}
\end{figure}

In the neighbourhood of the phase transition, the function 
$\overline{O}=\overline{O}(J)$ may be discontinuous. In
other words, there may exist a range of $\overline{O}$ values that do not 
correspond to any $J$ value, see Fig.~\ref{discont_maxwell}.

An effective potential associated with the observable $O$
can be defined as:
\begin{equation}
e^{-V\Gamma(\overline{O})}=\int \mathrm{d}[\psi]\,e^{-\beta S[\psi]}\,
\delta\left( \frac{1}{V}\int \mathrm{d^D}x\,O(x)-\overline{O} \right)\,,
\end{equation}
where $\delta$ is the Dirac delta function. Then it can be obtained
\begin{equation}
e^{-V\phi(\beta,\,J)}=\int \mathrm{d}\overline{O}\,e^{-V[\Gamma(\overline{O})+J\overline{O}]}\,.
\label{defi_phi}
\end{equation}
But for a large volume this integral is dominated by the saddle point
\begin{equation}
\frac{\partial \Gamma}{\partial \overline{O}}=-J\,,
\end{equation}
and we can conclude that $\phi(\beta,J)$ is the Legendre transform of $\Gamma$:
\begin{equation}
\phi(\beta,J)=\Gamma(\overline{O})+J\overline{O}\,.
\label{legendre}
\end{equation}

In Eq.~(\ref{legendre}), the $\overline{O}$  must be chosen that produces a
global minimum of $\Gamma(\overline{O})+J\overline{O}$. The contribution of
all other minima will be exponentially suppressed. In general $\Gamma(\overline{O})+J\overline{O}$ 
has two local minima. One will be the absolute minimum for $J<J_c$, and the
other will be that corresponding to $J>J_c$. Recall that $\overline{O}=\overline{O}(J)$ 
may be discontinuous.

The condition for a minimum located at $\overline{O}$ is
\begin{equation}
\frac{\partial}{\partial \overline{O}}\left( \Gamma(\overline{O})+J\overline{O}\right)=0\, .
\end{equation}
Let us define the two local minima as:
\begin{equation}
\min_{J<J_c} [\Gamma(\overline{O})+J\overline{O}] \equiv \overline{O}_A \,,
\end{equation}
\begin{equation}
\min_{J>J_c} [\Gamma(\overline{O})+J\overline{O}] \equiv \overline{O}_B \,.
\end{equation}

Therefore
\begin{equation}
\left. \frac{\partial \Gamma}{\partial \overline{O}}\right|_{\overline{O}_A}=
\left. \frac{\partial \Gamma}{\partial \overline{O}}\right|_{\overline{O}_B}=-J_c \,.
\end{equation}

\begin{figure}[!ht]
\begin{center}
\includegraphics[width=0.65\columnwidth]{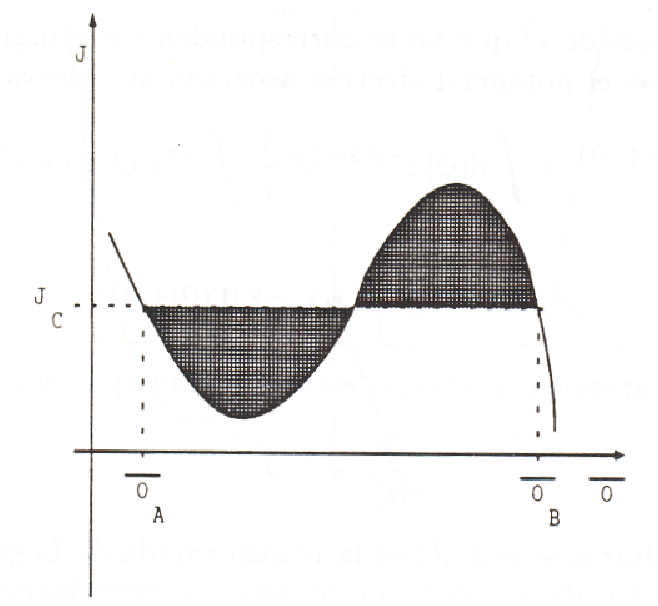}
\caption{Usual form of the function $J=J(\overline{O})$.  Also shown is the Maxwell construction.}
\label{const_maxwell}
\end{center}
\end{figure}

Just at $J_c$ both minima will result in the same value of $\Gamma(\overline{O})+J\overline{O}$, i.e.,
\begin{equation}
\Gamma(\overline{O}_A)+J_c\overline{O_A}=\Gamma(\overline{O}_B)+J_c\overline{O_B}\, .
\label{phiA_phiB}
\end{equation}
Given that
\begin{equation}
\Gamma(\overline{O})=-\int_\mathrm{Cte}^{\overline{O}} \mathrm{d}O\,J(O)\, ,
\end{equation}
Eq.~(\ref{phiA_phiB}) can be written as
\begin{equation}
\int_{\overline{O}_A}^{\overline{O}_B} \mathrm{d}O\,J(O) -J_c({\overline{O}_B}-{\overline{O}_A})=0\, ,
\end{equation}
which implies that the shaded areas in
Fig.~\ref{const_maxwell} must be equal (in absolute
value). This represents the well-known form of the Maxwell
construction.

\clearpage{\thispagestyle{empty}\cleardoublepage}

\vspace{4cm}
\chapter[Lee-Yang Zeros]{Lee-Yang Zeros}
\label{Appendix_LY}

In 1952 T.~D.~Lee and C.~N.~Yang, as part of their study of
the phenomenon of spontaneous symmetry breaking, wrote two
impressive papers~\cite{LEEYANG}. In analysing the behaviour
of a lattice gas (which is equivalent to an Ising model in a
magnetic field), they approached the problem of its phase
transition in an absolutely novel way focusing on finding
the zeros of the partition function in terms of an external
field allowed to take complex values. With this new
approach, the dimension, size, structure, and periodicity of
the lattice play no part at all in the main result.

Their starting point is that for a real (inverse) temperature, $\beta$, the partition
function of the finite system is a polynomial in the activity $\exp(-2h)$, where $h$ is
the external field. Since all the coefficients of the polynomial are positive,
none of their roots can be on the real positive axis, but are in general complex.

\section*{Lee-Yang Theorem}
\label{Appendix_LY_theorem}

The discovery of Lee and Yang is that the zeros of the partition function are all located
on the unit circle of the complex activity plane, or equivalently on the imaginary
$h$-axis. This was originally demonstrated for an Ising system with
ferromagnetic interactions (with no need for any limit to first nearest neighbours) although the result is valid
for more general models~\cite{LEEYANG_EXTENDED}.

The distribution of the zeros on the unit circle will
determine whether or not a phase transition exists. As the number
of spins, $N$, approaches infinity, if the zeros do not
condense onto the positive real axis the free energy, $F$,
will remain analytic and no phase transition exists at the
given $\beta$; otherwise, if the zeros condense onto the real
positive axis a phase transition will exist at this
$\beta=\beta_\text{c}$.

We will use the demonstration of the theorem described in~\cite{DROUFFE} including
the derivation of the magnetisation of the system from the angular distribution of roots on
the unit circle. We will focus on a general Ising model defined on a graph of $N$ sites 
(vertices of the graphs) with at most one link joining each pair of vertices, which are
then defined as neighbours. The total number of links is $L$.

The partition function for $N$ spins, $\sigma_i$, is:
\begin{equation}
Z_N=\frac{1}{2^N}\sum_{\sigma_i=\pm 1} \exp \left(\beta \sum_{(ij)}\sigma_i\sigma_j + \sum_i h_i \sigma_i \right)\,,
\label{Z_original}
\end{equation}
where $\sum_{(ij)}$ is a sum over all links and $h_i$
denotes the magnetic site-dependent field at the site $i$
(strictly the true external field is proportional to
$h_i/\beta$). Let us define
\begin{equation}
\rho_i=e^{-2h_i}\quad\quad ; \quad \quad \tau=e^{-2\beta} \,.
\label{LY_fugacity}
\end{equation}
 Then the partition function can be recast in the form
\begin{equation}
Z_N=\frac{1}{2^N} \exp \left(\beta L + \sum_i h_i \right) P(\tau, \rho_i) \,,
\label{Z_casted}
\end{equation}
where $P$ is a polynomial in $\tau$ and $\rho$ obtained as
\begin{equation}
P=\sum_{\sigma_i=\pm 1} \exp \left[ \beta \sum_{(ij)}(\sigma_i\sigma_j -1) + \sum_i h_i (\sigma_i -1) \right]\,.
\label{P_definition}
\end{equation}
For $\tau$ and $\rho_i$ real and positive, $P$ is always positive and cannot vanish.
We assume in the following $0<\tau<1$, i.e., we are in the ferromagnetic regime.

It is easy to find that the polynomials $P$ corresponding to the simplest graphs are:
\begin{eqnarray}
\ ^1\bullet\overline{\ \ \ \ \ }\,\bullet^2  \quad & P_{12} & =  1+\tau(\rho_1+\rho_2)+(\rho_1\rho_2) 
\label{P_2vertices}\\
\ ^1\bullet\overline{\ \ \ \ \ }\bullet^2 \overline{\ \ \ \ \ }\,\bullet^3 \quad & P_{123} & =  (1+\rho_1\tau)(1+\rho_3\tau)+\rho_2(\tau+\rho_1)(\tau+\rho_3) 
\label{P_3vertices}
\end{eqnarray}
It can then be seen that $P$ is a polynomial of degree one in each $\rho_i$ individually and of degree
$N$ in all of them.

We will analyse how the polynomials are generated by
building a graph step by step. First, one observes that for
any disjoint union of subsets of the graph, the polynomial
$P$ factorizes, $P=P_1 P_2$, where $P_1$ and $P_2$ are the
corresponding polynomials of the two separate subsets. With
this in hand, we generate a new polynomial by identifying a
site $a$ from the subset 1 with a site $b$ from subset 2. We
call $P_{(12)}$ the corresponding (contracted) polynomial of
the union.
\begin{figure}[!ht]
\begin{center}
\includegraphics[width=0.85\columnwidth]{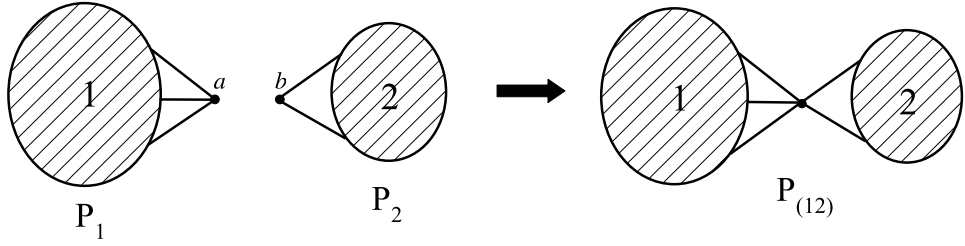}
\caption{The contraction process obtained by identifying $a$ and $b$ from two disjoint subsets of the whole graph.}
\label{contraction}
\end{center}
\end{figure}
Since $P_1$ is linear in $\rho_a$ and $P_2$ is linear in $\rho_b$ we can write
\begin{equation}
P_1 = A_{+} +\rho_a A_{-} \quad\quad;\quad\quad P_2 = B_{+} +\rho_b B_{-}  \,,
\label{P_divided}
\end{equation}
where $A_+(B_+)$ correspond to the contribution with spin up, $\sigma_a=+1\ (\sigma_b=+1)$, and
$A_-(B_-)$ correspond to the contribution with spin down, $\sigma_a=-1\ (\sigma_b=-1)$.
When the identification of $a$ and $b$ is made, a new activity variable, $\rho_{ab}$, is attached to the
new site and we have the following contraction process, see Fig.~\ref{contraction}:
\begin{align}
P_1P_2 \equiv A_{+} B_{+} +\rho_a A_{-} B_{+} + \rho_b A_{+} B_{-} + \rho_a \rho_b A_{-} B_{-} 
\quad \longrightarrow \label{P_contracted1} \\
P_{(12)} \equiv A_{+} B_{+} +\rho_{ab} A_{-} B_{-} \,. \qquad\qquad\qquad\label{P_contracted2}
\end{align}

Using this process, one can obtain for example the
expression of Eq.~(\ref{P_3vertices}) by joining two
unconnected two-vertex graphs, such as that of
Eq.~(\ref{P_2vertices}), and by identifying two of the four
points. It is a good exercise to obtain the expression for a
cyclic graph of three vertices by first obtaining that
corresponding to a graph of four vertices (by joining a
graph of two with a graph of three) which is
\begin{eqnarray}
\ ^1\bullet\overline{\ \ \ \ \ }\bullet^2 \overline{\ \ \ \ \ }\bullet^3 \overline{\ \ \ \ \ }\bullet^4  \quad  P_{1234} & = &  1+\tau\{\rho_1[1+\rho_2(1+\rho_3)]+\rho_4[1+\rho_3(1+\rho_2)]\}+\notag \\
& & \tau^2[\rho_1\rho_4(1+\rho_2+\rho_3)+\rho_2(1+\rho_3)+\rho_3]+ \notag \\
& & \tau^3(\rho_1\rho_3+\rho_2\rho_4)+ \rho_1\rho_2\rho_3\rho_4 \,,
\label{P_4vertices}
\end{eqnarray}
and then identifying the outer vertices (1 and 4 in the above equation) to obtain for the cyclic graph:
\begin{equation}
P_\text{cyclic} =  1+\tau^2(\rho_1+\rho_2+\rho_3+\rho_1\rho_2+\rho_1\rho_3+\rho_2\rho_3)+\rho_1\rho_2\rho_3 \,.
\label{P_cyclic}
\end{equation}

The contraction process can be applied to a single connected part, where at first the sites $a$ and $b$
are different, then they are identified as a single site $ab$ with activity variable $\rho_{ab}$ and
\begin{equation}
P\equiv A_{++} +A_{-+}\rho_a +A_{+-}\rho_b +A_{--}\rho_a\rho_b \quad \longrightarrow  \quad P_{ab}\equiv A_{++} +\rho_{ab}A_{--} \,.
\label{P_samesubset}
\end{equation}
As in the initial set, in the contracted graph no pair of vertices can be joined by more than
one link.

Thus we have demonstrated that the contraction process allows the 
polynomial $P$ of any arbitrary graph to be obtained by starting from the elementary result of the
simplest graph of two vertices joined by a link. In this case, from Eq.~(\ref{P_2vertices}),
the zero of the polynomial is obtained for:
\begin{equation}
\rho_1=-\frac{1+\tau\rho_2}{\tau+\rho_2} \,,
\label{zero_2vertices}
\end{equation}
which defines a one-to-one mapping between the complex planes $\rho_1$ and $\rho_2$. For $\tau$
real it leaves the unit circle invariant while for $0<\tau<1$ exchanges the interior
and the exterior of this circle. Therefore it can be stated that if $|\rho_1|<1$ and $|\rho_2|<1$
, or $|\rho_1|>1$ and $|\rho_2|>1$, the polynomial can not vanish.

The previous property for a graph of two vertices is generalised in the following: For an
arbitrary graph, if all $\rho_i$ lie inside, or all $\rho_i$ lie outside the unit circle,
$P$ is different from zero.

To demonstrate the foregoing statement, it is sufficient to
verify that the property survives the contraction
process. Let us assume that for a given graph $P(\rho_i)\neq
0$ when $|\rho_i|<1$ for all $i$, and apply the contraction
process: When the dependence on the two points ($a$ and $b$)
to be identified is made explicit, $P$ is given by
Eq.~(\ref{P_samesubset}), while if the identification is
already made $P_{ab}$ is a function only of the subset
$\{\rho_i\}-\{\rho_a,\rho_b\}$ and a new variable
$\rho_{ab}$. We fix all the $\rho_i$'s distinct from
$\rho_a$ and $\rho_b$ within the unit circle. We want to
show that $|A_{++}|>|A_{--}|$, in which case $P_{ab}$ will
be non-vanishing for $|\rho_{ab}|<1$.  Recalling that $P\neq
0$ for $|\rho_a|<1$, $|\rho_b|<1$, the polynomial
$A_{++}+\rho(A_{+-}+A_{-+})+\rho^2A_{--}$ must have its two
roots equal to or greater than unity, which means that
$|A_{++}|\geq|A_{--}|$. Thus $P_{ab}$ is different from zero
when all its $\rho$'s are within the unit circle, and this
property holds for any graph.

If we set now all $\rho_i\equiv\rho$, corresponding to a uniform external field, from the symmetry property
under reversal of the field $h\to -h\,$, $\rho \to \rho^{-1}\,$, and $Z(h)=Z(-h)$, we have:
\begin{equation}
\text{e}^{+Nh}P(\tau,\rho) = \text{e}^{-Nh}P(\tau,\rho^{-1}) \,,
\label{LY_eq1}
\end{equation}
or equivalently:
\begin{equation}
P(\tau,\rho) = \rho^{N}P(\tau,\rho^{-1}) \,.
\label{LY_eq2}
\end{equation}
As a result if $P\neq 0$ for $|\rho|<1$, it follows that also $P\neq 0$ for $|\rho|>1$. The Lee-Yang theorem
has thus been demonstrated -- the partition function can (and does) only vanish on the unit circle $|\rho|=1$.

\section*{Distribution of Roots on the Unit Circle}
\label{Appendix_LY_roots}

Using the definition of $P$ in Eq.~(\ref{P_definition}) we can obtain the 
two extreme cases of the distribution of the zeros on the unit circle:
\begin{itemize}
 \item For a system at infinite temperature ($\tau=1$):
\begin{equation}
P(\tau,\rho) = P(1,\rho) =(1+\rho)^{N} \,.
\label{P_tauinf}
\end{equation}

\item For a system at zero temperature ($\tau=0$):
\begin{equation}
P(\tau,\rho) = P(0,\rho) =1+\rho^{N} \,.
\label{P_tauzero}
\end{equation}
\end{itemize}

Then, decreasing the temperature from infinity, one goes from a degenerate zero with multiplicity $N$
located at $\rho=-1$ to a uniform distribution $\rho_k=\text{e}^{i\pi(2k+1)/N}$. In addition, if
$N$ remains finite, no zero will be located on the real positive axis ($\rho=1$, $h=0$).

Let us obtain the general result for a lattice of $N$ sites, with coordination number
$q$, see footnote in Appendix~\ref{Appendix_Harris},
when taking the infinite volume limit. According to Eq.~(\ref{Z_original}), the free 
energy per site in a uniform field is:
\begin{equation}
F \equiv \frac{1}{N} \ln Z= \frac{1}{2}q\beta +h + \ln 2 + \lim_{N\to\infty} \frac{1}{N} \ln P_N(\tau,\rho) \,,
\label{F_LY}
\end{equation}
where we have used that $L=Nq/2$, with $L$ being the total number of links in the lattice. $P_N$
can be factorized in terms of its roots
\begin{equation}
P_N =  \prod_{a=1}^N \left( 1-\frac{\rho}{\rho_a(\tau)} \right) \,.
\label{P_factorized}
\end{equation}
As was demonstrated in the previous section, the zeros will accumulate for $N\to\infty$ on
the unit circle, $\rho=e^{i\varphi}$, with a ($\tau$-dependent) angular density $N\mu(\varphi)$ satisfying
\begin{equation}
\mu(\varphi)=\mu(-\varphi)\geq 0 \quad\quad ;\quad\quad \int_{-\pi}^{+\pi} \text{d}\varphi\mu(\varphi) =1 \,,
\label{mu_props}
\end{equation}
where the above property is a consequence of the invariance under field reversal, $h\to-h$.
Making the change
\begin{equation}
 \frac{1}{N} \ln P_N(\tau,\rho) = \frac{1}{N} \sum_{a=1}^N \ln \left( 1-\frac{\rho}{\rho_a(\tau)} \right) \longrightarrow 
 \frac{1}{N} \int_{-\pi}^{\pi} \text{d}\varphi N \mu(\varphi) \ln \left( 1-\frac{\rho}{\rho(\varphi)} \right) \,,
\label{LY_change}
\end{equation}
and using the symmetry property of the angular distribution to join the contributions of the conjugate zeros we arrive at
\begin{equation}
F = \frac{1}{2}q\beta +h + \ln 2 + \frac{1}{2} \int_{-\pi}^{+\pi} \text{d}\varphi\mu(\varphi) \ln (1+\rho^2-2\rho\cos\varphi) \,,
\label{F_LY_2}
\end{equation}
which is valid over the whole range $0<\tau<1$. Below the
critical temperature, where the support of $\mu(\varphi)$ is
the full circle, this defines in general two different
analytic functions, one for $|\rho|<1$ and another for
$|\rho|>1$. At $\rho=1$ ($h=0$), $F$ is
continuous. Nevertheless, let us consider the behaviour of its
derivative with respect to $h$, i.e. the magnetisation, as
the temperature varies. By definition
\begin{equation}
M=\frac{\partial F}{\partial h}=1+ \int_{-\pi}^{+\pi} \text{d}\varphi \mu(\varphi) \frac{\partial }{\partial h}
\left[ \ln (1-\rho\text{e}^{-i\varphi}) \right] \,,
\label{LY_magne1}
\end{equation}
but
\begin{equation}
\frac{\partial }{\partial h}=\frac{\partial \rho}{\partial h}\frac{\partial }{\partial \rho}=
-2\rho \frac{\partial }{\partial \rho} \,,
\label{LY_change2}
\end{equation}
thus
\begin{eqnarray}
M & = & 1+ 2\int_{-\pi}^{+\pi} \text{d}\varphi \mu(\varphi) \frac{\rho}{\text{e}^{i\varphi}-\rho}=\int_{-\pi}^{+\pi} \text{d}\varphi \mu(\varphi)
 \frac{1+\rho\text{e}^{-i\varphi}}{1-\rho\text{e}^{-i\varphi}} \,, \\
& = & \int_{-\pi}^{+\pi} \text{d}\varphi \mu(\varphi)
 \frac{1-\rho^2}{1-2\rho\cos\varphi+\rho^2}\,,
\label{LY_magne2}
\end{eqnarray}
where we made use of the normalisation of $\mu(\varphi)$,
Eq.~(\ref{mu_props}). The magnetisation vanishes as $h\to0$
($\rho\to1$) for $\tau>\tau_\text{c}$ and becomes
discontinuous for $\tau<\tau_\text{c}$. The above equation
is not zero for $\rho=1$ only if $\cos\varphi=1$. Therefore
the value of the magnetisation is directly related to the
value of the density of zeros on the real positive axis
($\varphi=0$), as a simple contour integration summing the
residues shows:
\begin{eqnarray}
\tau<\tau_\text{c} \quad\quad ; \quad\quad M_{\pm} \equiv \lim_{h\to\pm 0} M= \pm 2\pi\mu(0)
\label{LY_magne3}
\end{eqnarray}
We have therefore demonstrated that in the Thermodynamic Limit the
spontaneous magnetisation is directly related to the
existence of zeros on the real positive axis.

\clearpage{\thispagestyle{empty}\cleardoublepage}

\vspace{4cm}
\chapter{IBERCIVIS}
\label{Appendix_Ibercivis}

During the year 2007, the BIFI (Institute for Biocomputation and Physics of Complex Systems) and the National Fusion Laboratory
of the CIEMAT (Centro de Investigaciones Energ\'eticas, Medioambientales y Tecnol\'ogicas), collaborating with
the city council of Zaragoza (Spain), leadered the ZIVIS project. The scope of the project was to
develop a volunteer supercomputing platform, based on individual computers located in both private homes and public
buildings, to be used by the scientific community in the University of Zaragoza. This network of individual
computers would make it possible to perform calculations as a single installation. The project
converted Zaragoza into the first city with a large and stable computing structure based only on the volunteer
effort of its citizens in a non-profit contribution to science and research.
The scientific goal of the project was the analysis of plasma trajectories in the next-generation nuclear fusion reactors.

The ZIVIS project was based on the BOINC (Berkeley Open
Infrastructure for Network Computing) software.  BOINC was
originally developed in the \emph{SETI@home} project for
analysing the electromagnetic radiation received from outer
space in the search for extra-terrestrial intelligence.  The
software every volunteer downloaded has the form of a
simple-to-instal program (client) that works as a
screensaver during idle times of the computer (on average,
some 80\% of total CPU time). This means that the user does
not notice any inconvenience when using the computer. A
server sends tasks via Internet to the clients, which return
the results of the computations as they are performed and
when their Internet connection is enabled.

ZIVIS was an impressive success that even took its promoters
aback. Around 3000 volunteers offered more than 5000
computers resulting in around 850\thinspace000 hours of CPU
time. It allowed the analysis of more than $4.2\times10^6$
ion trajectories. During the short life of the project (40
days), it achieved a performance comparable with a large
 ``traditional'' supercomputing facility (in fact it
became one of the five most important supercomputing
facilities in Spain).

After the marvellous experience of ZIVIS, and the interest
it raised in the rest of the country, an extension of the
project to a higher level was designed. The result was
called IBERCIVIS~\cite{IBERCIVIS1}. It was predicted that it
would include more than 50\thinspace000 nodes (only in
Spain) resulting in the largest computer of this kind all
around the world.  The project was officially launched in
June 2008 and it has helped the scientific community since
then. The project is becoming a major achievement both 
 scientifically (now six applications are
producing high-precision numerical results) and
socially (more than 6000 users are sharing their
computers and getting involved with scientific research
daily).  It differs from ZIVIS in some major points:
\begin{itemize}
 \item There is not just one scientific application running
 within IBERCIVIS. Scientists all around the world are
 invited to use the infrastructure created to run their
 programs.  The ``machine'' is an open structure where any
 research group could in principle execute their
 programs. The diagram of the process an application must
 follow to run in IBERCIVIS is outlined in
 Fig.~\ref{porting_ibercivis}. While at first IBERCIVIS
 started just running three different applications (plasma
 trajectories in fusion reactors, protein folding, and phase
 transitions in disordered systems), it is today running
 simultaneously six scientific programs (the former three
 plus neuronal amino acid simulations, adsorption in porous
 media, and light behaviour at the
 nanoscale)~\cite{IBERCIVIS1}.  Every volunteer can choose
 which application they prefer to run on their computer.
\begin{figure}[!ht]
\begin{center}
\includegraphics[width=1\columnwidth]{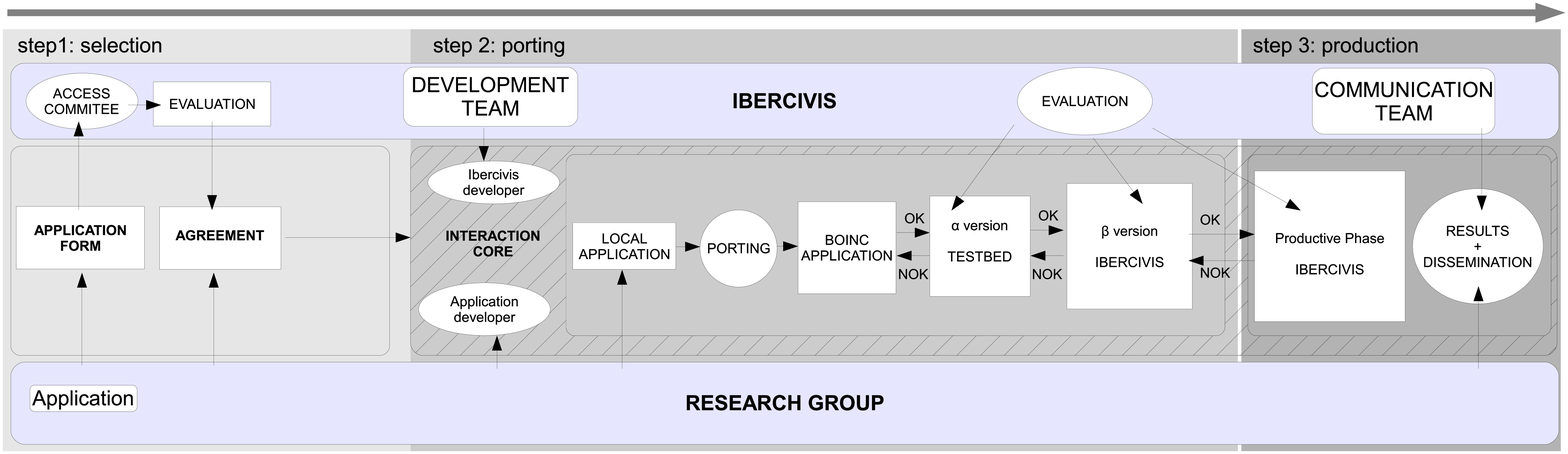}
\caption{An IBERCIVIS application road map from the beginning of the collaboration
up to the dissemination of the results.}
\label{porting_ibercivis}
\end{center}
\end{figure}

\item IBERCIVIS is not a temporary project (unlike ZIVIS or many other BOINC-based volunteer computing projects), 
so it will be possible to submit applications
indefinitely. This implies that IBERCIVIS is seen by
scientists as a stable structure like traditional
supercomputing centres on which they can run their programs via
a user-friendly interface with a typical queueing system,
launch their simulations from their personal computers, and
receive the output on a special server with high
storage capability.

\end{itemize}

\section*{Advantages and Disadvantages versus Other Structures}
\label{Appendix_iberciviscomparisin}

On the one hand, the strong points of the project are the following:
\begin{itemize}
 \item Apart from the huge scientific interest that nowadays every supercomputing facility produces,
this one has the additional
feature of its extremely low cost. As it mainly uses existing infrastructure (both computers and networks),
it only
requires the effort of the development and support of the specific software and servers, apart from an 
effort in publicity to persuade people to join the project (for example, there are periodically competitions
between individual clients or between teams of clients with prizes for the winners).

\item It provides an excellent way for bringing  science close to  people; the best way to make someone interested
in something is to get them involved with it. People feel
themselves to be part of the solution of a hard research
problem and learn about the subject.  Channels of
communication can be
established between volunteers and scientists through
blogs and social networks~\cite{IBERCIVIS1}.  In this way,
the most advanced scientific knowledge is spread to
society using modern information technologies.

\end{itemize}

On the other hand, the main objections to the project, and
generally to any kind of distributed volunteer computation,
are the following:
\begin{itemize}
 \item Every node of the supercomputing facility is located,
 in principle, away from other nodes making the
 communication between nodes very expensive. This fact makes
 direct parallelisation of the computing problem impossible
 if communication between nodes is a must. The range of
 applicability is therefore lower than for ``one-room''
 supercomputers.  Nevertheless, there exists a large class
 of problem for which communication between nodes is
 irrelevant. These are the so-called \emph{embarrassingly
 parallel problems}.  Within this class, problems can be
 split into independent simulations whose outputs can be
 joined later.

\item Given that all the data necessary for the job must be transferred to the volunteer's computer via the Internet,
the input/output of the program cannot be too massive. Otherwise, it would interfere with the volunteers'
network traffic and they would naturally become upset. Typically the
size of the transfered I/O files should not be greater than a
few megabytes.

\item Volunteers' computers are not as stable as computers within a traditional
supercomputing facility -- they are more likely to be
restarted or even turned off.  Therefore, in order to
increase the probability for a task to be finished, a long
computation (say of a week) must be divided into short
portions (say of one hour). This produces an increase both
in the network traffic and in the probability of corruption
of transferred data as they undergo several iterations from
computer to computer.  The implementation of a bulletproof
system for the validation of the output of each task
becomes crucial in order to ensure reliability of the final
data.

\end{itemize}

\section*{The Numbers of the Project}
\label{Appendix_numbersibercivis}

\begin{figure}[!ht]
\begin{center}
\includegraphics[width=1\columnwidth]{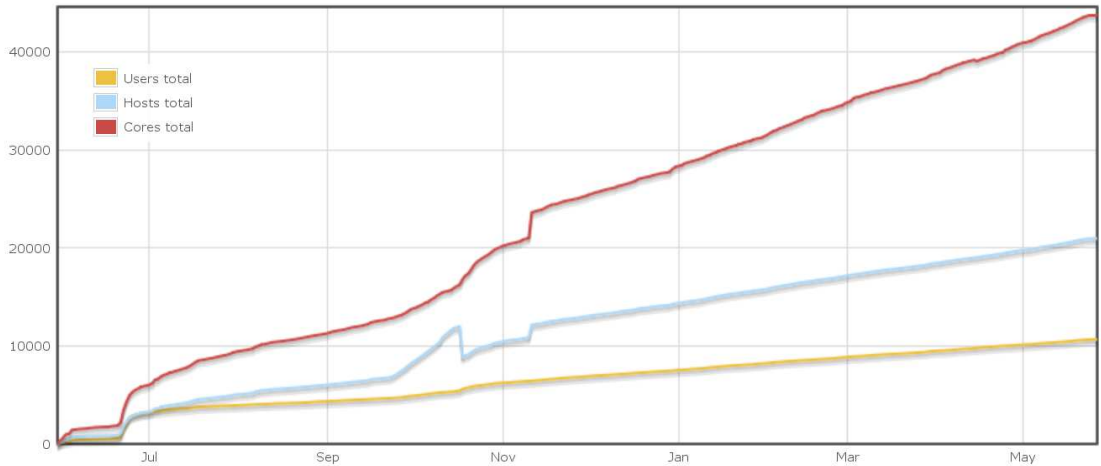}
\caption{Evolution in time (from June 2008 to June 2009) of the number of volunteers, computers, and cores involved in the project.}
\label{users_ibercivis}
\end{center}
\end{figure}

As was said before, the project is being a success both in scientific results and in citizen
contributions.  The number of users and of available nodes are both increasing continuously,
see Fig~\ref{users_ibercivis}. The number of registered users is today (May 2009) around 11\thinspace000 and
every day around 5500 of them share their computers with the scientists. The cluster equivalence 
of the IBERCIVIS structure is currently of 900 nodes located in a classical supercomputing 
facility.

Only during the project's first year it has produced around
eight millions hours of CPU time.  The total economic cost
has been around 270\thinspace000 euros, which is not too
much taking into account that the first year of life will be
surely the most expensive.

At present, six research groups are running simultaneously
on the platform, and three new applications are in the
porting process. This is indicative of the interest that
the project is producing in the scientific community.

The publicity campaign of the project has also been really
important. Apart from appearances in newspapers, magazines,
and TV~\cite{IBERCIVIS1}, more than 200\thinspace000 entries can
be found using the Google search engine for the word
``ibercivis''.

\section*{Our Experience}
\label{Appendix_ouribercivis}

In our case we have been using IBERCIVIS for
approximately a year to simulate disordered magnetic systems
defined in lattices through Monte Carlo methods.  In
particular, we have studied the three-dimensional
eight-state Potts model~\cite{Wu} in the presence of dilution.
The results of these simulations are presented in
Sec.~\ref{potts3D:eight-state}.

We started running on IBERCIVIS from its earliest stages
(around May 2008) so that we have followed all the
evolutional process of the project, thus suffering its
teething problems but also experiencing its gratifying
educational advances.

Our starting point was code written in C that we had been
running during the previous year on different supercomputing
facilities. The code was neither too complex nor did it make
use of any exotic C libraries -- facts which made the
porting process simpler, ensuring compatibility between the
different platforms of the volunteer's computers (Windows,
Mac, or Linux operating systems with 32 or 64 bit processor
architectures).

Our application is the perfect example of an embarrassingly
parallelisable one.  We parallel in four fully independent
ways: Firstly we have to simulate different system
sizes. Secondly, for each size, we have to simulate
different values of the dilution of the system. Thirdly, for
each dilution, we have to simulate different realizations of
the random spatial hole distribution (each one is called a
\emph{sample}).  And fourthly, we have to simulate each sample at
different values of its internal energy.

Our application does not have strong requirements of RAM
within the volunteer nodes (around 40 megabytes for the most
demanding case) or of disk storage (around 2 megabytes to
store the I/O configurations of the largest systems).  The
main problem of our simulations is that, as the system size
is increased, the run time grows
exponentially.  This fact forced us to design a continuity
system allowing the division of long simulations into small
(in terms of time) parts. In particular, the process that one
of our runs for a given dilution and system size follows is:
\begin{enumerate}
 \item For each sample (typically there will be around 1000
 of them), a random spatial hole configuration is generated
 depending on the system dilution. The holes will remain
 fixed in time (quenched disorder).

 \item For each energy (typically there will be around 30 of
 them), to each non-empty site of the lattice a spin
 variable is assigned. The assignation can be or a random Potts
 state, or a fixed one, or even a value depending on the
 system's energy. The result is called a configuration and saved
 into a file.

 \item Each configuration is sent to a volunteer's machine
 and updated there using a MC method. In addition,
 some measurements are taken into the system during the
 update process and saved into a measurement file. It has
 been calculated that the optimal time for a run on each
 volunteer machine (in order to minimise both the errors due
 to unexpected shutdowns and the web traffic) is around half
 an hour. Therefore the number of MC updates of the
 configuration must be set to last around this amount of
 time.

 \item When the specified number of MC updates have been
 made, both the output configuration and the measurement
 file are uploaded from the volunteer's computer to a
 server, where they are checked properly in order to avoid
 corruption.

 \item If the total desired simulation time is longer than
 half an hour, the continuity system takes over control: It
 will resend both the configuration and the measurement file
 to another volunteer (step 3), who will continue updating
 the system from the last configuration. The process is
 repeated until the total number of desired MC
 updates is performed. The number of continuity iterations
 of a given job must be specified a priori in our
 application.
\end{enumerate}

\begin{figure}[!ht]
\begin{center}
\includegraphics[width=1\columnwidth]{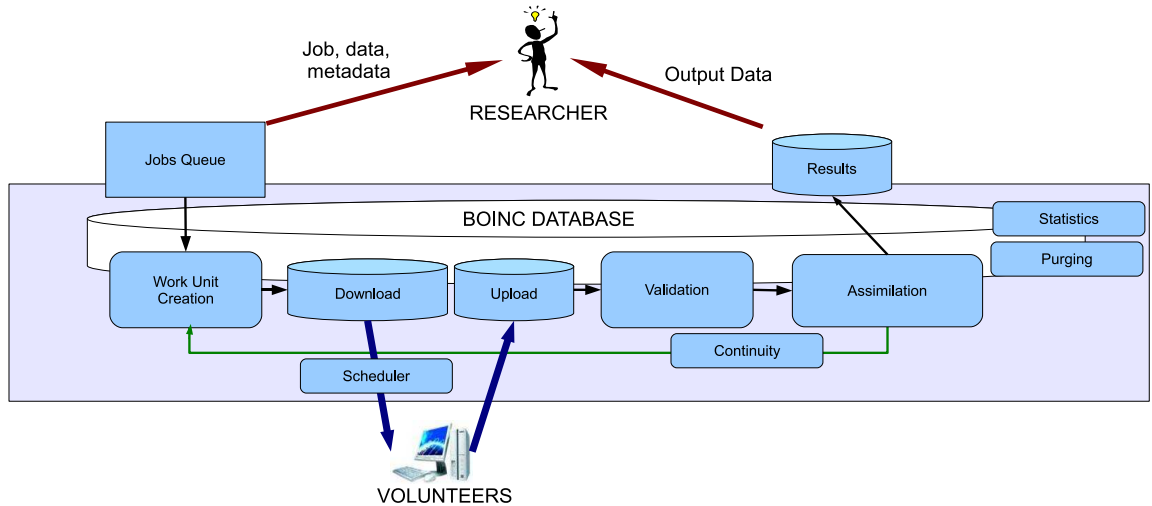}
\caption{Different stages of a simulation from the user submission to the results download.
The term ``work unit'' is used the define each small portion of the whole job that is sent to a single client
(each work unit should last around half an hour). Both validation and assimilation stages are crucial in order
to avoid data corruption. The continuity stage is only used if the required time for the job is much
longer than half an hour.}
\label{jobcycle_ibercivis}
\end{center}
\end{figure}

For example, for the smallest simulated system (with $24^3$
sites) the configurations travel only twice between the
IBERCIVIS server's and the volunteer's computers while for
the largest system (with $128^3$ sites) the continuity
process is repeated twenty times. See
Fig.~\ref{jobcycle_ibercivis} for a full sketch of the
process.

The implementation of the continuity systems is by no means
naive. Each file must be univocally identified in order to
avoid misdirections. If only a single file is lost or
misplaced, the rest of the continuity process will fail for
this configuration making the analysis of the corresponding
sample impossible. In addition, if there were a temporary
failure in the servers, some files would surely fail to be
transferred, resulting in a breakdown of the continuity
process.  This sometimes happened in the early stages of the
project: due to some server crashes, the continuity process
was unstable and simulation of the largest systems was
impossible. Finally, by building paranoid
assimilation-validation systems and by developing a
univocal nomenclature for each file we decreased the
failure rate of the continuity system to less than 0.5\%.

Using IBERCIVIS, we have simulated the site-diluted three-dimensional Potts model with a precision
never reached before. We obtained more than $2.5\times 10^6$ hours of CPU time.
A more detailed description of our usage of the infrastructure is given in Table~\ref{ournumbers_ibercivis}.
\begin{table}[ht]
\begin{center}
\begin{tabular}{|c|c|c|c|c|c|}\hline
$L$ & \# dilutions & \# samples & \# iterations & \# energies & CPU time ($\times 10^3$ hours) \\\hline\hline
24  &  14   & 7000  & 2  &  40  &  280  \\\cline{1-6}
32  &  13   & 6500  & 2  &  30  &  195 \\\cline{1-6}
48  &  12   & 24000 & 2  &  30  &  720 \\\cline{1-6}
64  &  8    & 10000 & 3  &  30  &  450 \\\cline{1-6}
128 &  4    & 2000  & 20 &  30  &  900 \\\hline
\end{tabular}
\caption{Approximate statistics of our simulations in IBERCIVIS for each linear lattice size $L$.  The fourth column
gives the number of continuity iterations used to
obtain enough MC time for each simulated energy.  }
\label{ournumbers_ibercivis}
\end{center}
\end{table}

\end{appendices}

\clearpage{\thispagestyle{empty}\cleardoublepage}



\end{document}